\documentclass[aps,a4paper,superscriptaddress,showpacs,preprintnumbers,amsmath,amssymb]{revtex4}

\usepackage{ulem}

\usepackage{psfrag} \usepackage{graphicx} \usepackage{dcolumn}
\usepackage{color} \usepackage{latexsym,amsfonts} \usepackage{bm}
\usepackage{amssymb}
\usepackage{hyperref} 
\begin{document}

\title{Radiative corrections of order $O(\alpha E_e/m_N)$\\ to
  Sirlin's radiative corrections of order $O(\alpha/\pi)$ to neutron
  lifetime}

\author{A. N. Ivanov}\email{ivanov@kph.tuwien.ac.at}
\affiliation{Atominstitut, Technische Universit\"at Wien, Stadionallee
  2, A-1020 Wien, Austria}
\author{R.~H\"ollwieser}\email{roman.hoellwieser@gmail.com}
\affiliation{Atominstitut, Technische Universit\"at Wien, Stadionallee
  2, A-1020 Wien, Austria}\affiliation{Department of Physics,
  Bergische Universit\"at Wuppertal, Gaussstr. 20, D-42119 Wuppertal,
  Germany} \author{N. I. Troitskaya}\email{natroitskaya@yandex.ru}
\affiliation{Atominstitut, Technische Universit\"at Wien, Stadionallee
  2, A-1020 Wien, Austria}
\author{M. Wellenzohn}\email{max.wellenzohn@gmail.com}
\affiliation{Atominstitut, Technische Universit\"at Wien, Stadionallee
  2, A-1020 Wien, Austria} \affiliation{FH Campus Wien, University of
  Applied Sciences, Favoritenstra\ss e 226, 1100 Wien, Austria}
\author{Ya. A. Berdnikov}\email{berdnikov@spbstu.ru}\affiliation{Peter
  the Great St. Petersburg Polytechnic University, Polytechnicheskaya
  29, 195251, Russian Federation}

\date{\today}

\begin{abstract}
We calculate the radiative corrections of order $O(\alpha E_e/m_N)$ as
next--to--leading order corrections in the large nucleon mass
expansion to Sirlin's radiative corrections of order $O(\alpha/\pi)$
to the neutron lifetime. The calculation is carried out within a
quantum field theoretic model of strong low--energy pion--nucleon
interactions described by the linear $\sigma$--model (L$\sigma$M) with
chiral $SU(2) \times SU(2)$ symmetry and electroweak hadron--hadron,
hadron--lepton and lepton--lepton interactions for the
electron--lepton family with $SU(2)_L \times U(1)_Y$ symmetry of the
Standard Electroweak Model (SEM). Such a quantum field theoretic model
is some kind a hadronized version of the Standard Model (SM). From a
gauge invariant set of the Feynman diagrams with one--photon exchanges
we reproduce Sirlin's radiative corrections of order $O(\alpha/\pi)$,
calculated to leading order in the large nucleon mass expansion, and
calculate next--to--leading corrections of order $O(\alpha
E_e/m_N)$. This confirms Sirlin's confidence level of the radiative
corrections $O(\alpha E_e/m_N)$. The contributions of the L$\sigma$M
are taken in the limit of the infinite mass of the scalar isoscalar
$\sigma$--meson. In such a limit the L$\sigma$M reproduces the results
of the current algebra (Weinberg, Phys. Rev. Lett. {\bf 18}, 188
(1967)) in the form of effective chiral Lagrangians of pion--nucleon
interactions with non--linear realization of chiral $SU(2)\times
SU(2)$ symmetry. In such a limit the L$\sigma$M is also equivalent to
Gasser--Leutwyler's chiral quantum field theory or chiral perturbation
theory (ChPT) with chiral $SU(2)\times SU(2)$ symmetry and the
exponential parametrization of a pion--field (Ecker,
Prog. Part. Nucl. Phys. {\bf 35}, 1 (1995)).
\end{abstract}

\maketitle

\section{Introduction}
\label{sec:introduction}

Nowadays the structure of the neutron in the $\beta^-$--decay
\cite{Abele2008,Nico2009} is investigated at the level of $10^{-3}$
related to the radiative corrections of order $O(\alpha/\pi)$, where
$\alpha$ is the fine--structure constant \cite{PDG2018}, and
corrections of order $O(E_e/m_N)$, caused by the weak magnetism and
proton recoil, where $E_e$ and $m_N$ are the electron energy and the
nucleon mass \cite{Gudkov2006, Ivanov2013, Ivanov2013a, Ivanov2017a,
  Ivanov2018a, Ivanov2018e, Ivanov2017}. The contributions of radiative
corrections of order $O(\alpha/\pi)$ has a long history
\cite{Berman1958}--\cite{Sirlin2013}. The contemporary shape of
radiative corrections to the neutron lifetime has been calculated by
Sirlin \cite{Sirlin1967} in the approximation of the one--photon
exchange and to leading order in the large nucleon mass expansion. The
contributions to the radiative corrections of the neutron lifetime,
caused electroweak boson exchanges and QCD corrections, have been
calculated by Marciano and Sirlin \cite{Sirlin1986,Sirlin2006} and
Czarnecki {\it et al.}  \cite{Sirlin2004}. In turn, the contemporary
shape of the radiative corrections to the correlation coefficients of
the electron--antineutrino 3--momentum correlations and correlations
between neutron spin and the electron 3--momentum has been calculated
by Shann \cite{Shann1971}. Recently radiative corrections of order
$O(\alpha/\pi)$ to leading order in the large nucleon mass expansion
have been calculated to the correlation coefficients of the neutron
$\beta^-$--decays with polarized neutron and electron and unpolarized
proton, and polarized electron and unpolarized neutron and proton
\cite{Ivanov2017a, Ivanov2018a, Ivanov2018e}. For the first time the
contributions of the weak magnetism and proton recoil of order
$O(E_e/m_N)$ to the neutron lifetime and correlation coefficients of
the neutron $\beta^-$--decay with polarized neutron and unpolarized
electron and proton have been calculated by Bilen'kii {\it et al.}
\cite{Bilenky1959} and then by Wilkinson \cite{Wilkinson1982}. To the
correlation coefficients of the neutron $\beta^-$--decays with
polarized neutron and electron and unpolarized proton, and with
polarized electron and unpolarized neutron and proton have been
calculated in \cite{Ivanov2017a, Ivanov2018a, Ivanov2018e}. At the
level of $10^{-3}$ the neutron as well as the proton has been treated
as a structureless particle. The contributions of strong low--energy
interactions to the $\beta^-$--decay of the structureless neutron with
a structureless decay proton are described by the axial coupling
constant $g_A$, and the isovector anomalous magnetic moment of the
nucleon $\kappa = \kappa_p - \kappa_n$, where $\kappa_p$ and
$\kappa_n$ are anomalous magnetic moments of the proton and neutron,
respectively, measured in the nuclear magneton \cite{PDG2018}. We
would like to remind that the axial coupling constant $g_A$ appears in
the standard $V - A$ theory of weak interactions
\cite{Feynman1958,Nambu1960,Marshak1969} as a trace of strong
low--energy interactions in the matrix element of the hadronic $n \to
p$ transition after renormalization of the matrix element of the
axial--vector hadronic current \cite{DeAlfaro1973}. As has been shown
by Sirlin \cite{Sirlin1967,Sirlin1978} the radiative corrections of
order $O(\alpha/\pi)$, calculated to leading order in the large
nucleon mass expansion, are independent of the axial coupling constant
$g_A$. In turn, the corrections of order $O(E_e/m_N)$, caused by the
weak magnetism and proton recoil, depend strongly on the axial
coupling constant $g_A$ and the isovector anomalous magnetic moment of
the nucleon $\kappa$ \cite{Bilenky1959,Wilkinson1982} (see also
\cite{Gudkov2006,Ivanov2013,Ivanov2017a,Ivanov2018a}).  The neutron
lifetime $\tau_n = 879.6(1.1)\,{\rm s}$, calculated in
\cite{Ivanov2013} at the axial coupling constant $g_A = 1.2750(9)$
\cite{Abele2008} (see also \cite{Mund2013}--\cite{Brown2018}), agrees
well with the neutron lifetime $\tau_n = 879.6(6)\,{\rm s}$, averaged
over the experimental values of the six bottle experiments
\cite{Mampe1993}--\cite{Arzumanov2015} included in the Particle Date
Group (PDG) \cite{PDG2018}. The values of the neutron lifetime $\tau_n
= 879.6(1.1)\,{\rm s}$ and axial coupling constant $g_A = 1.2750(9)$
agree also well with i) the values $\tau^{(\rm favoured)}_n =
879.6(4)\,{\rm s}$ and $g^{(\rm favoured)}_A = 1.2755(11)$, which have
been recommended by Czarnecki {\it et al.}  \cite{Sirlin2018} as {\it
  favoured} by a global analysis of the experimental data on the
neutron lifetime and the electron asymmetry of the neutron
$\beta^-$--decay with a polarized neutron and unpolarized proton and
electron, and ii) recent experimental value $g_A = 1.27641(45)_{\rm
  stat.}(33)_{\rm sys.}$ \cite{Abele2018}.

\begin{figure}
\includegraphics[height=0.15\textheight]{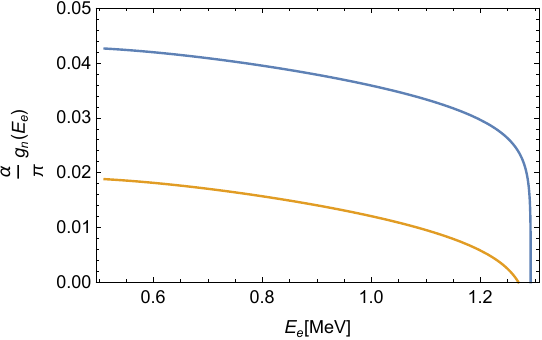}
  \caption{Radiative corrections $(\alpha/\pi)\,g_n(E_e)$ to the
    neutron lifetime in the electron energy region $m_e \le E_e <
    E_0$.  Blue and golden curves show the behaviour of the function
    $(\alpha/\pi)\,g_n(E_e)$ with and without contributions of
    electroweak--boson exchanges and QCD corrections, respectively,
    where QCD corrections make up of about $1.7\,\%$ of the
    contributions of electroweak--boson exchanges, which make up of
    about $60\,\%$ of the total radiative corrections to the neutron
    lifetime averaged over the electron--energy spectrum
    \cite{Sirlin2004}.}
\label{fig:fig1}
\end{figure}

For the first time deviations of the nucleon from a structureless
point--like particle in the neutron $\beta^-$--decay have been taken
into account by Wilkinson \cite{Wilkinson1982}. As has been shown in
\cite{Ivanov2018a} these corrections are of order $10^{-5}$. The
problem of non--trivial influence of hadronic structure of the
nucleon, caused by strong low--energy interactions, on gauge
properties of radiative corrections of order $O(\alpha^2/\pi^2)$ has
been pointed out in \cite{Ivanov2017b} within the standard $V - A$
effective theory of weak interactions.  As has been found in
\cite{Ivanov2017b} the interactions of real and virtual photons with
hadronic structure of the neutron and proton should provide not only
gauge invariance of radiative corrections of order $O(\alpha^2/\pi^2)$
but also non--trivial dependence of these corrections on the electron
$E_e$ and photon $\omega$ energies. This agrees well with Weinberg's
assertion that strong low--energy interactions play an important role
in weak decays \cite{Weinberg1957}. Hence, according to Weinberg
\cite{Weinberg1957}, contributions of strong low--energy interactions
beyond the axial coupling constant $g_A$ seem to be in principle
important for gauge invariant description of radiative corrections to
neutron $\beta^-$ decays to all orders in the fine--structure constant
expansion. However, as has been shown by Sirlin
\cite{Sirlin1967,Sirlin1968,Sirlin1978} the contribution of strong
low--energy interactions to the radiative corrections of order
$O(\alpha/\pi)$ to the neutron lifetime, calculated to leading order
in the large nucleon mass expansion, is a constant independent of the
electron energy. Because of such a property of strong low--energy
interactions their contributions to neutron $\beta^-$ decays have been
left at the level of the axial coupling constant $g_A$ and screened in
the radiative corrections \cite{Sirlin1967}--\cite{Sirlin2013} (see
also \cite{Gudkov2006, Ivanov2013, Ivanov2013a, Ivanov2017a,
  Ivanov2018a}). As has been shown in \cite{Ivanov2013} the
contributions of the weak magnetism and proton recoil of order
$O(E_e/m_N)$ to the neutron lifetime are much smaller than the
contributions of the radiative corrections. An enhancement of the
radiative corrections with respect to the corrections from the weak
magnetism and proton recoil is caused also by the contributions of the
electroweak--boson exchanges. The necessity to take into account
contributions of electroweak--boson exchanges \cite{Weinberg1971} for
the calculation of radiative corrections of order $O(\alpha/\pi)$ has
been pointed out by Sirlin
\cite{Sirlin1974, Sirlin1975b, Sirlin1978, Sirlin1982}. The analysis of
electroweak--boson exchanges and QCD corrections has been continued by
Marciano and Sirlin \cite{Sirlin1986, Sirlin2006}, Degrassi and Sirlin
\cite{Sirlin1992}, Czarnecki, Marciano and Sirlin \cite{Sirlin2004},
and Sirlin and Ferroglia \cite{Sirlin2013}. As has been shown by
Czarnecki {\it et al.} \cite{Sirlin2004} the contributions of
electroweak-boson exchanges change crucially the value of the
radiative corrections of order $O(\alpha/\pi)$. Indeed, the radiative
corrections to the neutron lifetime, averaged over the
electron--energy spectrum, are equal to $\langle
(\alpha/\pi)\,g_n(E_e)\rangle = 0.015056$ and $\langle
(\alpha/\pi)\,g_n(E_e)\rangle = 0.0390(8)$ without and with the
contributions of the electroweak-boson exchanges and QCD corrections,
respectively \cite{Sirlin2004}, where the function $g_n(E_e)$
describes the radiative corrections to the neutron lifetime in
notation \cite{Ivanov2013,Ivanov2017a}. In Fig.\,\ref{fig:fig1} the
function $g_n(E_e)$ is plotted without (golden line) and with (blue
line) the contributions of the electroweak--boson and QCD corrections.
It is important to emphasize that the contribution of QCD corrections,
caused by the quark structure of the neutron and proton and
gluon--exchanges, is by two orders of magnitude smaller than the
contribution of the electroweak--boson exchanges
\cite{Sirlin2004}.

For the correct gauge invariant calculation of radiative corrections
of order $O(\alpha^2/\pi^2)$ and as well as $O(\alpha E_e/m_N)$ to the
rate of the neutron radiative $\beta^-$--decay $n \to p + e^- +
\bar{\nu}_e + \gamma$ within the standard $V - A$ effective theory of
weak interactions an appearance of non--trivial contributions of
strong low--energy interactions dependent on the energies of decay
particles has been pointed out in \cite{Ivanov2017b}.
The problem of gauge invariant and infrared stable non--trivial
contributions of strong low--energy interactions to the radiative
corrections to neutron $\beta^-$ decays is closely related to the
analysis of corrections of order $10^{-5}$, calculated in the SM
\cite{Ivanov2017a,Ivanov2017b,Ivanov2018b,Ivanov2018c,Ivanov2018d}.

A contemporary level of relative accuracy of about $10^{-4}$ or even
better of the experimental analysis of the neutron $\beta^-$--decay
\cite{Abele2016} for searches of contributions of interactions beyond
the SM and contributions of second class currents
\cite{Lee1956}--\cite{Severijns2019} (see also \cite{Gudkov2006,
  Ivanov2013, Ivanov2018a, Ivanov2018e}) demands the SM theoretical
background of order $10^{-5}$, since "discovery" experiments with the
required 5$\sigma$ sensitivity will require experimental uncertainties
of a few parts in $10^{-5}$ \cite{Ivanov2017a, Ivanov2018a,
  Ivanov2018e}. The calculation of the radiative corrections of order
$O(\alpha E_e/m_N) \sim 10^{-5}$ as next--to--leading order
corrections in the large nucleon mass expansion to Sirlin's
corrections of order $O(\alpha/\pi) \sim 10^{-3}$, which we carry out
in this paper, is the first step to the calculation of the complete
set of the SM corrections of order $10^{-5}$ to the neutron lifetime
and correlation coefficients of the neutron $\beta^-$--decays with
different polarization states of the neutron and massive decay
fermions.

The paper is organized as follows. In section \ref{sec:sm} we discuss
shortly a low--energy hadronization of the Standard Model (SM), where
strong low--energy pion--nucleon interactions are described at the
hadronic level by the linear $\sigma$--model (L$\sigma$M) with linear
realization of chiral $SU(2)\times SU(2)$ symmetry. In section
\ref{sec:sigma} we outline the structure and properties of the
L$\sigma$M with chiral $SU(2)\times SU(2)$ symmetry. In section
\ref{sec:nlsigma} we demonstrate an equivalence at the Lagrangian
level between the L$\sigma$M, taken in the limit of the infinite mass
of the scalar isoscalar $\sigma$--meson $m_{\sigma} \to \infty$, and
chiral quantum field theories with non--linear realization of chiral
$SU(2)\times SU(2)$ symmetry by Weinberg and by Gasser and
Leutwyler. In section \ref{sec:sem} we propose a quantum field
theoretic model of strong low--energy and electroweak interactions
with electroweak $SU(2)_L \times U(1)_Y$ symmetry as a hadronized
version of the SM at low energies. Having switched off the electroweak
coupling constants this model reduces to the L$\sigma$M with chiral
$SU(2)\times SU(2)$ symmetry. In section \ref{sec:lorentzstr} we
calculate the matrix element of the hadronic $n \to p$ transition in
the neutron $\beta^-$--decay in the tree--approximation for
electroweak interactions and to one--hadron--loop approximation for
strong low--energy interactions in the quantum field theoretic model
proposed in section \ref{sec:sem} and described by the Lagrangian
Eq.(\ref{eq:44}). We show that the quantum field theoretic model,
described by the Lagrangian Eq.(\ref{eq:44}), reproduces well the
standard Lorentz structure of the matrix element of the hadronic $n
\to p$ transition with the axial coupling constant $g_A \neq 1$, the
isovector anomalous nucleon magnetic moment $\kappa$ and the
one--pion--pole contribution. The latter is important for gauge
invariance of the matrix element of the hadronic $n \to p$ transition
in the chiral limit $m_{\pi} \to 0$, where $m_{\pi}$ is a pion--meson
mass. Such a gauge invariance or an independence of a longitudinal
part of the propagator of the electroweak $W^-$--boson is required by
conservation of the axial--vector hadronic current in the chiral limit
$m_{\pi} \to 0$ \cite{Nambu1960}. Section \ref{sec:oneewloop} is
devoted to the analysis of the calculation of the radiative
corrections of order $O(\alpha E_e/m_N)$ to Sirlin's radiative
corrections of order $O(\alpha/\pi)$ to the neutron lifetime. We point
out that for the calculation of the radiative corrections of order
$O(\alpha E_e/m_N)$ as next--to--leading order corrections to Sirlin's
corrections of order $O(\alpha/\pi)$ calculated to leading order in
the large nucleon mass expansion, it is enough to analyze the
contribution of the Feynman diagrams with one--virtual--photon
exchanges in Fig.\,\ref{fig:fig6a}. Such a set of the Feynman diagrams
is gauge invariant, i.e. independent of a gauge parameter $\xi$ of a
longitudinal part of the photon propagator. Then, to leading order in
the large mass of the electroweak $W^-$--boson exchanges the Feynman
diagrams in Fig.\,\ref{fig:fig6a} reduce to the Feynman diagrams, used
by Sirlin for the calculation of the radiative corrections of order
$O(\alpha/\pi)$ to the neutron lifetime \cite{Sirlin1967}. The
calculation of the contributions of hadronic structure of the nucleon
to the radiative corrections of order $O(\alpha/\pi)$ and $O(\alpha
E_e/m_N)$, caused by one--virtual--photon exchanges and demanding the
analysis of two--loop Feynman diagrams, and the contributions of the
Feynman diagrams with electroweak $W$-- and $Z$--boson in the
one--electroweak--loop approximation goes beyond the scope of this
paper. We are planning to carry out these calculations in our
forthcoming publications. In section \ref{sec:conclusion} we discuss
the obtained results and perspectives of application of the quantum
field theoretic model of strong low--energy and electroweak
interactions, described by the Lagrangian Eq.(\ref{eq:44}), to the
analysis of neutron lifetime and correlation coefficients of the
neutron $\beta^-$--decays with different polarization states of the
neutron and massive decay fermions. In section \ref{sec:appendix} or
``Supplemental Material'' including Appendices A, B, C and D we give
detailed calculations of the matrix element of the hadronic $n \to p$
transition and radiative corrections of order $O(\alpha/\pi)$ and
$O(\alpha E_e/m_N)$ to the amplitude of the neutron $\beta^-$--decay,
respectively, and discuss gauge properties of the amplitude of the
neutron radiative $\beta^-$--decay. In Appendix A we give the
calculation of the matrix element of the hadronic $n \to p$ transition
to one--hadron--loop approximation in the quantum field theoretic
model described by the Lagrangian Eq.(\ref{eq:44}). In Appendices B
and C we give the analysis of gauge properties of the Feynman diagrams
in Fig.\,\ref{fig:fig6a} and the calculation of these diagrams in
details. We show that the Feynman diagrams in Fig.\,\ref{fig:fig6a}
are gauge invariant, i.e. independent of a gauge parameter $\xi$ of
the photon propagator, and renormalizable. In Appendix D we discuss
gauge properties of the amplitude of the neutron radiative
$\beta^-$--decay taken to leading order in the large mass of the
electroweak $W^-$--boson expansion.

\section{Low--energy dynamics of the Standard Model}
\label{sec:sm}

The Standard Model (SM) of particle physics is a quantum field theory
based on the $SU(3)_C \times SU(2)_L \times U(1)_Y$ gauge symmetry
group which describes strong, weak, and electromagnetic (or
electroweak) interactions among fundamental particles, which are i)
eight gluons $(g)$, mediating strong interactions between quarks with
six flavours $(q = u,d,s,c,b,t)$ and three colour degrees of freedom
each, electroweak bosons $(W^{\pm}, Z)$ and photon $(\gamma)$,
mediating weak and electromagnetic interactions between quarks and
three lepton families $(\ell^-,\nu_{\ell})$ for $\ell = e, \mu, \tau$
or electron, muon and tauon and electron-, muon-, and
tauon--neutrinos, and a Higgs boson $({\rm H})$ with mass $M_{\rm H} =
125\,{\rm GeV}$ coupled to quarks, leptons, electroweak bosons, photon
and gluons \cite{PDG2018,Bijnens2000a,Pich2007}. The part of the SM
invariant under $SU(3)_C$ gauge symmetry or Quantum Chromodynamics
(QCD) \cite{PDG2018,Bijnens2000a,Pich2007}, describing strong
interactions, was mainly formulated in
\cite{Fritsch1973}--\cite{Gross1974}. In turn, the Standard
Electroweak Model (SEM) or the part of the SM invariant under $SU(2)_L
\times U(1)_Y$ gauge symmetry has been formulated in
\cite{Glashow1961}--\cite{Kobayashi1973}. Renormalizability of the SM,
including a renormalizability of non--abelian massless and massive
Yang--Mills theories, has been proved in
\cite{Hooft1971,Hooft1971a}--\cite{Kraus1998}. The number of coloured
quarks and lepton families is constrained by a requirement of
renormalizability of the SEM to all orders of perturbation theory,
violation of which can occur because of Adler--Bell--Jackiw anomalies
\cite{Adler1969}--\cite{Bjorken1973}. Following Bijnens
\cite{Bijnens2000a} the SM Lagrangian we write as follows
\begin{eqnarray}\label{eq:1}
{\cal L}_{\rm SM} &=& {\cal L}_{\rm Higgs}(\phi) + {\cal L}_{\rm
  gauge}(g, W, Z, \gamma) + \!\!\!  \!\!\!\!\sum_{q =
  u,d,s,c,b,t}\!\!\!\!\!\!\!\!\bar{\psi}_qi\gamma^{\mu}D_{\mu} \psi_q
+ \!\!\!  \!\!\!\!\sum_{\ell = e, \mu,
  \tau}\!\!\!\!\bar{\psi}_{\ell}i\gamma^{\mu}D_{\mu} \psi_{\ell}
+\!\!\!  \!\!\!\!  \sum_{\nu_\ell = \nu_e, \nu_\mu,
  \nu_\tau}\!\!\!\!\!\!\!\!\bar{\psi}_{\nu_\ell}i\gamma^{\mu}D_{\mu}
\psi_{\nu_\ell}\nonumber\\ &-&\sum_{q =
  u,c,t}\!\!\!\!g_{qq}\big(\bar{\Psi}_{qL}\psi_{qR}\phi^c +
\phi^{c\dagger}\bar{\psi}_{qR}\Psi_{qL}\big) - \!\!\!
\!\!\!\!\!\!\!\!\sum_{(qq') = (ud),(cs),(tb)}\!\!\! \!\!\!\!
\!\!\!\!\!\!\!  g_{qq'}\big(\bar{\psi}_{qL}\psi_{q'R}\phi +
\phi^{\dagger}\bar{\psi}_{q'R}\psi_{qL}\big)\nonumber\\ &-&\sum_{\ell
  = e,\mu,\tau}\!\!\!\! g_\ell\big(\bar{\Psi}_{\ell L}\psi_{\ell
  R}\phi + \phi^{\dagger}\bar{\psi}_{\ell R}\Psi_{\ell L}\big),
\end{eqnarray}
where ${\cal L}_{\rm Higgs}(\phi)$ is the Lagrangian of the Higgs
field $\phi$, ${\cal L}_{\rm gauge}(g, W, Z, \gamma)$ is the
Lagrangian of the kinetic terms of gauge bosons, the third and fourth
terms in ${\cal L}_{\rm SM}$ are the kinetic terms and interactions of
quarks and leptons with gauge bosons, and the last three terms in
${\cal L}_{\rm SM}$ define Yukawa interactions of quarks and leptons
with the Higgs field. In the phase of the spontaneously broken
$SU(2)_L \times U(1)_Y$ symmetry these interactions produce masses of
charged fermions. Then, $\Psi_{u L}, \Psi_{c L}, \Psi_{t L}$ are quark
left--handed doublets with components $(P_L\psi_u, P_L\psi_d)$,
$(P_L\psi_c, P_L\psi_s)$ and $(P_L\psi_t, P_L\psi_b)$, respectively,
and $\Psi_{\ell L}$ are the lepton left--handed $SU(2)_L \times
U(1)_Y$ doublets with components $(P_L\psi_{\ell}, P_L
\psi_{\nu_\ell})$ for $\ell = e, \mu$ and $\tau$, respectively,
$\psi_{qR} = P_R\psi_q$ and $\psi_{\ell R} = P_R\psi_{\ell}$ are the
right--handed quark and charged lepton $SU(2)_L \times U(1)_Y$
singlets, where $P_{L,R} = (1 \mp \gamma^5)/2$ are the projection
operators $P^2_L = P_L$, $P^2_R = P_R$ and $P_LP_R = P_RP_L = 0$. For
$g_{qq} = g_{qq'} = 0$ the Lagrangian ${\cal L}_{\rm SM}$ is invariant
under chiral $SU(N_f) \times SU(N_f)$ transformations of the quark
fields \cite{Bijnens2000a,Pich2007}, where $N_f$ is the number of
quark fields.

Having integrated over gluon and quark degrees of freedom we arrive at
the effective Lagrangian for hadrons coupled to electroweak bosons
$(W,Z)$, photons $(\gamma)$ and the Higgs field $(\rm H)$, and
leptons. The strong low--energy interactions are described by the
effective Lagrangian. After the integration over the fields of baryons
with masses larger than $m_B > 1\,{\rm GeV}$ and of mesons with masses
larger than the $\pi$--meson mass $m_M > m_{\pi} = 0.14\,{\rm GeV}$ we
arrive at the effective quantum field theory for pions and nucleons
pions, described by the Chiral perturbation theory (ChPT) with a
non--linear realization of chiral $SU(2) \times SU(2)$ symmetry
\cite{Gasser1984}--\cite{Scherer2011}, based on the quantum field
theory of chiral dynamics developed by Weinberg
\cite{Weinberg1967a,Weinberg1968,Weinberg1979} and the general theory
of phenomenological or effective chiral Lagrangians
\cite{Coleman1969,Callan1969,Gasiorowicz1969}, which reproduce fully
(see Dashen and Weinstein \cite{Dashen1969}) the results of the
current algebra with partially conserved axial--vector hadronic
current (PCAC) \cite{Adler1968}--\cite{Treiman1985} (see also
\cite{DeAlfaro1973,Marshak1969}) on all possible soft--pion theorems
related to multi--pion production \cite{Nambu1962}. As has been shown
by Weinberg \cite{Weinberg1967a} the effective chiral Lagrangians with
non--linear realization of chiral $SU(2) \times SU(2)$ symmetry can be
derived from the linear $\sigma$--model (L$\sigma$M) with linear
realization of chiral $SU(2) \times SU(2)$ symmetry
\cite{GellMann1960} in the limit of the infinite mass of the scalar
$\sigma$--meson. Then, by applying pion--field redefinition one may
arrive at any form of effective chiral Lagrangian with non--linear
realization of chiral $SU(2) \times SU(2)$ symmetry
\cite{Weinberg1968}. Such an effective theory can be generalized by a
series of higher order terms of covariant derivatives of the
pion--field producing perturbative corrections to the current algebra
results, i.e. chiral perturbation theory \cite{Weinberg1979}. A
consistent realization of this idea has been carried out by Gasser and
Leutwyler \cite{Gasser1984} (see also
\cite{Gasser1987}--\cite{Scherer2011} and many other papers). The
L$\sigma$M with a linear realization of chiral $SU(2) \times SU(2)$
symmetry attracts strong attention by the following properties: i)
spontaneously broken chiral $SU(2)\times SU(2)$ symmetry, ii) the
partially conserved axial--vector hadronic current (PCAC) and the
Goldberger--Treiman relation \cite{Goldberger1958} at the quantum
field theoretic level, and iii) renormalizability
\cite{Bernstein1960}--\cite{Becchi1975}.

The analysis of the contributions of hadronic structure of the nucleon
or strong low--energy interactions to the neutron $\beta^-$--decay
within the L$\sigma$M has been carried out in
\cite{Ivanov2018b,Ivanov2018c,Ivanov2018d} in the standard $V - A$
effective theory of weak interactions
\cite{Feynman1958,Nambu1960,Marshak1969}. As has been shown in
\cite{Ivanov2018b} (see also \cite{Ivanov2018}) the contributions of
the L$\sigma$M, calculated to one--hadron--loop approximation,
reproduce well the Lorentz structure of the matrix element $\langle
p(\vec{k}_p, \sigma_p)|J^{(+)}_{\mu}(0)|n(\vec{k}_n,\sigma_n\rangle$
of the hadronic $n \to p$ transition, where $J^{(+)}_{\mu}(0) =
V^{(+)}_{\mu}(0) - A^{(+)}_{\mu}(0)$ is the charged weak hadronic
current \cite{Feynman1958,Nambu1960,Marshak1969}. According to the
analysis of contributions of hadronic structure of the nucleon to the
radiative corrections of the neutron lifetime, described by QED and
the L$\sigma$M in the standard $V - A$ effective theory of weak
interactions, the radiative corrections to order $O(\alpha/\pi)$ are
gauge invariant with contributions of strong low--energy interactions
described by the axial coupling constant $g_A$ to leading order in the
large nucleon mass $m_N$ expansion only.  This agrees well with the
analysis of radiative corrections carried out by Sirlin
\cite{Sirlin1967,Sirlin1978}. In other words in such an approximation
the neutron and proton can be treated as point--like particles.
Non--trivial contributions of hadronic structure of the nucleon to the
radiative corrections can appear only to order $O(\alpha E_e/m_N)$
\cite{Ivanov2018b}. However, these contributions are gauge
non--invariant and dependent on the ultra--violet cut--off, which
cannot be removed by renormalization. As has been pointed out in
\cite{Ivanov2018b, Ivanov2018c, Ivanov2018d} the problem of an
appearance of gauge non--invariant contributions and contributions,
violating renormalizability of the amplitude of the neutron
$\beta^-$--decays, to order $O(\alpha E_e/m_N)$ and even smaller, can
be explained as follows. Indeed, the effective $V - A$ vertex of weak
interactions is not the vertex of the combined quantum field theory
including the L$\sigma$M and QED. This implies that correct gauge
invariant contributions to the amplitude of the neutron
$\beta^-$--decays can be obtained in any loop approximation and
without violation of renormalizability only in the hadronized version
of the SEM with renormalizable quantum field theory of strong
low--energy interactions. In such a combined quantum field theory the
vertex of the effective $V - A$ weak interactions is defined by the
electroweak $W^-$--boson exchange. This should result in a gauge
invariant set of Feynman diagrams including electroweak bosons and
photons coupled to leptons, nucleon, and hadrons from hadronic
structure of the nucleon, described by a renormalizable quantum field
theory of strong low--energy interactions. Since the effective chiral
Lagrangians with non--linear realization of chiral $SU(2)\times SU(2)$
symmetry can be derived from the L$\sigma$M in the limit of the
infinite mass of the scalar isoscalar $\sigma$--meson
\cite{Weinberg1967a} (see also \cite{Gasiorowicz1969}) and by
redefinition of hadronic quantum fields \cite{Weinberg1968}, for the
description of strong low--energy interactions of the nucleon and
pions we choose the L$\sigma$M in the infinite limit of the scalar
isoscalar $\sigma$--meson mass. Because of the equivalence theorem
\cite{Kamefuchi1961}--\cite{Lam1973} such redefinitions of hadronic
quantum fields do not affect observable quantities, defined by matrix
elements of the ${\mathbb S}$--matrix on mass--shell of interacting
particles.

\section{Linear $\sigma$--model (L$\sigma$M) with chiral $SU(2) \times
 SU(2)$ symmetry \cite{Ivanov2018d}}
\label{sec:sigma}

\subsection{Chirally symmetric phase of the L$\sigma$M}

The L$\sigma$M with linear realization of chiral $SU(2)\times SU(2)$
symmetry describes strong low--energy nucleon--nucleon, pion--nucleon
and pion--pion interactions with a mediation of the scalar isoscalar
$\sigma$--meson \cite{GellMann1960}.  In the chirally symmetric phase
the Lagrangian of the L$\sigma$M is given by \cite{DeAlfaro1973}
\begin{eqnarray}\label{eq:2}
\hspace{-0.15in}{\cal L}_{\rm L\sigma M} =
\bar{\psi}_N\big(i\gamma^{\mu}\partial_{\mu} - g_{\pi N}(\tau_0 \sigma
+ i\, \gamma^5 \vec{\tau}\cdot \vec{\pi}\,)\big)\psi_N +
\frac{1}{2}\,\big(\partial_{\mu}\sigma\partial^{\mu}\sigma +
\partial_{\mu}\vec{\pi}\cdot \partial^{\mu}\vec{\pi}\,\big) +
\frac{1}{2}\,\mu^2\,\big(\sigma^2 + \vec{\pi}^{\,2}\big) -
\frac{1}{4}\,\gamma\,\big(\sigma^2 + \vec{\pi}^{\,2}\big)^2,
\end{eqnarray}
where $\psi_N$ is the isospin doublet of the nucleon field operator
with components $(\psi_p, \psi_n)$, where $\psi_p$ and $\psi_n$ are
the proton and neutron field operators, respectively, $\sigma$ and
$\vec{\pi} = (\pi^+, \pi^0, \pi^-)$ are the scalar isospin--scalar
(isoscalar) $\sigma$-- and pseudoscalar isospin--vector (isovector)
pion--meson field operators, $\mu^2$, $\gamma$ and $g_{\pi N}$ are
input parameters of the L$\sigma$M, $\vec{\tau} = (\tau_1,
\tau_2,\tau_3)$ are the isospin $2\times 2$ Pauli matrices, and
$\tau_0$ is the isospin $2\times 2$ unit matrix.

Under isospin--vector and isospin--axial--vector (or chiral)
infinitesimal transformations with parameters $\vec{\alpha}_V$ and
$\vec{\alpha}_A$, respectively, the nucleon and meson fields transform
as follows
\begin{eqnarray}\label{eq:3}
\hspace{-0.3in} \psi_N \stackrel{\vec{\alpha}_V}\longrightarrow
\psi'_N &=& \Big(1 + i\,\frac{1}{2}\,\vec{\alpha}_V\cdot
\vec{\tau}\,\Big) \psi_N \quad, \quad \bar{\psi}_N \stackrel{\vec{\alpha}_V}\longrightarrow \bar{\psi}'_N =
\bar{\psi}_N\Big(1 - i\,\frac{1}{2}\,\vec{\alpha}_V\cdot
\vec{\tau}\,\Big),\nonumber\\
\hspace{-0.3in}\sigma \stackrel{\vec{\alpha}_V}\longrightarrow \sigma'
&=& \sigma \quad,\quad \vec{\pi}
\stackrel{\vec{\alpha}_V}\longrightarrow \vec{\pi}^{\,'} =  \vec{\pi}
- \vec{\alpha}_V \times \vec{\pi},\nonumber\\
\hspace{-0.3in} \psi_N \stackrel{\vec{\alpha}_A}\longrightarrow
\psi'_N &=& \Big(1 + i\,\frac{1}{2}\,\gamma^5 \vec{\alpha}_A\cdot
\vec{\tau}\,\Big) \psi_N\quad, \quad \bar{\psi}_N
\stackrel{\vec{\alpha}_A}\longrightarrow \bar{\psi}'_N = \bar{\psi}_N
\Big(1 + i\,\frac{1}{2}\,\gamma^5 \vec{\alpha}_A\cdot
\vec{\tau}\,\Big),\nonumber\\
\hspace{-0.3in}\sigma \stackrel{\vec{\alpha}_A}\longrightarrow \sigma'
&=& \sigma + \vec{\alpha}_A \cdot \vec{\pi} \quad , \quad \vec{\pi}
\stackrel{\vec{\alpha}_A}\longrightarrow \vec{\pi}^{\,'} = \vec{\pi} -
\vec{\alpha}_A \sigma.
\end{eqnarray}
The Lagrangian Eq.(\ref{eq:2}) is invariant under global
transformations Eq.(\ref{eq:3}). Under local transformations
Eq.(\ref{eq:3}) the Lagrangian Eq.(\ref{eq:2}) acquires the following
corrections
\begin{eqnarray}\label{eq:4}
\delta {\cal L}_{\rm L\sigma M} = - \partial^{\mu}\vec{\alpha}_V\cdot
\Big(\bar{\psi}_N \gamma_{\mu}\,\frac{1}{2}\,\vec{\tau}\,\psi_N +
\vec{\pi} \times \partial_{\mu}\vec{\pi}\,\Big) -
\partial^{\mu}\vec{\alpha}_A \cdot \Big(\bar{\psi}_N
\gamma_{\mu}\,\gamma^5 \frac{1}{2}\,\vec{\tau}\,\psi_N + \big(\sigma
\,\partial_{\mu}\vec{\pi} -
\vec{\pi}\,\partial_{\mu}\sigma\,\big)\Big),
\end{eqnarray}
which allow to define the vector and axial--vector hadronic currents
\cite{Adler1968}
\begin{eqnarray}\label{eq:5}
\vec{V}_{\mu} &=& - \frac{\delta {\cal L}_{\rm L\sigma M}}{\delta
  \partial^{\mu}\vec{\alpha}_V} = \bar{\psi}_N
\gamma_{\mu}\,\frac{1}{2}\,\vec{\tau}\,\psi_N + \vec{\pi} \times
\partial_{\mu}\vec{\pi},\nonumber\\ \vec{A}_{\mu} &=& - \frac{\delta
  {\cal L}_{\rm L\sigma M}}{\delta \partial^{\mu}\vec{\alpha}_A} =
\bar{\psi}_N \gamma_{\mu}\,\gamma^5 \frac{1}{2}\,\vec{\tau}\,\psi_N +
\big( \sigma \,\partial_{\mu}\vec{\pi} -
\vec{\pi}\,\partial_{\mu}\sigma\,\big).
\end{eqnarray}
Using the equations of motion for the nucleon, scalar and pseudoscalar
fields one may show that in the chirally symmetric phase the
divergences of the vector and axial--vector hadronic currents vanish
$\partial^{\mu}\vec{V}_{\mu} = \partial^{\mu}\vec{A}_{\mu} = 0$.  This
means that in the chirally symmetric phase the vector and
axial--vector hadronic currents are locally conserved.

\subsection{Phase of spontaneously broken chiral symmetry}

We would like to notice that the nucleon, scalar and pseudoscalar
fields in Eq.(\ref{eq:2}) are unphysical. Indeed, the nucleon is
massless and the mass term of the scalar and pseudoscalar fields
enters with incorrect sign. Hence, physical hadronic states can appear
in the L$\sigma$M only in the phase of spontaneously broken chiral
symmetry \cite{GellMann1960}. In the L$\sigma$M the phase of
spontaneously broken chiral $SU(2) \times SU(2)$ symmetry can be
described by the Lagrangian \cite{DeAlfaro1973}
\begin{eqnarray}\label{eq:6}
\hspace{-0.3in}&&{\cal L}_{\rm L\sigma M} =
\bar{\psi}_N\big(i\gamma^{\mu}\partial_{\mu} - g_{\pi N}(\tau_0 \sigma
+ i\gamma^5 \vec{\tau}\cdot \vec{\pi}\,)\big) \psi_N +
\frac{1}{2}\,\big(\partial_{\mu}\sigma\partial^{\mu}\sigma +
\partial_{\mu}\vec{\pi}\cdot \partial^{\mu}\vec{\pi}\,\big) +
\frac{1}{2}\,\mu^2\,\big(\sigma^2 + \vec{\pi}^{\,2}\big) -
\frac{1}{4}\,\gamma\,\big(\sigma^2 + \vec{\pi}^{\,2}\big)^2 + a
\sigma,\nonumber\\
  \hspace{-0.3in}&&
\end{eqnarray}
where the last term $a \sigma$ is non--invariant under chiral
transformations Eq.(\ref{eq:3}).

The phase of spontaneously broken chiral symmetry characterizes by a
non--vanishing vacuum expectation value of the $\sigma$--field
$\langle \sigma \rangle = b \neq 0$. The transition to the fields of
physical hadronic states goes through the change of the
$\sigma$--field $\sigma \to \sigma + b$, where in the
right--hand--side (r.h.s.) the $\sigma$--field possesses a vanishing
vacuum expectation value. After such a change of the $\sigma$--field
the dynamics of physical hadronic states is described by the
Lagrangian
\begin{eqnarray}\label{eq:7}
\hspace{-0.3in}{\cal L}_{\rm L\sigma M} &=&
\bar{\psi}_N\big(i\gamma^{\mu}\partial_{\mu} - m_N - g_{\pi N}(\tau_0
\sigma + i\gamma^5 \vec{\tau}\cdot \vec{\pi}\,)\big)\, \psi_N +
\frac{1}{2}\,\big(\partial_{\mu}\sigma \partial^{\mu}\sigma -
m^2_{\sigma} \sigma^2\big) +
\frac{1}{2}\,\big(\partial_{\mu}\vec{\pi}\cdot \partial^{\mu}\vec{\pi}
- m^2_{\pi}\vec{\pi}^{\,2}\big)\nonumber\\
\hspace{-0.3in}&-&\gamma\,b\,\sigma\big(\sigma^2 +
\vec{\pi}^{\,2}\big) - \frac{1}{4}\,\gamma\,\big(\sigma^2 +
\vec{\pi}^{\,2}\big)^2,
\end{eqnarray}
where the masses of physical hadrons and coupling constants are
determined by
\begin{eqnarray}\label{eq:8}
\hspace{-0.3in}m_N = g_{\pi N} b\;,\; m^2_{\sigma} = 3\gamma b^2 -
\mu^2\;,\; m^2_{\pi} = \gamma b^2 - \mu^2\;,\,a = m^2_{\pi}b,
\end{eqnarray}
where $b = f_{\pi}$ with $f_{\pi}$ is the $\pi$--meson leptonic
coupling constant \cite{DeAlfaro1973}, and $\gamma = (m^2_{\sigma} -
m^2_{\pi})/2f^2_{\pi}$. In the phase of spontaneously broken chiral
symmetry the vector and axial--vector hadronic currents are equal to
\begin{eqnarray}\label{eq:9}
\vec{V}_{\mu} &=&\bar{\psi}_N \gamma_{\mu}\,\frac{1}{2}\,\vec{\tau}\,\psi_N +
\vec{\pi} \times \partial_{\mu}\vec{\pi},\nonumber\\ \vec{A}_{\mu} &=&
\bar{\psi}_N \gamma_{\mu}\,\gamma^5 \frac{1}{2}\,\vec{\tau}\,\psi_N + \big(
\sigma \,\partial_{\mu}\vec{\pi} -
\vec{\pi}\,\partial_{\mu}\sigma\,\big) + b\,\partial_{\mu}\vec{\pi}.
\end{eqnarray}
Using the equations of motion for the nucleon, scalar and pseudoscalar
fields one may show that the divergences of the vector and axial
vector hadronic currents are given by $\partial^{\mu}\vec{V}_{\mu} =
0$ and $\partial^{\mu}\vec{A}_{\mu} = - m^2_{\pi}
f_{\pi}\,\vec{\pi}$. This result agrees well with that by Adler and
Dashen \cite{Adler1968} (see Eq.(1.49) of Ref.\cite{Adler1968}). Thus,
the L$\sigma$M reproduces well the hypothesis of partial conservation
of the axial--vector hadronic current (the PCAC hypothesis) at the
quantum field theoretic level \cite{GellMann1960}.  Unlike the
axial--vector hadronic current the vector hadronic current is locally
conserved even in the phase of spontaneously broken chiral
symmetry. Conservation of the vector hadronic current in the
L$\sigma$M can be violated only by isospin symmetry breaking.

The mass of the scalar isoscalar $\sigma$--meson $m_{\sigma} = \sqrt{2
  f^2_{\pi}\gamma + m^2_{\pi}}$ is practically arbitrary because of an
arbitrariness of the coupling constant $\gamma$. Following Weinberg
\cite{Weinberg1967} one may take the limit $m_{\sigma} \to \infty$
corresponding to the limit $\gamma \to \infty$ at $\sqrt{\mu^2/\gamma}
= $ fixed. As has been pointed out by Weinberg \cite{Weinberg1967}
(see also \cite{Gasiorowicz1969}), in the limit $m_{\sigma} \to
\infty$ (or $\gamma \to \infty$) the L$\sigma$M reproduces the results
of the current algebra \cite{Adler1968}, and it is equivalent to
chiral quantum field theories of strong low--energy pion--nucleon
interactions with non--linear realizations of chiral $SU(2) \times
SU(2)$ symmetry.

For massless pions $m_{\pi} = 0$ or $a = 0$ the vacuum expectation
value of the $\sigma$--field is equal to $\langle \sigma \rangle =
\sqrt{\mu^2/\gamma} = f_{\pi}$. In this case the mass of the
$\sigma$--meson is $m_{\sigma} =\sqrt{3 \gamma}\,f_{\pi}$.

\subsection{Renormalization of the L$\sigma$M}

For the discussion of renormalization procedure in the L$\sigma$M we
rewrite the Lagrangian Eq.(\ref{eq:7}) as follows
\cite{Lee1969,Gervais1969,Mignaco1971}
\begin{eqnarray}\label{eq:10}
\hspace{-0.3in}{\cal L}^{(0)}_{\rm L\sigma M} &=&
\bar{\psi}^{(0)}_N\big(i\gamma^{\mu}\partial_{\mu} - m^{(0)}_N -
g^{(0)}_{\pi N}(\tau_0 \sigma^{(0)} + i \gamma^5 \vec{\tau}\cdot
\vec{\pi}^{\,(0)})\big)\,\psi^{(0)}_N\nonumber\\
\hspace{-0.3in} &+& \frac{1}{2}\,\big(\partial_{\mu}\sigma^{(0)}
\partial^{\mu}\sigma^{(0)} - m^{(0)2}_{\sigma} \sigma^{(0)2}\big) +
\frac{1}{2}\,\big(\partial_{\mu}\vec{\pi}^{\,(0)}\cdot
\partial^{\mu}\vec{\pi}^{\,(0)} -
m^{(0)2}_{\pi}\vec{\pi}^{\,(0)2}\big)\nonumber\\
\hspace{-0.3in}&+&\gamma^{(0)}\,f^{(0)}_{\pi}\,\sigma^{(0)}\big(\sigma^{(0)2}
+ \vec{\pi}^{\,(0)2}\big) - \frac{1}{4}\,\gamma^{(0)}\,\big(\sigma^{(0)2}
+ \vec{\pi}^{\,(0)2}\big)^2,
\end{eqnarray}
where $\psi^{(0)}_N$, $\sigma^{(0)}$ and $\vec{\pi}^{\,(0)}$ are {\it
  bare} hadronic fields, $m^{(0)}_N$, $m^{(0)}_{\sigma}$,
$m^{(0)}_{\pi}$ and $\gamma^{(0)}$, $f^{(0)}_{\pi}$ are {\it bare}
hadronic masses and coupling constants, respectively.  After the
calculation of hadron--loop contributions the dynamics of physical
fields is described by the Lagrangian
\begin{eqnarray}\label{eq:11}
\hspace{-0.3in}{\cal L}^{(r)}_{\rm L\sigma M} &=&
\bar{\psi}^{(r)}_N\big(i\gamma^{\mu}\partial_{\mu} - m^{(r)}_N -
g^{(r)}_{\pi N}(\tau_0 \sigma^{(r)} + i \gamma^5 \vec{\tau}\cdot
\vec{\pi}^{\,(r)}\,)\big)\,\psi^{(r)}_N +
\frac{1}{2}\,\big(\partial_{\mu}\sigma^{(r)}
\partial^{\mu}\sigma^{(r)} - m^{(r)2}_{\sigma}
(\sigma^{(r)})^2\big)\nonumber\\
\hspace{-0.3in}&+&
\frac{1}{2}\,\big(\partial_{\mu}\vec{\pi}^{\,(r)}\cdot
\partial^{\mu}\vec{\pi}^{\,(r)} -
m^{(r)2}_{\pi}(\vec{\pi}^{\,(r)})^2\big) +
\gamma^{(r)}\,f^{(r)}_{\pi}\,\sigma^{(r)}\big((\sigma^{(r)})^2 +
(\vec{\pi}^{\,(r)})^2\big) -
\frac{1}{4}\,\gamma^{(r)}\,\big((\sigma^{(r)})^2 +
(\vec{\pi}^{\,(r)})^2\big)^2\nonumber\\
\hspace{-0.3in}&+& {\cal L}^{(\rm CT)}_{\rm L\sigma M},
\end{eqnarray}
where the Lagrangian ${\cal L}^{(\rm CT)}_{\rm L\sigma M}$
is given by
\begin{eqnarray}\label{eq:12}
\hspace{-0.3in}&&{\cal L}^{(\rm CT)}_{\rm L\sigma M} = (Z_N
- 1)\bar{\psi}^{(r)}_N\big(i\gamma^{\mu}\partial_{\mu} -
m^{(r)}_N\big)\,\psi^{(r)}_N - Z_N \delta
m^{(r)}_N\,\bar{\psi}^{(r)}_N \psi^{(r)}_N - \big(Z_{M N} -
1\big)\,g^{(r)}_{\pi N}\bar{\psi}^{(r)}_N \big(\tau_0 \sigma^{(r)} +
i\gamma^5 \vec{\tau}\cdot \vec{\pi}^{\,(r)}\big)\,
\psi^{(r)}_N\nonumber\\
\hspace{-0.3in}&& +\big(Z_M -
1\big)\,\frac{1}{2}\,\big(\partial_{\mu}\sigma^{(r)}
\partial^{\mu}\sigma^{(r)} - m^{(r)2}_{\sigma} (\sigma^{(r)})^2\big) -
Z_M \delta m^{(r)2}_{\sigma}(\sigma^{(r)})^2 + \big(Z_M - 1\big)\,
\frac{1}{2}\,\big(\partial_{\mu}\vec{\pi}^{\,(r)}\cdot
\partial^{\mu}\vec{\pi}^{\,(r)} -
m^{(r)2}_{\pi}(\vec{\pi}^{\,(r)})^2\big)\nonumber\\
\hspace{-0.3in}&& - Z_M\delta m^{(r)2}_{\pi}(\vec{\pi}^{\,(r)})^2 +
\big(Z_{3M} -
1\big)\,\gamma^{(r)}\,f^{(r)}_{\pi}\,\sigma^{(r)}\big((\sigma^{(r)})^2
+ (\vec{\pi}^{\,(r)})^2\big) - \big(Z_{4M} -
1\big)\,\frac{1}{4}\,\gamma^{(r)}\,\big((\sigma^{(r)})^2 +
(\vec{\pi}^{\,(r)})^2\big)^2.
\end{eqnarray}
Here $Z_N$, $Z_M$ and $\delta m^{(r)}_N$, $\delta m^{(r)2}_{\sigma}$,
$\delta m^{(r)2}_{\pi}$ are renormalization constants of wave
functions and masses of the nucleon, scalar and pseudoscalar fields,
respectively. Then, $Z_{M N}$, $Z_{3 M}$ and $Z_{4 M}$ are
renormalization constants of the corresponding vertices of
meson--nucleon and meson--meson field interactions. The abbreviation
``CT'' means ``Counter--Terms''. If the fields, masses, coupling
constants and renormalization constants satisfy the relations
\begin{eqnarray}\label{eq:13}
\hspace{-0.3in}\psi^{(0)}_N &=& \sqrt{Z_N}\,\psi^{(r)}_N\;,\; \sigma^{(0)} =
\sqrt{Z_M}\,\sigma^{(r)}\;,\; \vec{\pi}^{\,(0)} =
\sqrt{Z_M}\,\vec{\pi}^{\,(r)},\nonumber\\
\hspace{-0.3in}m^{(0)}_N &=& m^{(r)}_N + \delta m^{(r)}_N\;,\;
m^{(0)2}_{\sigma} = m^{(r)2}_{\sigma} + \delta m^{(r)2}_{\sigma}\;,\;
m^{(0)2}_{\pi} = m^{(r)2}_{\pi} + \delta m^{(r)2}_{\pi},\nonumber\\
\hspace{-0.3in} g^{(0)}_{\pi N} &=& Z_{M N } Z^{-1}_N Z^{-1/2}_M
g^{(r)}_{\pi N}\;,\; f^{(0)}_{\pi} = Z_{3 M} Z^{-1}_{4 M}Z^{1/2}_M
f^{(r)}_{\pi} \;,\;\gamma^{(0)} = Z_{4 M}Z^{-2}_M \gamma^{(r)},\nonumber\\
Z_{3 M} &=& Z_{4 M}.
\end{eqnarray}
the Lagrangian Eq.(\ref{eq:11}) reduces to the Lagrangian
Eq.(\ref{eq:10}). The relation $Z_{3 M} = Z_{4 M}$ implies that the
pion decay constant $f^{(r)}_{\pi}$ is renormalized only by
renormalization of the wave function of the $\vec{\pi}$--meson,
i.e. $f^{(0)}_{\pi} = Z^{1/2}_M f^{(r)}_{\pi}$.

\section{Equivalence of the L$\sigma$M  to quantum field
 theories of strong low--energy pion--nucleon interactions with
 non--linear realization of chiral $SU(2) \times SU(2)$ symmetry}
\label{sec:nlsigma}

In this section we discuss an equivalence of the L$\sigma$M with a
linear realization of chiral $SU(2) \times SU(2)$ symmetry to quantum
field theories with non--linear realizations of chiral $SU(2) \times
SU(2)$ symmetry or chiral perturbation theory (ChPT). For this aim we
follow Ecker \cite{Ecker1995}. We introduce the fields
\begin{eqnarray}\label{eq:14}
U = \frac{1}{f_{\pi}}\,(\tau_0 \sigma + i \vec{\tau}\cdot
\vec{\pi}\,)\quad,\quad U^{\dagger} = \frac{1}{f_{\pi}}\,(\tau_0 \sigma - i
\vec{\tau}\cdot \vec{\pi}\,)
\end{eqnarray}
and rewrite the Lagrangian ${\cal L}_{\rm L\sigma M}$ in
Eq.(\ref{eq:6}) as follows
\begin{eqnarray}\label{eq:15}
{\cal L}_{\rm L\sigma M} &=&
\bar{\psi}_N\big(i\gamma^{\mu}\partial_{\mu} - g_{\pi N}(\tau_0 \sigma
+ i\gamma^5 \vec{\tau}\cdot \vec{\pi}\,)\big) \psi_N +
\frac{f^2_{\pi}}{4}\,\langle
\partial_{\mu}U^{\dagger}\partial^{\mu}U\rangle +
\frac{1}{4}\,m^2_{\pi}f^2_{\pi}\, \langle\big(U +
U^{\dagger}\big)\rangle\nonumber\\ &+&
\frac{1}{4}\,m^2_{\pi}f^2_{\pi}\langle\big(1 -
U^{\dagger}U\big)\rangle - \frac{f^4_{\pi}}{8}\,\gamma\,\langle \big(1
- U^{\dagger}U\big)^2\rangle,
\end{eqnarray}
where $\langle\ldots\rangle$ is a trace over isospin matrices
\cite{Ecker1995}. Taking the limit $\gamma \to \infty$ corresponding
to the infinite limit of the scalar isoscalar $\sigma$--meson mass
$m_{\sigma} \to \infty$, we get $U^{\dagger}U = 1$. This allows to
transcribe Eq.(\ref{eq:15}) into the form
\begin{eqnarray}\label{eq:16}
{\cal L}_{\rm ChPT} = \bar{\psi}_N\big(i\gamma^{\mu}\partial_{\mu} -
g_{\pi N}(\tau_0 \sigma + i\gamma^5 \vec{\tau}\cdot \vec{\pi}\,)\big)
\psi_N + \frac{f^2_{\pi}}{4}\,\langle
\partial_{\mu}U^{\dagger}\partial^{\mu}U\rangle +
\frac{1}{4}\,m^2_{\pi}f^2_{\pi}\, \langle\big(U +
U^{\dagger}\big)\rangle.
\end{eqnarray}
From the condition $U^{\dagger}U = 1$ we obtain $\sigma^2 +
\vec{\pi}^{\,2} = f^2_{\pi}$ \cite{Weinberg1968}.  Following again
Ecker \cite{Ecker1995} we rewrite Eq.(\ref{eq:16}) as follows
\begin{eqnarray}\label{eq:17}
{\cal L}_{\rm ChPT} &=&
\bar{\psi}_{NL}i\gamma^{\mu}\partial_{\mu}\psi_{NL} +
\bar{\psi}_{NR}i\gamma^{\mu}\partial_{\mu}\psi_{NR} - m_N
(\bar{\psi}_{NL}U\psi_{NR} +
\bar{\psi}_{NR}U^{\dagger}\psi_{NL})\nonumber\\ &+&
\frac{f^2_{\pi}}{4}\,\langle
\partial_{\mu}U^{\dagger}\partial^{\mu}U\rangle\rangle +
\frac{1}{4}\,m^2_{\pi}f^2_{\pi}\, \langle\big(U +
U^{\dagger}\big)\rangle,
\end{eqnarray}
where $\psi_{NL} = P_L \psi_N$ and $\psi_{NR} = P_R \psi_N$ are the
left- and right-handed nucleon fields, respectively, $g_{\pi N} =
m_N/f_{\pi}$ is the Goldberger--Treiman (GT) relation with the axial
coupling constant $g_A = 1$ \cite{Goldberger1958}. Then, we make
unitary transformations \cite{Ecker1995}
\begin{eqnarray}\label{eq:18}
\psi_{NL} = u\,\psi'_{NL}\quad,\quad \psi_{NR} =
  u^{\dagger}\,\psi'_{NR}\quad,\quad \bar{\psi}_{NL} =
  \bar{\psi}'_{NL}u^{\dagger} \quad,\quad \bar{\psi}_{NR} =
  \bar{\psi}'_{NR} u.
\end{eqnarray}
Plugging Eq.(\ref{eq:18}) into Eq.(\ref{eq:16}) we arrive at the
Lagrangian 
\begin{eqnarray}\label{eq:19}
{\cal L}_{\rm ChPT} &=&
\bar{\psi}'_{NL}i\gamma^{\mu}(\partial_{\mu} +
u^{\dagger}\partial_{\mu}u)\psi'_{NL} +
\bar{\psi}'_{NR}i\gamma^{\mu}(\partial_{\mu} +
u\partial_{\mu}u^{\dagger})\psi'_{NR} - m_N
(\bar{\psi}'_{NL}u^{\dagger} U u^{\dagger}\psi'_{NR} +
\bar{\psi}'_{NR}u U^{\dagger} u \psi'_{NL})\nonumber\\ &+&
\frac{f^2_{\pi}}{4}\,\langle
\partial_{\mu}U^{\dagger}\partial^{\mu}U\rangle\rangle +
\frac{1}{4}\,m^2_{\pi}f^2_{\pi}\, \langle\big(U +
U^{\dagger}\big)\rangle.
\end{eqnarray}
Setting $u^{\dagger} U u^{\dagger} = u U^{\dagger} u = 1$ that gives
$U = u^2$ we transcribe Eq.(\ref{eq:19}) into the form
\begin{eqnarray}\label{eq:20}
{\cal L}_{\rm ChPT} &=& \bar{\psi}'_N\Big(i\gamma^{\mu}
\partial_{\mu} +
i\gamma^{\mu}\frac{1}{2}\,[u^{\dagger},\partial_{\mu}u] -
i\gamma^{\mu}\gamma^5\frac{1}{2}\,\{u^{\dagger},\partial_{\mu}u\} -
m_N\Big)\psi'_N \nonumber\\ &+& \frac{f^2_{\pi}}{4}\,\langle
\partial_{\mu}U^{\dagger}\partial^{\mu}U\rangle\rangle +
\frac{1}{4}\,m^2_{\pi}f^2_{\pi}\, \langle\big(U +
U^{\dagger}\big)\rangle,
\end{eqnarray}
where we have used the relation $u\partial_{\mu}u^{\dagger} = -
\partial_{\mu}u u^{\dagger}$ \cite{Scherer2002} and denoted
$[u^{\dagger},\partial_{\mu}u] = u^{\dagger} \partial_{\mu}u -
\partial_{\mu}u u^{\dagger}$ and $\{u^{\dagger},\partial_{\mu}u\} =
u^{\dagger} \partial_{\mu}u + \partial_{\mu}u u^{\dagger} =
u^{\dagger} \partial_{\mu}U u^{\dagger}$. The Lagrangian
Eq.(\ref{eq:20}) can be written also in the following form
\cite{Gasser1987, Gasser1988, Bernard1995, Ecker1995, Scherer2002}
\begin{eqnarray}\label{eq:21}
{\cal L}_{\rm ChPT} = \bar{\psi}'_N\Big(i\gamma^{\mu} D_{\mu} -
i\gamma^{\mu}\gamma^5\frac{1}{2}\,\{u^{\dagger},\partial_{\mu}u\} -
m_N\Big)\psi'_N + \frac{f^2_{\pi}}{4}\,\langle
\partial_{\mu}U^{\dagger}\partial^{\mu}U\rangle\rangle +
\frac{1}{4}\,m^2_{\pi}f^2_{\pi}\, \langle\big(U +
U^{\dagger}\big)\rangle,
\end{eqnarray}
where $D_{\mu} = \partial_{\mu} + \Gamma_{\mu}$ is the covariant
derivative and $\Gamma_{\mu} = (1/2)[u^{\dagger},\partial_{\mu}u]$ has
a meaning of an affine connection \cite{Ecker1995,Scherer2002}.

\subsection{\bf Quantum field theory of strong low--energy 
pion--nucleon interactions with non--linear chiral $SU(2) \times SU(2)$
symmetry in Weinberg's parametrization}

In Weinberg's parametrization $u = (1 + i\vec{\tau}\cdot
\vec{\xi})/\sqrt{1 + \vec{\xi}^{\,2}}$, where $\vec{\xi} =
\vec{\pi}'/2f_{\pi}$ \cite{Weinberg1967}, the effective chiral
Lagrangian Eq.(\ref{eq:21}) takes the form
\begin{eqnarray}\label{eq:22}
\hspace{-0.15in}{\cal L}_{\rm ChPT} =
\bar{\psi}'_N\Big(i\gamma^{\mu}\partial_{\mu} - m_N -
\gamma^{\mu}\,\frac{1}{4f^2_{\pi}}\,\frac{\vec{\tau}\cdot
  (\vec{\pi}'\times \partial_{\mu}\vec{\pi}')}{\displaystyle 1 +
  \vec{\pi}'^2/4 f^2_{\pi}} +
\gamma^{\mu}\gamma^5\frac{1}{2f_{\pi}}\,\frac{\vec{\tau}\cdot
  \partial_{\mu}\vec{\pi}'}{\displaystyle 1 + \vec{\pi}'^2/4
  f^2_{\pi}} \Big)\psi'_N +
\frac{1}{2}\,\frac{\partial_{\mu}\vec{\pi}'\cdot
  \partial^{\mu}\vec{\pi}'- m^2_{\pi}\,\vec{\pi}'^2}{\displaystyle 1 +
  \vec{\pi}'^2/4 f^2_{\pi}}
\end{eqnarray}
and describes the quantum field theory of strong low--energy
pion--nucleon interactions with non--linear realization of chiral
$SU(2)\times SU(2)$ symmetry in Weinberg's chiral perturbation theory
\cite{Weinberg1967, Weinberg1968, Weinberg1979}. A deviation of the
axial coupling constant from unity $g_A > 1$ can be obtained in the
hadron--loop approximation.

\subsection{\bf Quantum field theory of strong low--energy 
pion--nucleon interactions with non--linear chiral $SU(2) \times SU(2)$
symmetry in Gasser--Leutwyler's parametrization}

In the exponential or Gasser--Leutwyler's parametrization $u =
e^{\,i\vec{\tau}\cdot \vec{\xi}}$, where $\vec{\xi} =
\vec{\pi}'/2f_{\pi}$, the effective chiral Lagrangian Eq.(\ref{eq:21})
retains its form with $U = u^2 = e^{\,i\vec{\tau}\cdot
  \vec{\pi}'/f_{\pi}}$ \cite{Gasser1984}--\cite{Scherer2011}
\begin{eqnarray}\label{eq:23}
{\cal L}_{\rm ChPT} = \bar{\psi}'_N\Big(i\gamma^{\mu} D_{\mu} - m_N
- i\gamma^{\mu}\gamma^5\frac{1}{2}\,u^{\dagger} \partial_{\mu}U
u^{\dagger}\Big)\psi'_N + \frac{f^2_{\pi}}{4}\,\langle
\partial_{\mu}U^{\dagger}\partial^{\mu}U\rangle\rangle +
\frac{1}{4}\,m^2_{\pi}f^2_{\pi}\, \langle\big(U +
U^{\dagger}\big)\rangle,
\end{eqnarray}
and describes the quantum field theory of strong low--energy
pion--nucleon interactions with non--linear realization of chiral
$SU(2)\times SU(2)$ symmetry in Gasser--Leutwyler's chiral
perturbation theory \cite{Gasser1984}--\cite{Scherer2011}. The
deviation of the axial coupling constant from unity $g_A > 1$ can be
obtained in the hadron--loop approximation
\cite{Gasser1987}--\cite{Scherer2011}. 

\subsection{\bf Quantum field theoretic model of strong low--energy 
pion--nucleon interactions for the description of hadronic structure
of the nucleon in neutron $\beta^-$--decays}

Since in the limit of infinite mass of the scalar isoscalar
$\sigma$--meson $m_{\sigma} \to \infty$ (or in the limit $\gamma \to
\infty$) the L$\sigma$M with linear realization of chiral $SU(2)\times
SU(2)$ symmetry is equivalent to chiral perturbation theory (ChPT)
with non--linear realization of chiral $SU(2) \times SU(2)$ symmetry
in Weinberg's and Gasser--Leutwyler's parametrizations, we shall use
the L$\sigma$M for the description of contributions of hadronic
structure of the nucleon to the neutron $\beta^-$--decays. We shall
calculate the corresponding Feynman diagrams for contributions of
strong low--energy interactions to the amplitude of the neutron
$\beta^-$--decays. We take the contributions of these Feynman diagrams
in the limit of the infinite scalar isoscalar $\sigma$--meson mass
$m_{\sigma} \to \infty$ (or in the limit $\gamma \to \infty$). After
renormalization the obtained expressions of the matrix elements of the
$\mathbb{S}$--matrix for the amplitudes of the neutron
$\beta^-$--decays should be in agreement with such properties of the
$\mathbb{S}$--matrix as analyticity, unitarity, cluster decomposition
and symmetry. This should imply that because of equivalence of the
L$\sigma$M in the infinite limit of the scalar isoscalar
$\sigma$--meson mass to quantum field theories of strong low--energy
pion--nucleon interactions with non--linear realization of chiral
$SU(2)\times SU(2)$ symmetry, the contributions of these Feynman
diagrams should be the same as the contributions of quantum field
theories with non--linear realization of chiral $SU(2) \times SU(2)$
symmetry and, correspondingly, current algebra. Such an assertion is
based on Weinberg's ``theorem'' \cite{Weinberg1979}. According to
Weinberg \cite{Weinberg1979}: ``The "theorem" says that although
individual quantum field theories have of course a good deal of
content, quantum field theory itself has no content beyond
analyticity, unitarity, cluster decomposition, and symmetry. This can
be put more precisely in the context of perturbation theory: if one
writes down the most general possible Lagrangian, including all terms
consistent with assumed symmetry principles, and then calculates
matrix elements with this Lagrangian to any given order of
perturbation theory, the result will simply be the most general
possible S-matrix consistent with analyticity, perturbative unitarity,
cluster decomposition and the assumed symmetry principles. As I said,
this has not been proved, but any counterexamples would be of great
interest, and I do not know of any.  With this "theorem", one can
obtain and justify the results of current algebra simply by writing
down the most general Lagrangian consistent with the assumed symmetry
principles, and then deriving low energy theorems by a direct study of
the Feynman graphs, without operator algebra. However, in order for
this to be a derivation and not merely a mnemonic, it is necessary to
include all possible terms in the Lagrangian, and take account of
graphs of all orders in perturbation theory.''  According to this
``theorem'', one may expect that the contributions of strong
low--energy interactions described by the L$\sigma$M to the neutron
$\beta^-$--decays are at Sirlin's confidence level of the description
of contributions of strong low--energy interactions to radiative
corrections for the neutron lifetime.

\section{\bf Quantum field theoretic model of strong low--energy 
pion--nucleon and electroweak interactions for the description of
neutron $\beta^-$--decays}
\label{sec:sem}

\subsection{ General properties of the Lagrangian for  quantum 
field theoretic model of strong low--energy and weak interactions of
pion--nucleon system coupled to electron and neutrino}

For the analysis of neutron $\beta^-$--decays within the quantum field
theoretic model of strong low--energy and electroweak interactions of
the pion--nucleon system coupled to electron and neutrino
(antineutrino) we propose to rewrite the Lagrangian of the L$\sigma$M
in the $SU(2)\times SU(2)$ symmetric phase, given by Eq.(\ref{eq:2}),
in term of the field operators
\begin{eqnarray}\label{eq:24}
\hspace{-0.3in}\Psi_{NL} &=& P_L \psi_N =
P_L\left(\begin{array}{c}\psi_p \\ \psi_n
\end{array}\right)\;,\; \psi_{pR} = P_R\psi_p\; , \; \psi_{nR} = P_R\psi_n, \nonumber\\ 
\hspace{-0.3in} \Phi &=& \frac{1}{\sqrt{2}}\,\left(\begin{array}{c}
  \sigma + i \pi^3\\ i(\pi^1 + i \pi^2)
\end{array}\right) = \frac{1}{\sqrt{2}}\,\left(\begin{array}{c} \sigma + i \pi^0\\ i\,\sqrt{2}\,\pi^-
\end{array}\right),\nonumber\\ 
\hspace{-0.3in} \Phi^c &=& - i\tau_{2L}\Phi^* = \frac{1}{\sqrt{2}}
\,\left(\begin{array}{c} i(\pi^1 - i\pi^2) \\ \sigma - i \pi^3
\end{array}\right) = \frac{1}{\sqrt{2}}\, \left(\begin{array}{c} i\, \sqrt{2}\,\pi^+
    \\ \sigma - i \pi^0
\end{array}\right),
\end{eqnarray}
where $\tau_{2L}$ is the Pauli $2\times 2$ matrix of the ``weak
isospin''. The field operators Eq.(\ref{eq:24}) have the following
properties under the $SU(2)_L \times U(1)_Y$ transformations

\begin{eqnarray}\label{eq:25}
&&\Psi_{NL} \stackrel{\vec{\alpha}_L\,,\, \alpha_Y}\longrightarrow
  \Psi'_{NL} = \Big(1 + i\,\frac{1}{2}\,\vec{\tau}_L\cdot
  \vec{\alpha}_L +
  i\,\frac{1}{2}\,Y\,\alpha_Y\Big)\Psi_{NL},\nonumber\\ &&\psi_{pR}
  \stackrel{\vec{\alpha}_L \,,\, \alpha_Y}\longrightarrow \psi'_{pR} =
  \Big(1 + i\,\frac{1}{2}\,Y\,\alpha_Y\Big)\psi_{pR} \;,\;\psi_{nR}
  \stackrel{\vec{\alpha}_L \,,\,\alpha_Y}\longrightarrow \psi'_{nR} =
  \Big(1 + i\,\frac{1}{2}\,Y\,\alpha_Y\Big) \psi_{nR},\nonumber\\ &&\Phi
  \stackrel{\vec{\alpha}_L \,,\,\alpha_Y}\longrightarrow \Phi' =
  \Big(1 + i\,\frac{1}{2}\,\vec{\tau}_L\cdot \vec{\alpha}_L +
  i\,\frac{1}{2}\,Y\,\alpha_Y\Big)\,\Phi,\nonumber\\ &&\Phi^c
  \stackrel{\vec{\alpha}_L \,,\,\alpha_Y}\longrightarrow {\Phi^{c}}' =
  \Big(1 + i\,\frac{1}{2}\,\vec{\tau}_L\cdot \vec{\alpha}_L +
  i\,\frac{1}{2}\,Y\,\alpha_Y\Big)\,\Phi^c,
\end{eqnarray}
where $\vec{I}_L = \frac{1}{2}\,\vec{\tau}_L$ and $Y$ are operators of
the ``weak isospin'' and ``weak hypercharge'', respectively,
$\vec{\alpha}_L$ and $\alpha_Y$ are infinitesimal parameters of the
$SU(2)_L$ and $U(1)_Y$ gauge group transformations, respectively. The
operators of the third component $I_{3L}$ of the ``weak isospin''
$\vec{I}_L$ and the ``weak hypercharge'' $Y$ are related by $Q =
I_{3L} + Y/2$ \cite{Weinberg1967,Weinberg1971} (see also
\cite{PDG2018}), where $Q$ is the operator of electric charge,
measured in $e$, which is the proton electric charge. The eigenvalues
of the third component of the ``weak isospin'' and ``weak
hypercharge'' are $((I_{3L})_{pL}, Y_{pL}) =(+1/2, +1)$,
$((I_{3L})_{nL}, Y_{nL}) =(-1/2, +1)$, $((I_{3L})_{pR},Y_{pR}) = (0,
+2)$, $((I_{3L})_{nR},Y_{nR}) = (0, 0)$, $((I_{3L})_{\sigma +
  i\pi^0},Y_{\Phi}) =(+1/2, -1)$, $((I_{3L})_{\pi^-},Y_{\Phi}) =(-1/2,
-1)$, $((I_{3L})_{\pi^+},Y_{\Phi^c}) =(+1/2, + 1)$, and
$((I_{3L})_{\sigma - i\pi^0},Y_{\Phi^c}) =(-1/2, + 1$,
respectively. In terms of the field operators Eq.(\ref{eq:24}) the
Lagrangian Eq.(\ref{eq:2}) takes the form
\begin{eqnarray}\label{eq:26}
\hspace{-0.3in} {\cal L}_{\rm L \sigma M} &=&
\bar{\Psi}_{NL}i\gamma^{\mu}\partial_{\mu}\Psi_{NL} +
\bar{\psi}_{pR}i\gamma^{\mu}\partial_{\mu}\psi_{pR} +
\bar{\psi}_{nR}i\gamma^{\mu}\partial_{\mu}\psi_{nR} - \sqrt{2}\,g_{\pi
  N}\big(\bar{\Psi}_{NL}\Phi\,\psi_{pR} +
\bar{\psi}_{pR}\Phi^{\dagger}\Psi_{NL}\big)\nonumber\\
\hspace{-0.3in} &-& \sqrt{2}\,g_{\pi N}\big(\bar{\Psi}_{NL}\Phi^c\,\psi_{nR} +
\bar{\psi}_{nR}\Phi^{c \dagger}\Psi_{NL}\big) + 
\partial_{\mu}\Phi^{\dagger}\partial^{\mu}\Phi + \mu^2\,
\Phi^{\dagger}\Phi - \frac{1}{2}\,\gamma\,
\big(\Phi^{\dagger}\Phi\big)^2.
\end{eqnarray}
The Lagrangian Eq.(\ref{eq:26}) is invariant under global $SU(2)_L
\times U(1)_Y$ transformations Eq.(\ref{eq:25}). Invariance under
local $SU(2)_L \times U(1)_Y$ transformations can be reached by the
inclusion of the interactions with gauge boson fields $\vec{W}_{\mu}$
and $B_{\mu}$ \cite{Weinberg1967,Weinberg1971}. This gives
\begin{eqnarray}\label{eq:27}
\hspace{-0.3in} {\cal L}_{\rm L \sigma M} &=&
\bar{\Psi}_{NL}i\gamma^{\mu}\Big(\partial_{\mu} +
i\,g\,\frac{1}{2}\,\vec{\tau}_L \cdot \vec{W}_{\mu} +
i\,g'\,\frac{1}{2}\,B_{\mu} \Big)\Psi_{NL} +
\bar{\psi}_{pR}i\gamma^{\mu}\big(\partial_{\mu} + i\,g'\,B_{\mu})\psi_{pR} +
\bar{\psi}_{nR}i\gamma^{\mu}\partial_{\mu}\psi_{nR}\nonumber\\
\hspace{-0.3in} &-& g_{\pi N}\big(\bar{\Psi}_{NL}\Phi\,\psi_{pR} +
\bar{\psi}_{pR}\Phi^{\dagger}\Psi_{NL}\big) - g_{\pi
  N}\big(\bar{\Psi}_{NL}\Phi^c\,\psi_{nR} + \bar{\psi}_{nR}\Phi^{c
  \dagger}\Psi_{NL}\big)\nonumber\\
\hspace{-0.3in} &+& \Big(\partial_{\mu}\Phi^{\dagger} -
i\,g\,\frac{1}{2}\,\Phi^{\dagger}\,\vec{\tau}_L\cdot \vec{W}_{\mu} +
i\,g'\,\frac{1}{2}\,\Phi^{\dagger} B_{\mu}
\Big)\Big(\partial^{\mu}\Phi + i\,g\,\frac{1}{2}\,\vec{\tau}_L\cdot
\vec{W}_{\mu}\Phi - i\,g'\,\frac{1}{2}\,B_{\mu}\Phi \Big) \nonumber\\
\hspace{-0.3in} &+& \mu^2\, \Phi^{\dagger}\Phi -
\frac{1}{2}\,\gamma\, \big(\Phi^{\dagger}\Phi\big)^2,
\end{eqnarray}
where $g$ and $g'$ are the electroweak coupling constants
\cite{Weinberg1967,Weinberg1971}. The gauge boson fields
$\vec{W}_{\mu}$ and $B_{\mu}$ have the following transformation
properties under the $SU(2)_L \times U(1)_Y$ local transformations
\begin{eqnarray}\label{eq:28}
&&\vec{W}_{\mu} \stackrel{\vec{\alpha}_L\,,\,\alpha_Y}\longrightarrow
  \vec{W}'_{\mu} = \vec{W}_{\mu} + \vec{W}_{\mu} \times \vec{\alpha}_L
  - \frac{1}{g}\,\partial_{\mu}\vec{\alpha}_L,\nonumber\\ &&B_{\mu}
  \stackrel{\vec{\alpha}_L\,,\,\alpha_Y}\longrightarrow B'_{\mu} =
  B_{\mu} - \frac{1}{g'}\,\partial_{\mu}\alpha_Y.
\end{eqnarray}
Having added to the Lagrangian Eq.(\ref{eq:27}) the kinetic terms of
the electroweak gauge boson fields, the interactions of the
electroweak gauge boson fields with the electron and neutrino fields
$(\Psi_{\ell L}, \psi_{eR})$ and the Higgs field $\phi$
\cite{Weinberg1967} we arrive at the Lagrangian
\begin{eqnarray}\label{eq:29}
\hspace{-0.15in} {\cal L}_{\rm L \sigma M + SEM} &=&
\bar{\Psi}_{NL}i\gamma^{\mu}\Big(\partial_{\mu} +
i\,g\,\frac{1}{2}\,\vec{\tau}_L \cdot \vec{W}_{\mu} +
i\,g'\,\frac{1}{2}\,B_{\mu} \Big)\Psi_{NL} +
\bar{\psi}_{pR}i\gamma^{\mu}\big(\partial_{\mu} +
i\,g'\,B_{\mu})\psi_{pR} +
\bar{\psi}_{nR}i\gamma^{\mu}\partial_{\mu}\psi_{nR}\nonumber\\
\hspace{-0.15in} &-& \sqrt{2}\,g_{\pi
  N}\big(\bar{\Psi}_{NL}\Phi\,\psi_{pR} +
\bar{\psi}_{pR}\Phi^{\dagger}\Psi_{NL}\big) - \sqrt{2}\, g_{\pi
  N}\big(\bar{\Psi}_{NL}\Phi^c\,\psi_{nR} + \bar{\psi}_{nR}\Phi^{c
  \dagger}\Psi_{NL}\big)\nonumber\\
\hspace{-0.15in} &+&  \Big(\partial_{\mu}\Phi^{\dagger} -
i\,g\,\frac{1}{2}\,\Phi^{\dagger}\,\vec{\tau}_L\cdot \vec{W}_{\mu} +
i\,g'\,\frac{1}{2}\,\Phi^{\dagger} B_{\mu}
\Big)\Big(\partial^{\mu}\Phi + i\,g\,\frac{1}{2}\,\vec{\tau}_L\cdot
\vec{W}_{\mu}\Phi - i\,g'\,\frac{1}{2}\,B_{\mu}\Phi \Big) \nonumber\\
\hspace{-0.15in} &+& \mu^2\, \Phi^{\dagger}\Phi -
\frac{1}{2}\,\gamma\, \big(\Phi^{\dagger}\Phi\big)^2 -
\frac{1}{4}\,\vec{W}_{\mu\nu}\cdot \vec{W}^{\mu\nu} -
\frac{1}{4}\,B_{\mu\nu} B^{\mu\nu} + \bar{\Psi}_{\ell
  L}i\gamma^{\mu}\Big(\partial_{\mu} +
i\,g\,\frac{1}{2}\,\vec{\tau}\cdot \vec{W}_{\mu} -
i\,g'\,\frac{1}{2}\,B_{\mu}\Big) \Psi_{\ell
  L}\nonumber\\ \hspace{-0.15in}&+& \bar{\psi}_{eR} i \gamma^{\mu}
\big(\partial_{\mu} - i\,g'\,B_{\mu}\big) \psi_{eR} -
\sqrt{2}\,g_e(\bar{\Psi}_{\ell L}\psi_{eR}\phi +
\phi^{\dagger}\bar{\psi}_{eR}\Psi_{\ell L}) +
\Big(\partial_{\mu}\phi^{\dagger} -
i\,g\,\frac{1}{2}\,\phi^{\dagger}\,\vec{\tau}\cdot \vec{W}_{\mu} -
i\,g'\,\frac{1}{2}\,\phi^{\dagger} B_{\mu}\Big)\nonumber\\
\hspace{-0.15in} &\times&\Big(\partial^{\mu}\phi +
i\,g\,\frac{1}{2}\,\vec{\tau}\cdot \vec{W}^{\,\mu}\phi +
i\,g'\,\frac{1}{2}\,B^{\,\mu}\phi\Big) + \tilde{\mu}^2\,
\phi^{\dagger}\phi - \tilde{\lambda}\,(\phi^{\dagger}\phi)^2
\end{eqnarray}
of the quantum field theoretic model of strong low--energy and
electroweak interactions, which we apply to the analysis of hadronic
structure of the nucleon in the neutron $\beta^-$--decays, where
$\vec{W}_{\mu\nu}$ and $B_{\mu\nu}$ are the operators of the field
strength tensors of the gauge boson $\vec{W}_{\mu}$ and $B_{\mu}$
fields
\begin{eqnarray}\label{eq:30}
\vec{W}_{\mu\nu} &=& \partial_{\mu}\vec{W}_{\nu} -
\partial_{\nu}\vec{W}_{\mu} - g\,\vec{W}_{\mu}\times
\vec{W}_{\nu},\nonumber\\ B_{\mu\nu}&=& \partial_{\mu}B_{\nu} -
\partial_{\nu}B_{\mu}
\end{eqnarray}
and the operators of the lepton and Higgs fields are defined by
\begin{eqnarray}\label{eq:31}
 \Psi_{\ell L} =  P_L\left(\begin{array}{c} 
\psi_{\nu_e} \\ \psi_e
\end{array}\right)\;,\; \psi_{eR} = P_R\psi_e \;,\;
 \phi  = \left(\begin{array}{c}\phi^+
    \\ \phi^0
\end{array}\right),
\end{eqnarray}
having the following properties under the $SU(2)_L \times U(1)_Y$
transformations
\begin{eqnarray}\label{eq:32}
&&\vec{W}_{\mu\nu} \stackrel{\vec{\alpha}_L \,,\,
    \alpha_Y}\longrightarrow \vec{W}\,'_{\mu\nu} = \vec{W}_{\mu\nu} +
  \vec{W}_{\mu\nu} \times \vec{\alpha}_L, \nonumber\\ &&B_{\mu\nu}
  \stackrel{\vec{\alpha}_L \,,\, \alpha_Y}\longrightarrow B'_{\mu\nu}
  = B_{\mu\nu},\nonumber\\ &&\Psi_{\ell L} \stackrel{\vec{\alpha}_L \,,\,
    \alpha_Y}\longrightarrow \Psi'_{\ell L} = \Big(1 +
  i\,\frac{1}{2}\,\vec{\tau}\cdot \vec{\alpha}_L +
  i\,\frac{1}{2}\,Y\,\alpha_Y\Big)\Psi_{\ell L},\nonumber\\ &&\psi_{eR}
  \stackrel{\vec{\alpha}_L \,,\,\alpha_Y}\longrightarrow \psi'_{eR} =
  \Big(1 +  i\,\frac{1}{2}\,Y\,\alpha_Y\Big)\psi_{eR},\nonumber\\ &&\phi
  \stackrel{\vec{\alpha}_L \,,\,\alpha_Y}\longrightarrow \phi' =
  \Big(1 + i\,\frac{1}{2}\,\vec{\tau}_L\cdot \vec{\alpha}_L +
  i\,\frac{1}{2}\,Y\,\alpha_Y\Big)\,\phi.
\end{eqnarray}
The eigenvalues of the third component of the ``weak isospin'' and
``weak hypercharge'' are $((I_{3L})_{eL}, Y_{eL}) =(-1/2, -1)$,
$((I_{3L})_{\nu_e L}, Y_{\nu_e L}) =(+1/2, -1)$,
$((I_{3L})_{pR},Y_{pR}) = (0, +2)$, $((I_{3L})_{eR},Y_{eR}) = (0,
-2)$, $((I_{3L})_{\phi^+},Y_{\phi}) =(+1/2, +1)$ and
$((I_{3L})_{\pi^0},Y_{\phi}) =(-1/2, + 1)$, respectively.  For the
derivation of the Lagrangians Eq.(\ref{eq:27}) and Eq.(\ref{eq:29}) we
have used the following standard definitions of the covariant
derivatives of the left--handed fermions and the Higgs field
$D_{L\mu}$ and the right--handed fermions $D_{R\mu}$ defined by
\cite{Weinberg1967}
\begin{eqnarray}\label{eq:33}
D_{L\mu} &=& \partial_{\mu} + i\,g\,\frac{1}{2}\,\vec{\tau}_L\cdot
\vec{W}_{\mu} + i\,g'\,\frac{1}{2}\, Y \,B_{\mu},\nonumber\\ D_{R\mu}
&=&\partial_{\mu} +
i\,g'\,\frac{1}{2}\,Y\,B_{\mu},
\end{eqnarray}
where $Y$ is the operator of the ``weak hypercharge''
\cite{Weinberg1967}.  In the physical phase or in the phase of
spontaneously broken $SU(2)_L \times U(1)_Y$ symmetry reduced to
$SU(2)_L \times U(1)_Y \to U(1)_{\rm em}$, where $U(1)_{\rm em}$ is a
gauge group of electromagnetic interactions, the components of the
Higgs field $\phi$ are equal to $\phi^+ = \phi^- = 0$ and $\phi^0 =
\phi^{0*} = (v + H)/\sqrt{2}$, respectively, where $v$ is the vacuum
expectation value $\langle \phi^0\rangle = \langle \phi^{0*}\rangle =
v/\sqrt{2}$ and $H$ is the observable scalar Higgs field with mass
$M_H = 125\,{\rm GeV}$ \cite{PDG2018}. In turn in the physical phase
the hadronic fields $\Phi$ and $\Phi^c$ are defined by
\begin{eqnarray}\label{eq:34}
  \hspace{-0.3in}\Phi  = \frac{1}{\sqrt{2}}\,\left(\begin{array}{c}
    \sigma + i \pi^0\\ i\,\sqrt{2}\,\pi^-
  \end{array}\right) \to  \frac{1}{\sqrt{2}}\,\left(\begin{array}{c}
    f _{\pi} + \sigma + i \pi^0\\ i\,\sqrt{2}\,\pi^-
  \end{array}\right)\;,\; \Phi^c  = \frac{1}{\sqrt{2}}\,
  \left(\begin{array}{c} i\, \sqrt{2}\,\pi^+
    \\ \sigma - i \pi^0
  \end{array}\right) \to \frac{1}{\sqrt{2}}\, \left(\begin{array}{c} i\,
    \sqrt{2}\,\pi^+ \\ f_{\pi} + \sigma - i \pi^0
\end{array}\right),
\end{eqnarray}
where the transition to the fields of physical hadronic states goes
through the change of the $\sigma$--field $\sigma \to f_{\pi} + \sigma
$ with a vanishing vacuum expectation value $\langle \sigma \rangle =
0$ of the $\sigma$--field in the right--hand--side (r.h.s.). In terms
of the fields of the physical states the Lagrangian Eq.(\ref{eq:29})
takes the form
\begin{eqnarray}\label{eq:35}
\hspace{-0.3in}&&{\cal L}_{\rm L\sigma M + SEM} =
\bar{\psi}_p\big(i\gamma^{\mu}\partial_{\mu} - m_N\big)\psi_p +
\bar{\psi}_n\big(i\gamma^{\mu}\partial_{\mu} - m_N\big)\psi_n +
\partial_{\mu}\pi^+\partial^{\mu}\pi^- + \frac{1}{2}\,
\partial_{\mu}\pi^0 \partial^{\mu}\pi^0 +
\frac{1}{2}\,\big(\partial_{\mu}\sigma \partial^{\mu}\sigma -
m^2_{\sigma} \sigma^2\big)\nonumber\\
\hspace{-0.3in}&& - \sqrt{2}\,g_{\pi N}\bar{\psi}_pi\gamma^5
\psi_n\,\pi^+ - \sqrt{2}\,g_{\pi N}\bar{\psi}_ni\gamma^5 \psi_p\,\pi^-
- g_{\pi N}\,\big( \bar{\psi}_pi\gamma^5 \psi_p -
\bar{\psi}_ni\gamma^5 \psi_n\big)\,\pi^0 - g_{\pi N}\,\big(
\bar{\psi}_p \psi_p + \bar{\psi}_n \psi_n\big)\,\sigma\nonumber\\
\hspace{-0.3in} && - \gamma\,f_{\pi}\,\sigma\big(\sigma^2 + 2 \pi^+
\pi^- + (\pi^0)^2\big) - \frac{1}{4}\,\gamma\,\big(\sigma^2 + 2 \pi^+
\pi^- + (\pi^0)^2\big)^2 - \frac{1}{2}\,W^+_{\mu\nu}W^{- \mu\nu} +
M^2_W W^+_{\mu}W^{-\mu} - \frac{1}{4}\,Z_{\mu\nu} Z^{\mu\nu}
\nonumber\\
\hspace{-0.3in} && + \frac{1}{2}\, M^2_Z Z_{\mu}Z^{\mu} -
\frac{1}{4}\,F_{\mu\nu}F^{\mu\nu} -
\frac{1}{2\xi}\,\big(\partial_{\mu}A^{\mu}\big)^2 +
\frac{1}{2}\,\partial_{\mu} H \partial^{\mu} H - \frac{1}{2}\,M^2_H
H^2 + \bar{\psi}_e \big(i\gamma^{\mu} \partial_{\mu} -
m_e\big)\,\psi_e + \bar{\psi}_{\nu L}
i\gamma^{\mu}\partial_{\mu}\psi_{\nu L}\nonumber\\
\hspace{-0.3in} && -
\frac{g}{2\sqrt{2}}\,\Big(\bar{\psi}_p\gamma^{\mu}(1 -
\gamma^5)\,\psi_n + i\,\sqrt{2}\,(\pi^0 \partial_{\mu} \pi^- -
\partial_{\mu}\pi^0 \pi^-) - \sqrt{2}\,(\sigma\, \partial_{\mu}\pi^- -
\partial_{\mu}\sigma\,\pi^-) - \sqrt{2}\,f_{\pi}\,\partial_{\mu}\pi^-
\Big)\,W^+_{\mu}\nonumber\\
\hspace{-0.3in} && -
\frac{g}{2\sqrt{2}}\,\Big(\bar{\psi}_n\gamma^{\mu}(1 -
\gamma^5)\,\psi_p + i\,\sqrt{2}\,(\pi^+ \partial_{\mu} \pi^0 - 
\partial_{\mu}\pi^+ \pi^0) - \sqrt{2}\,(\sigma\, \partial_{\mu}\pi^+ -
\partial_{\mu}\sigma\,\pi^+) - \sqrt{2}\,f_{\pi}\,\partial_{\mu}\pi^+
\Big)\,W^-_{\mu}\nonumber\\
\hspace{-0.3in} && -
\frac{g}{2\cos\theta_W}\,\Big(\frac{1}{2}\,\bar{\psi}_p\gamma^{\mu}\,\big(1
- 4\sin^2\theta_W - \gamma^5\big)\,\psi_p - \frac{1}{2}\,\bar{\psi}_n
\gamma^{\mu} \big(1 - \gamma^5)\,\psi_n + i\,(1 -
2\,\sin^2\theta_W)\,(\pi^+\partial_{\mu}\pi^- -
\partial_{\mu}\pi^+\,\pi^-)\nonumber\\
\hspace{-0.3in} && - (\sigma\,\partial_{\mu}\pi^0 -
\partial_{\mu}\sigma\,\pi^0) -
f_{\pi}\,\partial_{\mu}\pi^0\Big)\,Z_{\mu} -
e\,\Big(\bar{\psi}_p\gamma^{\mu}\,\psi_p +
i\,(\pi^-\partial_{\mu}\pi^+ - \partial_{\mu}\pi^-\,\pi^+)\Big)\,
A_{\mu} - \frac{g}{2\sqrt{2}}\,\bar{\psi}_e\gamma^{\mu} \big(1 -
\gamma^5\big)\,\psi_{\nu L}\,W^-_{\mu} \nonumber\\
\hspace{-0.3in} &&- \frac{g}{2\sqrt{2}}\,\bar{\psi}_{\nu L}\gamma^{\mu}
\big(1 - \gamma^5\big)\,\psi_e\,W^+_{\mu} +
\frac{g}{4\cos\theta_W}\,\bar{\psi}_e\,\gamma^{\mu}\, \big(1 -
4\sin^2\theta_W - \gamma^5\big)\,\psi_e\,Z_{\mu} -
\frac{g}{4\cos\theta_W}\,\bar{\psi}_{\nu L}\,\gamma^{\mu}\big(1 -
\gamma^5\big)\,\psi_{\nu L}\, Z_{\mu}\nonumber\\
\hspace{-0.3in} && + e\,\bar{\psi}_e\gamma^{\mu}\,\psi_e \,A_{\mu} +
\frac{1}{2}\,\Big(e g \,A_{\mu} + g^2 \cos\theta_W
\tan^2\theta_W\,Z_{\mu}\Big)\,\Big(i\,(f_{\pi} + \sigma)\,(\pi^+
W^{-\mu} - \pi^- W^{+\mu}) - \pi^0\,(\pi^+ W^{-\mu} + \pi^-
W^{+\mu})\Big)\nonumber\\
\hspace{-0.3in} && + \frac{1}{4}\,g^2\,\big(2\,f_{\pi}\,\sigma^2 +
(\pi^0)^2 + 2\pi^+ \pi^-\big)\,W^+_{\mu}W^{-\mu} +
\frac{1}{8}\,\frac{g^2}{\cos^2\theta_W}\,\big(2\,f_{\pi}\,\sigma^2 +
(\pi^0)^2\big)\,Z_{\mu}Z^{\mu} + \pi^+ \pi^- \Big(e\,A_{\mu} + 
\frac{g}{2\cos\theta_W}\nonumber\\
\hspace{-0.3in} &&\times \,(1 - 2\sin^2
\theta_W)\,Z_{\mu}\Big)\Big(e\,A^{\mu} + \frac{g}{2\cos\theta_W}\,(1 -
2\sin^2 \theta_W)\,Z^{\mu}\Big) +
i\,\frac{1}{2}\,e\,W^-_{\mu\nu}\big(W^{+\mu}A^{\nu} - A^{\mu}
W^{+\nu}\big) +
i\,\frac{1}{2}\,g\,\cos\theta_W\nonumber\\
\hspace{-0.3in}&&\times\,W^-_{\mu\nu}
\big(W^{+\mu}Z^{\nu} - Z^{\mu} W^{+\nu}\big) +
i\,\frac{1}{2}\,e\,W^+_{\mu\nu}\big( A^{\mu} W^{-\nu} -
W^{-\mu}A^{\nu}\big) + i\,\frac{1}{2}\,g\,\cos\theta_W\,W^+_{\mu\nu}
\big(Z^{\mu} W^{-\nu}
W^{-\mu}Z^{\nu}\big)\nonumber\\ \hspace{-0.3in}&& +
\frac{1}{2}\,e^2\,\big(W^+_{\mu}A_{\nu} - A_{\mu} W^+_{\nu}\big)\big(
A^{\mu} W^{-\nu} - W^{-\mu}A^{\nu}\big) +
\frac{1}{2}\,g^2\,\cos^2\theta_W \, \big(W^+_{\mu}Z_{\nu} - Z_{\mu}
W^+_{\nu}\big)\big(Z^{\mu} W^{-\nu} - W^{-\mu}Z^{\nu}\big)
\nonumber\\ \hspace{-0.3in}&& + \frac{1}{2}\,e\, g\,\cos\theta_W
\,\big(W^+_{\mu}A_{\nu} - A_{\mu} W^+_{\nu}\big)\big( Z^{\mu} W^{-\nu}
- W^{-\mu}Z^{\nu}\big) + \frac{1}{2}\,e\, g\,\cos\theta_W
\,\big(W^+_{\mu}Z_{\nu} - Z_{\mu} W^+_{\nu}\big)\big( A^{\mu} W^{-\nu}
- W^{-\mu}A^{\nu}\big)\nonumber\\ \hspace{-0.3in}&& +
\frac{1}{2}\,i\,e\,F_{\mu\nu}\big( W^{-\mu} W^{+\nu} - W^{+\mu}
W^{-\nu}\big) + \frac{1}{2}\,i\,g\,\cos\theta_W\,Z_{\mu\nu}\, \big(
W^{-\mu} W^{+\nu} - W^{+\mu} W^{-\nu}\big) + \frac{1}{4}\,g^2\,\big(
W^-_{\mu} W^+_{\nu} - W^+_{\mu}
W^-_{\nu}\big)\nonumber\\ \hspace{-0.3in}&&\times\,\big( W^{-\mu}
W^{+\nu} - W^{+\mu} W^{-\nu}\big) + \frac{M^2_W}{v} W^+_{\mu} W^{-\mu}
H + \frac{1}{4}\,\frac{M^2_W}{v^2}\, W^+_{\mu}W^{-\mu} H^2 +
\frac{1}{2}\,\frac{M^2_Z}{v}\, Z_{\mu}Z^{\mu}H +
\frac{1}{8}\,\frac{M^2_Z}{v^2}\, Z_{\mu}Z^{\mu}
H^2\nonumber\\ \hspace{-0.3in}&& -
\frac{m_e}{v}\,\bar{\psi}_e\psi_e\,H -
\frac{1}{2}\,\frac{M^2_H}{v}\,H^3 -
\frac{1}{8}\,\frac{M^2_H}{v^2}\,H^4,
\end{eqnarray}
where $\theta_W$ is the Weinberg angle defined by $\tan \theta_W =
g'/g$ \cite{PDG2018, Weinberg1967}, the field operators $W^{\pm} =
(W^1 \mp iW^2)/\sqrt{2}$ of the $W^{\pm}$--boson, $Z_{\mu} =
W^3_{\mu}\cos\theta_W - B_{\mu}\sin\theta_W$ of the $Z$--boson and
$A_{\mu} = B_{\mu} \cos\theta_W + W^3_{\mu} \sin\theta_W$ of the
electromagnetic fields, respectively, $e = g\,\sin\theta_W$ is the
proton electric charge. Then, we have denoted $X_{\mu\nu} =
\partial_{\mu}X_{\nu} - \partial_{\nu}X_{\mu}$ for $X = W^{\pm}, Z$
and $F_{\mu\nu} = \partial_{\mu} A_{\nu} - \partial_{\nu}A_{\mu}$ is
the electromagnetic field strength tensor. The term
$(1/2\xi)\,(\partial_{\mu}A^{\mu})^2$ fixes a gauge of the
electromagnetic field, where $\xi$ is a gauge parameter
\cite{Itzykson1980}. The massive fields of the $W^{\pm}$-- and
$Z$--electroweak bosons are defined in the physical gauge with masses
equal to
\begin{eqnarray}\label{eq:36}
M^2_W = \frac{1}{4}\,g^2\big(v^2 + f^2_{\pi}\big)\;,\; M^2_Z =
\frac{M^2_W}{\cos^2\theta_W}
\end{eqnarray}
with the hadronic contribution defined by the term proportional to
$f^2_{\pi}$. The vacuum expectation values $v$ and $f_{\pi}$ of the
Higgs and $\sigma$--meson fields are equal to $v =
\sqrt{\tilde{\mu}^2/\tilde{\lambda}}$ and $f_{\pi} =
\sqrt{\mu^2/\gamma}$, respectively. The masses of the hadrons,
electron and Higgs boson are given by
\begin{eqnarray}\label{eq:37}
m_N = g_{\pi N} f_{\pi}\;,\, m^2_{\pi} = 0 \;,\; m^2_{\sigma} =
2f^2_{\pi}\gamma\;,\; m_e = g_ev\;,\; M^2_H = 2 v^2 \tilde{\lambda}. 
\end{eqnarray}
The Lagrangian Eq.(\ref{eq:35}) as well as the Lagrangian
Eq.(\ref{eq:29}) is invariant under gauge $SU(2)_L \times U(1)_Y$
transformations Eqs.(\ref{eq:25}), (\ref{eq:28}) and
(\ref{eq:32}). Such an invariance is being retained as long as pions
$\vec{\pi} = (\pi^{\pm},\pi^0)$ are massless. 

The term violating chiral $SU(2)\times SU(2)$ invariance and providing
a non--vanishing pion mass is equal to $\delta {\cal L}_{\rm L \sigma M} =
m^2_{\pi} f_{\pi}\,\sigma$. This leads to the hadronic masses
\begin{eqnarray}\label{eq:38}
m_N = g_{\pi N} f_{\pi}\;,\, m^2_{\pi} = f^2_{\pi}\gamma - \mu^2 \;,\;
m^2_{\sigma} = 2f^2_{\pi}\gamma + m^2_{\pi}
\end{eqnarray}
and $f_{\pi} =\sqrt{(\mu^2 + m^2_{\pi})/\gamma} >
\sqrt{\mu^2/\gamma}$. Since the $\sigma$--field is a component of the
$SU(2)_L\times U(1)_Y$ doublet $\sigma = (\Phi_{+1/2} +
\Phi^c_{-1/2})/\sqrt{2}$, where $\Phi_{+1/2}$ and $\Phi^c_{-1/2}$ are
the {\it up} and {\it down} components of the $SU(2)_L\times U(1)_Y$
doublets $\Phi$ and $\Phi^c$, respectively, the term $\delta {\cal
  L}_{\rm L \sigma M} \to \delta {\cal L}_{\rm L \sigma M + SEM} =
m^2_{\pi} f_{\pi}\,\sigma = m^2_{\pi} f_{\pi}\,(\Phi_{+1/2} +
\Phi^c_{-1/2})/\sqrt{2}$ violates also invariance under $SU(2)_L\times
U(1)_Y$ transformations. Restoration of invariance under
$SU(2)_L\times U(1)_Y$ transformations can be reached following
Weinberg \cite{Weinberg1971} and introducing the interaction
\begin{eqnarray}\label{eq:39}
\delta {\cal L}_{\rm L \sigma M + SEM} =
\frac{m^2_{\pi}}{\sqrt{2}}\,\frac{f_{\pi}}{v}\,\big(\Phi^{c\dagger}\phi
+ \phi^{\dagger}\Phi^c\big).
\end{eqnarray}
This allows to deal with the term $\delta {\cal L}_{\rm L \sigma M} =
m^2_{\pi} f_{\pi}\,\sigma$ in the form invariant under $SU(2)_L\times
U(1)_Y$ transformations.  In the phase of spontaneously broken
$SU(2)_L\times U(1)_Y$ symmetry the interaction Eq.(\ref{eq:39})
acquires a form
\begin{eqnarray}\label{eq:40}
\delta {\cal L}_{\rm L \sigma M + SEM} = m^2_{\pi}f_{\pi}
\sigma\,\Big(1 + \frac{H}{v}\Big).
\end{eqnarray}
In the chirally broken phase, when  $\sigma \to f_{\pi} + \sigma$, the
contribution of the interaction Eq.(\ref{eq:40}) to the Lagrangian
${\cal L}_{\rm L \sigma M + SEM}$ in Eq.(\ref{eq:35}) is given by
\begin{eqnarray}\label{eq:41}
\delta {\cal L}_{\rm L \sigma M + SEM} \longrightarrow - m^2_{\pi}
\pi^+\pi^- - \frac{1}{2}\,m^2_{\pi} (\pi^0)^2 + \frac{m^2_{\pi}
  f_{\pi}}{v}\,\sigma\,H.
\end{eqnarray}
The terms linear in $\sigma$ and $H$, which appear in the $SU(2)_L
\times U(1)_Y$ symmetry broken phase, lead to a redefinition of the
vacuum expectation value $v$ of the Higgs field only. A relative
correction $\delta v/v_0 = f^2_{\pi}m^2_{\pi}/v^2_0 M^2_H$ to the
standard value $v_0 = \sqrt{\tilde{\mu}^2/\tilde{\lambda}} = 246\,{\rm
  GeV}$ \cite{PDG2018} is of about $10^{-13}$, calculated for the
Higgs--boson mass $M_H = 125\,{\rm GeV}$, $f_{\pi} = 92.4\,{\rm MeV}$
and $m_{\pi} = 140\,{\rm MeV}$ \cite{PDG2018}. We would like to
accentuate that the interaction Eq.(\ref{eq:41}) amends only
invariance under global $SU(2)_L \times U(1)_Y$ transformations but
not gauge ones. Indeed, a non--vanishing pion mass leads to
non--conservation (or partial conservation) of the hadronic
axial--vector current current, violating invariance under $SU(2)_L
\times U(1)_Y$ gauge transformations. Below we show this by example of
the neutron $\beta^-$--decays.

Together with the contribution of the interaction Eq.(\ref{eq:39}),
taken in the physical phase given by Eq.(\ref{eq:41}), the quantum
field theoretic model of strong low--energy pion--nucleon and
electroweak hadron--hadron, hadron--lepton and lepton--lepton
interactions, where leptons are an electron $e^-$ and neutrino
$\nu_e$, is described by the Lagrangian
\begin{eqnarray*}
\hspace{-0.30in}&&{\cal L}_{\rm L\sigma M + SEM} =
\bar{\psi}_p\big(i\gamma^{\mu}\partial_{\mu} - m_N\big)\psi_p +
\bar{\psi}_n\big(i\gamma^{\mu}\partial_{\mu} - m_N\big)\psi_n +
\big(\partial_{\mu}\pi^+\partial^{\mu}\pi^- - m^2_{\pi}\big) \pi^+
\pi^-\nonumber\\ \hspace{-0.3in}&& + \frac{1}{2}\,
\big(\partial_{\mu}\pi^0 \partial^{\mu}\pi^0 - m^2_{\pi}(\pi^0)^2\big)
+ \frac{1}{2}\,\big(\partial_{\mu}\sigma \partial^{\mu}\sigma -
m^2_{\sigma} \sigma^2\big) - \sqrt{2}\,g_{\pi N}\bar{\psi}_pi\gamma^5
\psi_n\,\pi^+ - \sqrt{2}\,g_{\pi N}\bar{\psi}_ni\gamma^5
\psi_p\,\pi^-\nonumber\\
\hspace{-0.3in}&& - g_{\pi N}\,\big( \bar{\psi}_pi\gamma^5 \psi_p -
\bar{\psi}_ni\gamma^5 \psi_n\big)\,\pi^0 - g_{\pi N}\,\big(
\bar{\psi}_p \psi_p + \bar{\psi}_n \psi_n\big)\,\sigma -
\gamma\,f_{\pi}\,\sigma\big(\sigma^2 + 2 \pi^+ \pi^- + (\pi^0)^2\big)
\nonumber\\
\hspace{-0.3in} && - \frac{1}{4}\,\gamma\,\big(\sigma^2 + 2 \pi^+
\pi^- + (\pi^0)^2\big)^2 - \frac{1}{2}\,W^+_{\mu\nu}W^{- \mu\nu} +
M^2_W W^+_{\mu}W^{-\mu} - \frac{1}{4}\,Z_{\mu\nu} Z^{\mu\nu} +
\frac{1}{2}\, M^2_Z Z_{\mu}Z^{\mu} - \frac{1}{4}\,F_{\mu\nu}F^{\mu\nu}
\nonumber\\
\hspace{-0.3in} &&-
\frac{1}{2\xi}\,\big(\partial_{\mu}A^{\mu}\big)^2 +
\frac{1}{2}\,\partial_{\mu} H \partial^{\mu} H - \frac{1}{2}\,M^2_H
H^2 + \bar{\psi}_e \big(i\gamma^{\mu} \partial_{\mu} -
m_e\big)\,\psi_e + \bar{\psi}_{\nu L}
i\gamma^{\mu}\partial_{\mu}\psi_{\nu L}\nonumber\\
\hspace{-0.3in} && -
\frac{g}{2\sqrt{2}}\,\Big(\bar{\psi}_p\gamma^{\mu}(1 -
\gamma^5)\,\psi_n + i\,\sqrt{2}\,(\pi^0 \partial_{\mu} \pi^- -
\partial_{\mu}\pi^0 \pi^-) - \sqrt{2}\,(\sigma\, \partial_{\mu}\pi^- -
\partial_{\mu}\sigma\,\pi^-) - \sqrt{2}\,f_{\pi}\,\partial_{\mu}\pi^-
\Big)\,W^+_{\mu}\nonumber\\
\hspace{-0.3in} && -
\frac{g}{2\sqrt{2}}\,\Big(\bar{\psi}_n\gamma^{\mu}(1 -
\gamma^5)\,\psi_p + i\,\sqrt{2}\,(\pi^+ \partial_{\mu} \pi^0 - 
\partial_{\mu}\pi^+ \pi^0) - \sqrt{2}\,(\sigma\, \partial_{\mu}\pi^+ -
\partial_{\mu}\sigma\,\pi^+) - \sqrt{2}\,f_{\pi}\,\partial_{\mu}\pi^+
\Big)\,W^-_{\mu}\nonumber\\
\hspace{-0.3in} && -
\frac{g}{2\cos\theta_W}\,\Big(\frac{1}{2}\,\bar{\psi}_p\gamma^{\mu}\,\big(1
- 4\sin^2\theta_W - \gamma^5\big)\,\psi_p - \frac{1}{2}\,\bar{\psi}_n
\gamma^{\mu} \big(1 - \gamma^5)\,\psi_n + i\,(1 -
2\,\sin^2\theta_W)\,(\pi^+\partial_{\mu}\pi^- -
\partial_{\mu}\pi^+\,\pi^-)\nonumber\\
\hspace{-0.3in} && - (\sigma\,\partial_{\mu}\pi^0 -
\partial_{\mu}\sigma\,\pi^0) -
f_{\pi}\,\partial_{\mu}\pi^0\Big)\,Z_{\mu} -
e\,\Big(\bar{\psi}_p\gamma^{\mu}\,\psi_p +
i\,(\pi^-\partial_{\mu}\pi^+ - \partial_{\mu}\pi^-\,\pi^+)\Big)\,
A_{\mu} - \frac{g}{2\sqrt{2}}\,\bar{\psi}_e\gamma^{\mu} \big(1 -
\gamma^5\big)\,\psi_{\nu L}\,W^-_{\mu} \nonumber\\
\hspace{-0.3in} &&- \frac{g}{2\sqrt{2}}\,\bar{\psi}_{\nu L}\gamma^{\mu}
\big(1 - \gamma^5\big)\,\psi_e\,W^+_{\mu} +
\frac{g}{4\cos\theta_W}\,\bar{\psi}_e\,\gamma^{\mu}\, \big(1 -
4\sin^2\theta_W - \gamma^5\big)\,\psi_e\,Z_{\mu} -
\frac{g}{4\cos\theta_W}\,\bar{\psi}_{\nu L}\,\gamma^{\mu}\big(1 -
\gamma^5\big)\,\psi_{\nu L}\, Z_{\mu}\nonumber\\
\hspace{-0.3in} && + e\,\bar{\psi}_e\gamma^{\mu}\,\psi_e \,A_{\mu} +
\frac{1}{2}\,\Big(e g \,A_{\mu} + g^2 \cos\theta_W
\tan^2\theta_W\,Z_{\mu}\Big)\,\Big(i\,(f_{\pi} + \sigma)\,(\pi^+
W^{-\mu} - \pi^- W^{+\mu}) - \pi^0\,(\pi^+ W^{-\mu} + \pi^-
W^{+\mu})\Big)\nonumber\\
\end{eqnarray*}
\begin{eqnarray}\label{eq:42}
\hspace{-0.3in} && + \frac{1}{4}\,g^2\,\big(2\,f_{\pi}\sigma +
\sigma^2 + (\pi^0)^2 + 2\pi^+ \pi^-\big)\,W^+_{\mu}W^{-\mu} +
\frac{1}{8}\,\frac{g^2}{\cos^2\theta_W}\,\big(2\,f_{\pi}\sigma +
\sigma^2 + (\pi^0)^2\big)\,Z_{\mu}Z^{\mu} + \pi^+ \pi^-
\Big(e\,A_{\mu} + \frac{g}{2\cos\theta_W}\nonumber\\
\hspace{-0.3in} &&\times \,(1 - 2\sin^2
\theta_W)\,Z_{\mu}\Big)\Big(e\,A^{\mu} + \frac{g}{2\cos\theta_W}\,(1 -
2\sin^2 \theta_W)\,Z^{\mu}\Big) +
i\,\frac{1}{2}\,e\,W^-_{\mu\nu}\big(W^{+\mu} A^{\nu} - A^{\mu}
W^{+\nu}\big) +
i\,\frac{1}{2}\,g\,\cos\theta_W\nonumber\\ \hspace{-0.3in}&&\times\,
W^-_{\mu\nu} \big(W^{+\mu}Z^{\nu} - Z^{\mu} W^{+\nu}\big) +
i\,\frac{1}{2}\,e\,W^+_{\mu\nu}\big( A^{\mu} W^{-\nu} -
W^{-\mu}A^{\nu}\big) + i\,\frac{1}{2}\,g\,\cos\theta_W\,W^+_{\mu\nu}
\big(Z^{\mu} W^{-\nu} -
W^{-\mu}Z^{\nu}\big)\nonumber\\ \hspace{-0.3in}&& +
\frac{1}{2}\,e^2\,\big(W^+_{\mu}A_{\nu} - A_{\mu} W^+_{\nu}\big)\big(
A^{\mu} W^{-\nu} - W^{-\mu}A^{\nu}\big) +
\frac{1}{2}\,g^2\,\cos^2\theta_W \, \big(W^+_{\mu}Z_{\nu} - Z_{\mu}
W^+_{\nu}\big)\big(Z^{\mu} W^{-\nu} - W^{-\mu}Z^{\nu}\big)
\nonumber\\ \hspace{-0.3in}&& + \frac{1}{2}\,e\, g\,\cos\theta_W
\,\big(W^+_{\mu}A_{\nu} - A_{\mu} W^+_{\nu}\big)\big( Z^{\mu} W^{-\nu}
- W^{-\mu}Z^{\nu}\big) + \frac{1}{2}\,e\, g\,\cos\theta_W
\,\big(W^+_{\mu}Z_{\nu} - Z_{\mu} W^+_{\nu}\big)\big( A^{\mu} W^{-\nu}
- W^{-\mu}A^{\nu}\big)\nonumber\\ \hspace{-0.3in}&& +
\frac{1}{2}\,i\,e\,F_{\mu\nu}\big( W^{-\mu} W^{+\nu} - W^{+\mu}
W^{-\nu}\big) + \frac{1}{2}\,i\,g\,\cos\theta_W\,Z_{\mu\nu}\, \big(
W^{-\mu} W^{+\nu} - W^{+\mu} W^{-\nu}\big) + \frac{1}{4}\,g^2\,\big(
W^-_{\mu} W^+_{\nu} - W^+_{\mu}
W^-_{\nu}\big)\nonumber\\ \hspace{-0.3in}&&\times\,\big( W^{-\mu}
W^{+\nu} - W^{+\mu} W^{-\nu}\big) + \frac{M^2_W}{v} W^+_{\mu} W^{-\mu}
H + \frac{1}{4}\,\frac{M^2_W}{v^2}\, W^+_{\mu}W^{-\mu} H^2 +
\frac{1}{2}\,\frac{M^2_Z}{v}\, Z_{\mu}Z^{\mu}H +
\frac{1}{8}\,\frac{M^2_Z}{v^2}\, Z_{\mu}Z^{\mu}
H^2\nonumber\\ \hspace{-0.3in}&& -
\frac{m_e}{v}\,\bar{\psi}_e\psi_e\,H -
\frac{1}{2}\,\frac{M^2_H}{v}\,H^3 -
\frac{1}{8}\,\frac{M^2_H}{v^2}\,H^4 + \frac{m^2_{\pi}
  f_{\pi}}{v}\,\sigma\,H.
\end{eqnarray}
We would like to emphasize that the $W^{\pm}$--bosons couple to the $V
- A$ hadronic currents, providing in the tree--approximation a
standard $V - A$ effective low--energy interactions for the
description of the neutron $\beta^-$--decays
\cite{Feynman1958,Nambu1960}. The vector and axial--vector hadronic
currents have baryonic and mesonic parts in agreement with
Eq.(\ref{eq:9}), which are necessary for conservation of vector and
partial conservation of axial--vector hadronic currents
\cite{Feynman1958, Nambu1960, Ivanov2018b, Ivanov2018}. A partial
conservation of the axial--vector hadronic current assumes a
proportionality of the divergence of the axial--vector hadronic
current to the squared pion mass \cite{Adler1968}. In the chiral
limit, i.e. in the limit of zero pion mass $m_{\pi} \to 0$, the
axial--vector hadronic current is conserved \cite{Nambu1960}. An
influence of partial conservation of the axial--vector hadronic
current on gauge invariance of radiative corrections, caused by
hadronic structure of the nucleon, we shall investigate below by
example of radiative corrections of order $O(\alpha E_e/ m_N)$ to the
neutron lifetime.

\subsection{Renormalization  of the quantum field theory of strong 
low--energy and electroweak interactions described by the Lagrangian
 Eq.(\ref{eq:42})}

For the discussion of renormalization procedure in the quantum field
theoretic model L$\sigma$M$+$SEM we rewrite the Lagrangian
Eq.(\ref{eq:42}) as follows
\begin{eqnarray*}
\hspace{-0.3in}&&{\cal L}^{(0)}_{\rm L\sigma M + SEM} =
\bar{\psi}^{(0)}_p\big(i\gamma^{\mu}\partial_{\mu} -
m^{(0)}_N\big)\psi^{(0)}_p +
\bar{\psi}^{(0)}_n\big(i\gamma^{\mu}\partial_{\mu} -
m^{(0)}_N\big)\psi^{(0)}_n +
\big(\partial_{\mu}\pi^{(0)+}\partial^{\mu}\pi^{(0)-} -
m^{(0)2}_{\pi}\big) \pi^{(0)+} \pi^{(0)-}\nonumber\\ \hspace{-0.15in}&&
+ \frac{1}{2}\, \big(\partial_{\mu}\pi^{(0)0} \partial^{\mu}\pi^{(0)0}
- m^{(0)2}_{\pi}(\pi^{(0)0})^2\big) +
\frac{1}{2}\,\big(\partial_{\mu}\sigma^{(0)}
\partial^{\mu}\sigma^{(0)} - m^{(0)2}_{\sigma} (\sigma^{(0)})^2\big) -
\sqrt{2}\,g^{(0)}_{\pi N}\bar{\psi}^{(0)}_pi\gamma^5
\psi^{(0)}_n\,\pi^{(0)+}\nonumber\\ \hspace{-0.15in}&& -
\sqrt{2}\,g^{(0)}_{\pi N}\bar{\psi}^{(0)}_ni\gamma^5
\psi_p\,\pi^{(0)-} - g^{(0)}_{\pi N}\,\big(
\bar{\psi}^{(0)}_pi\gamma^5 \psi^{(0)}_p - \bar{\psi}^{(0)}_ni\gamma^5
\psi^{(0)}_n\big)\,\pi^{(0)0} - g^{(0)}_{\pi N}\,\big(
\bar{\psi}^{(0)}_p \psi^{(0)}_p + \bar{\psi}^{(0)}_n
\psi^{(0)}_n\big)\,\sigma^{(0)}\nonumber\\
\hspace{-0.15in}&& -
\gamma^{(0)}\,f^{(0)}_{\pi}\,\sigma^{(0)}\big((\sigma^{(0)})^2 + 2
\pi^{(0)+} \pi^{(0)-} + (\pi^{(0)0})^2\big) -
\frac{1}{4}\,\gamma^{(0)}\,\big((\sigma{(0)})^2 + 2 \pi^{(0)+}
\pi^{(0)-} + (\pi^{(0)0})^2\big)^2 -
\frac{1}{2}\,W^{(0)+}_{\mu\nu}W^{(0)- \mu\nu}\nonumber\\
\hspace{-0.15in}&& + M^{(0)2}_W W^{(0)+}_{\mu}W^{(0)-\mu} -
\frac{1}{4}\,Z^{(0)}_{\mu\nu} Z^{(0)\mu\nu} + \frac{1}{2}\, M^{(0)2}_Z
Z^{(0)}_{\mu}Z^{(0)\mu} - \frac{1}{4}\,F^{(0)}_{\mu\nu}F^{(0)\mu\nu} -
\frac{1}{2\xi^{(0)}}\,\big(\partial_{\mu}A^{(0)\mu}\big)^2 +
\frac{1}{2}\,\partial_{\mu} H^{(0)} \partial^{\mu} H^{(0)}\nonumber\\
\hspace{-0.15in} && - \frac{1}{2}\,M^{(0)2}_H (H^{(0)})^2 +
\bar{\psi}^{(0)}_e \big(i\gamma^{\mu} \partial_{\mu} -
m^{(0)}_e\big)\,\psi^{(0)}_e + \bar{\psi}^{(0)}_{\nu L}
i\gamma^{\mu}\partial_{\mu}\psi^{(0)}_{\nu L} -
\frac{g^{(0)}}{2\sqrt{2}}\,\Big(\bar{\psi}^{(0)}_p\gamma^{\mu}(1 -
\gamma^5)\,\psi^{(0)}_n + i\,\sqrt{2}\,(\pi^{(0)0} \partial_{\mu}
\pi^{(0)-}\nonumber\\
\hspace{-0.15in} && - \partial_{\mu}\pi^{(0)0} \pi^{(0)-}) -
\sqrt{2}\,(\sigma^{(0)}\, \partial_{\mu}\pi^{(0)-} -
\partial_{\mu}\sigma^{(0)}\,\pi^{(0)-}) -
\sqrt{2}\,f^{(0)}_{\pi}\,\partial_{\mu}\pi^{(0)-}
\Big)\,W^{(0)+}_{\mu} -
\frac{g^{(0)}}{2\sqrt{2}}\,\Big(\bar{\psi}^{(0)}_n\gamma^{\mu}(1 -
\gamma^5)\,\psi^{(0)}_p\nonumber\\
\hspace{-0.15in}&& + i\,\sqrt{2}\,(\pi^{(0)+} \partial_{\mu} \pi^{(0)0}
- \partial_{\mu}\pi^{(0)+} \pi^{(0)0}) - \sqrt{2}\,(\sigma^{(0)}\,
\partial_{\mu}\pi^{(0)+} - \partial_{\mu}\sigma^{(0)}\,\pi^{(0)+}) -
\sqrt{2}\,f^{(0)}_{\pi}\,\partial_{\mu}\pi^{(0)+}
\Big)\,W^{(0)-}_{\mu}\nonumber\\
\hspace{-0.15in}&& -
\frac{g^{(0)}}{2\cos\theta_W}\,\Big(\frac{1}{2}\,\bar{\psi}^{(0)}_p\gamma^{\mu}\,\big(1
- 4\sin^2\theta_W - \gamma^5\big)\,\psi^{(0)}_p -
\frac{1}{2}\,\bar{\psi}^{(0)}_n \gamma^{\mu} \big(1 -
\gamma^5)\,\psi^{(0)}_n + i\,(1 -
2\,\sin^2\theta_W)\,(\pi^{(0)+}\partial_{\mu}\pi^{(0)-}\nonumber\\
\hspace{-0.15in} && - \partial_{\mu}\pi^{(0)+}\,\pi^{(0)-}) -
(\sigma^{(0)}\,\partial_{\mu}\pi^{(0)0} -
\partial_{\mu}\sigma^{(0)}\,\pi^{(0)0}) -
f^{(0)}_{\pi}\,\partial_{\mu}\pi^{(0)0}\Big)\,Z^{(0)}_{\mu} -
e^{(0)}\,\Big(\bar{\psi}^{(0)}_p\gamma^{\mu}\,\psi^{(0)}_p +
i\,(\pi^{(0)-}\partial_{\mu}\pi^{(0)+}\nonumber\\
\hspace{-0.15in} && - \partial_{\mu}\pi^{(0)-}\,\pi^{(0)+})\Big)\,
A^{(0)}_{\mu} -
\frac{g^{(0)}}{2\sqrt{2}}\,\bar{\psi}^{(0)}_e\gamma^{\mu} \big(1 -
\gamma^5\big)\,\psi^{(0)}_{\nu L}\,W^{(0)-}_{\mu} -
\frac{g^{(0)}}{2\sqrt{2}}\,\bar{\psi}^{(0)}_{\nu L} \gamma^{\mu}
\big(1 - \gamma^5\big)\,\psi^{(0)}_e\,W^{(0)+}_{\mu} +
\frac{g^{(0)}}{4\cos\theta_W}\nonumber\\
\end{eqnarray*}
\begin{eqnarray}\label{eq:43}
\hspace{-0.15in} &&\times\, \bar{\psi}^{(0)}_e\,\gamma^{\mu}\, \big(1 -
4\sin^2\theta_W - \gamma^5\big)\,\psi^{(0)}_e\,Z^{(0)}_{\mu} -
\frac{g^{(0)}}{4\cos\theta_W}\,\bar{\psi}^{(0)}_{\nu
  L}\,\gamma^{\mu}\big(1 - \gamma^5\big)\,\psi^{(0)}_{\nu L}\,
Z^{(0)}_{\mu} + e^{(0)}\,\bar{\psi}^{(0)}_e\gamma^{\mu}\,\psi^{(0)}_e
\,A^{(0)}_{\mu}\nonumber\\
\hspace{-0.15in} && 
+ \frac{1}{2}\,\Big(e^{(0)} g^{(0)} \,A^{(0)}_{\mu} + g^{(0)2}
\cos\theta_W
\tan^2\theta_W\,Z^{(0)}_{\mu}\Big)\,\Big(i\,(f^{(0)}_{\pi} +
\sigma^{(0)})\,(\pi^{(0)+} W^{(0)-\mu} - \pi^{(0)-}
W^{(0)+\mu})\nonumber\\
\hspace{-0.15in} && - \pi^{(0)0}\,(\pi^{(0)+} W^{(0)-\mu} + \pi^{(0)-}
W^{(0)+\mu})\Big) +
\frac{1}{4}\,g^{(0)2}\,\big(2\,f^{(0)}_{\pi} \sigma^{(0)} + (\sigma^{(0)})^2 +
(\pi^{(0)0})^2 + 2\pi^{(0)+}
\pi^{(0)-}\big)\,W^{(0)+}_{\mu}W^{(0)-\mu}\nonumber\\
\hspace{-0.15in} && + \frac{1}{8}\,\frac{g^{(0)2}}{\cos^2\theta_W}\,
\big(2\,f^{(0)}_{\pi}\sigma^{(0)} + (\sigma^{(0)})^2 +
(\pi^{(0)0})^2\big)\,Z^{(0)}_{\mu} Z^{(0)\mu} + \pi^{(0)+} \pi^{(0)-}
\Big(e^{(0)}\,A^{(0)}_{\mu} + \frac{g^{(0)}}{2\cos\theta_W}\,(1 -
2\sin^2 \theta_W)\,Z^{(0)}_{\mu}\Big)\nonumber\\
\hspace{-0.15in} && \times \Big(e^{(0)}\,A^{(0)\mu} +
\frac{g^{(0)}}{2\cos\theta_W}\,(1 - 2\sin^2 \theta_W)\,Z^{(0)\mu}\Big)
+ i\,\frac{1}{2}\,e^{(0)}\,W^{(0)-}_{\mu\nu}\big(W^{(0)+\mu}A^{(0)\nu}
- A^{(0)\mu} W^{(0)+\nu}\big) +
i\,\frac{1}{2}\,g^{(0)}\,\cos\theta_W\nonumber\\
\hspace{-0.15in}&&\times\,
W^{(0)-}_{\mu\nu} \big(W^{(0)+\mu}Z^{(0)\nu} - Z^{(0)\mu}
W^{(0)+\nu}\big) + i\,\frac{1}{2}\,e^{(0)}\,W^{(0)+}_{\mu\nu}\big(
A^{(0)\mu} W^{(0)-\nu} - W^{(0)-\mu}A^{(0)\nu}\big) +
i\,\frac{1}{2}\,g^{(0)}\,\cos\theta_W\,W^{(0)+}_{\mu\nu}\nonumber\\
\hspace{-0.15in}&&\times
\big(Z^{(0)\mu} W^{(0)-\nu} - W^{(0)-\mu}Z^{(0)\nu}\big) +
\frac{1}{2}\,e^{(0)2}\,\big(W^{(0)+}_{\mu}A^{(0)}_{\nu} -
A^{(0)}_{\mu} W^{(0)+}_{\nu}\big)\big( A^{(0)\mu} W^{(0)-\nu} -
W^{(0)-\mu}A^{(0)\nu}\big) +
\frac{1}{2}\,g^{(0)2}\nonumber\\
\hspace{-0.15in}&&\times\,\cos^2\theta_W
\, \big(W^{(0)+}_{\mu}Z^{(0)}_{\nu} - Z^{(0)}_{\mu}
W^{(0)+}_{\nu}\big)\big(Z^{(0)\mu} W^{(0)-\nu} -
W^{(0)-\mu}Z^{(0)\nu}\big) + \frac{1}{2}\,e^{(0)}\,
g^{(0)}\,\cos\theta_W \,\big(W^{(0)+}_{\mu}A^{(0)}_{\nu} -
A^{(0)}_{\mu} W^{(0)+}_{\nu}\big)\nonumber\\
\hspace{-0.15in}&&\times
\big( Z^{(0)\mu} W^{(0)-\nu} - W^{(0)-\mu}Z^{(0)\nu}\big) +
\frac{1}{2}\,e^{(0)}\, g^{(0)}\,\cos\theta_W
\,\big(W^{(0)+}_{\mu}Z^{(0)}_{\nu} - Z^{(0)}_{\mu}
W^{(0)+}_{\nu}\big)\big( A^{(0)\mu} W^{(0)-\nu} -
W^{(0)-\mu}A^{(0)\nu}\big)\nonumber\\
\hspace{-0.15in}&& +
\frac{1}{2}\,i\,e^{(0)}\,F^{(0)}_{\mu\nu}\big( W^{(0)-\mu} W^{(0)+\nu}
- W^{(0)+\mu} W^{(0)-\nu}\big) +
\frac{1}{2}\,i\,g^{(0)}\,\cos\theta_W\,Z^{(0)}_{\mu\nu}\, \big(
W^{(0)-\mu} W^{(0)+\nu} - W^{(0)+\mu}
W^{(0)-\nu}\big)\nonumber\\
\hspace{-0.15in}&& +
\frac{1}{4}\,g^{(0)2}\,\big( W^{(0)-}_{\mu} W^{(0)+}_{\nu} -
W^{(0)+}_{\mu} W^{(0)-}_{\nu}\big)\,\big( W^{(0)-\mu} W^{(0)+\nu} -
W^{(0)+\mu} W^{(0)-\nu}\big) + \frac{M^{(0)2}_W}{v^{(0)}}
W^{(0)+}_{\mu} W^{(0)-\mu} H^{(0)} \nonumber\\
\hspace{-0.15in}&& +
\frac{1}{4}\,\frac{M^{(0)2}_W}{v^{(0)2}}\, W^{(0)+}_{\mu}W^{(0)-\mu}
(H^{(0)})^2 + \frac{1}{2}\,\frac{M^{(0)2}_Z}{v^{(0)}}\,
Z^{(0)}_{\mu}Z^{(0)\mu}H^{(0)} +
\frac{1}{8}\,\frac{M^{(0)2}_Z}{v^{(0)2}}\, Z^{(0)}_{\mu}Z^{(0)\mu}
(H^{(0)})^2 -
\frac{m^{(0)}_e}{v^{(0)}}\,\bar{\psi}^{(0)}_e\psi^{(0)}_e\,H^{(0)}\nonumber\\
\hspace{-0.15in}&&
- \frac{1}{2}\,\frac{M^{(0)2}_H}{v^{(0)}}\,(H^{(0)})^3 -
\frac{1}{8}\,\frac{M^{(0)2}_H}{v^{(0)2}}\,(H^{(0)})^4 +
\frac{m^{(0)2}_{\pi} f^{(0)}_{\pi}}{v^{(0)}}\,\sigma^{(0)}\,H^{(0)},
\end{eqnarray}
where the subscript $(0)$ denotes {\it bare} fields and their {\it
  bare} masses and coupling constants, respectively.  After the
calculation of loop--contributions the dynamics of strong low--energy
and electroweak interactions of physical fields is described in the
quantum field theoretic model L$\sigma$M$+$SEM by the Lagrangian
\begin{eqnarray*}
\hspace{-0.3in}&&{\cal L}^{(r)}_{\rm L\sigma M + SEM} =
\bar{\psi}^{(r)}_p\big(i\gamma^{\mu}\partial_{\mu} -
m^{(r)}_N\big)\psi^{(r)}_p +
\bar{\psi}^{(r)}_n\big(i\gamma^{\mu}\partial_{\mu} -
m^{(r)}_N\big)\psi^{(r)}_n +
\big(\partial_{\mu}\pi^{(r)+}\partial^{\mu}\pi^{(r)-} -
m^{(r)2}_{\pi}\big) \pi^{(r)+}
\pi^{(r)-}\nonumber\\ \hspace{-0.15in}&& + \frac{1}{2}\,
\big(\partial_{\mu}\pi^{(r)0} \partial^{\mu}\pi^{(r)0} -
m^{(r)2}_{\pi}(\pi^{(r)0})^2\big) +
\frac{1}{2}\,\big(\partial_{\mu}\sigma^{(r)}
\partial^{\mu}\sigma^{(r)} - m^{(r)2}_{\sigma} (\sigma^{(r)})^2\big) -
\sqrt{2}\,g^{(r)}_{\pi N}\bar{\psi}^{(r)}_pi\gamma^5
\psi^{(r)}_n\,\pi^{(r)+}\nonumber\\ \hspace{-0.15in}&& -
\sqrt{2}\,g^{(r)}_{\pi N}\bar{\psi}^{(r)}_ni\gamma^5
\psi_p\,\pi^{(r)-} - g^{(r)}_{\pi N}\,\big(
\bar{\psi}^{(r)}_pi\gamma^5 \psi^{(r)}_p - \bar{\psi}^{(r)}_ni\gamma^5
\psi^{(r)}_n\big)\,\pi^{(r)0} - g^{(r)}_{\pi N}\,\big(
\bar{\psi}^{(r)}_p \psi^{(r)}_p + \bar{\psi}^{(r)}_n
\psi^{(r)}_n\big)\,\sigma^{(r)}\nonumber\\
\hspace{-0.15in}&& -
\gamma^{(r)}\,f^{(r)}_{\pi}\,\sigma^{(r)}\big((\sigma^{(r)})^2 + 2
\pi^{(r)+} \pi^{(r)-} + (\pi^{(r)0})^2\big) -
\frac{1}{4}\,\gamma^{(r)}\,\big((\sigma{(r)})^2 + 2 \pi^{(r)+}
\pi^{(r)-} + (\pi^{(r)0})^2\big)^2 -
\frac{1}{2}\,W^{(r)+}_{\mu\nu}W^{(r)- \mu\nu}\nonumber\\
\hspace{-0.15in}&& + M^{(r)2}_W W^{(r)+}_{\mu}W^{(r)-\mu} -
\frac{1}{4}\,Z^{(r)}_{\mu\nu} Z^{(r)\mu\nu} + \frac{1}{2}\, M^{(r)2}_Z
Z^{(r)}_{\mu}Z^{(r)\mu} - \frac{1}{4}\,F^{(r)}_{\mu\nu}F^{(r)\mu\nu} -
\frac{1}{2\xi^{(r)}}\,\big(\partial_{\mu}A^{(r)\mu}\big)^2 +
\frac{1}{2}\,\partial_{\mu} H^{(r)} \partial^{\mu} H^{(r)}\nonumber\\
\hspace{-0.15in} && - \frac{1}{2}\,M^{(r)2}_H (H^{(r)})^2 +
\bar{\psi}^{(r)}_e \big(i\gamma^{\mu} \partial_{\mu} -
m^{(r)}_e\big)\,\psi^{(r)}_e + \bar{\psi}^{(r)}_{\nu L}
i\gamma^{\mu}\partial_{\mu}\psi^{(r)}_{\nu L} -
\frac{g^{(r)}}{2\sqrt{2}}\,\Big(\bar{\psi}^{(r)}_p\gamma^{\mu}(1 -
\gamma^5)\,\psi^{(r)}_n + i\,\sqrt{2}\,(\pi^{(r)0} \partial_{\mu}
\pi^{(r)-}\nonumber\\
\hspace{-0.15in} && - \partial_{\mu}\pi^{(r)0} \pi^{(r)-}) -
\sqrt{2}\,(\sigma^{(r)}\, \partial_{\mu}\pi^{(r)-} -
\partial_{\mu}\sigma^{(r)}\,\pi^{(r)-}) -
\sqrt{2}\,f^{(r)}_{\pi}\,\partial_{\mu}\pi^{(r)-}
\Big)\,W^{(r)+}_{\mu} -
\frac{g^{(r)}}{2\sqrt{2}}\,\Big(\bar{\psi}^{(r)}_n\gamma^{\mu}(1 -
\gamma^5)\,\psi^{(r)}_p\nonumber\\
\hspace{-0.15in}&& + i\,\sqrt{2}\,(\pi^{(r)+} \partial_{\mu} \pi^{(r)0}
- \partial_{\mu}\pi^{(r)+} \pi^{(r)0}) - \sqrt{2}\,(\sigma^{(r)}\,
\partial_{\mu}\pi^{(r)+} - \partial_{\mu}\sigma^{(r)}\,\pi^{(r)+}) -
\sqrt{2}\,f^{(r)}_{\pi}\,\partial_{\mu}\pi^{(r)+}
\Big)\,W^{(r)-}_{\mu}\nonumber\\
\hspace{-0.15in}&& -
\frac{g^{(r)}}{2\cos\theta_W}\,\Big(\frac{1}{2}\,\bar{\psi}^{(r)}_p\gamma^{\mu}\,\big(1
- 4\sin^2\theta_W - \gamma^5\big)\,\psi^{(r)}_p -
\frac{1}{2}\,\bar{\psi}^{(r)}_n \gamma^{\mu} \big(1 -
\gamma^5)\,\psi^{(r)}_n + i\,(1 -
2\,\sin^2\theta_W)\,(\pi^{(r)+}\partial_{\mu}\pi^{(r)-}\nonumber\\
\hspace{-0.15in} && - \partial_{\mu}\pi^{(r)+}\,\pi^{(r)-}) -
(\sigma^{(r)}\,\partial_{\mu}\pi^{(r)0} -
\partial_{\mu}\sigma^{(r)}\,\pi^{(r)0}) -
f^{(r)}_{\pi}\,\partial_{\mu}\pi^{(r)0}\Big)\,Z^{(r)}_{\mu} -
e^{(r)}\,\Big(\bar{\psi}^{(r)}_p\gamma^{\mu}\,\psi^{(r)}_p +
i\,(\pi^{(r)-}\partial_{\mu}\pi^{(r)+}\nonumber\\
\hspace{-0.15in} && - \partial_{\mu}\pi^{(r)-}\,\pi^{(r)+})\Big)\,
A^{(r)}_{\mu} -
\frac{g^{(r)}}{2\sqrt{2}}\,\bar{\psi}^{(r)}_e\gamma^{\mu} \big(1 -
\gamma^5\big)\,\psi^{(r)}_{\nu L}\,W^{(r)-}_{\mu} -
\frac{g^{(r)}}{2\sqrt{2}}\,\bar{\psi}^{(r)}_{\nu L} \gamma^{\mu}
\big(1 - \gamma^5\big)\,\psi^{(r)}_e\,W^{(r)+}_{\mu} +
\frac{g^{(r)}}{4\cos\theta_W}\nonumber\\
\end{eqnarray*}
\begin{eqnarray}\label{eq:44}
\hspace{-0.15in} &&\times\, \bar{\psi}^{(r)}_e\,\gamma^{\mu}\, \big(1 -
4\sin^2\theta_W - \gamma^5\big)\,\psi^{(r)}_e\,Z^{(r)}_{\mu} -
\frac{g^{(r)}}{4\cos\theta_W}\,\bar{\psi}^{(r)}_{\nu
  L}\,\gamma^{\mu}\big(1 - \gamma^5\big)\,\psi^{(r)}_{\nu L}\,
Z^{(r)}_{\mu} + e^{(r)}\,\bar{\psi}^{(r)}_e\gamma^{\mu}\,\psi^{(r)}_e
\,A^{(r)}_{\mu}\nonumber\\
\hspace{-0.15in} && 
+ \frac{1}{2}\,\Big(e^{(r)} g^{(r)} \,A^{(r)}_{\mu} + g^{(r)2}
\cos\theta_W
\tan^2\theta_W\,Z^{(r)}_{\mu}\Big)\,\Big(i\,(f^{(r)}_{\pi} +
\sigma^{(r)})\,(\pi^{(r)+} W^{(r)-\mu} - \pi^{(r)-}
W^{(r)+\mu})\nonumber\\
\hspace{-0.15in} && - \pi^{(r)0}\,(\pi^{(r)+} W^{(r)-\mu} + \pi^{(r)-}
W^{(r)+\mu})\Big) +
\frac{1}{4}\,g^{(r)2}\,\big(2\,f^{(r)}_{\pi}\sigma^{(r)} + (\sigma^{(r)})^2 +
(\pi^{(r)0})^2 + 2\pi^{(r)+}
\pi^{(r)-}\big)\,W^{(r)+}_{\mu}W^{(r)-\mu}\nonumber\\
\hspace{-0.15in} && + \frac{1}{8}\,\frac{g^{(r)2}}{\cos^2\theta_W}\,
\big(2\,f^{(r)}_{\pi}\sigma^{(r)} + (\sigma^{(r)})^2 +
(\pi^{(r)0})^2\big)\,Z^{(r)}_{\mu} Z^{(r)\mu} + \pi^{(r)+} \pi^{(r)-}
\Big(e^{(r)}\,A^{(r)}_{\mu} + \frac{g^{(r)}}{2\cos\theta_W}\,(1 -
2\sin^2 \theta_W)\,Z^{(r)}_{\mu}\Big)\nonumber\\
\hspace{-0.15in} && \times \Big(e^{(r)}\,A^{(r)\mu} +
\frac{g^{(r)}}{2\cos\theta_W}\,(1 - 2\sin^2 \theta_W)\,Z^{(r)\mu}\Big)
+ i\,\frac{1}{2}\,e^{(r)}\,W^{(r)-}_{\mu\nu}\big(W^{(r)+\mu}A^{(r)\nu}
- A^{(r)\mu} W^{(r)+\nu}\big) +
i\,\frac{1}{2}\,g^{(r)}\,\cos\theta_W\nonumber\\ \hspace{-0.15in}&&\times\,
W^{(r)-}_{\mu\nu} \big(W^{(r)+\mu}Z^{(r)\nu} - Z^{(r)\mu}
W^{(r)+\nu}\big) + i\,\frac{1}{2}\,e^{(r)}\,W^{(r)+}_{\mu\nu}\big(
A^{(r)\mu} W^{(r)-\nu} - W^{(r)-\mu}A^{(r)\nu}\big) +
i\,\frac{1}{2}\,g^{(r)}\,\cos\theta_W\,W^{(r)+}_{\mu\nu}\nonumber\\ \hspace{-0.15in}&&\times
\big(Z^{(r)\mu} W^{(r)-\nu} - W^{(r)-\mu}Z^{(r)\nu}\big) +
\frac{1}{2}\,e^{(r)2}\,\big(W^{(r)+}_{\mu}A^{(r)}_{\nu} -
A^{(r)}_{\mu} W^{(r)+}_{\nu}\big)\big( A^{(r)\mu} W^{(r)-\nu} -
W^{(r)-\mu}A^{(r)\nu}\big) +
\frac{1}{2}\,g^{(r)2}\nonumber\\ \hspace{-0.15in}&&\times\,\cos^2\theta_W
\, \big(W^{(r)+}_{\mu}Z^{(r)}_{\nu} - Z^{(r)}_{\mu}
W^{(r)+}_{\nu}\big)\big(Z^{(r)\mu} W^{(r)-\nu} -
W^{(r)-\mu}Z^{(r)\nu}\big) + \frac{1}{2}\,e^{(r)}\,
g^{(r)}\,\cos\theta_W \,\big(W^{(r)+}_{\mu}A^{(r)}_{\nu} -
A^{(r)}_{\mu} W^{(r)+}_{\nu}\big)\nonumber\\ \hspace{-0.15in}&&\times
\big( Z^{(r)\mu} W^{(r)-\nu} - W^{(r)-\mu}Z^{(r)\nu}\big) +
\frac{1}{2}\,e^{(r)}\, g^{(r)}\,\cos\theta_W
\,\big(W^{(r)+}_{\mu}Z^{(r)}_{\nu} - Z^{(r)}_{\mu}
W^{(r)+}_{\nu}\big)\big( A^{(r)\mu} W^{(r)-\nu} -
W^{(r)-\mu}A^{(r)\nu}\big)\nonumber\\
\hspace{-0.15in}&& +
\frac{1}{2}\,i\,e^{(r)}\,F^{(r)}_{\mu\nu}\big( W^{(r)-\mu} W^{(r)+\nu}
- W^{(r)+\mu} W^{(r)-\nu}\big) +
\frac{1}{2}\,i\,g^{(r)}\,\cos\theta_W\,Z^{(r)}_{\mu\nu}\, \big(
W^{(r)-\mu} W^{(r)+\nu} - W^{(r)+\mu}
W^{(r)-\nu}\big)\nonumber\\
\hspace{-0.15in}&& +
\frac{1}{4}\,g^{(r)2}\,\big( W^{(r)-}_{\mu} W^{(r)+}_{\nu} -
W^{(r)+}_{\mu} W^{(r)-}_{\nu}\big)\,\big( W^{(r)-\mu} W^{(r)+\nu} -
W^{(r)+\mu} W^{(r)-\nu}\big) + \frac{M^{(r)2}_W}{v^{(r)}}
W^{(r)+}_{\mu} W^{(r)-\mu} H^{(r)} \nonumber\\
\hspace{-0.15in}&& +
\frac{1}{4}\,\frac{M^{(r)2}_W}{v^{(r)2}}\, W^{(r)+}_{\mu}W^{(r)-\mu}
(H^{(r)})^2 + \frac{1}{2}\,\frac{M^{(r)2}_Z}{v^{(r)}}\,
Z^{(r)}_{\mu}Z^{(r)\mu}H^{(r)} +
\frac{1}{8}\,\frac{M^{(r)2}_Z}{v^{(r)2}}\, Z^{(r)}_{\mu}Z^{(r)\mu}
(H^{(r)})^2 -
\frac{m^{(r)}_e}{v^{(r)}}\,\bar{\psi}^{(r)}_e\psi^{(r)}_e\,H^{(r)}\nonumber\\
\hspace{-0.15in}&&
- \frac{1}{2}\,\frac{M^{(r)2}_H}{v^{(r)}}\,(H^{(r)})^3 -
\frac{1}{8}\,\frac{M^{(r)2}_H}{v^{(r)2}}\,(H^{(r)})^4 +
\frac{m^{(r)2}_{\pi} f^{(r)}_{\pi}}{v^{(r)}}\,\sigma^{(r)}\,H^{(r)} +{\cal L}^{(\rm CT)}_{\rm L\sigma M + SEM},
\end{eqnarray}
where the Lagrangian ${\cal L}^{(\rm CT)}_{\rm L\sigma M + SEM}$
contains the contributions of the counter--terms. We define it
following \cite{Ivanov2018b,Ivanov2018c,Ivanov2018d} and
\cite{BWLee1972}--\cite{BWLee1977}
\begin{eqnarray*}
\hspace{-0.15in}&&{\cal L}^{(\rm CT)}_{\rm L\sigma M + SEM} = \big(Z_N
\tilde{Z}^{(N)}_2 Z^{(p)}_2 -
1)\,\bar{\psi}_p\big(i\gamma^{\mu}\partial_{\mu} - m_N\big)\psi_p +
\big(Z_N \tilde{Z}^{(N)}_2 -
1)\,\bar{\psi}_n\big(i\gamma^{\mu}\partial_{\mu} - m_N\big)\psi_n -
Z_N \tilde{Z}^{(N)}_2 Z^{(p)}_2\delta
m^{(r)}_N\nonumber\\ \hspace{-0.15in}&&\times
\,\bar{\psi}^{(r)}_p\,\psi^{(r)}_p - Z_N \tilde{Z}^{(N)}_2\delta
m^{(r)}_N\,\bar{\psi}^{(r)}_n\,\psi^{(r)}_n + \big(Z_M Z^{(\pi)}_2
\tilde{Z}^{(M)}_2 -
1\big)\,\big(\partial_{\mu}\pi^{(r)+}\partial^{\mu}\pi^{(r)-} -
(m^{(r)}_{\pi})^2\pi^{(r)+}\pi^{(r)-}\big) - Z_M Z^{(\pi)}_2
\tilde{Z}^{(M)}_2\nonumber\\ \hspace{-0.15in}&&\times\, \delta
m^{(r)2}_{\pi}\pi^{(r)+}\pi^{(r)-} + \frac{1}{2}\,\big(Z_M
\tilde{Z}^{(M)}_2 - 1\big)
\,\big(\partial_{\mu}\pi^{(r)0}\partial^{\mu}\pi^{(r)0} -
(m^{(r)}_{\pi})^2(\pi^{(r)0})^2\big) - \frac{1}{2}\,Z_M
\tilde{Z}^{(M)}_2 \delta m^{(r)2}_{\pi}
(\pi^{(r)0})^2\nonumber\\ \hspace{-0.15in}&&+ \frac{1}{2}\, \big(Z_M
\tilde{Z}^{(M)}_2 - 1\big)\,\big(\partial_{\mu}
\sigma^{(r)}\partial^{\mu}\sigma^{(r)} - (m^{(r)}_{\sigma})^2
(\sigma^{(r)})^2\big) - \frac{1}{2}\,Z_M \tilde{Z}^{(M)}_2\delta
m^{(r)2}_{\sigma} (\sigma^{(r)})^2 - \big(Z_{MN} \tilde{Z}^{(N)}_2
\sqrt{Z^{(p)}_2 Z^{(\pi)}_2 \tilde{Z}^{(M)}_2} -
1\big)\nonumber\\ \hspace{-0.15in}&&\times \, \sqrt{2}\,g^{(r)}_{\pi
  N}\bar{\psi}^{(r)}_pi\gamma^5 \psi^{(r)}_n\,\pi^{(r)+} - \big(Z_{MN}
\tilde{Z}^{(N)}_2 \sqrt{Z^{(p)}_2 Z^{(\pi)}_2 \tilde{Z}^{(M)}_2} -
1\big) \,\sqrt{2}\,g^{(r)}_{\pi N}\bar{\psi}^{(r)}_ni\gamma^5
\psi_p\,\pi^{(r)-} - \big(Z_{MN} \tilde{Z}^{(N)}_2
Z^{(p)}_2\sqrt{\tilde{Z}^{(M)}_2} -
1\big)\nonumber\\ \hspace{-0.15in}&&\times\, g^{(r)}_{\pi N}\,
\bar{\psi}^{(r)}_pi\gamma^5 \psi^{(r)}_p\,\pi^{(r)0} + \big(Z_{MN}
\tilde{Z}^{(N)}_2 \sqrt{\tilde{Z}^{(M)}_2} - 1\big) g^{(r)}_{\pi
  N}\,\bar{\psi}^{(r)}_ni\gamma^5 \psi^{(r)}_n \pi^{(r)0} -
\big(Z_{MN} \tilde{Z}^{(N)}_2 Z^{(p)}_2\sqrt{\tilde{Z}^{(M)}_2} -
1\big) g^{(r)}_{\pi N} \bar{\psi}^{(r)}_p \psi^{(r)}_p
\sigma^{(r)}\nonumber\\ \hspace{-0.15in}&& - \big(Z_{MN}
\tilde{Z}^{(N)}_2 \sqrt{\tilde{Z}^{(M)}_2} - 1\big)\,g^{(r)}_{\pi N}\,
\bar{\psi}^{(r)}_n \psi^{(r)}_n \sigma^{(r)} - \big(Z_{3M}
(\tilde{Z}^{(M)}_2)^{3/2} -
1\big)\,\gamma^{(r)}\,f^{(r)}_{\pi}\,(\sigma^{(r)})^3 - \big(Z_{3M}
(\tilde{Z}^{(M)}_2)^{3/2} Z^{(\pi)}_2 -
1\big)\nonumber\\ \hspace{-0.15in}&&\times\,
2\,\gamma^{(r)}\,f^{(r)}_{\pi}\,\sigma^{(r)} \pi^{(r)+} \pi^{(r)-} -
\big(Z_{3M}(\tilde{Z}^{(\sigma)}_2)^{3/2} - 1\big)\,
\gamma^{(r)}\,f^{(r)}_{\pi}\,\sigma^{(r)}(\pi^{(r)0})^2 -
\big(Z_{4M}(\tilde{Z}^{(M)}_2)^2 -
1\big)\,\frac{1}{4}\,\gamma^{(r)}\,(\sigma^{(r)})^4
\nonumber\\
\hspace{-0.15in}&& - \big(Z_{4M}(\tilde{Z}^{(M)}_2
Z^{(\pi)}_2)^2 -
1\big)\,\gamma^{(r)}\,\big(\pi^{(r)+}\pi^{(r)-}\big)^2 -
\big(Z_{4M}(\tilde{Z}^{(M)}_2)^2 - 1\big) \,\frac{1}{4}\,
\gamma^{(r)}\, (\pi^{(r)0})^4 - \big(Z_{4M}(\tilde{Z}^{(M)}_2)^2
Z^{(\pi)}_2 -
1\big)\,\gamma^{(r)}\,(\sigma^{(r)})^2\nonumber\\
\hspace{-0.15in}&&\times\,
\pi^{(r)+}\pi^{(r)-} - \big(Z_{4M}(\tilde{Z}^{(M)}_2)^2 -
1\big)\,\frac{1}{2}\,\gamma^{(r)}\,(\sigma^{(r)})^2(\pi^{(r)0})^2 -
\big(Z_{4M} (\tilde{Z}^{(M)}_2)^2 \, Z^{(\pi)}_2 -
1\big)\,\gamma^{(r)}\pi^{(r)+}\pi^{(r)-} (\pi^{(r)0})^2
\nonumber\\
\hspace{-0.15in}&& + \big(Z^{(W)}_3 -
1\big)\,\Big(- \frac{1}{2}\,W^{(r)+}_{\mu\nu}W^{(r)- \mu\nu} +
M^{(r)2}_W W^{(r)+}_{\mu}W^{(r)-\mu}\Big) + Z^{(W)}_3\delta M^{(r)2}_W
W^{(r)+}_{\mu}W^{(r)-\mu} + \big(Z^{(Z)}_3 - 1\big) \Big(- \frac{1}{4}
Z^{(r)}_{\mu\nu} Z^{(r)\mu\nu}\nonumber\\
\hspace{-0.15in}&& +
\frac{1}{2}\, M^{(r)2}_Z Z^{(r)}_{\mu}Z^{(r)\mu}\Big) + Z^{(Z)}_3
\frac{1}{2}\, \delta M^{(r)2}_Z Z^{(r)}_{\mu}Z^{(r)\mu} -
\big(Z^{(\gamma)}_3 -
1\big)\,\frac{1}{4}\,F^{(r)}_{\mu\nu} F^{(r)\mu\nu} -
\frac{Z^{(\gamma)}_3 - 1}{Z_{\xi}}
\frac{1}{2\xi^{(r)}}\,\big(\partial_{\mu} A^{(r)\mu}\big)^2 +
\frac{1}{2}\big(Z^{(H)}_2 - 1
\big)\nonumber\\
\end{eqnarray*}
\begin{eqnarray*}
\hspace{-0.15in}&&\times\,\Big(\partial_{\mu} H^{(r)}
\partial^{\mu} H^{(r)} - M^{(r)2}_H (H^{(r)})^2\Big) -
\frac{1}{2}\,Z^{(H)}_2 \,\delta M^{(r)2}_H (H^{(r)})^2 +
\big(Z^{(e)}_2 \tilde{Z}^{(e)}_2 - 1\big)\,\bar{\psi}^{(r)}_e
\big(i\gamma^{\mu} \partial_{\mu} - m^{(r)}_e\big)\,\psi^{(r)}_e -
Z^{(e)}_2 \tilde{Z}^{(e)}_2
\nonumber\\
\hspace{-0.15in}&&\times\,\delta m^{(r)}_e\,
\bar{\psi}^{(r)}_e \psi^{(r)}_e + \big(\tilde{Z}^{(\ell)}_2 - 1\big)\,
\bar{\psi}^{(r)}_{\nu L} i\gamma^{\mu}\partial_{\mu}\psi^{(r)}_{\nu L}
- \big(\tilde{Z}^{(N)}_1 Z_N \sqrt{Z^{(p)}_2} -
1\big)\,\frac{g^{(r)}}{2\sqrt{2}}\,\bar{\psi}^{(r)}_p\gamma^{\mu}(1 -
\gamma^5)\,\psi^{(r)}_n\,W^{(r)+}_{\mu} - \big(\tilde{Z}^{(M)}_1\,Z_M
\nonumber\\
\hspace{-0.15in}&&\times\,\sqrt{Z^{(\pi)}_2} - 1\big) \,
\frac{g^{(r)}}{2\sqrt{2}}\,i\,\sqrt{2}(\pi^{(r)0} \partial_{\mu}
\pi^{(r)-} - \partial_{\mu}\pi^{(r)0} \pi^{(r)-})\, W^{(r)+}_{\mu} -
\big(\tilde{Z}^{(M)}_1\,Z_M \sqrt{Z^{(\pi)}_2} - 1\big) \,
\frac{g^{(r)}}{2\sqrt{2}}\, \sqrt{2}\,\big(\sigma^{(r)}\,
\partial_{\mu}\pi^{(r)-} -
\partial_{\mu}\sigma^{(r)}\nonumber\\
\hspace{-0.15in}&&\times
\,\pi^{(r)-}\big)\, W^{(r)+}_{\mu} -
(\tilde{Z}^{(M)}_1Z_M\sqrt{Z^{(\pi)}_2/\tilde{Z}^{(M)}_2} -
1) \frac{g^{(r)}}{2\sqrt{2}}\,\sqrt{2}\,f^{(r)}_{\pi}\,\partial_{\mu}\pi^{(r)-}
\,W^{(r)+}_{\mu} - \big(\tilde{Z}^{(N)}_1 Z_N \sqrt{Z^{(p)}_2} -
1\big) \frac{g^{(r)}}{2\sqrt{2}} \bar{\psi}^{(r)}_n\gamma^{\mu}(1 -
\gamma^5)\nonumber\\
\hspace{-0.15in}&&\times\,\psi^{(r)}_p\,W^{(r)-}_{\mu}
- \big(\tilde{Z}^{(M)}_1\,Z_M\,\sqrt{Z^{(\pi)}_2} - 1\big) \,
\frac{g^{(r)}}{2\sqrt{2}}\, i\,\sqrt{2}\,(\pi^{(r)+} \partial_{\mu}
\pi^{(r)0} - \partial_{\mu}\pi^{(r)+} \pi^{(r)0}) \,W^{(r)-}_{\mu} -
\big(\tilde{Z}^{(M)}_1\,Z_M \sqrt{Z^{(\pi)}_2} - 1\big) \,
\frac{g^{(r)}}{2\sqrt{2}}\nonumber\\
\hspace{-0.15in}&&\times\,\sqrt{2}\,
\big(\sigma^{(r)}\, \partial_{\mu}\pi^{(r)+} -
\partial_{\mu}\sigma^{(r)}\,\pi^{(r)+}\big)\,W^{(r)-}_{\mu} -
\big(\tilde{Z}^{(M)}_1Z_M\sqrt{Z^{(\pi)}_2/\tilde{Z}^{(M)}_2} -
1\big)\, \sqrt{2}\,f^{(r)}_{\pi}\,\partial_{\mu}\pi^{(r)+}
\,W^{(r)-}_{\mu} - \big(\tilde{Z}^{(N)}_1 Z_N Z^{(p)}_2 -
1\big) \nonumber\\
\hspace{-0.15in}&& \times\,
\frac{g^{(r)}}{2\cos\theta_W}\,\Big(\frac{1}{2}\,\bar{\psi}^{(r)}_p
\gamma^{\mu}\,\big(1 - 4\sin^2\theta_W -
\gamma^5\big)\,\psi^{(r)}_p\,Z^{(r)}_{\mu} + \big(\tilde{Z}^{(N)}_1
Z_N - 1\big)\, \frac{g^{(r)}}{2\cos\theta_W}\,
\frac{1}{2}\,\bar{\psi}^{(r)}_n \gamma^{\mu} \big(1 -
\gamma^5)\,\psi^{(r)}_n \,Z^{(r)}_{\mu} - \big(\tilde{Z}^{(M)}_1 Z_M\nonumber\\
\hspace{-0.15in}&&\times\, Z^{(\pi)}_2 - 1\big)\,
\frac{g^{(r)}}{2\cos\theta_W}\,(1 - 2\,\sin^2\theta_W)\,
i\,(\pi^{(r)+}\partial_{\mu}\pi^{(r)-} -
\partial_{\mu}\pi^{(r)+}\,\pi^{(r)-})\, Z^{(r)}_{\mu} + \big(\tilde{Z}^{(M)}_1 Z_M -
1\big)\, \frac{g^{(r)}}{2\cos\theta_W}\,
(\sigma^{(r)}\,\partial_{\mu}\pi^{(r)0}\nonumber\\
\hspace{-0.15in}&& -
\partial_{\mu}\sigma^{(r)}\,\pi^{(r)0})\,Z^{(r)}_{\mu} +
\big(\tilde{Z}^{(M)}_1Z_M(\tilde{Z}^{(M)}_2)^{-1/2} -
1\big)\,\frac{g^{(r)}}{2\cos\theta_W}\,
f^{(r)}_{\pi}\,\partial_{\mu}\pi^{(r)0} \,Z^{(r)}_{\mu} -
\big(Z^{(p)}_1 Z_N \tilde{Z}^{(N)}_2 -
1\big)\,e^{(r)}\,\bar{\psi}^{(r)}_p\gamma^{\mu}\,\psi^{(r)}_p\,A^{(r)}_{\mu}
\nonumber\\
\hspace{-0.15in}&& - \big(Z^{(\pi)}_1 Z_M \tilde{Z}^{(M)}_2 - 1\big)
\,e^{(r)}\,i\,(\pi^{(r)-}\partial_{\mu}\pi^{(r)+} -
\partial_{\mu}\pi^{(r)-}\,\pi^{(r)+})\, A^{(r)}_{\mu} -
\big(\tilde{Z}^{(\ell)}_1 \sqrt{Z^{(e)}_2} -
1\big)\,\frac{g^{(r)}}{2\sqrt{2}}\,\bar{\psi}^{(r)}_e\gamma^{\mu}
\big(1 - \gamma^5\big)\,\psi^{(r)}_{\nu L}\,W^{(r)-}_{\mu}\nonumber\\
\hspace{-0.15in}&& - \big(\tilde{Z}^{(\ell)}_1 \sqrt{Z^{(e)}_2} -
1\big)\,\frac{g^{(r)}}{2\sqrt{2}}\,\bar{\psi}^{(r)}_{\nu L}
\gamma^{\mu} \big(1 - \gamma^5\big)\,\psi^{(r)}_e\,W^{(r)+}_{\mu} +
\big(\tilde{Z}^{(\ell)}_1 Z^{(e)}_2\sqrt{Z^{(Z)}_3/Z^{(W)}_3} -
1\big)\,\frac{g^{(r)}}{4\cos\theta_W}\,
\bar{\psi}^{(r)}_e \gamma^{\mu} \big(1 - 4\sin^2\theta_W -
\gamma^5\big)\nonumber\\
\hspace{-0.15in}&&\times\,\psi^{(r)}_e\,Z^{(r)}_{\mu} -
\big(\tilde{Z}^{(\ell)}_1 \sqrt{Z^{(Z)}_3/Z^{(W)}_3} -
1\big)\,\frac{g^{(r)}}{4\cos\theta_W}\,\bar{\psi}^{(r)}_{\nu
  L}\,\gamma^{\mu}\big(1 - \gamma^5\big)\,\psi^{(r)}_{\nu L}\,
Z^{(r)}_{\mu} + \big(Z^{(e)}_1 \tilde{Z}^{(\ell)}_2 - 1\big)\,
e^{(r)}\,\bar{\psi}^{(r)}_e\gamma^{\mu}\,\psi^{(r)}_e \,A^{(r)}_{\mu}\nonumber\\
\hspace{-0.15in} && + \big(Z^{(\pi)}_1 \tilde{Z}^{(M)}_1 Z_M -
1\big)\, \frac{1}{2}\,i\,f^{(r)}_{\pi}\,e^{(r)} g^{(r)}
\,A^{(r)}_{\mu}\,(\pi^{(r)+} W^{(r)-\mu} - \pi^{(r)-} W^{(r)+\mu}) +
\big(Z^{(\pi)}_1 \tilde{Z}^{(M)}_1 Z_M - 1\big)\,\frac{1}{2}\,e^{(r)}
g^{(r)} \,A^{(r)}_{\mu} \nonumber\\
\hspace{-0.15in} &&\times \,\Big(i \,\sigma^{(r)}\,(\pi^{(r)+}
W^{(r)-\mu} - \pi^{(r)-} W^{(r)+\mu}) - \pi^{(r)0}\,(\pi^{(r)+}
W^{(r)-\mu} + \pi^{(r)-} W^{(r)+\mu})\Big)  + \big((\tilde{Z}^{(M)}_1)^2 Z_M
\sqrt{Z^{(\pi)}_2/ Z^{(W)}_3}\nonumber\\
\hspace{-0.15in} &&\times \, (\tilde{Z}^{(M)}_2)^{-3/2} - 1\big)\,
\frac{1}{2}\, g^{(r)2} \cos\theta_W\, \tan^2\theta_W\,
i\,f^{(r)}_{\pi}\,Z^{(r)}_{\mu} (\pi^{(r)+} W^{(r)-\mu} - \pi^{(r)-}
W^{(r)+\mu}) + \big((\tilde{Z}^{(M)}_1)^2 Z_M \sqrt{Z^{(\pi)}_2/
  Z^{(W)}_3}\nonumber\\
\hspace{-0.15in} &&\times \,(\tilde{Z}^{(M)}_2)^{-1} - 1\big)\,
\frac{1}{2}\, g^{(r)2} \cos\theta_W\, \tan^2\theta_W \,Z^{(r)}_{\mu}
\,\big(i\, \sigma^{(r)}\,(\pi^{(r)+} W^{(r)-\mu} - \pi^{(r)-}
W^{(r)+\mu}) - \pi^{(r)0}\,(\pi^{(r)+} W^{(r)-\mu}\nonumber\\
\hspace{-0.15in} && + \pi^{(r)-} W^{(r)+\mu})\big) +
\big((\tilde{Z}^{(M)}_1)^2 Z_M (\tilde{Z}^{(M)}_2)^{-1/2} -
1\big)\,\frac{1}{4}\,g^{(r)2}\,2\,f^{(r)}_{\pi}\sigma^{(r)}
\,W^{(r)+}_{\mu}W^{(r)-\mu} + \big((\tilde{Z}^{(M)}_1)^2 Z_M -
1\big)\,\frac{1}{4}\,g^{(r)2}\nonumber\\
\hspace{-0.15in} &&\times \,\big((\sigma^{(r)})^2 +
(\pi^{(r)0})^2\big)\,W^{(r)+}_{\mu}W^{(r)-\mu} +
\big((\tilde{Z}^{(M)}_1)^2 Z_M Z^{(\pi)}_2 -
1\big)\,\frac{1}{4}\,g^{(r)2}\, 2\,\pi^{(r)+}
\pi^{(r)-}\big)\,W^{(r)+}_{\mu}W^{(r)-\mu}\nonumber\\
\hspace{-0.15in} &&+ \big((\tilde{Z}^{(M)}_1)^2
Z_MZ^{(Z)}_3(Z^{(W)}_3)^{-1}(\tilde{Z}^{(M)}_2)^{-1/2} - 1\big)\,
\frac{1}{8}\,\frac{g^{(r)2}}{\cos^2\theta_W}\,2\,f^{(r)}_{\pi}
\sigma^{(r)}\, Z^{(r)}_{\mu} Z^{(r)\mu} + \big((\tilde{Z}^{(M)}_1)^2
Z_MZ^{(Z)}_3(Z^{(W)}_3)^{-1} - 1\big)\nonumber\\
\hspace{-0.15in} &&\times \,\frac{1}{8}\,
\frac{g^{(r)2}}{\cos^2\theta_W} \,\big((\sigma^{(r)})^2 +
(\pi^{(r)0})^2\big)\,Z^{(r)}_{\mu} Z^{(r)\mu} +
\big((Z^{(\pi)}_1)^2(Z^{(\pi)}_2)^{-1} - 1\big)\, e^{(r)2}\,\pi^{(r)+}
\pi^{(r)-}\,A^{(r)}_{\mu}A^{(r)\mu} +
\big(Z^{(\pi)}_1\tilde{Z}^{(M)}_1\nonumber\\
\hspace{-0.15in} &&\times\, \sqrt{Z^{(Z)}_3/Z^{(W)}_3} - 1\big)\,
\frac{2 e^{(r)}g^{(r)}}{2\cos\theta_W}\, (1 - 2\sin^2
\theta_W)\,\pi^{(r)+} \pi^{(r)-}\, A^{(r)}_{\mu}\, Z^{(r)\mu} +
\big((Z^{(M)}_1)^2 Z_M Z^{(\pi)}_2 Z^{(Z)}_3/Z^{(W)}_3 - 1\big)\,
\frac{g^{(r)2}}{4\cos^2\theta_W} \nonumber\\
\hspace{-0.15in} &&\times\, (1 - 2\sin^2 \theta_W)^2\,\pi^{(r)+}
\pi^{(r)-}\, Z^{(r)}_{\mu}\, Z^{(r)\mu} + \big(Z^{(W)}_1 - 1\big)\,
i\,\frac{1}{2}\,e^{(r)}\,W^{(r)-}_{\mu\nu}\big(W^{(r)+\mu}A^{(r)\nu} -
A^{(r)\mu} W^{(r)+\nu}\big) + \big(Z^{(W)}_1\sqrt{Z^{(Z)}_3}\nonumber\\
\hspace{-0.15in} &&\times\,(Z^{(W)}_3)^{-1/2} - 1\big)\,
i\,\frac{1}{2}\,g^{(r)}\,\cos\theta_W \, W^{(r)-}_{\mu\nu}
\big(W^{(r)+\mu}Z^{(r)\nu} - Z^{(r)\mu} W^{(r)+\nu}\big) +
\big(Z^{(W)}_3 - 1\big)\,
i\,\frac{1}{2}\,e^{(r)}\,W^{(r)+}_{\mu\nu}\big( A^{(r)\mu} W^{(r)-\nu}\nonumber\\
\hspace{-0.15in} && - W^{(r)-\mu}A^{(r)\nu}\big) + \big(Z^{(W)}_1
\sqrt{Z^{(Z)}_3/Z^{(W)}_3} - 1\big)\,
i\,\frac{1}{2}\,g^{(r)}\,\cos\theta_W\,W^{(r)+}_{\mu\nu}\,
\big(Z^{(r)\mu} W^{(r)-\nu} - W^{(r)-\mu}Z^{(r)\nu}\big) +
\big(Z^{(W)}_3 - 1\big)\nonumber\\
\hspace{-0.15in} && \times\,
\frac{1}{2}\,e^{(r)2}\,\big(W^{(r)+}_{\mu}A^{(r)}_{\nu} -
A^{(r)}_{\mu} W^{(r)+}_{\nu}\big)\big( A^{(r)\mu} W^{(r)-\nu} -
W^{(r)-\mu}A^{(r)\nu}\big) + \big((Z^{(W)}_1)^2
Z^{(Z)}_3(Z^{(W)}_3)^{-2} - 1\big)\,
\frac{1}{2}\,g^{(r)2}\,\cos^2\theta_W
\nonumber\\ \hspace{-0.15in}&&\times \,
\big(W^{(r)+}_{\mu}Z^{(r)}_{\nu} - Z^{(r)}_{\mu}
W^{(r)+}_{\nu}\big)\big(Z^{(r)\mu} W^{(r)-\nu} -
W^{(r)-\mu}Z^{(r)\nu}\big)+ \big(Z^{(W)}_1\sqrt{Z^{(Z)}_3/Z^{(W)}_3} -
1\big)\, \frac{1}{2}\,e^{(r)}\, g^{(r)}\,\cos\theta_W
\,\big(W^{(r)+}_{\mu}A^{(r)}_{\nu}\nonumber\\
\hspace{-0.15in}&& -
A^{(r)}_{\mu} W^{(r)+}_{\nu}\big)\, \big( Z^{(r)\mu} W^{(r)-\nu} -
W^{(r)-\mu}Z^{(r)\nu}\big) + \big(Z^{(W)}_1\sqrt{Z^{(Z)}_3/Z^{(W)}_3}
- 1\big)\, \frac{1}{2}\,e^{(r)}\, g^{(r)}\,\cos\theta_W
\,\big(W^{(r)+}_{\mu}Z^{(r)}_{\nu} - Z^{(r)}_{\mu}
W^{(r)+}_{\nu}\big)\nonumber\\
\hspace{-0.15in}&&\times \, \big(
A^{(r)\mu} W^{(r)-\nu} - W^{(r)-\mu}A^{(r)\nu}\big) + \big(Z^{(W)}_3 -
1\big)\, \frac{1}{2}\,i\,e^{(r)}\,F^{(r)}_{\mu\nu}\big( W^{(r)-\mu}
W^{(r)+\nu} - W^{(r)+\mu} W^{(r)-\nu}\big) +
\big(Z^{(W)}_1\sqrt{Z^{(Z)}_3/Z^{(W)}_3}\nonumber\\ \hspace{-0.15in}&&
- 1\big)\, \frac{1}{2}\,i\,g^{(r)}\,\cos\theta_W\,Z^{(r)}_{\mu\nu}\,
\big( W^{(r)-\mu} W^{(r)+\nu} - W^{(r)+\mu} W^{(r)-\nu}\big) +
\big((Z^{(W)}_1)^2\sqrt{Z^{(W)}_3} - 1\big)\,
\frac{1}{4}\,g^{(r)2}\,\big( W^{(r)-}_{\mu} W^{(r)+}_{\nu} -
W^{(r)+}_{\mu}\nonumber\\
\end{eqnarray*}
\begin{eqnarray}\label{eq:45}
\hspace{-0.15in}&&\times \,
W^{(r)-}_{\nu}\big)\,\big( W^{(r)-\mu} W^{(r)+\nu} - W^{(r)+\mu}
W^{(r)-\nu}\big) + \big(Z^{(W)}_3 \sqrt{Z^{(H)}_2}\,Z^{-1}_v -
1\big)\,\frac{M^{(r)2}_W}{v^{(r)}} W^{(r)+}_{\mu} W^{(r)-\mu} H^{(r)}
+ Z^{(W)}_3
\sqrt{Z^{(H)}_2}\,Z^{-1}_v\nonumber\\ \hspace{-0.15in}&&\times
\,\frac{\delta M^{(r)2}_W}{v^{(r)}} W^{(r)+}_{\mu} W^{(r)-\mu} H^{(r)}
+ \big(Z^{(W)}_3 Z^{(H)}_2\,Z^{-2}_v - 1\big)\,
\frac{1}{4}\,\frac{M^{(r)2}_W}{v^{(r)2}}\, W^{(r)+}_{\mu}W^{(r)-\mu}
(H^{(r)})^2 + Z^{(W)}_3 Z^{(H)}_2\,Z^{-2}_v \,
\frac{1}{4}\,\frac{\delta
  M^{(r)2}_W}{v^{(r)2}}\nonumber\\
\hspace{-0.15in}&&\times \,
W^{(r)+}_{\mu}W^{(r)-\mu} (H^{(r)})^2 + \big(Z^{(Z)}_3
\sqrt{Z^{(H)}_2}\,Z^{-1}_v -
1\big)\,\frac{1}{2}\,\frac{M^{(r)2}_Z}{v^{(r)}}\,
Z^{(r)}_{\mu}Z^{(r)\mu}H^{(r)} + Z^{(Z)}_3
\sqrt{Z^{(H)}_2}\,Z^{-1}_v\,\frac{1}{2}\,\frac{\delta
  M^{(r)2}_Z}{v^{(r)}}\, Z^{(r)}_{\mu}Z^{(r)\mu}
H^{(r)}\nonumber\\ \hspace{-0.15in}&& + \big(Z^{(Z)}_3
Z^{(H)}_2\,Z^{-2}_v - 1\big)\,
\frac{1}{8}\,\frac{M^{(r)2}_Z}{v^{(r)2}}\, Z^{(r)}_{\mu}Z^{(r)\mu}
(H^{(r)})^2 + Z^{(Z)}_3 Z^{(H)}_2\,Z^{-2}_v \,
\frac{1}{8}\,\frac{\delta M^{(r)2}_Z}{v^{(r)2}}\,
Z^{(r)}_{\mu}Z^{(r)\mu} (H^{(r)})^2 - \big(Z^{(e)}_2
\tilde{Z}^{(\ell)}_2 \sqrt{Z^{(H)}_2} -
1\big)\nonumber\\ \hspace{-0.15in}&&\times \,
\frac{m^{(r)}_e}{v^{(r)}}\, \bar{\psi}^{(r)}_e\psi^{(r)}_e\,H^{(r)} -
\big((Z^{(H)}_2)^{3/2}Z^{-1}_v - 1\big)\,
\frac{1}{2}\,\frac{M^{(r)2}_H}{v^{(r)}}\,(H^{(r)})^3 -
(Z^{(H)}_2)^{3/2}Z^{-1}_v\, \frac{1}{2}\,\frac{\delta
  M^{(r)2}_H}{v^{(r)}}\,(H^{(r)})^3 - \big((Z^{(H)}_2)^2 Z^{-2}_v -
1\big)\nonumber\\ \hspace{-0.15in}&&\times\,\frac{1}{8}\,\frac{M^{(r)2}_H}{v^{(r)2}}\,(H^{(r)})^4
- (Z^{(H)}_2)^2 Z^{-2}_v \,\frac{1}{8}\,\frac{\delta
  M^{(r)2}_H}{v^{(r)2}}\,(H^{(r)})^4 + \big(Z_M
\sqrt{Z^{(H)}_2}\,Z^{-1}_v - 1\big)\, \frac{m^{(r)2}_{\pi}
  f^{(r)}_{\pi}}{v^{(r)}}\,\sigma^{(r)}\,H^{(r)} + Z_M
\sqrt{Z^{(H)}_2} \,Z^{-1}_v\nonumber\\\hspace{-0.15in}&&\times\,
\frac{\delta m^{(r)2}_{\pi}
  f^{(r)}_{\pi}}{v^{(r)}}\,\sigma^{(r)}\,H^{(r)},
\end{eqnarray}
where $Z_{MN}$, $Z_N$, $Z_M$, $Z^{(a)}_j$ and $\tilde{Z}^{(a')}_{j'}$
are renormalization constants of the field operators and vertices of
strong and electroweak interactions. Then, $Z_v$ is a renormalization
constant of the vacuum expectation value $v^{(r)}$, and $\delta
m^{(r)}_N$, $\delta m^{(r)2}_{\pi}$, $\delta m^{(r)2}_{\sigma}$,
$\delta M^{(r)2}_W$ and so on are the counter--terms of
mass--renormalization. Rescaling the field operators and the coupling
constants
\begin{eqnarray}\label{eq:46}
\hspace{-0.3in}&& \psi^{(0)}_p = \sqrt{Z_N Z^{(p)}_2
  \tilde{Z}^{(N)}_2}\,\psi^{(r)}_p\;,\;\psi^{(0)}_n = \sqrt{Z_N
  \tilde{Z}^{(N)}_2}\,\psi^{(r)}_n\;,\;\pi^{(0)\pm} = \sqrt{Z_M
  Z^{(\pi)}_2 \tilde{Z}^{(M)}_2}\,\pi^{(r)\pm}\;,\;\pi^{(0)0} =
\sqrt{Z_M \tilde{Z}^{(M)}_2}\,\pi^{(r)0},\nonumber\\
\hspace{-0.3in}&&\sigma^{(0)} = \sqrt{Z_M
  \tilde{Z}^{(M)}_2}\,\sigma^{(r)}\;,\; A^{(0)}_{\mu} =
\sqrt{Z^{(\gamma)}_3}\,A^{(r)}_{\mu}\;,\;W^{(0)\pm}_{\mu} =
\sqrt{Z^{(W)}_3}\,W^{(r)\pm}_{\mu}\;,\; Z^{(0)}_{\mu} =
\sqrt{Z^{(Z)}_3}\,Z^{(r)}_{\mu}\;,\; H^{(0)} =
\sqrt{Z^{(H)}_2}\,H^{(r)},\nonumber\\
\hspace{-0.3in}&& \psi^{(0)}_e = \sqrt{Z^{(e)}_2
  \tilde{Z}^{(\ell)}_2}\,\psi^{(r)}_e\;,\;\psi^{(0)}_{\nu L} = \sqrt{
  \tilde{Z}^{(\ell)}_2}\,\psi^{(r)}_{\nu L},\nonumber\\
\hspace{-0.3in}&&g^{(0)}_{\pi N} = Z_{MN}Z^{-1}_N
Z^{-1/2}_M\, g^{(r)}_{\pi N}\;,\; f^{(0)}_{\pi} = 
Z^{1/2}_M\,f^{(r)}_{\pi}\;,\; \gamma^{(0)} =
Z_{3M}Z^{-2}_M\,\gamma^{(r)},\nonumber\\
\hspace{-0.3in}&&e^{(0)} = Z^{(p)}_1 Z^{(p)-1}_2 Z^{(\gamma) -
  1/2}_3\,e^{(r)} = Z^{(\pi)}_1 Z^{(\pi)-1}_2 Z^{(\gamma)-
  1/2}_3\,e^{(r)} = Z^{(e)}_1 Z^{(e)-1}_2 Z^{(\gamma)-1/2}_3\,e^{(r)}
= Z^{(\gamma) - 1/2}_3\,e^{(r)},\nonumber\\
\hspace{-0.3in}&&g^{(0)} = \tilde{Z}^{(N)}_1 \tilde{Z}^{(N)-1}_2
Z^{(W) - 1/2}_3\,g^{(r)} = \tilde{Z}^{(M)}_1 \tilde{Z}^{(M)-1}_2
Z^{(W) - 1/2}_3\,g^{(r)} = \tilde{Z}^{(\ell)}_1 \tilde{Z}^{(\ell)-1}_2
Z^{(W) - 1/2}_3\,g^{(r)} =\nonumber\\
\hspace{-0.3in}&&= Z^{(W)}_1 Z^{(W) - 3/2}_3\,g^{(r)}\;,\; v^{(0)} =
Z_v v^{(r)}\;,\;\xi^{(0)} = Z_{\xi} \xi^{(r)},
\end{eqnarray}
where we have set $Z_{4M} = Z_{3M}$ (see Eq.(\ref{eq:13})), and using
the relations
\begin{eqnarray}\label{eq:47}
\hspace{-0.3in}&& m^{(0)}_N = m^{(r)}_N + \delta m^{(r)}_N\;,\;
m^{(0)2}_{\pi} = m^{(r)2}_{\pi} + \delta
m^{(r)2}_{\pi}\;,\;m^{(0)2}_{\sigma} = m^{(r)2}_{\sigma} + \delta
m^{(r)2}_{\sigma},\nonumber\\
\hspace{-0.3in}&& M^{(0)2}_W = M^{(r)2}_W + \delta M^{(r)2}_W \;,\;
M^{(0)2}_Z = M^{(r)2}_Z + \delta M^{(r)2}_Z\;,\; M^{(0)2}_H =
M^{(r)2}_H + \delta M^{(r)2}_H,\nonumber\\
\hspace{-0.3in}&&m^{(0)}_e = m^{(r)}_e + \delta m^{(r)}_e
\end{eqnarray}
we transcribe the Lagrangian in Eq.(\ref{eq:44}) into the Lagrangian
in Eq.(\ref{eq:43}).

\section{\bf Matrix element of the hadronic $n \to p$ transition in
  the neutron $\beta^-$--decay $n \to p + e^- + \bar{\nu}_e$}
\label{sec:lorentzstr}

The amplitude of the neutron $\beta^-$--decay is defined by
\cite{Ivanov2018,Ivanov2018b}
\begin{eqnarray}\label{eq:48}
M(n \to p e^-\bar{\nu}_e) = \big\langle {\rm out},
\bar{\nu}_e(\vec{k}_{\bar{\nu}},+ \frac{1}{2}), e^-(\vec{k}_e, \sigma_e),
p(\vec{k}_p,\sigma_p) \big|n(\vec{k}_n,
\sigma_n), {\rm in}\big\rangle,
\end{eqnarray}
where $\langle {\rm out}, \chi(\vec{k}_{\chi},\sigma_{\chi})|$ and
$|{\rm in}, n(\vec{k}_n, \sigma_n)\rangle$ are the wave functions of
the free antineutrino, electron and proton ($\chi = \bar{\nu}_e, e^-,
p$) in the final state (i.e. out--state at $t \to + \infty$) and the
free neutron in the initial state (i.e. in--state at $t \to - \infty$)
\cite{Itzykson1980}. Using the relation $\langle {\rm out}, 
\prod_{\chi} \chi(\vec{k}_{\chi},\sigma_{\chi})| = \langle {\rm in},
\prod_{\chi} \chi(\vec{k}_{\chi},\sigma_{\chi})|{\mathbb S}$, where
${\mathbb S}$ is the S--matrix, we rewrite the matrix element
Eq.(\ref{eq:48}) as follows
\begin{eqnarray}\label{eq:49}
M(n \to p e^-\bar{\nu}_e) = \big\langle {\rm in},
\bar{\nu}_e(\vec{k}_{\bar{\nu}},+ \frac{1}{2}), e^-(\vec{k}_e, \sigma_e),
p(\vec{k}_p,\sigma_p) \big|{\mathbb S}\big|
n(\vec{k}_n, \sigma_n), {\rm in}\big\rangle.
\end{eqnarray}
 The corresponding S--matrix is determined by
 \cite{Itzykson1980,Ivanov2018,Ivanov2018b}
\begin{eqnarray}\label{eq:50}
{\mathbb S} = {\rm T}e^{\textstyle i\int d^4x\,{\cal L}_{\rm 
    L\sigma M + SEM}(x)},
\end{eqnarray}
where ${\rm T}$ is a time--ordering operator and ${\cal L}_{\rm
  L\sigma M + SEM}$ is given by Eq.(\ref{eq:44}). Plugging
Eq.(\ref{eq:50}) into Eq.(\ref{eq:49}) we get
\cite{Ivanov2018,Ivanov2018b}
\begin{eqnarray}\label{eq:51}
M(n \to p e^-\bar{\nu}_e) = \big\langle {\rm in},
\bar{\nu}_e(\vec{k}_{\bar{\nu}},+ \frac{1}{2}), e^-(\vec{k}_e,
\sigma_e), p(\vec{k}_p,\sigma_p) \big|{\rm T}e^{\textstyle i\int
  d^4x\,{\cal L}_{\rm L\sigma M + SEM}(x)}\big| n(\vec{k}_n,
\sigma_n), {\rm in} \big\rangle.
\end{eqnarray}
The wave functions of fermions we determine in terms of the operators
of creation (annihilation)
 \begin{eqnarray}\label{eq:52}
|n(\vec{k}_n, \sigma_n), {\rm in}\rangle &=& a^{\dagger}_{n,\rm
  in}(\vec{k}_n, \sigma_n)|0\rangle,\nonumber\\ \big\langle {\rm in},
\bar{\nu}_e(\vec{k}_{\bar{\nu}},+ \frac{1}{2}), e^-(\vec{k}_e, \sigma_e),
p(\vec{k}_p,\sigma_p) &=& \langle 0|b_{\bar{\nu}_e,\rm in}(\vec{k}_{\bar{\nu}},+
\frac{1}{2})a_{e,\rm in} (\vec{k}_e, \sigma_e) a_{p,\rm
  in}(\vec{k}_p,\sigma_p).
\end{eqnarray}
The operators of creation (annihilation) obey standard anticommutation
relations \cite{Itzykson1980,Ivanov2018}.  

\subsection{Neutron beta decay in the tree--approximation 
for strong low--energy and electroweak interactions described by the
Lagrangian Eq.(\ref{eq:44})}

\begin{figure}
\includegraphics[height=0.11\textheight]{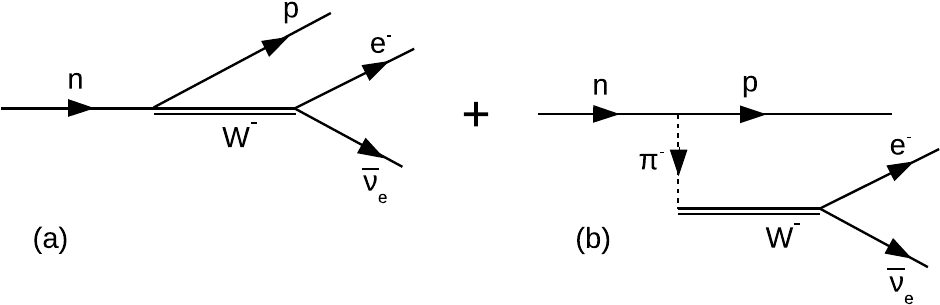}
  \caption{The Feynman diagrams, defining the amplitude of the neutron
    $\beta^-$--decay $n \to p + e^- + \bar{\nu}_e$ in the
    tree--approximation in the quantum field theoretic model of strong
    low--energy and electroweak interactions described by the
    Lagrangian Eq.(\ref{eq:44}).}
\label{fig:fig1a}
\end{figure}

In the tree--approximation for the electroweak $W^-$--boson exchange
and strong low--energy interactions the amplitude of neutron
$\beta^-$-decay $n \to p + e^- + \bar{\nu}_e$ is defined by the
Feynman diagrams in Fig.\,\ref{fig:fig1a}
\begin{eqnarray}\label{eq:53}
\hspace{-0.15in}M(n \to p e^-\bar{\nu}_e) &=& G_V\langle
p(\vec{k}_p,\sigma_p)|J^+_{\mu}(0)|n(\vec{k}_n, \sigma_n)\rangle_{\rm
  Fig.\,\ref{fig:fig1a}}\,\frac{M^2_W}{M^2_W - q^2 - i0}\, \Big(-
\eta^{\mu\nu} + \frac{q^{\mu}q^{\nu}}{M^2_W}\Big)\nonumber\\ &&\times
\, \Big[\bar{u}_e\big(\vec{k}_e, \sigma_e\big) \gamma_{\nu}\big(1 -
  \gamma^5\big)v_{\bar{\nu}}\big(\vec{k}_{\bar{\nu}}, +
  \frac{1}{2}\big)\Big],
\end{eqnarray}
where $G_V = g^2/8 M^2_W$, $J^+_{\mu}(0) = V^+_{\mu}(0) -
A^+_{\mu}(0)$ is the $V - A$ charged hadronic current
\cite{Feynman1958,Nambu1960}, appearing naturally in our model caused
by the electroweak $W^-$--boson exchanges (see Eq.(\ref{eq:43}) and
Eq.(\ref{eq:44})), where the vector and axial--vector current possess
both baryonic and mesonic parts (see Eq.(\ref{eq:9})). Then,
$\bar{u}_e$ and $v_{\nu}$ are Dirac wave functions of the free
electron and electron antineutrino, respectively, a momentum
transferred of the decay is equal to $q = k_p - k_n = - k_e -
k_{\nu}$. Then, since strong low--energy interactions give the
contributions to the matrix element of the charged hadronic current
only, we have denoted $\langle {\rm in}, p(\vec{k}_p,\sigma_p)|{\rm
  T}\big(e^{\textstyle i\int d^4x\,{\cal L}_{\rm L\sigma
    M}(x)}J^+_{\mu}(0)\big)|n(\vec{k}_n, \sigma_n),{\rm in}\rangle =
\langle p(\vec{k}_p,\sigma_p)|J^+_{\mu}(0)|n(\vec{k}_n,
\sigma_n)\rangle$. This matrix element describes the hadronic $n \to
p$ transition in the neutron $\beta^-$--decay
\cite{Leitner2006,Ivanov2018,Ivanov2018b}. The matrix element of the
hadronic $V - A$ current $\langle
p(\vec{k}_p,\sigma_p)|J^+_{\mu}(0)|n(\vec{k}_n, \sigma_n)\rangle$,
calculated in the tree--approximation (see Fig.\,\ref{fig:fig1a}), is
equal to (see also \cite{Ivanov2018,Ivanov2018b})
\begin{eqnarray}\label{eq:54}
\langle p(\vec{k}_p,\sigma_p)|J^+_{\mu}(0)|n(\vec{k}_n,
\sigma_n)\rangle_{\rm Fig. \ref{fig:fig1a}} = \bar{u}_p\big(\vec{k}_p,
\sigma_p\big)\Big(\gamma_{\mu}\big(1 - \gamma^5\big) - \frac{2\,g_{\pi
    N}\,f_{\pi}}{m^2_{\pi} - q^2}\,
q_{\mu}\,\gamma^5\Big)\,u_n\big(\vec{k}_n, \sigma_n\big),
\end{eqnarray}
where $\bar{u}_p$ and $u_n$ are the Dirac wave functions of the free
proton and neutron.  The matrix element of the divergence of the
charged hadronic current $\partial^{\mu} J^+_{\mu}$ is equal to
\begin{eqnarray}\label{eq:55}
\langle p(\vec{k}_p,\sigma_p)|\partial^{\mu}J^+_{\mu}(0)|n(\vec{k}_n,
\sigma_n)\rangle_{\rm Fig. \ref{fig:fig1a}} = i
\bar{u}_p\big(\vec{k}_p, \sigma_p\big)\Big(\big(-2 m_N + 2g_{\pi N}
f_{\pi}\big)\,\gamma^5 - 2\,g_{\pi N}\,f_{\pi}\frac{m^2_{\pi}
}{m^2_{\pi} - q^2}\, \gamma^5\Big)\,u_n\big(\vec{k}_n,
\sigma_n\big),
\end{eqnarray}
Because of the Goldberger--Treiman (GT) relation $g_{\pi N} =
m_N/f_{\pi}$ \cite{Goldberger1958} (see also
\cite{GellMann1960,Bernstein1960,Nambu1960,Strubbe1972}), which
appears naturally in the L$\sigma$M (see Eq.(\ref{eq:8})) at $b =
f_{\pi}$ with the axial coupling constant $g_A$ equal to $g_A = 1$, we
get
\begin{eqnarray}\label{eq:56}
\langle p(\vec{k}_p,\sigma_p)|\partial^{\mu}J^+_{\mu}(0)|n(\vec{k}_n,
\sigma_n)\rangle_{\rm Fig. \ref{fig:fig1a}} = - 2\,g_{\pi N}\,f_{\pi}
\frac{m^2_{\pi} }{m^2_{\pi} - q^2}\, \bar{u}_p\big(\vec{k}_p,
\sigma_p\big)\,i\, \gamma^5\,u_n\big(\vec{k}_n, \sigma_n\big).
\end{eqnarray}
Because of conservation of the charged hadronic vector current
$\partial^{\mu}V^+_{\mu} = 0$ \cite{Feynman1958, Ivanov2018b,
  Ivanov2018, Leitner2006} leading to 
\begin{eqnarray}\label{eq:57}
\langle
p(\vec{k}_p,\sigma_p)|\partial^{\mu}V^+_{\mu}(0)|n(\vec{k}_n,
\sigma_n)\rangle_{\rm Fig. \ref{fig:fig1a}} = i q^{\mu} \langle
p(\vec{k}_p,\sigma_p)|V^+_{\mu}(0)|n(\vec{k}_n, \sigma_n)\rangle_{\rm
  Fig. \ref{fig:fig1a}} = 0,
\end{eqnarray}
the right--hand--side (r.h.s.) of
Eq.(\ref{eq:56}) is fully defined by the divergence of the charged
hadronic axial--vector current
\begin{eqnarray}\label{eq:58}
\langle p(\vec{k}_p,\sigma_p)|\partial^{\mu}A^+_{\mu}(0)|n(\vec{k}_n,
\sigma_n)\rangle_{\rm Fig. \ref{fig:fig1a}} = 2\,g_{\pi N}\,f_{\pi}
\frac{m^2_{\pi} }{m^2_{\pi} - q^2}\, \bar{u}_p\big(\vec{k}_p,
\sigma_p\big)\,i\, \gamma^5\,u_n\big(\vec{k}_n, \sigma_n\big).
\end{eqnarray}
Such a matrix element is caused by the partial conservation of the
axial--vector hadronic current (PCAC) $\partial^{\mu}A^+_{\mu} = -
\sqrt{2}\,m^2_{\pi} f_{\pi}\,\pi^+$ \cite{GellMann1960,Adler1968}.
Plugging the GT--relation $g_{\pi N} = m_N/f_{\pi}$ into
Eq.(\ref{eq:37}) we arrive at the matrix element of the charged $V -
A$ hadronic current, calculated in the tree--approximation in the
L$\sigma$M + SEM (see also \cite{Ivanov2018})
\begin{eqnarray}\label{eq:59}
\langle p(\vec{k}_p,\sigma_p)|J^+_{\mu}(0)|n(\vec{k}_n,
\sigma_n)\rangle_{\rm Fig.\,\ref{fig:fig1}} = \bar{u}_p\big(\vec{k}_p,
  \sigma_p\big)\Big(\gamma_{\mu}\big(1 - \gamma^5\big) - \frac{2\,m_N
  }{m^2_{\pi} - q^2}\, q_{\mu}\, \gamma^5\Big)\,u_n\big(\vec{k}_n,
  \sigma_n\big).
\end{eqnarray}
The matrix element of the charged hadronic current Eq.(\ref{eq:59})
has the standard Lorentz structure with the vector, axial--vector and
pseudoscalar form factors equal to unity
\cite{Nambu1960,Marshak1969,Leitner2006} (see also \cite{Ivanov2018}).
The amplitude of the neutron $\beta^-$--decay in the
tree--approximation is equal to
\begin{eqnarray}\label{eq:60}
\hspace{-0.15in}M(n \to p e^-\bar{\nu}_e) &=&
G_V\,\bar{u}_p\big(\vec{k}_p, \sigma_p\big)\Big(\gamma_{\mu}\big(1 -
\gamma^5\big) - \frac{2\,m_N }{m^2_{\pi} - q^2}\, q_{\mu}
\gamma^5\Big)\,u_n\big(\vec{k}_n, \sigma_n\big)\,\frac{M^2_W}{M^2_W -
  q^2 - i0}\,\Big(- \eta^{\mu\nu} +
\frac{q^{\mu}q^{\nu}}{M^2_W}\Big)\nonumber\\ \hspace{-0.15in}&&\times
\, \Big[\bar{u}_e\big(\vec{k}_e, \sigma_e\big) \gamma_{\nu}\big(1 -
  \gamma^5\big)v_{\bar{\nu}}\big(\vec{k}_{\bar{\nu}}, + \frac{1}{2}\big)\Big],
\end{eqnarray}
As a consequence of the PCAC the longitudinal part of the electroweak
$W^-$--boson propagator, proportional to $q^{\mu}q^{\nu}/M^2_W$, does
not vanish. This violates gauge invariance, as we have pointed out
above. The contribution of such a violation of gauge invariance to the
amplitude of the neutron $\beta^-$--decay is of order $O(2m_N
m_e/M^2_W) \sim 1.5\times 10^{-7}$.  This is two orders of magnitude
smaller the corrections of order $O(\alpha E_e/ m_N) \sim 10^{-5}$,
which we are searching for. In the chiral limit $m_{\pi} \to 0$ the
r.h.s. of Eq.(\ref{eq:58}) and, correspondingly, Eq.(\ref{eq:56})
vanish that leads to local conservation of the charged axial--vector
hadronic current $\partial^{\mu}A^+_{\mu} = 0$, providing gauge
invariance of the amplitude of the neutron $\beta^-$--decay,
i.e. independence of the longitudinal part of the electroweak
$W^-$--boson propagator.

\subsection{Neutron beta decay in the tree--approximation for 
electroweak interactions and to one--hadron--loop approximation for
strong low--energy interactions described by the Lagrangian
Eq.(\ref{eq:44})}

\begin{figure}
\centering \includegraphics[height=0.20\textheight]{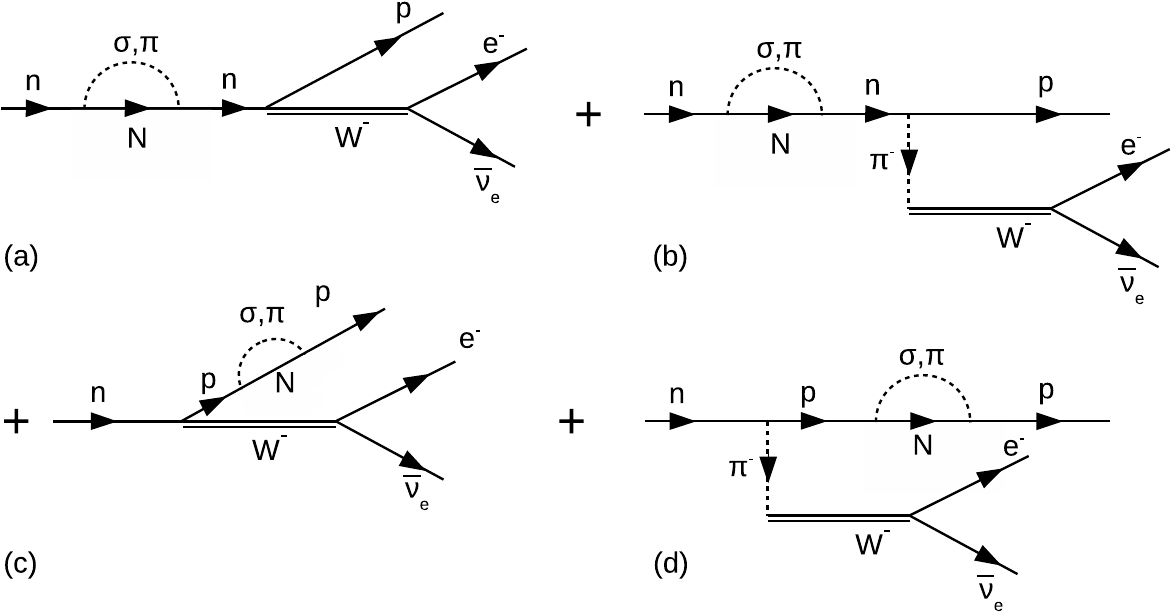}
  \caption{The Feynman diagrams, describing the contributions to the
    amplitude of the neutron $\beta^-$--decay of the self--energy
    corrections to the neutron and proton states in the
    one--hadron--loop approximation in the L$\sigma$M $\&$ SEM
    described by the Lagrangian Eq.(\ref{eq:44}).}
\label{fig:fig2a}
\end{figure}

\begin{figure}
\centering\includegraphics[height=0.12\textheight]{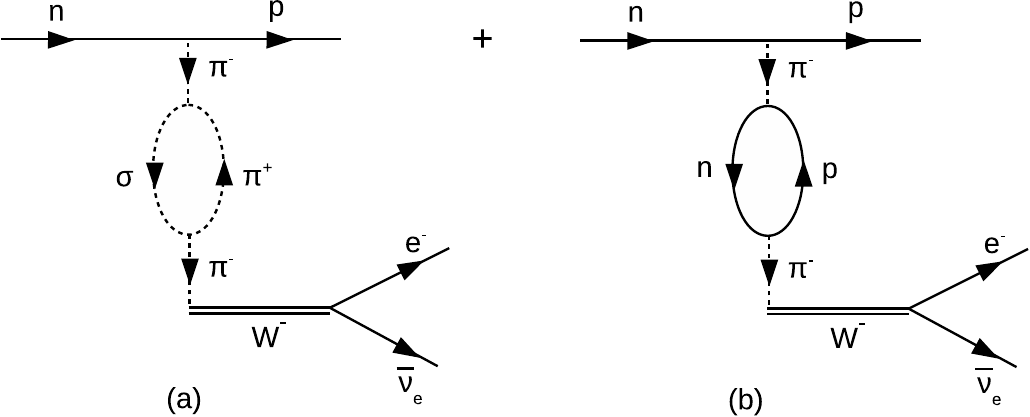}
  \caption{The Feynman diagrams, describing self--energy corrections
    to the $\pi^-$--meson state in the one--hadron--loop approximation
    in the L$\sigma$M $\&$ SEM described by the Lagrangian
    Eq.(\ref{eq:44}).}
\label{fig:fig3a}
\end{figure}

\begin{figure}
\centering \includegraphics[height=0.29\textheight]{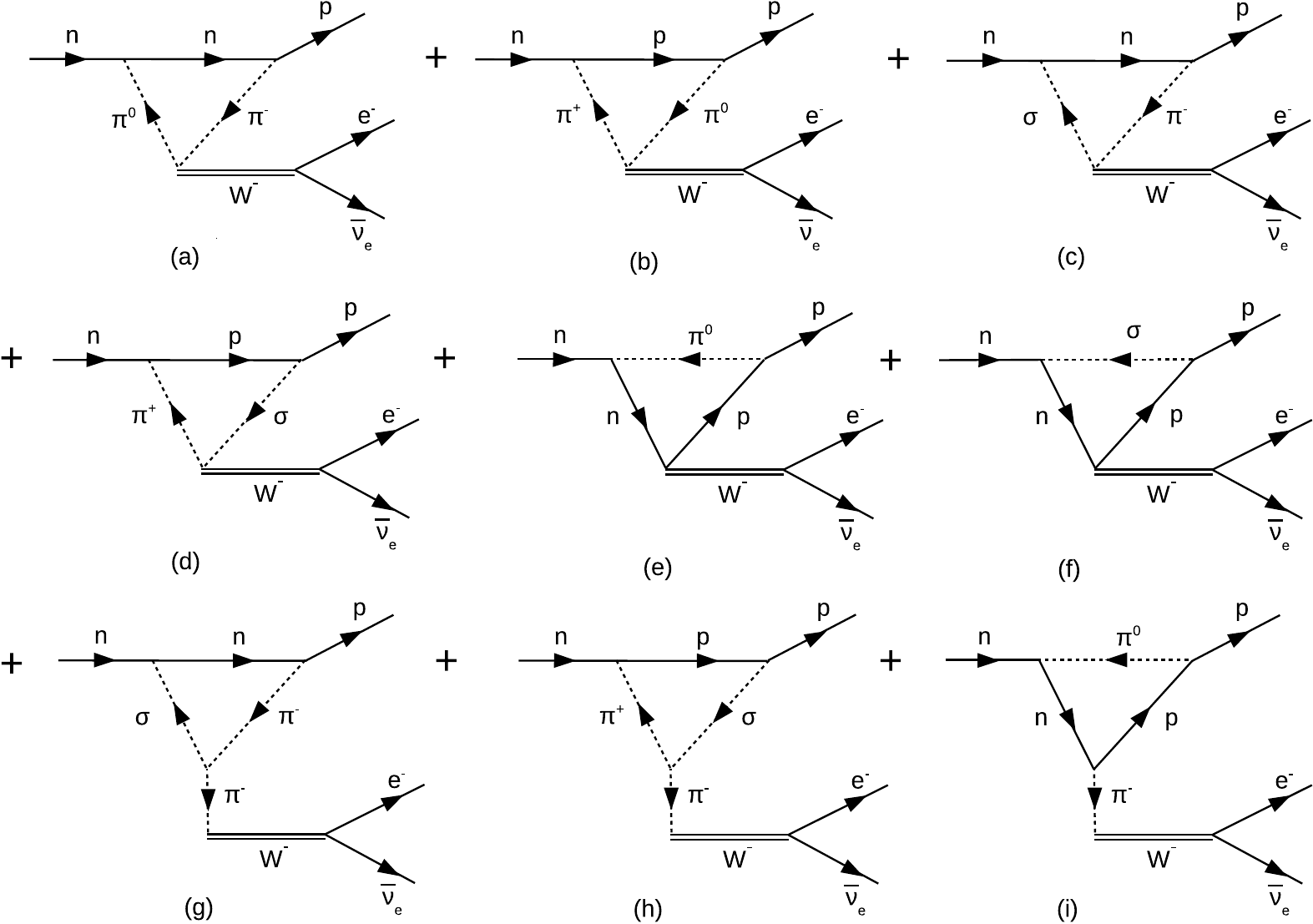}
\includegraphics[height=0.11\textheight]{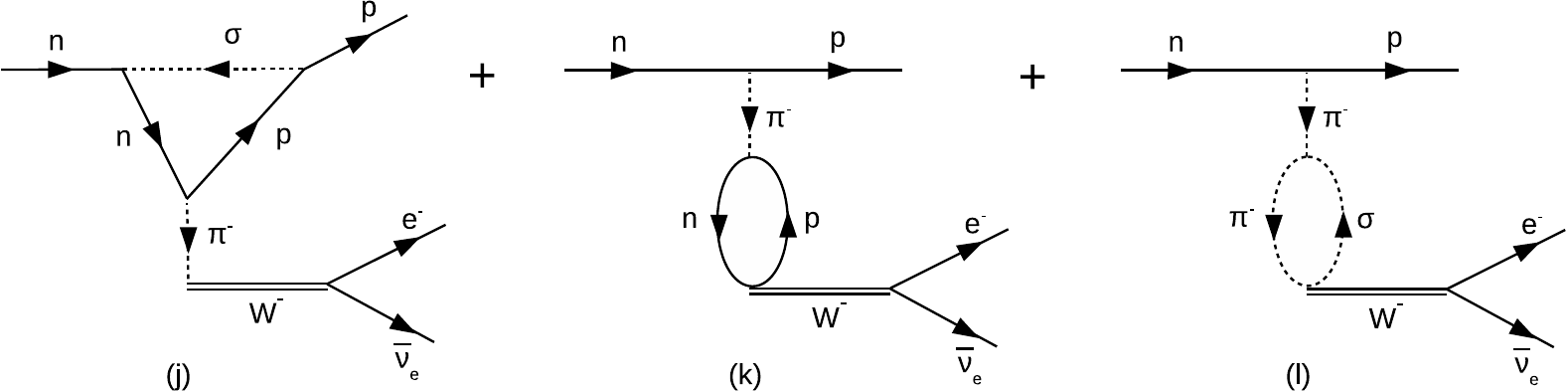}
  \caption{The Feynman diagrams, describing the contributions of the
    hadronic structure of the neutron and proton, and the
    $\pi^-$--meson to the amplitude of the neutron $\beta^-$--decay in
    the one--hadron--loop approximation in the L$\sigma$M $\&$ SEM
    described by the Lagrangian Eq.(\ref{eq:44}).}
\label{fig:fig4a}
\end{figure}

The amplitude of the neutron $\beta^-$--decay in the
tree--approximation for the electroweak $W^-$--boson exchange and to
one--hadron--loop approximation can be taken in the following form
\cite{Ivanov2018b}
\begin{eqnarray}\label{eq:61}
\hspace{-0.15in}M(n \to p e^-\bar{\nu}_e) &=& G_V\langle
p(\vec{k}_p,\sigma_p)|J^+_{\mu}(0)|n(\vec{k}_n, \sigma_n)\rangle_{\rm
  Fig.\,\ref{fig:fig1a} + \ldots +
  Fig.\,\ref{fig:fig4a}} \, \frac{M^2_W}{M^2_W - q^2 - i0}\,\Big(-
\eta^{\mu\nu} + \frac{q^{\mu}q^{\nu}}{M^2_W}\Big)\nonumber\\ &&\times
\, \Big[\bar{u}_e\big(\vec{k}_e, \sigma_e\big) \gamma_{\nu}\big(1 -
  \gamma^5\big)v_{\bar{\nu}}\big(\vec{k}_{\bar{\nu}}, + \frac{1}{2}\big)\Big],
\end{eqnarray}
For the calculation of the one--hadron--loop corrections we shall use
the normal ordered form of the Lagrangians Eq.(\ref{eq:44}) and
Eq.(\ref{eq:45}), respectively \cite{Lee1969a}. This allows to avoid
the tadpole--contributions. Using the normal ordered form of the
Lagrangians Eq.(\ref{eq:44}) and Eq.(\ref{eq:45}) the Feynman
diagrams, defining the one--hadron--loop contributions to the
amplitude of the neutron $\beta^-$--decay, are shown in
Fig.\ref{fig:fig2a}, Fig.\ref{fig:fig3a} and Fig.\ref{fig:fig4a},
respectively.  The Feynman diagrams in Fig.\,\ref{fig:fig2a} and
Fig.\,\ref{fig:fig3a} define the contributions of the self--energy
corrections, caused by strong low--energy interactions to the neutron
and proton, and $\pi^-$--states, respectively. It is obvious that
after normalization the contributions of these diagrams to matrix
element of the hadronic $n \to p$ transition vanish
\cite{Ivanov2017b,Ivanov2018b}. The non--trivial structure of the
matrix element of the hadronic $n \to p$ transition is caused by the
contributions of the Feynman diagrams in Fig.\,\ref{fig:fig4a}
\cite{Ivanov2018b}.

The matrix element of the hadronic $n \to p$ transition, calculated in
the one--hadron--loop approximation, is equal to (see Appendix A in
section \ref{sec:appendix})
\begin{eqnarray}\label{eq:62}
\hspace{-0.15in}&&\langle p(\vec{k}_p,\sigma_p) |J^+_{\mu}(0)|
n(\vec{k}_n, \sigma_n)\rangle_{\rm Fig.\,\ref{fig:fig1a} + \ldots +
  Fig.\,\ref{fig:fig4a}} = \bar{u}_p\big(\vec{k}_p,
\sigma_p\big)\,\Big\{\Big[1 + \big(\tilde{Z}^{(N)}_1 - 1\big) +
  \big(Z_N - 1\big) + \frac{g^2_{\pi N}}{8\pi^2} \Big({\ell
    n}\frac{\Lambda^2}{m^2_N} - \frac{1}{4}\,{\ell
    n}\frac{m^2_{\sigma}}{m^2_N}\Big)\Big]\gamma_{\mu}\nonumber\\
\hspace{-0.15in}&& - \Big[1 + \big(\tilde{Z}^{(N)}_1 - 1\big) +
  \big(Z_N - 1\big) + \frac{g^2_{\pi N}}{8\pi^2} \Big( \frac{5}{4}
      {\ell n}\frac{m^2_{\sigma}}{m^2_N}- {\ell
        n}\frac{\Lambda^2}{m^2_N} \Big)\Big] \gamma_{\mu} \gamma^5 +
\frac{5 g^2_{\pi N}}{16\pi^2} \frac{i \sigma_{\mu\nu}q^{\nu}}{2 m_N} -
\frac{2 m_N q_{\mu}}{m^2_{\pi} - q^2 - i0}\,\gamma^5\, \Big[1 +
  \big(Z_{MN} - 1\big)\nonumber\\
\hspace{-0.15in}&& + \big(\tilde{Z}^{(N)}_2 - 1\big)+
\big(\tilde{Z}^{(M)}_1 - 1\big) + (Z_M - 1) + \frac{g^2_{\pi
    N}}{8\pi^2}\Big( {\ell n}\frac{m^2_{\sigma}}{m^2_N} + {\ell
  n}\frac{\Lambda^2}{m^2_N}\Big)\Big] \Big\}\,u_n\big(\vec{k}_n,
\sigma_n\big).
\end{eqnarray}
Using Eq.(\ref{eq:A.6}) we arrive at the matrix element of the
hadronic $n \to p$ transition, calculated to the one--hadron--loop
approximation. We get
\begin{eqnarray}\label{eq:63}
\hspace{-0.15in}&&\langle
p(\vec{k}_p,\sigma_p)|J^+_{\mu}(0)|n(\vec{k}_n, \sigma_n)\rangle_{\rm
  Fig.\,\ref{fig:fig1a} + \ldots + Fig.\,\ref{fig:fig4a}} =
\bar{u}_p\big(\vec{k}_p, \sigma_p\big)\,\Big\{\Big[1 +
  \big(\tilde{Z}^{(N)}_1 - 1\big) - \big(\tilde{Z}^{(N)}_2 -
  1\big)\Big]\gamma_{\mu} - \Big[1 + \big(\tilde{Z}^{(N)}_1 - 1\big)
  \nonumber\\
\hspace{-0.15in}&& - \big(\tilde{Z}^{(N)}_2 - 1\big) + \frac{g^2_{\pi
    N}}{8\pi^2} \Big( \frac{3}{2}\, {\ell
  n}\frac{m^2_{\sigma}}{m^2_N}- 2\,{\ell n}\frac{\Lambda^2}{m^2_N}
\Big)\Big]\, \gamma_{\mu} \gamma^5 + \frac{5 g^2_{\pi N}}{16\pi^2}
\frac{i \sigma_{\mu\nu}q^{\nu}}{2 m_N} - \frac{2 m_N
  q_{\mu}}{m^2_{\pi} - q^2 - i0}\,\gamma^5\, \Big[1 + \Big(\big(Z_{MN}
  - 1\big) \nonumber\\
\hspace{-0.15in}&& - \big(Z_N - 1\big) - \frac{Z_M - 1}{2}\Big) +
\big(\tilde{Z}^{(M)}_1 - 1\big) - \frac{3}{2}\,\big(\tilde{Z}^{(M)}_2
- 1\big) - \frac{g^2_{\pi N}}{8\pi^2}\Big(3 \,{\ell
  n}\frac{\Lambda^2}{m^2_N} - \frac{5}{4}\, {\ell
  n}\frac{m^2_{\sigma}}{m^2_N}\Big)\Big] \Big\}\,u_n\big(\vec{k}_n,
\sigma_n\big).
\end{eqnarray}
Because of gauge invariance and Ward identities $\tilde{Z}^{(N)}_2 =
\tilde{Z}^{(N)}_1$ and $\tilde{Z}^{(M)}_2 = \tilde{Z}^{(M)}_1$ we
transcribe the r.h.s. of Eq.(\ref{eq:63}) into the form
\begin{eqnarray}\label{eq:64}
\hspace{-0.15in}&&\langle
p(\vec{k}_p,\sigma_p)|J^+_{\mu}(0)|n(\vec{k}_n, \sigma_n)\rangle_{\rm
  Fig.\,\ref{fig:fig1a} + \ldots + Fig.\,\ref{fig:fig4a}} =
\bar{u}_p\big(\vec{k}_p, \sigma_p\big)\,\Big\{\gamma_{\mu} - \Big[1 +
  \frac{g^2_{\pi N}}{8\pi^2} \Big( \frac{3}{2}\, {\ell
    n}\frac{m^2_{\sigma}}{m^2_N}- 2\,{\ell n}\frac{\Lambda^2}{m^2_N}
  \Big)\Big]\, \gamma_{\mu} \gamma^5 \nonumber\\
\hspace{-0.15in}&& + \frac{5 g^2_{\pi N}}{16\pi^2}
\frac{i \sigma_{\mu\nu}q^{\nu}}{2 m_N} - \frac{2 m_N q_{\mu}}{m^2_{\pi} - q^2 -
  i0}\,\gamma^5 \Big[1 + \Big(\big(Z_{MN} - 1\big) - \big(Z_N - 1\big)
  - \frac{Z_M - 1}{2}\Big) - \frac{1}{2}\,\big(\tilde{Z}^{(M)}_2 -
  1\big) - \frac{g^2_{\pi N}}{8\pi^2}\nonumber\\
\hspace{-0.15in}&&\times \Big(3 \,{\ell n}\frac{\Lambda^2}{m^2_N} -
\frac{5}{4}\, {\ell n}\frac{m^2_{\sigma}}{m^2_N}\Big)\Big] \Big\}
u_n\big(\vec{k}_n, \sigma_n\big).
\end{eqnarray}
Since the counter--term $Z_{MN}Z^{-1}_NZ^{-1/2}_M$ renormalizes the
pion--nucleon coupling constant $g_{\pi N}$, we set
\begin{eqnarray}\label{eq:65}
\hspace{-0.15in}\big(Z_{MN} - 1\big) - \big(Z_N - 1\big)
  - \frac{Z_M - 1}{2} = g_A - 1,
\end{eqnarray}
where $g_A \neq 1$ is the axial coupling constant, defining a finite
non--trivial renormalization of the pion--nucleon coupling constant
$g_{\pi N}$ \cite{DeAlfaro1973}. Setting then the relation
\begin{eqnarray}\label{eq:66}
\hspace{-0.15in}\tilde{Z}^{(M)}_2 - 1 = - \frac{g^2_{\pi N}}{8\pi^2}
\Big(3 \,{\ell n}\frac{\Lambda^2}{m^2_N} - \frac{5}{4}\, {\ell
  n}\frac{m^2_{\sigma}}{m^2_N}\Big)
\end{eqnarray}
we arrive at the following matrix element of the hadronic $n \to p$
transition
\begin{eqnarray}\label{eq:67}
\hspace{-0.15in}&&\langle
p(\vec{k}_p,\sigma_p)|J^+_{\mu}(0)|n(\vec{k}_n, \sigma_n)\rangle_{\rm
  Fig.\,\ref{fig:fig1a} + \ldots + Fig.\,\ref{fig:fig4a}} =\nonumber\\
\hspace{-0.15in}&&= \bar{u}_p\big(\vec{k}_p,
\sigma_p\big)\,\Big\{\gamma_{\mu} - \Big[1 + \frac{g^2_{\pi
      N}}{8\pi^2} \Big( \frac{3}{2}\, {\ell
    n}\frac{m^2_{\sigma}}{m^2_N}- 2\,{\ell n}\frac{\Lambda^2}{m^2_N}
  \Big)\Big]\, \gamma_{\mu} \gamma^5 + \frac{5 g^2_{\pi N}}{16\pi^2}
\frac{i \sigma_{\mu\nu}q^{\nu}}{2 m_N} - \frac{2 m_N g_A
  q_{\mu}}{m^2_{\pi} - q^2 - i0}\,\gamma^5 \Big\} u_n\big(\vec{k}_n,
\sigma_n\big).
\end{eqnarray}
In the chiral limit $m_{\pi} \to 0$ the matrix element
Eq.(\ref{eq:67}) should obey the requirement of conservation of the
axial--vector hadronic current \cite{Nambu1960}, i.e.
\begin{eqnarray}\label{eq:68}
\hspace{-0.15in}q^{\mu}\lim_{m_{\pi} \to 0}\langle
p(\vec{k}_p,\sigma_p)|J^+_{\mu}(0)|n(\vec{k}_n, \sigma_n)\rangle_{\rm
  Fig.\,\ref{fig:fig1a} + \ldots + Fig.\,\ref{fig:fig4a}} = 0.
\end{eqnarray}
This allows to impose the following relation
\begin{eqnarray}\label{eq:69}
\hspace{-0.15in}g_A = 1 + \frac{g^2_{\pi N}}{8\pi^2} \Big(
\frac{3}{2}\, {\ell n}\frac{m^2_{\sigma}}{m^2_N}- 2\,{\ell
  n}\frac{\Lambda^2}{m^2_N} \Big).
\end{eqnarray}
The axial coupling constant $g_A$ defines a finite renormalization of
the axial--vector hadronic current \cite{DeAlfaro1973}. This means
that the r.h.s. of Eq.(\ref{eq:69}) should be finite even in the limit
$\Lambda \to \infty$ and $m_{\sigma} \to \infty$. This can be reached
if $m^2_{\sigma} = (\Lambda^2/m^2_N)^{4/3}M^2$, where $M$ is a finite
scale parameter. That fact that the mass of the $\sigma$--meson tends
to infinity faster than the ultra--violet cut--off does not contradict
our analysis of the equivalence of the L$\sigma$M to the chiral
quantum field theories with non--linear realizations of chiral $SU(2)
\times SU(2)$ symmetry (see section \ref{sec:nlsigma}). As a result,
the matrix element of the hadronic $n \to p$ transition takes the
standard form \cite{Leitner2006}
  \begin{eqnarray}\label{eq:70}
\hspace{-0.15in}&&\langle
p(\vec{k}_p,\sigma_p)|J^+_{\mu}(0)|n(\vec{k}_n, \sigma_n)\rangle_{\rm
  Fig.\,\ref{fig:fig1a} + \ldots + Fig.\,\ref{fig:fig4a}} =
\bar{u}_p\big(\vec{k}_p, \sigma_p\big)\,\Big\{\gamma_{\mu}\big(1 - g_A
\gamma^5\big) + \frac{\kappa}{2 m_N}\,i
  \sigma_{\mu\nu}q^{\nu} - \frac{2 m_N g_Aq_{\mu}}{m^2_{\pi} -
  q^2 - i0}\,\gamma^5 \Big\}\,u_n\big(\vec{k}_n,
\sigma_n\big).\nonumber\\
\hspace{-0.15in}&&
\end{eqnarray}
where $\kappa = 5g^2_{\pi N}/16\pi^2$ is the isovector anomalous n
magnetic moment of the nucleon defining of the intensity of the
so--cald weak magnetism \cite{Marshak1969}. The experimental value of
the isovector anomalous magnetic moment of the nucleon is equal to
$\kappa = \kappa_p - \kappa_n = 3.70589$ with $\kappa_p = 1.7928473$
and $\kappa_n = - 1.9130427$, where $\kappa_p$ and $\kappa_n$ are
anomalous magnetic moments of the proton and neutron, respectively
\cite{PDG2018}. Setting $\kappa = 3.70589$ one may estimate the value
of the pion--nucleon coupling constant $g_{\pi N} =
\sqrt{\kappa\,16\pi^2/5} = 10.82$. This defines the leptonic decay (or
the PCAC) constant of pion $f_{\pi} = 86.8\,{\rm MeV}$ at $m_N = (m_p
+ m_n)/2 = 939\,{\rm MeV}$ \cite{PDG2018}, which agrees well with the
definition of a {\it bare} leptonic decay constant of pion
\cite{Scherer2002}.  In our analysis a {\it bare} leptonic decay
constant of pion $f_{\pi} = 86.8\,{\rm MeV}$ deviates from the
observable value of the pion--leptonic constant is equal to $f^{(\rm
  obs.)}_{\pi} = 92.4\,{\rm MeV}$ \cite{PDG2018} by about $6\,\%$.

Thus, we have shown that the matrix element of the hadronic $n \to p$
transition, calculated to one--hadron--loop approximation in the
quantum field theoretic model of strong low--energy and electroweak
interactions described by the Lagrangian Eq.(\ref{eq:44}), possesses a
standard Lorentz structure, where contributions of strong low--energy
interactions are defined by the axial coupling constant $g_A \neq 1$,
the isovector anomalous magnetic moment of the nucleon $\kappa$ and
the one--pion--pole exchange. In the chiral limit the matrix element
of the hadronic $ n\to p$ transition provides independence of the
amplitude of the neutron $\beta^-$--decay of the longitudinal part of
the electroweak $W^-$--boson propagator. This agrees well with a
requirement of conservation of the axial--vector hadronic current in
the chiral limit \cite{Nambu1960}.

Using the experimental value of the axial coupling constant
$g^{(\exp)}_A = 1.27641(45)_{\rm stat.}(33)_{\rm syst.}$, measured
recently by the spectrometer PERKEO III \cite{Abele2018}, we estimate
the value of the scale parameter $M \simeq 1\,{\rm GeV}$, agreeing
well with a scale $\Lambda_{\chi} \sim 1\,{\rm GeV}$ of spontaneous
breakdown of chiral symmetry \cite{Scherer2002}.

\begin{figure}
\centering \includegraphics[height=0.20\textheight]{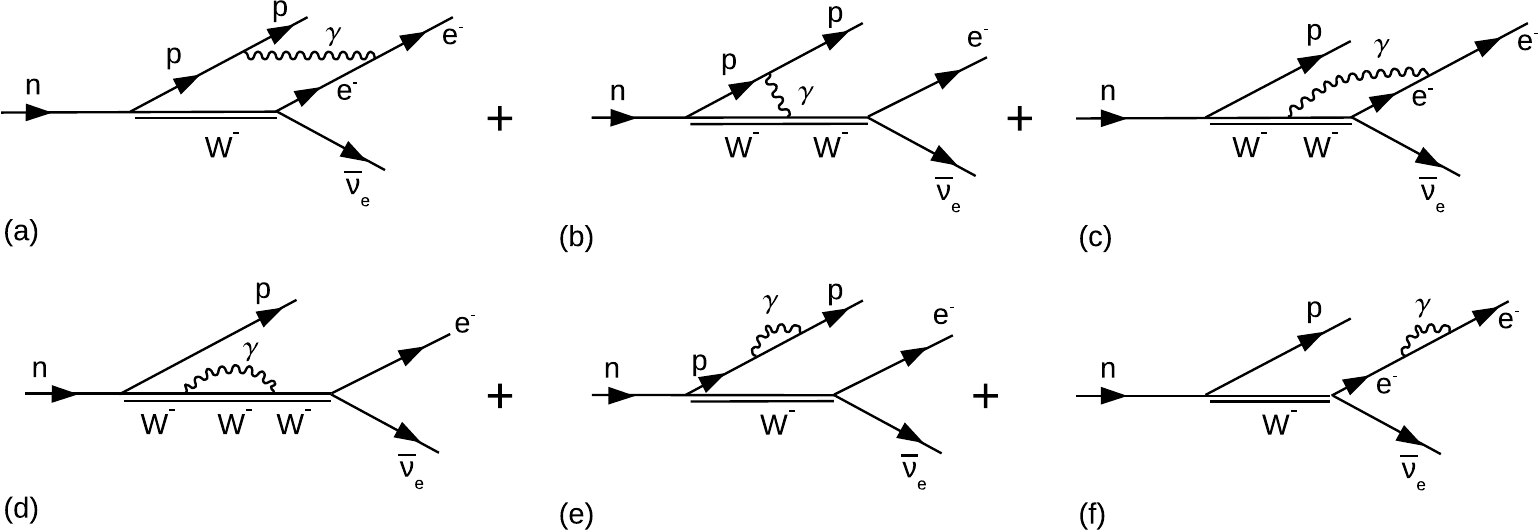}
  \caption{The Feynman diagrams, describing the one--photon--loop
    radiative corrections to the part of the amplitude of the neutron
    $\beta^-$--decay, described by the Feynman diagram in
    Fig.\,\ref{fig:fig1a}a.}
\label{fig:fig6a}
\end{figure}

\section{\bf Radiative one--loop electromagnetic corrections to the
 neutron $\beta^-$--decay $n \to p + e^- + \bar{\nu}_e$ }
\label{sec:oneewloop}

In this section we proceed to the calculation of the radiative
corrections of order $O(\alpha/\pi)$ and its next--to--leading order
corrections of order $O(\alpha E_e/m_N)$ to the neutron
$\beta^-$--decay, caused by one--virtual--photon exchanges. For this
aim we start with the analysis of the radiative electromagnetic
corrections to the amplitude of the neutron $\beta^-$--decay taken in
the tree--approximation for strong low--energy interactions at $g_A =
1$ and described by the Feynman diagrams in Fig.\,\ref{fig:fig6a}. We
show that the set of Feynman diagrams in Fig.\,\ref{fig:fig6a} is gauge
invariant, i.e. independent of a gauge parameter $\xi$ of the photon
propagator. Then, we show that gauge properties of the Feynman
diagrams in Fig.\,\ref{fig:fig6a} are not changed even for $g_A \neq
1$ and calculate these diagrams setting $g_A \neq 1$. This allows us
to take contributions of strong low--energy interactions at Sirlin's
confidence level \cite{Sirlin1967, Sirlin1978}.

\subsection{\bf Amplitude of the neutron $\beta^-$--decay 
 in the tree--approximation for strong low--energy interactions and to
 one--loop approximation for electromagnetic interactions described by
 the Lagrangian Eq.(\ref{eq:44})}

The amplitude of the neutron $\beta^-$--decay, calculated in the
tree--approximation for strong low--energy and electroweak
interactions is described by the Feynman diagrams in
Fig.\,\ref{fig:fig1a}. The radiative corrections to this part of the
amplitude of the neutron $\beta^-$--decay, caused by
one--virtual--photon exchanges and described by the Lagrangian
Eq.(\ref{eq:44}), are defined by the Feynman diagrams in
Fig.\,\ref{fig:fig6a}. We do not analyze the radiative corrections to
the one--pion--pole exchange, since as has been shown in
\cite{Ivanov2018d} the radiative corrections to the one--pion--pole
exchange are of order $10^{-9}$ and can be neglected in comparison
with the radiative corrections of order $10^{-5}$, which we are
searching for in this paper. The details of the calculation of the
Feynman diagrams in Fig.\,\ref{fig:fig6a} one may find in Appendices B
and C in section \ref{sec:appendix}, where we show that the Feynman
diagrams in Fig.\,\ref{fig:fig6a} are gauge invariant and do not
depend on a gauge parameter $\xi$ of the photon propagator. We show
also that gauge invariance of the Feynman diagrams in
Fig.\,\ref{fig:fig6a} retains even for the axial coupling constant
$g_A \neq 1$. As has been shown in section \ref{sec:lorentzstr} (see
Eq.(\ref{eq:70}), the axial coupling constant $g_A \neq$ in the
amplitude of the neutron $\beta^-$--decay appears as contribution of
one--hadron--loop diagrams, caused by strong low--energy interactions
described by the Lagrangian Eq.(\ref{eq:44}). Gauge invariance of the
Feynman diagrams in Fig.\,\ref{fig:fig6a} for $g_A \neq 1$ allows us
to take into account partly the contributions of strong low--energy
interactions to the one--hadron--loop approximation. As we have shown
in Appendices B and C in section \ref{sec:appendix} the amplitude of
the neutron $\beta^-$--decay with radiative corrections of order
$O(\alpha/\pi)$ and $O(\alpha E_e/m_N)$ after renormalization takes
the form

\begin{eqnarray}\label{eq:71}
\hspace{-0.3in}&&M(n\to p\,e^-\bar{\nu}_e) = - 2 m_N G_V\Big\{\Big[1
  + \frac{\alpha}{2\pi}\Big(f_{\beta^-_c}(E_e,\mu) +
  \frac{E_e}{m_N}\,f_V(E_e)\Big)\Big]
       [\varphi^{\dagger}_p\varphi_n][\bar{u}_e\gamma^0(1 -
         \gamma^5)v_{\bar{\nu}}]\nonumber\\
\hspace{-0.3in}&&+ g_A \Big[1 + \frac{\alpha}{2\pi}\Big(
  f_{\beta^-_c}(E_e,\mu) + \frac{E_e}{m_N} f_A(E_e) + \frac{5}{2}
  \frac{m^2_N}{M^2_W} {\ell n}\frac{M^2_W}{m^2_N}\Big)\Big]
       [\varphi^{\dagger}_p\vec{\sigma}\,\varphi_n]\cdot
       [\bar{u}_e\vec{\gamma}\, (1 - \gamma^5)v_{\bar{\nu}}]\nonumber\\
\hspace{-0.3in}&&+ \frac{\alpha}{2\pi}\Big[
  [\varphi^{\dagger}_p\varphi_n][\bar{u}_e (1 -
    \gamma^5)v_{\bar{\nu}}]\Big(- \frac{\sqrt{1 -
      \beta^2}}{2\beta}{\ell n}\Big(\frac{1 + \beta}{1 - \beta}\Big) +
  \frac{E_e}{m_N} f_S(E_e)\Big) + g_A
       [\varphi^{\dagger}_p\vec{\sigma}\,\varphi_n] \cdot
       [\bar{u}_e\gamma^0 \vec{\gamma} \,(1 -
         \gamma^5)v_{\bar{\nu}}]\nonumber\\
\hspace{-0.3in}&&\times \Big(- \frac{\sqrt{1 -
    \beta^2}}{2\beta}\,{\ell n}\Big(\frac{1 + \beta}{1 - \beta}\Big) +
\frac{E_e}{m_N}\,f_T(E_e)\Big) +
\Big([\varphi^{\dagger}_p\frac{\vec{k}_e\cdot \vec{\sigma}}{E_e}
  \varphi_n]\frac{E_e}{m_N}\,g_S(E_e) +
    [\varphi^{\dagger}_p\frac{\vec{k}_{\bar{\nu}}\cdot
        \vec{\sigma}}{E_e} \varphi_n ]\frac{E_e}{m_N}\,h_S(E_e)\Big)
    \nonumber\\
\hspace{-0.3in}&&\times [\bar{u}_e(1 - \gamma^5)v_{\bar{\nu}}] +
       [\varphi^{\dagger}_p \frac{\vec{k}_e\cdot \vec{\sigma}}{E_e}
         \varphi_n][\bar{u}_e\gamma^0(1 - \gamma^5)v_{\bar{\nu}}]
       \frac{E_e}{m_N}g_V(E_e) +
            [\varphi^{\dagger}_p\frac{(\vec{k}_e\cdot
                \vec{\sigma}\,)\vec{\sigma}}{E_e} \varphi_n]\cdot
            [\bar{u}_e \vec{\gamma}\,(1 -
              \gamma^5)v_{\bar{\nu}}]\frac{E_e}{m_N}
            h_A(E_e)\Big]\Big\},
\end{eqnarray}
where the functions $f_{\beta^-_c}(E_e, \mu)$, $f_V(E_e)$, $f_A(E_e)$,
$f_S(E_e)$ and so on are calculated in Appendices B and C in section
\ref{sec:appendix} and are given in Eq.(\ref{eq:C.4}). The function
$f_{\beta^-_c}(E_e, \mu)$, where $\mu$ is an infinitesimal photon
mass, realizing relativistic covariant infrared regularization of the
radiative corrections caused by one--photon loop exchanges
\cite{Sirlin1967}, has been calculated by Sirlin \cite{Sirlin1967}
(the details of the calculation one may find in
\cite{Ivanov2013}). This function together with the terms (see
Eq.(\ref{eq:C.5})), which survive to leading order in the large
nucleon mass expansion, define the famous Sirlin's function
$\bar{g}(E_e)$ \cite{Sirlin1967}, describing radiative corrections to
the neutron lifetime. The functions $f_V(E_e)$, $f_A(E_e)$,
$f_S(E_e)$, $f_T(E_e)$, $g_S(E_e)$, $h_S(E_e)$, $g_V(E_e)$ and
$h_A(E_e)$ (see Eq.(\ref{eq:C.4})) are related to the radiative
corrections of order $O(\alpha E_e/m_N)$. The term
$(\alpha/\pi)(5m^2_N/2M^2_W){\ell n}(M^2_W/m^2_N) \sim 10^{-5}$,
calculated at $m_N = 0.939\,{\rm GeV}$ and $M_W = 80.379\,{\rm GeV}$
\cite{PDG2018}, is the rest of the contributions of the virtual
electroweak $W^-$--boson exchanges (see Feynman diagrams in
Fig.\,\ref{fig:fig6a}) after renormalization (see Appendices B and C in
section \ref{sec:appendix}).

\subsection{\bf Rate of the neutron $\beta^-$--decay 
$n \to p + e^- + \bar{\nu}_e$ described by the amplitude
  Eq.(\ref{eq:71}). Corrections of order $O(\alpha E_e/m_N)$ to
  Sirlin's function}

First, following \cite{Ivanov2013} we calculate the electron--energy
and angular distribution of the neutron $\beta^-$--decay $n \to p +
e^- + \bar{\nu}_e$ with unpolarized massive fermions, described the
amplitude Eq.(\ref{eq:71}). We get
\begin{eqnarray*}
\hspace{-0.15in}&&\frac{d^5\lambda_{\beta^-_c}(E_e, \vec{k}_e,
  \vec{k}_{\bar{\nu}}, \mu)}{d E_e d\Omega_e d\Omega_{\bar{\nu}}}= (1
+ 3 g^2_A)\,\frac{|G_V|^2}{16\pi^5}\,\Big\{1 +
\frac{\alpha}{\pi}\,\Big(f_{\beta^-_c}(E_e, \mu) - \frac{1 -
  \beta^2}{2\beta}\,{\ell n}\Big(\frac{1 + \beta}{1 - \beta}\Big)\Big)
+ \frac{1 - g^2_A}{1 + 3 g^2_A}\,\Big(1 +
\frac{\alpha}{\pi}\,f_{\beta^-_c}(E_e, \mu)\Big)\nonumber\\
\end{eqnarray*}
\begin{eqnarray}\label{eq:72}
\hspace{-0.15in}&&\times \,\frac{\vec{k}_e\cdot
  \vec{k}_{\bar{\nu}}}{E_e E_{\bar{\nu}}} +
\frac{\alpha}{\pi}\,\Big[\frac{1}{1 + 3
    g^2_A}\,\frac{E_e}{m_N}\,\Big(f_V(E_e) + \sqrt{1 -
    \beta^2}\,f_S(E_e) + \beta^2 (1 + 2 g_A) h_A(E_e) + g_A \beta^2
  g_V(E_e) + g_A \frac{E_0 - E_e}{E_e}\nonumber\\
\hspace{-0.15in}&& \times \,\sqrt{1 - \beta^2}\,h_S(E_e)\Big) +
\frac{3g^2_A}{1 + 3 g^2_A}\,\frac{E_e}{m_N}\,\Big(f_A(E_e) + \sqrt{1 -
  \beta^2}\,f_T(E_e)\Big) + \frac{3g^2_A}{1 + 3
  g^2_A}\,\frac{5}{2}\,\frac{m^2_N}{M^2_W}\, {\ell
  n}\frac{M^2_W}{m^2_N}\Big] + \frac{\alpha}{\pi}\,\Big[\frac{1}{1 + 3
    g^2_A} \nonumber\\
\hspace{-0.15in}&&\times \,\frac{E_e}{m_N}\,\Big(f_V(E_e) + (1 - 2
g_A) h_A(E_e) + g_A \sqrt{1 - \beta^2}\,g_S(E_e) + g_A g_V(E_e)\Big) -
\frac{g^2_A}{1 + 3 g^2_A} \frac{E_e}{m_N} f_A(E_e) - \frac{g^2_A}{1 +
  3 g^2_A} \frac{5}{2}\,\frac{m^2_N}{M^2_W}\nonumber\\
\hspace{-0.15in}&& \times\, {\ell n}\frac{M^2_W}{m^2_N}\Big]
\frac{\vec{k}_e\cdot \vec{k}_{\bar{\nu}}}{E_e
  E_{\bar{\nu}}}\Big\}\sqrt{E^2_e - m^2_e} E_e F(E_e, Z = 1),
\end{eqnarray}
where $d\Omega_e$ and $d\Omega_{\bar{\nu}}$ are infinitesimal solid
angles in the directions of the electron and antineutrino 3--momenta,
$F(E_e, Z = 1)$ is the well--known relativistic Fermi function,
describing electron--proton Coulomb final--state interaction (see, for
example, \cite{Wilkinson1982}). The rate $\lambda_{\beta^-_c}(\mu)$ of
the neutron $\beta^-$--decay $n \to p + e^- + \bar{\nu}_e$ is defined
by the integral
\begin{eqnarray}\label{eq:73}
\hspace{-0.15in}&&\lambda_{\beta^-_c}(\mu) = (1 + 3
g^2_A)\,\frac{|G_V|^2}{\pi^3}\int^{E_0}_{m_e}\,\Big\{1 +
\frac{\alpha}{\pi}\,\Big(f_{\beta^-_c}(E_e, \mu) - \frac{1 -
  \beta^2}{2\beta}\,{\ell n}\Big(\frac{1 + \beta}{1 - \beta}\Big)\Big)
+ \frac{\alpha}{\pi}\,\Big[\frac{1}{1 + 3
    g^2_A}\,\frac{E_e}{m_N}\,\Big(f_V(E_e)\nonumber\\
\hspace{-0.15in}&&+ \sqrt{1 - \beta^2}\,f_S(E_e) + \beta^2 (1 + 2 g_A)
h_A(E_e) + g_A \beta^2 g_V(E_e) + g_A \frac{E_0 - E_e}{E_e}\,\sqrt{1 -
  \beta^2}\,h_S(E_e)\Big) + \frac{3g^2_A}{1 + 3
  g^2_A}\,\frac{E_e}{m_N}\nonumber\\
\hspace{-0.15in}&& \times \,\Big(f_A(E_e) + \sqrt{1 -
  \beta^2}\,f_T(E_e)\Big) + \frac{3g^2_A}{1 + 3
  g^2_A}\,\frac{5}{2}\,\frac{m^2_N}{M^2_W}\, {\ell
  n}\frac{M^2_W}{m^2_N}\Big]\Big\}\sqrt{E^2_e - m^2_e}\,E_e F(E_e, Z =
1)\,dE_e,
\end{eqnarray}
where $E_0 = (m^2_n - m^2_p + m^2_e)/2m_n = 1.2927\,{\rm MeV}$ is the
end--point energy of the electron--energy spectrum
\cite{Ivanov2013}. In the integrand the first term of order
$O(\alpha/\pi)$ reproduces fully Sirlin's result \cite{Sirlin1967},
calculated to leading order in the large nucleon mass expansion.

For the cancellation of the infrared divergence in the rate
$\lambda_{\beta^-_c}(\mu)$ of the neuron $\beta^-$--decay $n \to p +
e^- + \bar{\nu}_e$ and to calculate the total rate of the neutron
$\beta^-$--decays we have to take into account the rate
$\lambda_{\beta^-_c\gamma}(\mu)$ of the neutron radiative
$\beta^-$--decay $n \to p + e^- + \bar{\nu}_e + \gamma$, where
$\gamma$ is a real photon \cite{Berman1958}--\cite{Abers1968}.
\begin{figure}
\centering \includegraphics[height=0.23\textheight]{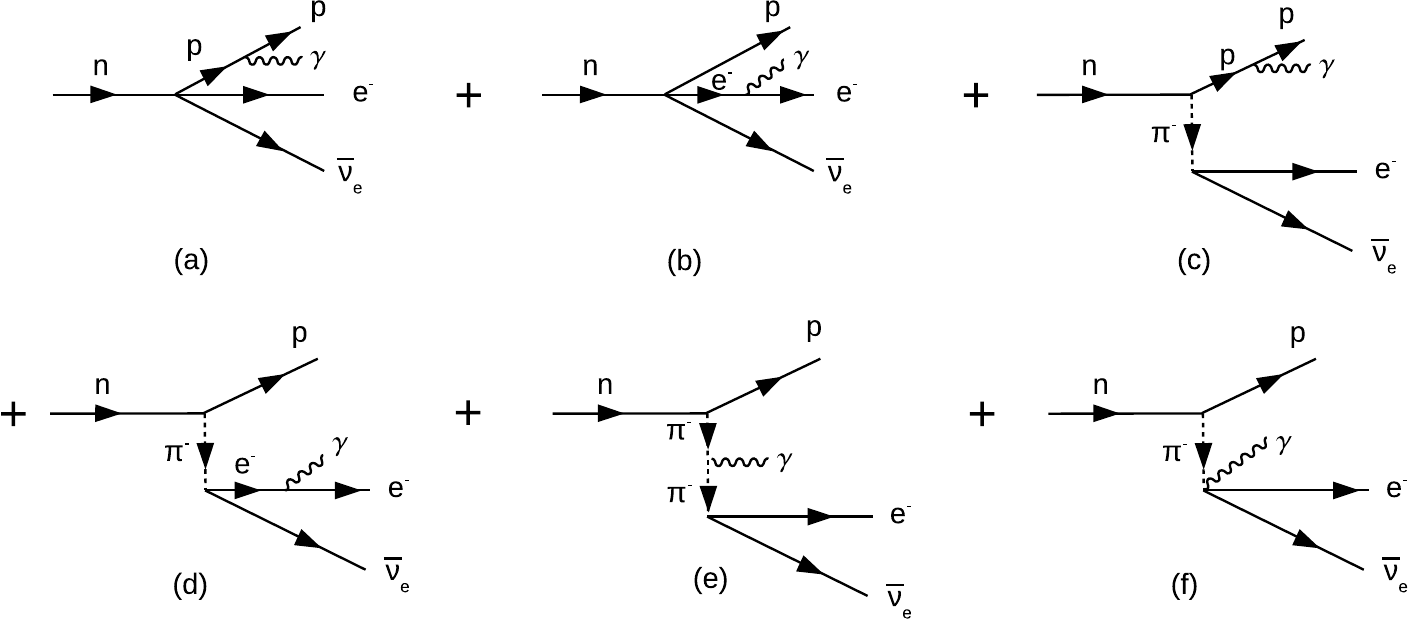}
  \caption{The Feynman diagrams, defining in the tree--approximation
    for the electroweak, electromagnetic and strong low--energy
    interactions the amplitude of the neutron radiative
    $\beta^-$--decay calculated with the Lagrangian
    Eq.(\ref{eq:44}). The Feynman diagrams are drawn to leading order
    in the large mass $M_W$ of the electroweak $W^-$--boson expansion
    at the neglect of the Feynman diagram with the vertex
    $W^-W^-\gamma$, the contribution of which is suppressed by the
    factor $q\cdot k/M^2_W$, where $k$ is a 4--momentum of a real
    photon.}
\label{fig:fig7a}
\end{figure}
The Feynman diagrams, describing the amplitude of the neutron
radiative $\beta^-$--decay in the tree--approximation for electroweak,
electromagnetic and strong low--energy interactions , are shown in
Fig.\,\ref{fig:fig7a}. The Feynman diagrams are drawn to leading order
in the large mass $M_W$ of the electroweak $W^-$--boson expansion at
the neglect of the Feynman diagram with the vertex $W^-W^-\gamma$, the
contribution of which is suppressed by the factor $q\cdot k/M^2_W$,
where $k$ is a 4--momentum of a real photon. The Feynman diagrams in
Fig.\,\ref{fig:fig7a}c -- Fig.\,\ref{fig:fig7a}f are caused by the
mesonic part of the charged hadronic axial-vector current. The
calculation of the Feynman diagrams in Fig.\,\ref{fig:fig7a} has been
carried out in \cite{Ivanov2018b} (see also Appendix D in section
\ref{sec:appendix}).

\begin{figure}
\centering \includegraphics[height=0.22\textheight]{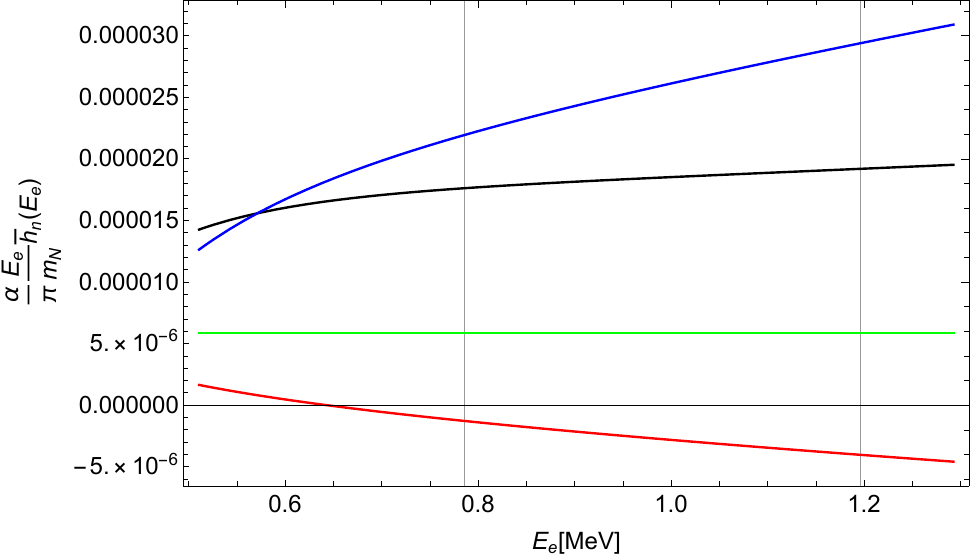}
  \caption{The radiative corrections of order $O(\alpha E_e/m_N)$,
    which are described by the function $\bar{h}_n(E_e)$ defining
    next--to--leading order corrections in the large nucleon mass
    expansion to Sirlin's function $\bar{g}_n(E_e)$, calculated to
    leading order in the large nucleon mass expansion (see
    Eq.(\ref{eq:77})), where i) black, ii) red, iii) blue and iv)
    green curves are defined by the contributions of i) all three
    terms, ii) of the first term, iii) of the last two terms and iv)
    of the last term of Eq.(\ref{eq:77}), respectively. The radiative
    corrections $O(\alpha E_e/m_N)$ are calculated in the
    electron--energy region $m_e < E_e < E_0$.}
\label{fig:fig8a}
\end{figure}

For the calculation of the neutron lifetime $\tau_n$, related to the
rate $\tau_n = 1/\lambda_n$, where $\lambda_n =
\lambda_{\beta^-_c}(\mu) + \lambda_{\beta^-_c \gamma}(\mu)$ and
$\lambda_{\beta^-_c\gamma}(\mu)$ is the rate of the neutron radiative
$\beta^-$--decay $n \to p + e^- + \bar{\nu}_e + \gamma$, we may use $
\lambda_{\beta\gamma}(\mu)$, calculated to leading order in the large
nucleon mass expansion. Using the results, obtained in Appendix B of
Ref.\cite{Ivanov2013}, we get the rate
$\lambda_{\beta^-_c\gamma}(\mu)$ of the neutron radiative
$\beta^-$--decay
\begin{eqnarray}\label{eq:74}
\hspace{-0.15in}&&\lambda_{\beta^-_c \gamma}(\mu) = (1 + 3
g^2_A)\,\frac{\alpha}{\pi}\,\frac{|G_V|^2}{\pi^3}\int^{E_0}_{m_e}\,
\Big\{\Big[2{\ell n}\Big(\frac{2(E_0 - E_e)}{\mu}\Big) - 3 +
  \frac{2}{3}\,\frac{E_0 - E_e}{E_e}\, \Big(1 + \frac{1}{8} \frac{E_0
    - E_e}{E_e} \Big)\Big]\Big[\frac{1}{2\beta}\,{\ell n}\Big(\frac{1
    + \beta}{1 - \beta}\Big) - 1\Big] \nonumber\\
\hspace{-0.3in}&& + 1 + \frac{1}{12} \frac{(E_0 - E_e)^2}{E^2_e}+
\frac{1}{2\beta}\,{\ell n}\Big(\frac{1 + \beta}{1 - \beta}\Big) -
\frac{1}{4\beta}\,{\ell n}^2\Big(\frac{1 + \beta}{1 - \beta}\Big) -
\frac{1}{\beta}\,{\rm Li}_2\Big(\frac{2 \beta}{1 + \beta}
\Big)\Big\}\sqrt{E^2_e - m^2_e}\,E_e F(E_e, Z = 1)\,dE_e
\end{eqnarray}
and the total rate $\lambda_n = \lambda_{\beta^-_c}(\mu) +
\lambda_{\beta^-_c \gamma}(\mu)$ of the neutron $\beta^-$--decay
\begin{eqnarray}\label{eq:75}
\hspace{-0.15in}&&\lambda_n = (1 + 3
g^2_A)\,\frac{|G_V|^2}{\pi^3}\int^{E_0}_{m_e}\,\Big(1 +
\frac{\alpha}{\pi}\,\bar{g}_n(E_e) +
\frac{\alpha}{\pi}\,\frac{E_e}{m_N}\,\bar{h}_n(E_e)\Big)\, \sqrt{E^2_e -
  m^2_e}\,E_e F(E_e, Z = 1)\,dE_e,
\end{eqnarray}
where the function $\bar{g}_n(E_e)$ is Sirlin's function equal to
\cite{Sirlin1967} (see also Appendix D of Ref. \cite{Ivanov2013})
\begin{eqnarray}\label{eq:76}
\hspace{-0.3in}\bar{g}_n(E_e) &=& \frac{3}{4}\,{\ell
  n}\Big(\frac{m^2_N}{m_e}\Big) - \frac{3}{8} +
\Big[\frac{1}{\beta} \,{\ell n}\Big(\frac{1 + \beta}{1 -
    \beta}\Big) - 2\Big]\Big[{\ell n}\Big(\frac{2(E_0 -
    E_e)}{m_e}\Big) - \frac{3} {2} + \frac{1}{3}\,\frac{E_0 -
  E_e}{E_e}\Big]\nonumber\\
\hspace{-0.3in}&-& \frac{2}{\beta}\,{\rm Li}_2\Big(\frac{2\beta}{1 +
  \beta}\Big) + \frac{1}{2\beta}{\ell n}\Big(\frac{1 + \beta}{1 -
  \beta}\Big)\,\Big[(1 + \beta^2) + \frac{1}{12} \frac{(E_0 -
    E_e)^2}{E^2_e} - {\ell n}\Big(\frac{1 + \beta}{1 -
    \beta}\Big)\Big],
\end{eqnarray}
and the function $\bar{h}_n(E_e)$, defining gauge invariant radiative
corrections of order $O(\alpha E_e/m_N)$ to Sirlin's function
$\bar{g}_n(E_e)$, is given by 
\begin{eqnarray}\label{eq:77}
\hspace{-0.15in}\bar{h}_n(E_e) &=& \frac{1}{1 + 3
  g^2_A}\,\Big(f_V(E_e) + \sqrt{1 - \beta^2}\,f_S(E_e) + \beta^2 (1 +
2 g_A) h_A(E_e) + g_A \beta^2 g_V(E_e) + g_A \frac{E_0 -
  E_e}{E_e}\nonumber\\
\hspace{-0.15in}&& \times \,\sqrt{1 - \beta^2}\,h_S(E_e)\Big) +
\frac{3g^2_A}{1 + 3 g^2_A}\,\Big(f_A(E_e) + \sqrt{1 -
  \beta^2}\,f_T(E_e)\Big) + \frac{3g^2_A}{1 + 3 g^2_A}
\,\frac{5}{2}\,\frac{m_N}{E_e}\,\frac{m^2_N}{M^2_W}\, {\ell
  n}\frac{M^2_W}{m^2_N},
\end{eqnarray}
where the functions $f_V(E_e)$, $f_V(E_e)$, $f_S(E_e)$, $f_T(E_e)$,
$g_S(E_e)$, $g_V(E_e)$, $h_S(E_e)$ and $h_A(E_e)$ are adduced in
Eq.(\ref{eq:C.4}) of Appendix C in section \ref{sec:appendix}.  Thus,
we have reproduced fully Sirlin's radiative corrections of order
$O(\alpha/\pi) \sim 10^{-3}$ to the neutron lifetime, calculated to
leading order in the large nucleon mass expansion, and obtained
radiative corrections of order $O(\alpha E_e/m_N) \sim 10^{-5}$ or
corrections to Sirlin's function $\bar{g}_n(E_e)$ in the gauge
invariant and renormalizable quantum field theoretic model of strong
low--energy and electroweak interactions, described by the Lagrangians
Eq.(\ref{eq:44}) and Eq.(\ref{eq:45}). In Fig.\,\ref{fig:fig8a} we
plot the function $(\alpha/\pi)\,(E_e/m_N)\,\bar{h}_n(E_e)$, where i)
the black curve is defined by the contributions of all three terms in
Eq.(\ref{eq:77}), ii) the red curve is given by the contribution of
only the first term, iii) the blue curve is defined by the
contributions of the last three terms, and iv) the green line
determines the contribution of the last term, caused by the
contribution of the electroweak $W^-$--boson exchanges. The function
$(\alpha/\pi)\,(E_e/m_N)\,\bar{h}_n(E_e)$ depends strongly on the
axial coupling constant $g_A$. The curves in Fig.\,\ref{fig:fig8a} are
calculated at $g_A = 1.2764$ \cite{Abele2018}, $m_e = 0.511\,{\rm
  MeV}$, $m_N = (m_n + m_p)/2 = 939\,{\rm MeV}$ and $M_W = 80379\,{\rm
  MeV}$, respectively \cite{PDG2018}, in the electron--energy region
$m_e < E_e < E_0$.

\section{\bf Discussion}
\label{sec:conclusion}

We have calculated radiative corrections of order $O(\alpha E_e/m_N)
\sim 10^{-5}$ as next--to--leading order corrections in the large
nucleon mass expansion to Sirlin's radiative corrections of order
$O(\alpha/\pi)$, calculated to leading order in the large nucleon mass
expansion to the neutron lifetime \cite{Sirlin1967}. For the extension
of Sirlin's result on the contributions of order $O(\alpha E_e/m_N)$
we have followed the assertion pointed out in \cite{Ivanov2018b,
  Ivanov2018c, Ivanov2018d} that for the analysis of corrections of
order $O(\alpha E_e/m_N)$ as next--to--leading order corrections in
the large nucleon mass expansion to Sirlin's radiative corrections of
order $O(\alpha/\pi)$ one has to deal with a combined quantum field
theoretic model at the hadronic level for strong low--energy
pion--nucleon interactions and electroweak interactions of the
Standard Electroweak Model with $SU(2)_L \times U(1)_Y$
symmetry. Thus, for the calculation of radiative corrections of order
$O(\alpha E_e/m_N) \sim 10^{-5}$ as next--to--leading order
corrections in the large nucleon mass expansion to Sirlin's radiative
corrections of order $O(\alpha/\pi)$ we have proposed a gauge
invariant quantum field theoretic model of strong low--energy
pion--nucleon interactions and electroweak pion--nucleon--lepton
interactions with electroweak $SU(2)_L \times U(1)_Y$ gauge symmetry,
described by the Lagrangian Eq.(\ref{eq:44}).  In the limit of
vanishing electroweak coupling constants such a quantum field
theoretic model reduces to the linear $\sigma$--model (L$\sigma$M) of
strong low--energy pion--nucleon interactions with chiral $SU(2)
\times SU(2)$ symmetry, which is treated as a hadronized version of
low-energy QCD. The latter is justified by an equivalence of the
L$\sigma$M with a linear realization of chiral $SU(2) \times SU(2)$
symmetry to Gasser--Leutwyler's chiral perturbation theory (ChPT) with
non--linear realization of chiral $SU(2) \times SU(2)$ symmetry in the
limit of the infinite mass $m_{\sigma} \to \infty$ of the scalar
isoscalar $\sigma$--meson (see section \ref{sec:nlsigma} and
\cite{Ecker1995}). We have shown that the quantum field theoretic
model of strong low--energy and electroweak interactions, described by
the Lagrangian Eq.(\ref{eq:44}), reproduces well in the
tree--approximation for electroweak $W^-$--boson exchanges and to
one--hadron--loop approximation, calculated in the limit of the
infinitely heavy scalar isoscalar $\sigma$--meson, a correct Lorentz
structure of the matrix element of the hadronic $n \to p$ transition
in the amplitude of the neutron $\beta^-$--decay. The contributions of
strong low--energy interactions are presented in the matrix element of
the hadronic $n \to p$ transition in terms of the axial coupling
constant $g_A \neq 1$, the anomalous isovector magnetic moment of the
nucleon $\kappa$, and the one--pion--pole exchange. In the chiral
limit $m_{\pi} \to 0$ such a matrix element does not depend on a
longitudinal part of the electroweak $W^-$--boson propagator. This
agrees well with the analysis of weak decays within effective standard
$V - A$ theory of weak interactions, carried out by Feynman and
Gell--Mann \cite{Feynman1958} and Nambu \cite{Nambu1960} (see also
\cite{Marshak1969}, \cite{Leitner2006} and \cite{Ivanov2018}).

In the quantum field theoretic model, described by the Lagrangian
Eq.(\ref{eq:44}), the radiative corrections of order $O(\alpha/\pi)$
are defined by the one--photon--loop Feynman diagrams in the
tree--approximation for strong low--energy hadronic interactions and
by two--loop Feynman diagrams with one--virtual--photon and --hadron
exchanges.  After renormalization these Feynman diagrams define the
radiative corrections of order $O(\alpha/\pi)$ to the neutron
$\beta^-$--decay with the traces of strong low--energy hadronic
interactions in terms of the axial coupling constant $g_A \neq 1$ and
the contributions of hadronic structure of the nucleon, which do not
reduce to the axial coupling constant.

As a first step towards a calculation of radiative corrections of
order $O(\alpha/\pi)$ valid to any order in the large nucleon mass
expansion and an understanding of gauge properties of these
corrections in dependence of one--virtual--photon exchanges with
hadronic structure of the neutron and proton, we have investigated the
contributions of one--photon--loop Feynman diagrams in the
tree--approximation for strong low--energy hadronic interactions,
which are shown in Fig.\,\ref{fig:fig6a}. To leading order in the
large electroweak $W^-$--boson exchanges these diagrams reduce to the
set of one--photon--loop Feynman diagrams, defined by
Fig.\,\ref{fig:fig6a}a, e and f, with point--like neutron and proton,
defined within the standard $V - A$ effective theory of weak
interactions and Quantum Electrodynamics (QED). Such a reduced set of
Feynman diagrams has been investigated by Sirlin \cite{Sirlin1967} to
leading order in the large nucleon mass expansion for the calculation
of the radiative corrections to the neutron lifetime, defined by the
function $(\alpha/\pi)\,\bar{g}_n(E_e)$. As has been pointed out by
Sirlin \cite{Sirlin1967}, these Feynman diagrams with one--virtual
photon coupled to point--like proton and electron is not gauge
invariant, and for gauge invariant set of Feynman diagrams defining
observable radiative corrections of order $O(\alpha/\pi)$ one has to
take into account Feynman diagrams of one--virtual--photon exchanges
with hadronic structure of the neutron and proton. Keeping only the
leading order contributions in the large nucleon mass expansion Sirlin
obtained that the observable radiative corrections of order
$O(\alpha/\pi)$ to the neutron lifetime do not depend on the axial
coupling constant $g_A \neq 1$ and the contributions of hadronic
structure of the nucleon coupled to one--virtual photon, responsible
for gauge invariance of radiative corrections of order
$O(\alpha/\pi)$, do not depend on the electron energy $E_e$
\cite{Sirlin1967, Sirlin1978}. Thus, the analysis of of the Feynman
diagrams in Fig.\,\ref{fig:fig6a}, taken in the tree--approximation
for strong low--energy interactions, only for the first step in the
calculation of radiative corrections of order $O(\alpha/\pi)$ should
shed light on the influence of hadronic structure of the nucleon on
gauge invariance of radiative corrections of order $O(\alpha/\pi)$
valid to any order in the large nucleon mass expansion.

The analytical expressions for these Feynman diagrams, adduced in
Appendix B in section \ref{sec:appendix}, can be obtained by using the
Lagrangian Eq.(\ref{eq:44}) with the axial coupling constant $g_A$
equal to $g_A = 1$. As we have shown in Appendix B in section
\ref{sec:appendix} these Feynman diagrams are gauge invariant and do
not depend on a gauge parameter $\xi$ of the longitudinal part of the
photon propagator. Having noticed that such an independence of a gauge
parameter $\xi$ retains also for $g_A \neq 1$, we have calculated the
contributions of the Feynman diagrams in Fig.\,\ref{fig:fig6a} at $g_A
\neq 1$. This has allowed us to take into account partly the
contributions of strong low--energy interactions in terms of the axial
coupling constant and to deal with gauge invariant radiative
contributions of order $O(\alpha/\pi)$ valid to any order in the large
nucleon mass expansion at Sirlin's confidence level \cite{Sirlin1967}.
The latter is very important for the calculation of next--to--leading
corrections in the large nucleon mass expansion to Sirlin's radiative
corrections, calculated to leading order in the large nucleon mass
expansion. After renormalization of the one--photon--loop
contributions we have obtained the radiative corrections to the
amplitude of the neutron $\beta^-$--decay of order $O(\alpha/\pi) \sim
10^{-3}$, agreeing fully with Sirlin's result \cite{Sirlin1967}( see
also Appendices C and D of Ref.\cite{Ivanov2013}) calculated to
leading order in the large nucleon mass expansion, and have taken into
account next--to--leading order corrections in the large nucleon mass
expansion of order $O(\alpha E_e/m_N) \sim 10^{-5}$ (see
Eq.(\ref{eq:70})). The amplitude of the neutron $\beta^-$--decay,
given by Eq.(\ref{eq:70}) and supplemented by next--to--leading order
$1/m_N$ proton recoil corrections and contributions of the weak
magnetism, might be used for the analysis of the neutron lifetime and
correlation coefficients of the neutron $\beta^-$--decays with
different polarization states of the neutron and massive decay
fermions to order $10^{-5}$. We are planning to carry out such an
analysis in our forthcoming publications.

The $O(\alpha E_e/m_N)$ corrections, defined by the function
$\bar{h}_n(E_ee)$ (see Eq.(\ref{eq:77})), to Sirlin's function
$\bar{g}_n(E_e)$ is plotted in Fig.\,\ref{fig:fig8a} in the
electron--energy region $m_e < E_e < E_0$. The order of the $O(\alpha
E_e/m_N)$ corrections is of about $10^{-5}$. Unlike Sirlin's
corrections $(\alpha/\pi) \bar{g}_n(E_e)$ of order $O(\alpha/\pi) \sim
10^{-3}$ , which do not depend on the axial coupling constant $g_A$,
the corrections of order $O(\alpha E_e/m_N) \sim 10^{-5}$ depend
strongly on the axial coupling constant or on strong low--energy
interactions. It is important to emphasize that the term
$(\alpha/\pi)(5/2)(m^2_N/M^2_W){\ell n}(M^2_W/m^2_N) \simeq 5\times
10^{-6}$ does not depend on the electron energy $E_e$. Such a
contribution comes from the Feynman diagrams in Fig.\,\ref{fig:fig6a}b
and c, which are important for gauge invariance of the
one--photon--loop exchanges, and agrees well with Sirlin's assertion
that the contribution of Feynman diagrams restoring gauge invariance
of the Feynman diagrams with one--virtual photon exchanges, when a
virtual photon emitted by the proton is hooked by the electron and
self--energy proton and electron Feynman diagrams, do not depend on
the electron energy. So one may assert that the radiative corrections
of order $O(\alpha E_e/m_N) \sim 10^{-5}$ calculated as
next--to--leading order corrections to Sirlin's radiative corrections
of order $O(\alpha/\pi)$, are defined at Sirlin's confidence level of
radiative corrections of order $O(\alpha/\pi)$. In addition the
calculation of the radiative corrections of order $O(\alpha/\pi)$
being valid to any order in the large nucleon mass expansion and
defined by a gauge invariant set of Feynman diagrams in
Fig.\,\ref{fig:fig6a} testifies that the contributions of one--virtual
photon interactions with hadronic structure of the nucleon should be
described by a set of Feynman diagrams, which are self--gauge
invariant. The shape of radiative corrections of order $O(\alpha/\pi)$
as functions of the electron energy $E_e$, caused by one--virtual
photon coupled to hadronic structure of the nucleon, is to some extent
model--dependent and can be calculated within the quantum field
theoretic model of strong low--energy and electroweak interactions
defined by the Lagrangian Eq.(\ref{eq:44}).  We would like also to
notice that the radiative corrections of order $O(\alpha E_e/m_N)$ to
the amplitude of the neutron $\beta^-$--decay Eq.(\ref{eq:71}) can be
also used for the calculation of radiative corrections of order
$10^{-5}$ to the proton recoil distribution of the neutron
$\beta^-$--decay \cite{Gluck1992}--\cite{Gluck1998b}.

Thus, concluding our discussion of the radiative corrections of order
$O(\alpha E_e/m_N)$ as next--to--leading order corrections in the
large nucleon mass expansion to Sirlin's radiative corrections of
order $O(\alpha/\pi)$, calculated to leading order in the large
nucleon mass expansion, we may argue that there are else three
problems, the analysis of which goes beyond the scope of this
paper. They are i) the radiative corrections to two--loop
approximation with one--hadron-- and one--photon--loop exchanges, the
contributions of which do not reduce to the axial coupling constant,
ii) the contribution of the electroweak $W$-- and $Z$--boson exchanges
to one--virtual electroweak boson approximation and iii) the radiative
corrections to two--loop approximation with one--hadron-- and
one--electroweak--boson--loop exchanges.

The contributions of the electroweak $W$-- and $Z$--boson exchanges
are defined by more than 24 Feynman diagrams with intermediate $W$--
and $Z$--boson virtual exchanges. Practically, they have been
calculated by Sirlin with co--workers (see, for example,
\cite{Sirlin2004}). We have to show that such a result can be obtained
in the quantum field theoretic model of strong low--energy and
electroweak interactions, described by the Lagrangian
Eq.(\ref{eq:44}). According to \cite{Sirlin2004}, the contribution of
the electroweak $W$-- and $Z$--boson exchanges do not depend on the
electron energy $E_e$.  Our analysis of the Feynman diagrams in
Appendices B and C in section \ref{sec:appendix}, where we have shown
that the contributions of Feynman diagrams with virtual electroweak
$W^-$--boson exchanges do not depend on the electron energy $E_e$,
agrees well with independence of the electron energy the corrections
caused by the electroweak $W$-- and $Z$--boson exchanges. So that the
contribution of the Feynman diagrams with $W$-- and $Z$--boson virtual
exchanges should not add any corrections of order $O(\alpha
E_e/m_N)$. As next--to--leading order corrections to the result
obtained in \cite{Sirlin2004} we may expect the corrections of order
$(m^2_N/M^2_X){\ell n}(M^2_X/m^2_N) \sim 10^{-5}$ for $X = W$ or $Z$,
respectively.  We are planning to take into account the contributions
of the virtual electroweak $W$-- and $Z$--boson exchanges in our
forthcoming publication.

Then, the contributions of the two--loop Feynman diagrams with virtual
hadron and photon exchanges, which cannot be reduced to the
contribution of the axial coupling constant, and electroweak $W$-- and
$Z$--boson exchanges are defined by a huge number of Feynman diagrams
which could in principle give some non--trivial but finite independent
of the electron energy $E_e$ contributions to the radiative
corrections of order $O(\alpha/\pi)$ and, correspondingly, to
next--to--leading order corrections in the large nucleon mass
expansion. However, according to Sirlin's analysis \cite{Sirlin1967,
  Sirlin1978}, the corrections of hadronic structure of the nucleon to
order $O(\alpha/\pi)$ and taken to leading in the large nucleon mass
expansion do not depend on the electron energy $E_e$ and can be
removed by renormalization of the Fermi and axial coupling constants.
Nevertheless, we are planning to carry out the investigation of the
problem of contributions of hadronic structure of the nucleon coupled
to one--virtual photon and virtual electroweak bosons, which cannot be
reduced to the axial coupling constant $g_A$, in our forthcoming
publications.

Finally we would like to notice that in the quantum field theoretic
model of strong low--energy and electroweak interactions strong
low--energy interactions are described by the L$\sigma$M with a linear
realization of chiral $SU(2)\times SU(2)$ symmetry. We calculate the
contributions of strong low--energy interactions in the limit of the
infinite mass of the scalar isoscalar $\sigma$--meson. In such a limit
the L$\sigma$M is equivalent to the quantum field theory with a
non--linear realization of chiral $SU(2)\times SU(2)$ symmetry. In the
exponential parametrization of the pion field the L$\sigma$M reduces
to ChPT (see \cite{Ecker1995}), which is accepted as a low--energy QCD
\cite{Ecker1996,Scherer2002}). Since the use of the L$\sigma$M as a
quantum field theoretic model of strong low--energy pion--nucleon
interactions makes to some extent the results of our analysis of
contributions of strong low--energy interactions to the neutron
$\beta^-$--decays model--dependent, we are planning to reformulate the
quantum field theoretic model of strong low--energy and electroweak
interactions, presented by the Lagrangian Eq.(\ref{eq:29}) and,
correspondingly, Eq.(\ref{eq:44}) with the sector of strong
low--energy pion--nucleon interactions, described by ChPT with a
non--linear realization of chiral $SU(2) \times SU(2)$
symmetry. However, first of all we would like to investigate the
problems mentioned above for subsequent investigations of an influence
of strong low--energy interactions on gauge properties and
renormalizability of radiative corrections to the neutron
$\beta^-$--decays in the quantum field theoretic model of strong
low--energy and electroweak interactions with the sector of strong
low--energy interactions described by the L$\sigma$M. According to
Weinberg's ``theorem'' \cite{Weinberg1979} (see also subsection C in
section \ref{sec:sigma}), because of an equivalence of the L$\sigma$M
to the ChPT \cite{Ecker1995} (see also section \ref{sec:nlsigma}) the
results obtained in the L$\sigma$M and in the ChPT as quantum field
theoretic models of strong low--energy interactions in the neutron
$\beta^-$--decays should be in principle the same. A comparison of the
results, obtained within these to quantum field theoretic models of
strong low--energy hadronic interactions with chiral $SU(2) \times
SU(2)$ symmetry in the neutron $\beta^-$--decays, should be to some
extent a good verification of Weinberg's ``theorem'', which is
required by Weinberg \cite{Weinberg1979}.

\section{Acknowledgements}

We thank Hartmut Abele for discussions stimulating the work under this
paper as a first step towards the analysis of the SM corrections of
order $10^{-5}$ to the neutron lifetime and correlation coefficients
of the neutron $\beta^-$--decays with different polarization states of
the neutron and massive decay fermions.  The work of A. N. Ivanov was
supported by the Austrian ``Fonds zur F\"orderung der
Wissenschaftlichen Forschung'' (FWF) under contracts P31702-N27,
P26781-N20 and P26636-N20 and ``Deutsche F\"orderungsgemeinschaft''
(DFG) AB 128/5-2. The work of R. H\"ollwieser was supported by the
Deutsche Forschungsgemeinschaft in the SFB/TR 55. The work of
M. Wellenzohn was supported by the MA 23 (FH-Call 16) under the
project ``Photonik - Stiftungsprofessur f\"ur Lehre''.

\newpage

\section{\bf Supplemental material}
\label{sec:appendix}

\section*{Appendix A: Analytical expressions for the Feynman diagrams 
in Fig.\,\ref{fig:fig2a} $-$ Fig.\,\ref{fig:fig4a}}
\renewcommand{\theequation}{A-\arabic{equation}}
\setcounter{equation}{0}

Since the Feynman diagrams in Fig.\,\ref{fig:fig2a} and
Fig.\,\ref{fig:fig3a} give the contributions to the amplitude of the
neutron $\beta^-$--decay only in terms of the matrix element of the
hadronic $n \to p$ transition, the contributions of these diagrams we
may write down as follows
\begin{eqnarray}\label{eq:A.1}
\hspace{-0.3in}\langle
p(\vec{k}_p,\sigma_p)|J^+_{\mu}(0)|n(\vec{k}_n, \sigma_n)\rangle_{\rm
  Fig. \ref{fig:fig2a}a + Fig. \ref{fig:fig2a}b} &=&
\Big[\bar{u}_p\big(\vec{k}_p, \sigma_p\big)\,\gamma_{\mu}(1 -
  \gamma^5)\,\frac{1}{m_N - \hat{k}_n - i 0}\,
  \Sigma_n(k_n)\,u_n\big(\vec{k}_n,
  \sigma_n\big)\Big]\nonumber\\
\hspace{-0.3in}&-&\frac{2 m_N q_{\mu}}{m^2_{\pi} - q^2 -
  i0}\Big[\bar{u}_p\big(\vec{k}_p,
  \sigma_p\big)\,\gamma^5\,\frac{1}{m_N - \hat{k}_n - i 0}\,
  \Sigma_n(k_n)\,u_n\big(\vec{k}_n,
  \sigma_n\big)\Big],\nonumber\\ \hspace{-0.3in}\langle
p(\vec{k}_p,\sigma_p)|J^+_{\mu}(0)|n(\vec{k}_n, \sigma_n)\rangle_{\rm
  Fig.\,\ref{fig:fig2a}c + Fig.\,\ref{fig:fig2a}d} &=&
\Big[\bar{u}_p\big(\vec{k}_p,
  \sigma_p\big)\,\Sigma_p(k_p)\,\frac{1}{m_N - \hat{k}_p - i
    0}\,\gamma_{\mu}(1 - \gamma^5)\,u_n\big(\vec{k}_n,
  \sigma_n\big)\Big]\nonumber\\ \hspace{-0.3in}&-&\frac{2 m_N
  q_{\mu}}{m^2_{\pi} - q^2 - i0}\Big[\bar{u}_p\big(\vec{k}_p,
  \sigma_p\big)\, \Sigma_p(k_p)\,\frac{1}{m_N - \hat{k}_p - i
    0}\,\gamma^5\,u_n\big(\vec{k}_n,
  \sigma_n\big)\Big],\nonumber\\ \hspace{-0.3in}\langle
p(\vec{k}_p,\sigma_p)|J^+_{\mu}(0)|n(\vec{k}_n, \sigma_n)\rangle_{\rm
  Fig.\,\ref{fig:fig3a}} &=& - \,\frac{2 m_N q_{\mu}}{m^2_{\pi} - q^2
  - i0}\,\Sigma_{\pi}(q)\,\frac{1}{m^2_{\pi} - q^2 -
  i0}\big[\bar{u}_p\big(\vec{k}_p,
  \sigma_p\big)\,\gamma^5\,u_n\big(\vec{k}_n, \sigma_n\big)\big],
\end{eqnarray}
where the self--energy correction $\Sigma_n(k_n)$ to the
neutron state is equal to
\begin{eqnarray}\label{eq:A.2}
\hspace{-0.3in}\Sigma_n(k_n) &=& - \delta m_N -
\Big((Z_N - 1) + (\tilde{Z}^{(N)}_2 - 1)\Big)\,(m_N - \hat{k}_n) + g^2_{\pi N}\int
\frac{d^4p}{(2\pi)^4 i}\,\frac{1}{m_N - \hat{p} - \hat{k}_n - i
  0}\,\frac{1}{m^2_{\sigma} - p^2 - i 0}
\nonumber\\ \hspace{-0.3in}&-& 3 g^2_{\pi N}\int \frac{d^4p}{(2\pi)^4
  i}\,\gamma^5\,\frac{1}{m_N - \hat{p} - \hat{k}_n - i
  0}\,\gamma^5\,\frac{1}{m^2_{\pi} - p^2 - i 0}.
\end{eqnarray}
The self--energy corrections to the proton state
$\Sigma_p(k_p)$ is defined by Eq.(\ref{eq:A.2}) with a
replacement of indices $n \to p$. In turn, the self--energy correction
$\Sigma_{\pi}(q)$ to the $\pi^-$--meson state is
\begin{eqnarray}\label{eq:A.3}
\hspace{-0.3in}\Sigma_{\pi}(q) &=& - \delta m^2_{\pi} +
\Big((Z_M - 1) + (\tilde{Z}^{(M)}_2 - 1)\Big)\,(m^2_{\pi} - q^2) + 4\gamma^2 f^2_{\pi}\int
\frac{d^4p}{(2\pi)^4 i}\frac{1}{m^2_{\pi} - (p + q)^2 - i
  0}\,\frac{1}{m^2_{\sigma} - p^2 - i 0}\nonumber\\ \hspace{-0.3in}&+&
2 g^2_{\pi N}\int \frac{d^4p}{(2\pi)^4 i}{\rm tr}\Big\{\gamma^5
\frac{1}{m_N - \hat{p} + \hat{q} - i 0} \gamma^5 \frac{1}{m_N -
  \hat{p} - i 0}\Big\}.
\end{eqnarray}
The contributions of the counter--terms, defined by the Lagrangian
Eq.(\ref{eq:45}), are equal to
\begin{eqnarray}\label{eq:A.4}
\hspace{-0.15in}\langle
p(\vec{k}_p,\sigma_p)|J^+_{\mu}(0)|n(\vec{k}_n, \sigma_n)\rangle_{\rm
  Fig. \ref{fig:fig4a}} &=&\bar{u}_p\big(\vec{k}_p,
\sigma_p\big)\,\Big\{\Big(\big(\tilde{Z}^{(N)}_1 - 1\big) + \big(Z_N -
1\big)\Big)\, \gamma_{\mu}(1 - \gamma^5) - \Big(\big(Z_{MN} - 1\big) +
\big(\tilde{Z}^{(N)}_2 - 1\big)\nonumber\\ \hspace{-0.15in}&& +
\big(\tilde{Z}^{(M)}_1 - 1\big) + \big(Z_M - 1\big)\Big)\,\frac{2 m_N
  q_{\mu}}{m^2_{\pi} - q^2 - i0}\, \gamma^5\Big\}\,u_n\big(\vec{k}_n,
\sigma_n\big),
\end{eqnarray}
where we have used the GT--relation $g_{\pi N} = m_N/f_{\pi}$. The
analytical expressions of the Feynman diagrams in
Fig.\,\ref{fig:fig4a}, defining a non--trivial Lorentz structure of
the matrix element of the hadronic $n \to p$ transition, are given by
\begin{eqnarray*}
\hspace{-0.3in}&&\langle
p(\vec{k}_p,\sigma_p)|J^+_{\mu}(0)|n(\vec{k}_n, \sigma_n)\rangle_{\rm
  Fig.\,\ref{fig:fig4a}a + Fig.\,\ref{fig:fig4a}b}
=\nonumber\\ \hspace{-0.3in}&& = \Big[\bar{u}_p\big(\vec{k}_p,
  \sigma_p\big)\,\Big\{4\,g^2_{\pi N}\int \frac{d^4p}{(2\pi)^4
    i}\,\gamma^5\,\frac{(2 p - q)_{\mu}}{m_N - \hat{p} - \hat{k}_n - i
    0}\,\gamma^5\, \frac{1}{m^2_{\pi} - (p - q)^2 - i
    0}\,\frac{1}{m^2_{\pi} - p^2 - i 0}\,\Big\}\,u_n\big(\vec{k}_n,
  \sigma_n\big)\Big], \nonumber\\ \hspace{-0.3in}&&\langle
p(\vec{k}_p,\sigma_p)|J^+_{\mu}(0)|n(\vec{k}_n, \sigma_n)\rangle_{\rm
  Fig.\,\ref{fig:fig4a}c + Fig.\,\ref{fig:fig4a}d}
=\nonumber\\ \hspace{-0.3in}&& = \Big[\bar{u}_p\big(\vec{k}_p,
  \sigma_p\big)\,\Big\{2\,g^2_{\pi N} \int \frac{d^4p}{(2\pi)^4
    i}\,\gamma^5\,\frac{(2 p + q)_{\mu}}{m_N - \hat{p} - \hat{k}_p - i
    0}\, \frac{1}{m^2_{\pi} - p^2 - i 0}\,\frac{1}{m^2_{\sigma} - (p +
    q)^2 - i 0}\Big\}\,u_n\big(\vec{k}_n, \sigma_n\big)\nonumber\\ &&
  -\bar{u}_p\big(\vec{k}_p, \sigma_p\big)\,\Big\{ 2\,g^2_{\pi N}\int
  \frac{d^4p}{(2\pi)^4 i}\,\frac{(2 p - q)_{\mu}}{m_N - \hat{p} -
    \hat{k}_n - i 0}\,\gamma^5\, \frac{1}{m^2_{\sigma} - (p - q)^2 - i
    0}\,\frac{1}{m^2_{\pi} - p^2 - i 0}\,\Big\}\, u_n\big(\vec{k}_n,
  \sigma_n\big)\Big],\nonumber\\ \hspace{-0.3in}&&\langle
p(\vec{k}_p,\sigma_p)|J^+_{\mu}(0)|n(\vec{k}_n, \sigma_n)\rangle_{\rm
  Fig.\,\ref{fig:fig4a}e + Fig.\,\ref{fig:fig4a}f} = \nonumber\\ &&=
\Big[\bar{u}_p\big(\vec{k}_p, \sigma_p\big)\,\Big\{ g^2_{\pi N}\int
  \frac{d^4p}{(2\pi)^4 i}\, \gamma^5\, \frac{1}{m_N - \hat{p} -
    \hat{k}_p - i0}\,\gamma_{\mu}(1 - \gamma^5)\,\frac{1}{m_N -
    \hat{p} - \hat{k}_n - i0}\,\gamma^5\,\frac{1}{m^2_{\pi} - p^2 -
    i0}\Big\}\, u_n\big(\vec{k}_n, \sigma_n\big)
  \nonumber\\
\end{eqnarray*}
\begin{eqnarray}\label{eq:A.5}
  \hspace{-0.3in}&&+ \bar{u}_p\big(\vec{k}_p,
  \sigma_p\big)\,\Big\{ g^2_{\pi N}\int \frac{d^4p}{(2\pi)^4
    i}\,\frac{1}{m_N - \hat{p} - \hat{k}_p - i0}\,\gamma_{\mu}(1 -
  \gamma^5)\,\frac{1}{m_N - \hat{p} - \hat{k}_n -
    i0}\,\frac{1}{m^2_{\sigma} - p^2 - i0}\Big\}\, u_n\big(\vec{k}_n,
  \sigma_n\big)\Big],\nonumber\\
\hspace{-0.3in}&&\langle
p(\vec{k}_p,\sigma_p)|J^+_{\mu}(0)|n(\vec{k}_n, \sigma_n)\rangle_{\rm
  Fig.\,\ref{fig:fig4a}g + Fig.\,\ref{fig:fig4a}h +
  Fig.\,\ref{fig:fig4a}i + Fig.\,\ref{fig:fig4a}j +
  Fig.\,\ref{fig:fig4a}k + Fig.\,\ref{fig:fig4a}\ell +
  Fig.\,\ref{fig:fig4a}m} = \frac{q_{\mu}}{m^2_{\pi} - q^2 -
  i0}\nonumber\\ \hspace{-0.3in}&&\times \Big[\bar{u}_p\big(\vec{k}_p,
  \sigma_p\big)\,\Big\{(- 4)\,g^2_{\pi N}\gamma f^2_{\pi} \int
  \frac{d^4p}{(2\pi)^4 i}\,\gamma^5 \frac{1}{m_N - \hat{p} - \hat{k}_p
    - i 0}\, \frac{1}{m^2_{\pi} - p^2 - i
    0}\,\frac{1}{m^2_{\sigma} - (p + q)^2 - i 0}\Big\}\,u_n\big(\vec{k}_n,
  \sigma_n\big) \nonumber\\ \hspace{-0.3in}&&+
  \bar{u}_p\big(\vec{k}_p, \sigma_p\big)\,\Big\{(- 4)\,g^2_{\pi
    N}\gamma f^2_{\pi} \int \frac{d^4p}{(2\pi)^4 i}\,\frac{1}{m_N -
    \hat{p} - \hat{k}_n - i 0}\,\gamma^5\, \frac{1}{m^2_{\sigma} - (p
    - q)^2 - i 0}\,\frac{1}{m^2_{\pi} - p^2 - i 0}
  \Big\}\,u_n\big(\vec{k}_n,
  \sigma_n\big)\nonumber\\ \hspace{-0.3in}&&+ \bar{u}_p\big(\vec{k}_p,
  \sigma_p\big)\,\Big\{ 2g^3_{\pi N} f_{\pi}\int \frac{d^4p}{(2\pi)^4
    i}\, \gamma^5\, \frac{1}{m_N - \hat{p} - \hat{k}_p -
    i0}\,\gamma^5\,\frac{1}{m_N - \hat{p} - \hat{k}_n -
    i0}\,\gamma^5\,\frac{1}{m^2_{\pi} - p^2 - i0}\Big\}\,
  u_n\big(\vec{k}_n, \sigma_n\big) \nonumber\\
\hspace{-0.3in}&&+ \bar{u}_p\big(\vec{k}_p, \sigma_p\big)\,\Big\{
2g^3_{\pi N}f_{\pi} \int \frac{d^4p}{(2\pi)^4 i}\, \frac{1}{m_N -
  \hat{p} - \hat{k}_p - i0}\,\gamma^5\,\frac{1}{m_N - \hat{p} -
  \hat{k}_n - i0}\,\frac{1}{m^2_{\sigma} - p^2 - i0}\Big\}\,
u_n\big(\vec{k}_n, \sigma_n\big)\Big] \nonumber\\ && +
\frac{1}{m^2_{\pi} - q^2 - i0}\,\Big[\bar{u}_p\big(\vec{k}_p,
  \sigma_p\big)\,\Big\{2 g^2_{\pi N}\int \frac{d^4p}{(2\pi)^4 i}\,{\rm
    tr}\Big\{\gamma^5\, \frac{1}{m_N - \hat{p} - \hat{q}-
    i0}\,\gamma_{\mu}(1 - \gamma^5)\, \frac{1}{m_N - \hat{p} -
    i0}\Big\}\,\gamma^5\, u_n\big(\vec{k}_n, \sigma_n\big) \nonumber\\
\hspace{-0.3in}&&+ \bar{u}_p\big(\vec{k}_p, \sigma_p\big)\,\Big\{(- 4)
g_{\pi N} \gamma f_{\pi} \int \frac{d^4p}{(2\pi)^4 i}\,\frac{(2 p -
  q)_{\mu}}{m^2_{\pi} - (p - q)^2 - i0}\,\frac{1}{m^2_{\sigma} - p^2 -
  i0}\Big\}\,\gamma^5\, u_n\big(\vec{k}_n, \sigma_n\big)\Big].
\end{eqnarray}
The result of the calculation of the integrals in Eq.(\ref{eq:A.2})
and Eq.(\ref{eq:A.3}) can be represented in the following general form
as $a_n + b_n(m_N - \hat{k}_n)$ and $a_{\pi} + b_{\pi}(m^2_{\pi} -
q^2)$, respectively. The contributions of $a_n$ and $a_{\pi}$ are
absorbed by the mass--renormalization $\delta m_N$ and $\delta
m^2_{\pi}$, respectively. In turn, the contributions of $b_n$ and
$b_{\pi}$ can be removed by the wave function renormalisations
$\tilde{Z}^{(N)}_2$, $Z_N$ and $\tilde{Z}^{(M)}_2$, respectively. For
the counter--terms $\tilde{Z}^{(N)}_2$, $Z_N$ and $\tilde{Z}^{(M)}_2$
we get the following expressions
\begin{eqnarray}\label{eq:A.6}
\hspace{-0.3in}\big(Z_N - 1) + \big(\tilde{Z}^{(N)}_2 - 1\big) &=& -
\frac{g^2_{\pi N}}{8\pi^2}\,\Big({\ell n}\frac{\Lambda^2}{m^2_N} -
\frac{1}{4}\,{\ell n}\frac{m^2_{\sigma}}{m^2_N}\Big),\nonumber\\
\hspace{-0.3in}\big(Z_M - 1\big) + \big(\tilde{Z}^{(M)}_2 - 1\big) &=&
- \frac{g^2_{\pi N}}{4\pi^2}\,{\ell n}\frac{\Lambda^2}{m^2_N},
\end{eqnarray}
where we have used the dimensional regularization with a replacement
$\Gamma(2 - n/2) \to {\ell n}(\Lambda^2/m^2_N)$ \cite{Itzykson1980},
where $n$ is a dimension of a momentum space and $\Lambda$ is an
ultra--violet cut--off.  The Feynman diagrams in Fig.\,\ref{fig:fig4a}
give the following contributions, calculated at $m_{\sigma} \gg m_N$,
to the matrix element of the hadronic $n \to p$ transition
\begin{eqnarray}\label{eq:A.7}
\hspace{-0.15in}\langle
p(\vec{k}_p,\sigma_p)|J^+_{\mu}(0)|n(\vec{k}_n, \sigma_n)\rangle_{\rm
  Fig.\,\ref{fig:fig4a}a + Fig.\,\ref{fig:fig4a}b} &=&
\bar{u}_p\big(\vec{k}_p, \sigma_p\big)\,\Big\{\frac{g^2_{\pi
    N}}{8\pi^2} {\ell n}\frac{\Lambda^2}{m^2_N}\, \gamma_{\mu} +
\frac{g^2_{\pi N}}{4\pi^2} \frac{i \sigma_{\mu\nu}q^{\nu}}{2
  m_N}\Big\} u_n\big(\vec{k}_n, \sigma_n\big),\nonumber\\
\hspace{-0.15in}\langle
p(\vec{k}_p,\sigma_p)|J^+_{\mu}(0)|n(\vec{k}_n, \sigma_n)\rangle_{\rm
  Fig.\,\ref{fig:fig4a}c + Fig.\,\ref{fig:fig4a}d} &=&
\bar{u}_p\big(\vec{k}_p, \sigma_p\big)\,\Big\{\frac{g^2_{\pi
    N}}{8\pi^2} \Big({\ell n}\frac{\Lambda^2}{m^2_N} - {\ell
  n}\frac{m^2_{\sigma}}{m^2_N}\Big) \gamma_{\mu}
\gamma^5\Big\} u_n\big(\vec{k}_n, \sigma_n\big),\nonumber\\
\hspace{-0.15in}\langle p(\vec{k}_p,\sigma_p)|J^+_{\mu}(0)|n(\vec{k}_n,
\sigma_n)\rangle_{\rm Fig.\,\ref{fig:fig4a}e + Fig.\,\ref{fig:fig4a}f}
&=& \bar{u}_p\big(\vec{k}_p, \sigma_p\big)\,\Big\{- \frac{g^2_{\pi N}}{8\pi^2}\, \frac{1}{4}\,{\ell
  n}\frac{m^2_{\sigma}}{m^2_N} \gamma_{\mu} -
\frac{g^2_{\pi N}}{8\pi^2}\, \frac{1}{4}\,{\ell
  n}\frac{m^2_{\sigma}}{m^2_N}\, \gamma_{\mu}
\gamma^5 \nonumber\\
\hspace{-0.15in}&+& \frac{g^2_{\pi N}}{16\pi^2} \frac{i
  \sigma_{\mu\nu}q^{\nu}}{2 m_N}\Big\} u_n\big(\vec{k}_n,
\sigma_n\big),\nonumber\\
\hspace{-0.15in}\langle p(\vec{k}_p,\sigma_p)|J^+_{\mu}(0)|n(\vec{k}_n,
\sigma_n)\rangle_{\rm Fig.\,\ref{fig:fig4a}g + Fig.\,\ref{fig:fig4a}h}
&=& - \frac{2 m_N q_{\mu}}{m^2_{\pi} - q^2 -
  i0}\,\bar{u}_p\big(\vec{k}_p, \sigma_p\big)\,\Big\{\frac{g^2_{\pi
    N}}{8\pi^2}\, \frac{1}{2}\,{\ell
  n}\frac{m^2_{\sigma}}{m^2_N}\Big\}\,\gamma^5\,u_n\big(\vec{k}_n,
\sigma_n\big),\nonumber\\
\hspace{-0.15in}\langle p(\vec{k}_p,\sigma_p)|J^+_{\mu}(0)|n(\vec{k}_n,
\sigma_n)\rangle_{\rm Fig.\,\ref{fig:fig4a}i + Fig.\,\ref{fig:fig4a}j}
&=& - \frac{2 m_N q_{\mu}}{m^2_{\pi} - q^2 -
  i0}\,\bar{u}_p\big(\vec{k}_p, \sigma_p\big)\,\Big\{\frac{g^2_{\pi
    N}}{8\pi^2}\,\Big(- {\ell n}\frac{\Lambda^2}{m^2_N}
+ \frac{1}{2}\,{\ell n}\frac{m^2_{\sigma}}{m^2_N}\Big)\Big\} \gamma^5
u_n\big(\vec{k}_n, \sigma_n\big),\nonumber\\
\hspace{-0.15in}\langle
p(\vec{k}_p,\sigma_p)|J^+_{\mu}(0)|n(\vec{k}_n, \sigma_n)\rangle_{\rm
  Fig.\,\ref{fig:fig4a}k + Fig.\,\ref{fig:fig4a}l} &=& - \frac{2 m_N
  q_{\mu}}{m^2_{\pi} - q^2 - i0}\,\bar{u}_p\big(\vec{k}_p,
\sigma_p\big)\,\Big\{ \frac{g^2_{\pi N}}{4\pi^2}\,{\ell
  n}\frac{\Lambda^2}{m^2_N} \Big\} \gamma^5 u_n\big(\vec{k}_n,
\sigma_n\big).
\end{eqnarray}
Thus, the contributions of the Feynman diagrams in Fig.\,\ref{fig:fig4a}
to the matrix element of the hadronic $n \to p$ transition is equal
to
\begin{eqnarray}\label{eq:A.8}
\hspace{-0.15in}\langle
p(\vec{k}_p,\sigma_p)|J^+_{\mu}(0)|n(\vec{k}_n, \sigma_n)\rangle_{\rm
  Fig.\,\ref{fig:fig4a}} &=& \bar{u}_p\big(\vec{k}_p,
\sigma_p\big)\,\Big\{ \frac{g^2_{\pi N}}{8\pi^2}\,\Big({\ell
  n}\frac{\Lambda^2}{m^2_N} - \frac{1}{4}\,{\ell
  n}\frac{m^2_{\sigma}}{m^2_N}\Big) \gamma_{\mu} -
\frac{g^2_{\pi N}}{8\pi^2} \Big(- {\ell n}\frac{\Lambda^2}{m^2_N} +
\frac{5}{4}\, {\ell n}\frac{m^2_{\sigma}}{m^2_N}\Big)
\gamma_{\mu} \gamma^5\nonumber\\
\hspace{-0.15in}&+& \frac{5 g^2_{\pi N}}{16\pi^2} \frac{i
  \sigma_{\mu\nu}q^{\nu}}{2 m_N} - \frac{2 m_N q_{\mu}}{m^2_{\pi} -
  q^2 - i0}\,\gamma^5\, \Big[\frac{g^2_{\pi N}}{8\pi^2}\,\Big( {\ell
    n}\frac{m^2_{\sigma}}{m^2_N} + {\ell
    n}\frac{\Lambda^2}{m^2_N}\Big)\Big] \Big\}\,u_n\big(\vec{k}_n,
\sigma_n\big).
\end{eqnarray}
The contributions of the Feynman diagrams in Fig.\,\ref{fig:fig3a} and
Fig.\,\ref{fig:fig4a} are calculated within the dimensional
regularization \cite{Itzykson1980} with the replacement $\Gamma(2 -
n/2) \to {\ell n}(\Lambda^2/m^2_N)$.

\section*{Appendix B: Analytical expressions for the Feynman diagrams 
in Fig.\,\ref{fig:fig6a} defining
one--photon--loop corrections to the amplitude of the neutron
$\beta^-$--decay, calculated in the tree--approximation for strong
low--energy interactions}
\renewcommand{\theequation}{B-\arabic{equation}}
\setcounter{equation}{0}

In this Appendix we calculate one--photon--loop radiative corrections
to the part of the amplitude of the neutron $\beta^-$--decay, given by
the Feynman diagram in Fig.\,\ref{fig:fig1a}a. The Feynman diagrams,
describing one--photon--loop radiative corrections to this part of the
amplitude of the neutron $\beta^-$--decay with the analytical
expression given in Eq.(\ref{eq:60}), are shown in
Fig.\,\ref{fig:fig6a}.  Since the radiative corrections to the part of
the amplitude of the neutron $\beta^-$--decay, induced by the
one--pion--pole contribution, is of order $10^{-9}$ \cite{Ivanov2018c}
we neglect it in comparison with the corrections of order $10^{-5}$,
which we are searching for in this paper.

\subsection*{Analytical calculation of the one--loop--photon exchange 
radiative corrections to the part of the amplitude of the neutron
$\beta^-$--decay, described by the Feynman diagram in
Fig.\,\ref{fig:fig6a}}

The analytical expressions for the Feynman diagrams in
Fig.\,\ref{fig:fig6a} are given by
\begin{eqnarray}\label{eq:B.1}
\hspace{-0.3in}&&M^{(W\gamma)}(n\to p\,e^-\bar{\nu}_e)_{\rm
  Fig.\,\ref{fig:fig6a}a} = (-e^2 G_V M^2_W)\int
\frac{d^4p}{(2\pi)^4i}\,\Big[\bar{u}_p\big(\vec{k}_p,
  \sigma_p\big)\,\gamma^{\alpha}\frac{1}{m_N - \hat{p} - \hat{k}_p -
    i0}\gamma^{\mu}(1 - \gamma^5)u_n\big(\vec{k}_n,
  \sigma_n\big)\Big]\nonumber\\
\hspace{-0.3in}&&\times\, D^{(W)}_{\mu\nu}(- p -
q)D^{(\gamma)}_{\alpha\beta}(p)\Big[\bar{u}_e\big(\vec{k}_e,
  \sigma_e\big)\,\gamma^{\beta}\frac{1}{m_e + \hat{p} - \hat{k}_e -
    i0} \gamma^{\nu}\big(1 - \gamma^5\big)v_{\bar{\nu}}\big(\vec{k}_{\bar{\nu}},
  + \frac{1}{2}\big)\Big]
\end{eqnarray}
and
\begin{eqnarray}\label{eq:B.2}
\hspace{-0.3in}&&M^{(W\gamma)}(n\to p\,e^-\bar{\nu}_e)_{\rm
  Fig.\,\ref{fig:fig6a}b} = (+ e^2 G_V M^2_W)\int
\frac{d^4p}{(2\pi)^4i}\,\Big[\bar{u}_p\big(\vec{k}_p,
  \sigma_p\big)\,\gamma^{\alpha}\frac{1}{m_N - \hat{p} - \hat{k}_p -
    i0}\gamma^{\mu}(1 - \gamma^5)u_n\big(\vec{k}_n,
  \sigma_n\big)\Big]\nonumber\\
\hspace{-0.3in}&&\times\, \big[(p + q)^{\varphi}\eta^{\rho\beta} - (p
  + 2q)^{\rho}\eta^{\varphi\beta} + q^{\beta}\eta^{\rho\varphi}\big]
D^{(W)}_{\mu\beta}(-p - q) D^{(\gamma)}_{\alpha\rho}(p)
D^{(W)}_{\varphi\nu}(-q)\Big[\bar{u}_e\big(\vec{k}_e, \sigma_e\big)
  \gamma^{\nu}\big(1 - \gamma^5\big)v_{\bar{\nu}}\big(\vec{k}_{\bar{\nu}}, +
  \frac{1}{2}\big)\Big],
\end{eqnarray}
and
\begin{eqnarray}\label{eq:B.3}
\hspace{-0.3in}&&M^{(W\gamma)}(n\to p\,e^-\bar{\nu}_e)_{\rm
  Fig.\,\ref{fig:fig6a}c} = (- e^2 G_V M^2_W)\int
\frac{d^4p}{(2\pi)^4i}\,\Big[\bar{u}_p\big(\vec{k}_p,
  \sigma_p\big)\,\gamma^{\mu}(1 - \gamma^5)u_n\big(\vec{k}_n,
  \sigma_n\big)\Big] D^{(W)}_{\mu\varphi}(- q)D^{(W)}_{\beta\nu}(- p -
q) D^{(\gamma)}_{\alpha\rho}(p)\nonumber\\
\hspace{-0.3in}&&\times\, \big[(p + q)^{\varphi}\eta^{\rho\beta} - (p
  + 2q)^{\rho}\eta^{\varphi\beta} +
  q^{\beta}\eta^{\rho\varphi}\big]\Big[\bar{u}_e\big(\vec{k}_e,
  \sigma_e\big)\,\gamma^{\alpha}\,\frac{1}{m_e + \hat{p} - \hat{k}_e -
    i0}\, \gamma^{\nu}\big(1 - \gamma^5\big)
 v_{\bar{\nu}}\big(\vec{k}_{\bar{\nu}}, + \frac{1}{2}\big)\Big]
\end{eqnarray}
and
\begin{eqnarray}\label{eq:B.4}
\hspace{-0.3in}&&M^{(W\gamma)}(n\to p\,e^-\bar{\nu}_e)_{\rm
  Fig.\,\ref{fig:fig6a}d} = (+ e^2 G_V M^2_W)\Big[\bar{u}_p\big(\vec{k}_p,
  \sigma_p\big)\,\gamma^{\mu}(1 - \gamma^5) u_n\big(\vec{k}_n,
  \sigma_n\big)\Big]\Big[\bar{u}_e\big(\vec{k}_e, \sigma_e\big)
  \gamma^{\nu}\big(1 - \gamma^5\big)v_{\bar{\nu}}\big(\vec{k}_{\bar{\nu}}, +
  \frac{1}{2}\big)\Big] \nonumber\\
\hspace{-0.3in}&&\times \int \frac{d^4p}{(2\pi)^4i}\,\big[(p +
  q)^{\omega}\eta^{\alpha\varphi} - (p + 2q)^{\alpha}\eta^{\varphi
    \omega} + q^{\varphi} \eta^{\alpha\omega}\big]
D^{(W)}_{\mu\omega}(-q) D^{(W)}_{\varphi\beta}(-p - q) D^{(W)}_{\kappa
  \nu}(- q) D^{(\gamma)}_{\alpha\rho}(p)\nonumber\\
\hspace{-0.3in}&&\times\, \big[(p + q)^{\kappa}\eta^{\beta\rho} - (p +
  2q)^{\rho} \eta^{\beta\kappa} + q^{\beta} \eta^{\kappa\rho}\big]
\end{eqnarray}
and 
\begin{eqnarray}\label{eq:B.5}
\hspace{-0.3in}&&M^{(W\gamma)}(n\to p\,e^-\bar{\nu}_e)_{\rm
  Fig.\,\ref{fig:fig6a}e} = (+ e^2 G_V M^2_W) \int
\frac{d^4p}{(2\pi)^4i}\, \Big[\bar{u}_p\big(\vec{k}_p,
  \sigma_p\big)\,\gamma^{\alpha}\,\frac{1}{m_N - \hat{p} - \hat{k}_p -
    i0}\,\gamma^{\rho}\,D^{(\gamma)}_{\alpha\rho}(p) \nonumber\\
\hspace{-0.3in}&&\times\,\frac{1}{m_N - \hat{k}_p -
  i0}\,\gamma^{\mu}(1 - \gamma^5)\,u_n\big(\vec{k}_n,
\sigma_n\big)\Big] D^{(W)}_{\mu\nu}(- q)\Big[\bar{u}_e\big(\vec{k}_e,
  \sigma_e\big) \gamma^{\nu}\big(1 - \gamma^5\big)
 v_{\bar{\nu}}\big(\vec{k}_{\bar{\nu}}, + \frac{1}{2}\big)\Big]
\end{eqnarray}
and 
\begin{eqnarray}\label{eq:B.6}
\hspace{-0.3in}&&M^{(W\gamma)}(n\to p\,e^-\bar{\nu}_e)_{\rm
  Fig.\,\ref{fig:fig6a}f} = (+ e^2 G_V M^2_W)\,
\Big[\bar{u}_p\big(\vec{k}_p, \sigma_p\big)\,\gamma^{\mu}(1 -
  \gamma^5)\,u_n\big(\vec{k}_n, \sigma_n\big)\Big] D^{(W)}_{\mu\nu}(-
q) \nonumber\\
\hspace{-0.3in}&&\times \int
\frac{d^4p}{(2\pi)^4i}\Big[\bar{u}_e\big(\vec{k}_e,
  \sigma_e\big)\,\gamma^{\alpha}\,\frac{1}{m_e + \hat{p} - \hat{k}_e -
    i0}\,\gamma^{\rho}\,D^{(\gamma)}_{\alpha\rho}(p)\,\frac{1}{m_e -
    \hat{k}_e - i0}\, \gamma^{\nu}\big(1 - \gamma^5\big)
 v_{\bar{\nu}}\big(\vec{k}_{\bar{\nu}}, + \frac{1}{2}\big)\Big],
\end{eqnarray}
where $D^{(W)}_{\alpha\beta}(k)$ and $D^{(\gamma)}_{\alpha\beta}(p)$
are the propagators of the electroweak $W^-$--boson and photon,
respectively, defined by
\begin{eqnarray}\label{eq:B.7}
D^{(W)}_{\alpha\beta}(k) = \frac{1}{M^2_W - k^2 - i0}\,\Big( -
\eta_{\alpha\beta} + \frac{k_{\alpha}k_{\beta}}{M^2_W}\Big)\quad,
\quad D^{(\gamma)}_{\alpha\beta}(p) = \frac{1}{p^2 +
  i0}\,\Big(\eta_{\alpha\beta} - \xi\,\frac{p_{\alpha}
  p_{\beta}}{p^2 + i0}\Big).
\end{eqnarray}
The propagator of the electroweak $W^-$--boson is taken in the
physical gauge, whereas the photon propagator depends on the gauge
parameter $\xi$. For the analysis of independence of the Feynman
diagrams of the gauge parameter $\xi$ we have to define the inverse
propagator of the electroweak $W^-$--boson $D^{(W)-1}_{\alpha\beta}$
equal to
\begin{eqnarray}\label{eq:B.8}
D^{(W)-1}_{\alpha\beta}(k) = - k_{\alpha}k_{\beta} - (M^2_W -
k^2)\,\eta_{\alpha\beta}
\end{eqnarray}
and obeying the constraint $D^{(W)-1}_{\alpha\rho}(k)
D^{(W)\rho}{}_{\beta}(k) = \eta_{\alpha\beta}$.

\subsection*{Gauge invariance of the Feynman diagrams
 in Fig.\,\ref{fig:fig6a} or their independence of a gauge parameter
 $\xi$}

For the analysis of independence of the Feynman diagrams in
Fig.\,\ref{fig:fig6a} of a gauge parameter $\xi$ we consider the
following expressions
\begin{eqnarray}\label{eq:B.9}
\hspace{-0.3in}&& \frac{\partial}{\partial \xi}M^{(W\gamma)}(n\to
p\,e^-\bar{\nu}_e)_{\rm Fig.\,\ref{fig:fig6a}a} = - e^2 G_V
M^2_W\Big[\bar{u}_p\big(\vec{k}_p, \sigma_p\big)\,\gamma^{\mu}(1 -
  \gamma^5)u_n\big(\vec{k}_n,
  \sigma_n\big)\Big]\Big[\bar{u}_e\big(\vec{k}_e,
  \sigma_e\big)\,\gamma^{\nu}\big(1 - \gamma^5\big)
 v_{\bar{\nu}}\big(\vec{k}_{\bar{\nu}}, + \frac{1}{2}\big)\Big]\nonumber\\
\hspace{-0.3in}&&\times\,\int
\frac{d^4p}{(2\pi)^4i}\,\frac{1}{(p^2 + i0)^2}\, D^{(W)}_{\mu\nu}(- p -
q)
\end{eqnarray}
and 
\begin{eqnarray}\label{eq:B.10}
\hspace{-0.3in}&& \frac{\partial}{\partial \xi}M^{(W\gamma)}(n\to
p\,e^-\bar{\nu}_e)_{\rm Fig.\,\ref{fig:fig6a}b} = + e^2 G_V
M^2_W\Big[\bar{u}_p\big(\vec{k}_p, \sigma_p\big)\,\gamma^{\mu}(1 -
  \gamma^5)u_n\big(\vec{k}_n,
  \sigma_n\big)\Big]\Big[\bar{u}_e\big(\vec{k}_e,
  \sigma_e\big)\,\gamma^{\nu}\big(1 - \gamma^5\big)
 v_{\bar{\nu}}\big(\vec{k}_{\bar{\nu}}, + \frac{1}{2}\big)\Big]\nonumber\\
\hspace{-0.3in}&&\times\,\int \frac{d^4p}{(2\pi)^4i}\,\frac{1}{(p^2 +
  i0)^2}\,\big[D^{(W)-1}_{\beta\varphi}(- q) -
  D^{(W)-1}_{\beta\varphi}(- p - q)\big] D^{(W)\beta}{}_{\mu}(- p - q)
D^{(W)\varphi}{}_{\nu}(- q) =\nonumber\\
\hspace{-0.3in}&&= + e^2 G_V
M^2_W\Big[\bar{u}_p\big(\vec{k}_p, \sigma_p\big)\,\gamma^{\mu}(1 -
  \gamma^5)u_n\big(\vec{k}_n,
  \sigma_n\big)\Big]\Big[\bar{u}_e\big(\vec{k}_e,
  \sigma_e\big)\,\gamma^{\nu}\big(1 - \gamma^5\big)
 v_{\bar{\nu}}\big(\vec{k}_{\bar{\nu}}, + \frac{1}{2}\big)\Big]\nonumber\\
\hspace{-0.3in}&&\times\,\int \frac{d^4p}{(2\pi)^4i}\,\frac{1}{(p^2 +
  i0)^2}\,\big[D^{(W)}_{\mu\nu}(- p- q) -
  D^{(W)}_{\mu\nu}(- q)\big]
\end{eqnarray}
and 
\begin{eqnarray}\label{eq:B.11}
\hspace{-0.3in}&& \frac{\partial}{\partial \xi}M^{(W\gamma)}(n\to
p\,e^-\bar{\nu}_e)_{\rm Fig.\,\ref{fig:fig6a}c} = + e^2 G_V
M^2_W\Big[\bar{u}_p\big(\vec{k}_p, \sigma_p\big)\,\gamma^{\mu}(1 -
  \gamma^5)u_n\big(\vec{k}_n,
  \sigma_n\big)\Big]\Big[\bar{u}_e\big(\vec{k}_e,
  \sigma_e\big)\,\gamma^{\nu}\big(1 - \gamma^5\big)
 v_{\bar{\nu}}\big(\vec{k}_{\bar{\nu}}, + \frac{1}{2}\big)\Big]\nonumber\\
\hspace{-0.3in}&&\times\,\int \frac{d^4p}{(2\pi)^4i}\,\frac{1}{(p^2 +
  i0)^2}\,\big[D^{(W)}_{\mu\nu}(- p- q) -
  D^{(W)}_{\mu\nu}(- q)\big]
\end{eqnarray}
and 
\begin{eqnarray}\label{eq:B.12}
\hspace{-0.3in}&& \frac{\partial}{\partial \xi}M^{(W\gamma)}(n\to
p\,e^-\bar{\nu}_e)_{\rm Fig.\,\ref{fig:fig6a}d} = - e^2 G_V M^2_W
\Big[\bar{u}_p\big(\vec{k}_p, \sigma_p\big)\,\gamma^{\mu}(1 -
  \gamma^5)u_n\big(\vec{k}_n,
  \sigma_n\big)\Big]\Big[\bar{u}_e\big(\vec{k}_e,
  \sigma_e\big)\,\gamma^{\nu}\big(1 - \gamma^5\big)
 v_{\bar{\nu}}\big(\vec{k}_{\bar{\nu}}, + \frac{1}{2}\big)\Big]\nonumber\\
\hspace{-0.3in}&&\times\,\int \frac{d^4p}{(2\pi)^4i}\,\frac{1}{(p^2 +
  i0)^2}\,\big[D^{(W)}_{\mu\nu}(- p- q) - 2 D^{(W)}_{\mu\nu}(- q) +
  D^{(W)}_{\mu}{}^{\beta}(- q) D^{(W)-1}_{\beta\kappa}(- p - q)
  D^{(W)\kappa}{}_{\nu}(-q)\big]
\end{eqnarray}
and 
\begin{eqnarray}\label{eq:B.13}
\hspace{-0.3in}&& \frac{\partial}{\partial \xi}M^{(W\gamma)}(n\to
p\,e^-\bar{\nu}_e)_{\rm Fig.\,\ref{fig:fig6a}e} = (+ e^2 G_V M^2_W)
\int \frac{d^4p}{(2\pi)^4i}\, \Big[\bar{u}_p\big(\vec{k}_p,
  \sigma_p\big)\,\frac{\hat{p}}{(p^2 + i0)^2}\,\frac{1}{m_N - \hat{k}_p -
  i0}\,\gamma^{\mu}(1 - \gamma^5)\,u_n\big(\vec{k}_n,
\sigma_n\big)\Big]\nonumber\\
\hspace{-0.3in}&&\times\, D^{(W)}_{\mu\nu}(-
q)\Big[\bar{u}_e\big(\vec{k}_e, \sigma_e\big) \gamma^{\nu}\big(1 -
  \gamma^5\big)v_{\bar{\nu}}\big(\vec{k}_{\bar{\nu}}, + \frac{1}{2}\big)\Big]
\end{eqnarray}
and 
\begin{eqnarray}\label{eq:B.14}
\hspace{-0.3in}&&\frac{\partial}{\partial \xi} M^{(W\gamma)}(n\to
p\,e^-\bar{\nu}_e)_{\rm Fig.\,\ref{fig:fig6a}f} = - e^2 G_V M^2_W\,
\Big[\bar{u}_p\big(\vec{k}_p, \sigma_p\big)\,\gamma^{\mu}(1 -
  \gamma^5)\,u_n\big(\vec{k}_n, \sigma_n\big)\Big] D^{(W)}_{\mu\nu}(-
q) \nonumber\\
\hspace{-0.3in}&&\times \int
\frac{d^4p}{(2\pi)^4i}\Big[\bar{u}_e\big(\vec{k}_e,
  \sigma_e\big)\,\frac{\hat{p}}{(p^2 + i0)^2}\,\frac{1}{m_e -
    \hat{k}_e - i0}\, \gamma^{\nu}\big(1 - \gamma^5\big)
 v_{\bar{\nu}}\big(\vec{k}_{\bar{\nu}}, + \frac{1}{2}\big)\Big].
\end{eqnarray}
Summing up the contributions we arrive at the expression
\begin{eqnarray}\label{eq:B.15}
\hspace{-0.3in}&& \frac{\partial}{\partial \xi}M^{(\gamma)}(n\to
p\,e^-\bar{\nu}_e)_{\rm Fig.\,\ref{fig:fig6a}} = e^2 G_V M^2_W
\Big[\bar{u}_p\big(\vec{k}_p, \sigma_p\big)\,\gamma^{\mu}(1 -
  \gamma^5)u_n\big(\vec{k}_n,
  \sigma_n\big)\Big]\Big[\bar{u}_e\big(\vec{k}_e,
  \sigma_e\big)\,\gamma^{\nu}\big(1 - \gamma^5\big)
 v_{\bar{\nu}}\big(\vec{k}_{\bar{\nu}}, + \frac{1}{2}\big)\Big]\nonumber\\
\hspace{-0.3in}&&\times\, D^{(W)\beta}{}_{\mu}(- q) \Big\{\int
\frac{d^4p}{(2\pi)^4i}\,\frac{1}{(p^2 +
  i0)^2}\,D^{(W)-1}_{\beta\kappa}(- p - q)\Big\}\,
D^{(W)\kappa}{}_{\nu}(-q)\big],
\end{eqnarray}
where the contributions of Eq.(\ref{eq:B.13}) and Eq.(\ref{eq:B.14})
vanish because of integration over directions of the 4--momentum
$p$. Using a dimensional regularization we get
\begin{eqnarray}\label{eq:B.16}
\hspace{-0.3in}&& \frac{\partial}{\partial \xi}M^{(W\gamma)}(n\to
p\,e^-\bar{\nu}_e)_{\rm Fig.\,\ref{fig:fig6a}} = e^2 G_V M^2_W
\Big[\bar{u}_p\big(\vec{k}_p, \sigma_p\big)\,\gamma^{\mu}(1 -
  \gamma^5)u_n\big(\vec{k}_n,
  \sigma_n\big)\Big]\Big[\bar{u}_e\big(\vec{k}_e,
  \sigma_e\big)\,\gamma^{\nu}\big(1 - \gamma^5\big)
 v_{\bar{\nu}}\big(\vec{k}_{\bar{\nu}}, + \frac{1}{2}\big)\Big]\nonumber\\
\hspace{-0.3in}&&\times\, \Big\{D^{(W)\beta}{}_{\mu}(- q)
D^{(W)}_{\beta \nu}(-q)\,\Big(1 - \frac{1}{n}\Big)\int
\frac{d^np}{(2\pi)^ni}\,\frac{1}{p^2 + i0} + D^{(W)}_{\mu\nu}(-q) \int
\frac{d^np}{(2\pi)^ni}\,\frac{1}{(p^2 + i0)^2} \Big\}.
\end{eqnarray}
Since in dimensional regularization these integrals are equal to zero
\cite{t'Hooft1973,Capper1972,Capper1973,Capper1974}, the r.h.s. of
Eq.(\ref{eq:B.16}) vanishes and we obtain
\begin{eqnarray}\label{eq:B.17}
\hspace{-0.3in}&& \frac{\partial}{\partial \xi}M^{(W\gamma)}(n\to
p\,e^-\bar{\nu}_e)_{\rm Fig.\,\ref{fig:fig6a}} = 0
\end{eqnarray}
that confirms independence of the Feynman diagrams in
Fig.\,\ref{fig:fig6a} of the gauge parameter $\xi$ of the photon
propagator or gauge invariance of these diagrams
\cite{Ivanov2019,Gerstein1969}. We would like to accentuate that gauge
invariance of the Feynman diagrams in Fig.\,\ref{fig:fig6a} retains
even for $g_A \neq 1$. As result, the calculation of the radiative
corrections, described by the Feynman diagrams in
Fig.\,\ref{fig:fig6a} with the replacement $\gamma^{\mu}(1 - \gamma^5)
\to \gamma^{\mu}(1 - g_A\gamma^5)$, we perform within dimensional
regularization using the Feynman gauge for the photon
propagator. Then, it is convenient to represent the electroweak
$W^-$--boson propagator $D^{(W)}_{\alpha\beta}(- p - q)$ in the
following form
\begin{eqnarray}\label{eq:B.18}
\hspace{-0.3in}D^{(W)}_{\alpha\beta}(- p - q) = -
\frac{1}{M^2_W}\,\eta_{\alpha\beta} - \frac{1}{M^2_W}\,\frac{(p + q)^2
  \eta_{\alpha\beta} - (p + q)_{\alpha} (p + q)_{\beta}}{M^2_W - (p +
  q)^2 - i0}.
\end{eqnarray}
We would like to emphasize that after the calculation of the Feynman
diagrams the contribution of the second term is not negligible in
comparison with the contribution of the first one. 

\subsection*{\bf Analytical calculation  of the Feynman diagram in 
Fig.\, \ref{fig:fig6a}a}

The analytical expression for the Feynman diagram in
Fig.\,\ref{fig:fig6a}a, taken in the Feynman gauge for the photon
propagator, is
\begin{eqnarray}\label{eq:B.19}
\hspace{-0.3in}&&M^{(W\gamma)}(n\to p\,e^-\bar{\nu}_e)_{\rm
  Fig.\,\ref{fig:fig6a}a} = - G_V\Big\{(-e^2)\int
\frac{d^4p}{(2\pi)^4i}\,\frac{\eta_{\mu\nu}}{p^2 +
  i0}\,\Big[\bar{u}_p\big(\vec{k}_p,
  \sigma_p\big)\,\gamma^{\alpha}\frac{1}{m_N - \hat{p} - \hat{k}_p -
    i0}\gamma^{\mu}(1 - g_A\gamma^5)u_n\big(\vec{k}_n, \sigma_n\big)\Big]
\nonumber\\
\hspace{-0.3in}&&\times\, \Big[\bar{u}_e\big(\vec{k}_e,
  \sigma_e\big)\,\gamma_{\alpha}\frac{1}{m_e + \hat{p} - \hat{k}_e -
    i0} \gamma^{\nu}\big(1 - \gamma^5\big)v_{\bar{\nu}}\big(\vec{k}_{\bar{\nu}},
  + \frac{1}{2}\big)\Big] + (- e^2)\int
\frac{d^4p}{(2\pi)^4i}\,\frac{1}{p^2 + i0}\,\frac{(p +
  q)^2 \eta_{\mu\nu} - (p + q)_{\mu} (p + q)_{\nu}}{M^2_W - (p + q)^2
  - i0}\nonumber\\
\hspace{-0.3in}&&\times\,\Big[\bar{u}_p\big(\vec{k}_p,
  \sigma_p\big)\,\gamma^{\alpha}\frac{1}{m_N - \hat{p} - \hat{k}_p -
    i0}\gamma^{\mu}(1 - \gamma^5)u_n\big(\vec{k}_n, \sigma_n\big)\Big]
\Big[\bar{u}_e\big(\vec{k}_e,
  \sigma_e\big)\,\gamma_{\alpha}\frac{1}{m_e + \hat{p} - \hat{k}_e -
    i0} \gamma^{\nu}\big(1 -
  \gamma^5\big)v_{\bar{\nu}}\big(\vec{k}_{\bar{\nu}}, +
  \frac{1}{2}\big)\Big]\Big\}.\nonumber\\
\hspace{-0.3in}&&
\end{eqnarray}
The first integral in Eq.(\ref{eq:B.19}) was calculated by Sirlin
\cite{Sirlin1967} to leading order in the large nucleon mass $m_N$
expansion. The details of this calculation one may find in
\cite{Ivanov2013}. Here we calculate this integral by taking into
account next--to--leading contributions in the large nucleon mass
$m_N$ expansion. Following \cite{Ivanov2013} we rewrite the first
integral in the r.h.s. of Eq.(\ref{eq:B.19}) as follows
\begin{eqnarray}\label{eq:B.20}
\hspace{-0.3in}{\cal M}^{(\Box W\gamma)}_{1 + 2} = -\, \frac{\alpha}{4\pi}
\int\frac{d^4p}{\pi^2i}\,\frac{[\bar{u}_p\gamma^{\alpha}(m_p +
    \hat{k}_p + \hat{p})\gamma^{\mu}(1 - g_A\gamma^5)
    u_n]\,[\bar{u}_e\gamma_{\alpha} (m_e + \hat{k}_e -
    \hat{p})\gamma_{\mu}(1 - \gamma^5) v_{\bar{\nu}}]}{(p^2 + i0)(p^2
  + 2 p\cdot k_p + i0)((p^2 - 2 p\cdot k_e + i0)},
\end{eqnarray}
where we have set $e^2 = 4\pi \alpha$. According to \cite{Ivanov2013},
we transcribe the numerator into the form
\begin{eqnarray}\label{eq:B.21}
\hspace{-0.3in}&&[\bar{u}_p\gamma^{\alpha}(m_p + \hat{k}_p +
  \hat{p})\gamma^{\mu}(1 - g_A \gamma^5)
  u_n]\,[\bar{u}_e\gamma_{\alpha} (m_e + \hat{k}_e -
  \hat{p})\gamma_{\mu}(1 - \gamma^5) v_{\bar{\nu}}] = - (p^2 - 2 p
\cdot k_e) [\bar{u}_p\gamma^{\mu}(1 - g_A \gamma^5) u_n]\nonumber\\
\hspace{-0.3in}&&\times\,[\bar{u}_e \gamma_{\mu}(1 - \gamma^5)
  v_{\bar{\nu}}]- (p^2 + 2 p \cdot k_p) [\bar{u}_p\gamma^{\mu}(1 - g_A
  \gamma^5) u_n]\,[\bar{u}_e \gamma_{\mu}(1 - \gamma^5) v_{\bar{\nu}}]
- 2 p^2[\bar{u}_p\gamma^{\mu}(1 - g_A
  \gamma^5)u_n]\nonumber\\
\hspace{-0.3in}&&\times\, [\bar{u}_e\gamma_{\mu}(1 -
  \gamma^5)v_{\bar{\nu}}] + 4(k_e\cdot k_p)[\bar{u}_p\gamma^{\mu}(1 -
  g_A \gamma^5) u_n]\,[\bar{u}_e \gamma_{\mu}(1 - \gamma^5)
  v_{\bar{\nu}}] - 2(1 - g_A)[\bar{u}_p\hat{p}(1 +
  \gamma^5)u_n][\bar{u}_e\hat{p}(1 - \gamma^5)v_{\bar{\nu}}]\nonumber\\
\hspace{-0.3in}&& + 2(1 -
g_A)p^2 [\bar{u}_p\gamma^{\mu}(1 +
  \gamma^5)u_n][\bar{u}_e\gamma_{\mu}(1 -
  \gamma^5)v_{\bar{\nu}}] +
2i[\bar{u}_p\sigma_{\alpha\beta}p^{\alpha}k^{\beta}_e \gamma^{\mu}(1 -
  g_A \gamma^5) u_n]\,[\bar{u}_e \gamma_{\mu}(1 - \gamma^5)
  v_{\bar{\nu}}]\nonumber\\
\hspace{-0.3in}&& - 2i[\bar{u}_p\gamma^{\mu}(1 - g_A\gamma^5)
  u_n]\,[\bar{u}_e \sigma_{\alpha\beta}p^{\alpha}k^{\beta}_p
  \gamma_{\mu}(1 - \gamma^5) v_{\bar{\nu}}]
\end{eqnarray}
and represent ${\cal M}^{(\Box W\gamma)}_{1 + 2}$ as ${\cal M}^{(\Box
  W\gamma)}_{1 + 2} = {\cal M}^{(\Box W\gamma)}_1 + {\cal M}^{(\Box
  W\gamma)}_2$, where ${\cal M}^{(\Box W\gamma)}_1$ and ${\cal
  M}^{(\Box W\gamma)}_2$ are given by
\begin{eqnarray}\label{eq:B.22}
\hspace{-0.3in}&& {\cal M}^{(\Box W\gamma)}_1 =
\frac{\alpha}{4\pi}\,[\bar{u}_p\gamma^{\mu}(1 - g_A
  \gamma^5)u_n][\bar{u}_e\gamma_{\mu}(1 -
  \gamma^5)v_{\bar{\nu}}]\Big\{\int \frac{d^4p}{\pi^2i}\frac{1}{p^2 +
  i0}\frac{1}{p^2 + 2 p \cdot k_p + i0} + \int \frac{d^4p}{\pi^2i}
\frac{1}{p^2 + i0}\frac{1}{p^2 - 2 p \cdot k_e + i0}\nonumber\\
\hspace{-0.3in}&&+ 2\int \frac{d^4p}{\pi^2i}\frac{1}{p^2 + 2 p \cdot
  k_p + i0}\frac{1}{p^2 - 2 p\cdot k_e + i0} - 4(k_e\cdot k_p) \int
\frac{d^4p}{\pi^2i}\frac{1}{p^2 + i0}\frac{1}{p^2 + 2 p \cdot k_p +
  i0} \frac{1}{p^2 - 2 p \cdot k_e + i0}\Big\}.
\end{eqnarray}
and
\begin{eqnarray*}
\hspace{-0.3in}&&{\cal M}^{(\Box W\gamma)}_2 =
\frac{\alpha}{4\pi}\,\Big\{- 2(1 - g_A)\, [\bar{u}_p\gamma^{\mu}(1 +
  \gamma^5)u_n][\bar{u}_e\gamma_{\mu}(1 - \gamma^5)v_{\bar{\nu}}] \int
\frac{d^4p}{\pi^2i} \frac{1}{p^2 + 2 p \cdot k_p + i0}\frac{1}{p^2 - 2
  p \cdot k_e + i0}\nonumber\\
\hspace{-0.3in}&& + 2(1 - g_A)\int \frac{d^4p}{\pi^2i}\frac{1}{p^2 +
  i0}\frac{1}{p^2 + 2 p \cdot k_p + i0}\frac{1}{p^2 - 2 p \cdot k_e +
  i0}\,[\bar{u}_p\hat{p}(1 + \gamma^5)u_n][\bar{u}_e\hat{p}(1 -
  \gamma^5)v_{\bar{\nu}}] \nonumber\\
\end{eqnarray*}
\begin{eqnarray}\label{eq:B.23}
\hspace{-0.3in}&& + 2i\int \frac{d^4p}{\pi^2i}\frac{1}{p^2 +
  i0}\frac{1}{p^2 + 2 p \cdot k_p + i0}\frac{1}{p^2 - 2 p \cdot k_e +
  i0}\Big([\bar{u}_p\sigma_{\alpha\beta}
  p^{\alpha}k^{\beta}_e\gamma^{\mu}(1 - g_A
  \gamma^5)u_n][\bar{u}_e\gamma_{\mu}(1 -
  \gamma^5)v_{\bar{\nu}}]\nonumber\\
\hspace{-0.3in}&& - [\bar{u}_p\gamma^{\mu}(1 - g_A
  \gamma^5)u_n][\bar{u}_e\sigma_{\alpha\beta}p^{\alpha}k^{\beta}_p\gamma_{\mu}(1
  - \gamma^5)v_{\bar{\nu}}]\Big)\Big\},
\end{eqnarray}
respectively. Using the dimensional regularization and the results,
obtained in \cite{Ivanov2013} (see Appendix C of
Ref.\cite{Ivanov2013}), we get for ${\cal M}^{(\Box W\gamma)}_1$ the
following expression
\begin{eqnarray}\label{eq:B.24}
{\cal M}^{(\Box W\gamma)}_1 &=&
\frac{\alpha}{2\pi}\,\Big\{\Big[2\,{\ell n}\frac{\Lambda^2}{m^2_N} + 4
  + \frac{1}{2}\,{\ell n}\Big(\frac{m^2_N}{m^2_e}\Big) + 2{\ell
    n}\Big(\frac{\mu}{m_e}\Big) + 2{\ell
    n}\Big(\frac{\mu}{m_e}\Big)\,\Big[\frac{1}{2\beta}\,{\ell
      n}\Big(\frac{1 + \beta}{1 - \beta}\Big) - 1\Big] -
  \frac{1}{4\beta}\,{\ell n}^2\Big(\frac{1 + \beta}{1 - \beta}\Big)
  \nonumber\\
\hspace{-0.3in}&-& \frac{1}{\beta}\,{\rm Li}_2\Big(\frac{2\beta}{1 +
  \beta}\Big)\Big] + \frac{E_e}{m_N}\Big[1 +\frac{1 -
    \beta^2}{\beta}\,{\ell n}\Big(\frac{1 + \beta}{1 -
    \beta}\Big)\Big]\Big\}[\bar{u}_p\gamma^{\mu}(1 - g_A
  \gamma^5)u_n][\bar{u}_e\gamma_{\mu}(1 - \gamma^5)v_{\bar{\nu}}],
\end{eqnarray}
where ${\rm Li}_2(x)$ is the Polylogarithmic function of a real
argument $x \le 1$ \cite{Mitchel1949}. For the calculation of the
terms of order $O(E_e/m_N)$ we have used the technique of the
calculation of the integrals Eq.(C-5) and Eq.(D-38) of
Ref.\cite{Ivanov2013}, and the relations for the Polylogarithmic
functions obtained by Mitchel \cite{Mitchel1949}. In turn, for $ {\cal
  M}^{(\Box W\gamma)}_2 $ after the integration over the virtual
momentum we obtain the following expression
\begin{eqnarray}\label{eq:B.25}
\hspace{-0.3in}&&{\cal M}^{(\Box W\gamma)}_2 =
\frac{\alpha}{4\pi}\,\Big\{- 2(1 - g_A)\, [\bar{u}_p\gamma^{\mu}(1 +
  \gamma^5)u_n][\bar{u}_e\gamma_{\mu}(1 - \gamma^5)v_{\bar{\nu}}]
\Big\{{\ell n}\frac{\Lambda^2}{m^2_N} - \int^1_0dx\,{\ell
  n}\frac{p^2(x)}{m^2_N}\Big\}\nonumber\\
\hspace{-0.3in}&& + 2(1 - g_A) [\bar{u}_p\gamma^{\mu}(1 +
  \gamma^5)u_n] [\bar{u}_e\gamma^{\nu}(1 -
  \gamma^5)v_{\bar{\nu}}]\Big\{\Big[\frac{1}{4}\,{\ell n}\frac{\Lambda^2}{m^2_N} + \frac{1}{4} - \frac{1}{4}\int^1_0dx\,{\ell
    n}\frac{p^2(x)}{m^2_N}\Big]\,\eta_{\mu\nu}\nonumber\\
\hspace{-0.3in}&& - \frac{1}{2} \int^1_0dx\,
\frac{p_{\mu}(x)p_{\nu}(x)}{p^2(x)}\Big\} +
2i[\bar{u}_p\sigma_{\alpha\beta} k^{\alpha}_pk^{\beta}_e\gamma^{\mu}(1
  - g_A \gamma^5)u_n][\bar{u}_e\gamma_{\mu}(1 -
  \gamma^5)v_{\bar{\nu}}]\int^1_0dx\, \frac{(1 -
  x)}{p^2(x)}\nonumber\\
\hspace{-0.3in}&& - 2i[\bar{u}_p\gamma^{\mu}(1 - g_A \gamma^5)u_n]
       [\bar{u}_e\sigma_{\alpha\beta}k^{\alpha}_pk^{\beta}_e\gamma_{\mu}(1
         - \gamma^5)v_{\bar{\nu}}]\int^1_0dx\, \frac{x}{p^2(x)}\Big\},
\end{eqnarray}
where $p(x) = k_ex - k_p(1 - x)$ and $p^2(x) = m^2_ex^2 + m^2_N(1 -
x)^2 - 2m_N E_e x(1 - x)$ \cite{Ivanov2013}. Then, it is convenient to
rewrite Eq.(\ref{eq:B.25}) as follows
\begin{eqnarray}\label{eq:B.26}
\hspace{-0.3in}&&{\cal M}^{(\Box W\gamma)}_2 =
\frac{\alpha}{4\pi}\,\Big\{- 2(1 - g_A)\, [\bar{u}_p\gamma^{\mu}(1 +
  \gamma^5)u_n][\bar{u}_e\gamma_{\mu}(1 - \gamma^5)v_{\bar{\nu}}]
\Big\{{\ell n}\frac{\Lambda^2}{m^2_N} - \int^1_0dx\,{\ell
  n}\frac{p^2(x)}{m^2_N}\Big\}\nonumber\\
\hspace{-0.3in}&& + 2(1 - g_A) [\bar{u}_p\gamma^{\mu}(1 +
  \gamma^5)u_n] [\bar{u}_e\gamma^{\nu}(1 -
  \gamma^5)v_{\bar{\nu}}]\Big\{\Big[\frac{1}{4}\,{\ell n}\frac{\Lambda^2}{m^2_N} + \frac{1}{4} - \frac{1}{4}\int^1_0dx\,{\ell
    n}\Big(\frac{p^2(x)}{m^2_N}\Big)\Big]\,\eta_{\mu\nu}\nonumber\\
\hspace{-0.3in}&& - \frac{1}{2}\,k_{p\mu}k_{p\nu} \int^1_0dx\,
\frac{(1 - x)^2}{p^2(x)} + \frac{1}{2}\,\big(k_{p\mu}k_{e\nu} +
k_{p\nu}k_{e\mu}\big) \int^1_0dx\, \frac{x(1 - x)}{p^2(x)} -
\frac{1}{2}\,k_{e\mu}k_{e\nu} \int^1_0dx\, \frac{x^2}{p^2(x)}\Big\}\nonumber\\
\hspace{-0.3in}&& + 2 m_N\, [\bar{u}_p\hat{k}_e \gamma^{\mu}(1 - g_A
  \gamma^5)u_n] [\bar{u}_e\gamma_{\mu}(1 -
  \gamma^5)v_{\bar{\nu}}]\int^1_0dx\, \frac{(1 - x)}{p^2(x)} + 2 m_e\,
       [\bar{u}_p\gamma^{\mu}(1 -
         g_A\gamma^5)u_n][\bar{u}_e\hat{k}_p\gamma_{\mu}(1 -
         \gamma^5)v_{\bar{\nu}}]\nonumber\\
\hspace{-0.3in}&&\times \int^1_0dx\, \frac{x}{p^2(x)} - 2 k_e\cdot
k_p[\bar{u}_p \gamma^{\mu}(1 - g_A
  \gamma^5)u_n][\bar{u}_e\gamma_{\mu}(1 -
  \gamma^5)v_{\bar{\nu}}]\int^1_0dx\, \frac{1}{p^2(x)}\Big\}.
\end{eqnarray}
The integrals over the Feynman parameter $x$ are equal to
\begin{eqnarray}\label{eq:B.27}
\hspace{-0.3in}&&\int^1_0dx\,{\ell n}\Big(\frac{p^2(x)}{m^2_N}\Big) =
- 2 + \frac{E_e}{m_N}\Big[{\ell n}\Big(\frac{m^2_N}{m^2_e}\Big) -
  \beta\,{\ell n}\Big(\frac{1 + \beta}{1 - \beta}\Big)\Big] +
O\Big(\frac{1}{m^2_N}\Big) + \ldots,\nonumber\\
\hspace{-0.3in}&&\int^1_0dx\, \frac{(1 - x)^2}{p^2(x)} =
\frac{1}{m^2_N}\,\Big\{1 - 2\,\frac{E_e}{m_N}\Big[1 -
  \frac{1}{2}\,{\ell n}\Big(\frac{m^2_N}{m^2_e}\Big) + \frac{1 +
    \beta^2}{4\beta}\,{\ell n}\Big(\frac{1 + \beta}{1 -
    \beta}\Big)\Big]\Big\} + O\Big(\frac{1}{m^4_N}\Big) +
\ldots,\nonumber\\
\hspace{-0.3in}&&\int^1_0dx\, \frac{x(1 - x)}{p^2(x)} = -
\frac{1}{m^2_N}\,\Big\{1 - \frac{1}{2}\,{\ell
  n}\Big(\frac{m^2_N}{m^2_e}\Big) - \frac{1}{2\beta}\,{\ell
  n}\Big(\frac{1 + \beta}{1 - \beta}\Big)\Big]\Big\}  +
O\Big(\frac{1}{m^3_N}\Big) +
  \ldots,\nonumber\\
\hspace{-0.3in}&&\int^1_0dx\, \frac{x^2}{p^2(x)} = - \frac{1}{2 m_N
  E_e \beta}\,{\ell n}\Big(\frac{1 + \beta}{1 - \beta}\Big)  +
O\Big(\frac{1}{m^2_N}\Big) +
\ldots,\nonumber\\
\hspace{-0.3in}&&\int^1_0dx\, \frac{(1 - x)}{p^2(x)} =
\frac{1}{m^2_N}\,\Big[\frac{1}{2}\,{\ell
    n}\Big(\frac{m^2_N}{m^2_e}\Big) - \frac{1}{2 \beta}\,{\ell
    n}\Big(\frac{1 + \beta}{1 - \beta}\Big)\Big] +
O\Big(\frac{1}{m^3_N}\Big) + \ldots,\nonumber\\
\hspace{-0.3in}&&\int^1_0dx\, \frac{x}{p^2(x)} = - \frac{1}{2 m_N E_e
  \beta}\,{\ell n}\Big(\frac{1 + \beta}{1 - \beta}\Big) -
\frac{1}{m^2_N}\,\Big[\frac{1}{2}\,{\ell
    n}\Big(\frac{m^2_N}{m^2_e}\Big) - \frac{1}{2 \beta}\,{\ell
    n}\Big(\frac{1 + \beta}{1 - \beta}\Big)\Big] +
O\Big(\frac{1}{m^3_N}\Big) + \ldots,\nonumber\\
\hspace{-0.3in}&&\int^1_0dx\, \frac{1}{p^2(x)} = - \frac{1}{2 m_N E_e
  \beta}\,{\ell n}\Big(\frac{1 + \beta}{1 - \beta}\Big) +
O\Big(\frac{1}{m^3_N}\Big) + \ldots
\end{eqnarray}
Plugging Eq.(\ref{eq:B.27}) into Eq.(\ref{eq:B.26}) we arrive at the
following analytical expression for ${\cal M}^{(\Box W\gamma)}_2$
\begin{eqnarray}\label{eq:B.28}
\hspace{-0.3in}&&{\cal M}^{(\Box W \gamma)}_2 =
\frac{\alpha}{2\pi}\,\Big\{(g_A - 1)\, [\bar{u}_p\gamma^{\mu}(1 +
  \gamma^5)u_n][\bar{u}_e\gamma_{\mu}(1 - \gamma^5)v_{\bar{\nu}}]
\Big\{\frac{3}{4}\,{\ell n}\frac{\Lambda^2}{m^2_N} + \frac{5}{4} -
\frac{3}{4}\,\frac{E_e}{m_N}\Big[{\ell n}\Big(\frac{m^2_N}{m^2_e}\Big)
  - \beta\,{\ell n}\Big(\frac{1 + \beta}{1 - \beta}\Big)\Big]\Big\}
\nonumber\\
\hspace{-0.3in}&& - (g_A - 1) [\bar{u}_p\gamma^{\mu}(1 +
  \gamma^5)u_n] [\bar{u}_e\gamma^{\nu}(1 -
  \gamma^5)v_{\bar{\nu}}]\Big\{-
\frac{1}{4}\,\frac{k_{p\mu}k_{p\nu}}{m^2_N}\Big\{1 -
2\,\frac{E_e}{m_N}\Big[1 - \frac{1}{2}\,{\ell
    n}\Big(\frac{m^2_N}{m^2_e}\Big) + \frac{1 +
    \beta^2}{4\beta}\,{\ell n}\Big(\frac{1 + \beta}{1 -
    \beta}\Big)\Big]\Big\}\nonumber\\
\hspace{-0.3in}&& - \frac{1}{4}\,\frac{k_{p\mu}k_{e\nu} +
  k_{p\nu}k_{e\mu}}{m^2_N}\Big\{1 - \frac{1}{2}\,{\ell
  n}\Big(\frac{m^2_N}{m^2_e}\Big) - \frac{1}{2\beta}\,{\ell
  n}\Big(\frac{1 + \beta}{1 - \beta}\Big)\Big]\Big\} +
  \frac{1}{4}\,\frac{k_{e\mu}k_{e\nu}}{E_e
    m_N}\,\frac{1}{2\beta}\,{\ell n}\Big(\frac{1 + \beta}{1 -
    \beta}\Big) \Big\} + [\bar{u}_p \frac{\hat{k}_e}{m_N}
    \gamma^{\mu}(1 - g_A \gamma^5)u_n] \nonumber\\
\hspace{-0.3in}&&\times [\bar{u}_e\gamma_{\mu}(1 -
  \gamma^5)v_{\bar{\nu}}]\Big[\frac{1}{2}\,{\ell
    n}\Big(\frac{m^2_N}{m^2_e}\Big) - \frac{1}{2 \beta}\,{\ell
    n}\Big(\frac{1 + \beta}{1 - \beta}\Big)\Big] - \frac{m_e}{2E_e}\,
       [\bar{u}_p\gamma^{\mu}(1 -
         g_A\gamma^5)u_n][\bar{u}_e\frac{\hat{k}_p}{m_N}\gamma_{\mu}(1
         - \gamma^5)v_{\bar{\nu}}]\Big\{\frac{1}{\beta}\,{\ell
         n}\Big(\frac{1 + \beta}{1 - \beta}\Big)\nonumber\\
\hspace{-0.3in}&& + \frac{E_e}{m_N}\,\Big[{\ell
    n}\Big(\frac{m^2_N}{m^2_e}\Big) - \frac{1}{\beta}\,{\ell
    n}\Big(\frac{1 + \beta}{1 - \beta}\Big) \Big]\Big\} + [\bar{u}_p
  \gamma^{\mu}(1 - g_A \gamma^5)u_n][\bar{u}_e\gamma_{\mu}(1 -
  \gamma^5)v_{\bar{\nu}}]\,\frac{1}{2\beta}\,{\ell n}\Big(\frac{1 +
  \beta}{1 - \beta}\Big) \Big\}.
\end{eqnarray}
Summing up Eq.(\ref{eq:B.24}) and Eq.(\ref{eq:B.28}) we obtain the
analytical expression for the amplitude ${\cal M}^{(\Box W\gamma)}_{1 + 2}$. 

For the calculation of the second integral in Eq.(\ref{eq:B.19}) we
represent it in the following form ${\cal M}^{(\Box W\gamma)}_{3+4+5}
= {\cal M}^{(\Box W\gamma)}_3 + {\cal M}^{(\Box W\gamma)}_4 + {\cal
  M}^{(\Box W\gamma)}_5$, where ${\cal M}^{(\Box W\gamma)}_j$ for $j
=3,4,5$ are equal to
\begin{eqnarray}\label{eq:B.29}
\hspace{-0.3in}{\cal M}^{(\Box W\gamma)}_3 &=&
\frac{\alpha}{2\pi}\,\big[\bar{u}_p \gamma^{\mu}(1 - g_A \gamma^5)
  u_n\big]\big[\bar{u}_e \gamma^{\nu}(1 -
  \gamma^5)v_{\bar{\nu}}\big]\,2 k_e\cdot k_p \int
\frac{d^4p}{\pi^2i}\,\frac{1}{p^2 + i0} \frac{(p + q)^2 \eta_{\mu\nu}
  - (p + q)_{\mu} (p + q)_{\nu}}{(p + q)^2 - M^2_W + i0}\nonumber\\
\hspace{-0.15in}&&\times \frac{1}{p^2 + 2p\cdot k_p +
  i0}\,\frac{1}{p^2 - 2p\cdot k_e + i0}
\end{eqnarray}
and 
\begin{eqnarray}\label{eq:B.30}
\hspace{-0.3in}{\cal M}^{(\Box W\gamma)}_4 &=& 
\frac{\alpha}{2\pi} \int \frac{d^4p}{\pi^2i}\,\frac{1}{p^2 +
  i0}\,\frac{(p + q)^2 \eta_{\mu\nu} - (p + q)_{\mu} (p + q)_{\nu}}{(p
  + q)^2 - M^2_W + i0}\,\frac{\big[\bar{u}_p \hat{k}_e\hat{p}
    \gamma^{\mu}(1 - g_A
    \gamma^5) u_n\big]\big[\bar{u}_e\,\gamma^{\nu}(1 -
    \gamma^5)v_{\bar{\nu}}\big]}{[p^2 + 2p\cdot k_p + i0][p^2 -
    2 p\cdot k_e + i0]}\nonumber\\
\hspace{-0.15in}&-& \frac{\alpha}{2\pi}\int
\frac{d^4p}{\pi^2i}\,\frac{1}{p^2 + i0}\,\frac{(p + q)^2 \eta_{\mu\nu}
  - (p + q)_{\mu} (p + q)_{\nu}}{(p + q)^2 - M^2_W + i0}
\frac{\big[\bar{u}_p \gamma^{\mu}(1 - g_A \gamma^5)
    u_n\big]\big[\bar{u}_e \hat{k}_p\hat{p}\gamma^{\nu}(1 -
    \gamma^5)v_{\bar{\nu}}\big]}{[p^2 + 2p\cdot k_p + i0][p^2 -
    2p\cdot k_e + i0]}
\end{eqnarray}
and
\begin{eqnarray}\label{eq:B.31}
\hspace{-0.3in}{\cal M}^{(\Box W\gamma)}_5 = - \frac{\alpha}{4\pi} \int
\frac{d^4p}{\pi^2i}\,\frac{1}{p^2 + i0}\,\frac{(p + q)^2 \eta_{\mu\nu}
  - (p + q)_{\mu} (p + q)_{\nu}}{(p + q)^2 - M^2_W +
  i0}\,\frac{\big[\bar{u}_p \gamma^{\alpha}\hat{p} \gamma^{\mu}(1 -
    g_A \gamma^5) u_n\big]\big[\bar{u}_e \gamma_{\alpha}\hat{p}
    \gamma^{\nu}(1 - \gamma^5)v_{\bar{\nu}}\big]}{[p^2 + 2p\cdot k_p +
    i0][p^2 - 2 p\cdot k_e + i0]}.
\end{eqnarray}
We merge denominators by using the Feynman parameters
\begin{eqnarray}\label{eq:B.32}
\hspace{-0.3in}&&\frac{1}{(p + q)^2 - M^2_W + i0}\frac{1}{p^2 +
  i0}\frac{1}{p^2 + 2p\cdot k_p + i0}\frac{1}{p^2 - 2 p\cdot k_e + i0}
= \nonumber\\
\hspace{-0.15in}&& =\int^1_0dx\int^1_0dy\, 2 y\int^1_0dz\,3
z^2\frac{1}{[(p - Q)^2 - (M^2_W - q^2)(1-z) - Q^2 + i0]^4},
\end{eqnarray}
where $Q = p(x)yz - q(1 - z)$ and $p(x) = k_e x - k_p (1 - x)$. After
the integration over the virtual momentum we arrive at the following
expression
\begin{eqnarray}\label{eq:B.33}
\hspace{-0.3in}{\cal M}^{(\Box W\gamma)}_3 &=&
\frac{\alpha}{2\pi}\,\big[\bar{u}_p \gamma^{\mu}(1 - g_A \gamma^5)
  u_n\big]\big[\bar{u}_e \gamma^{\nu}(1 -
  \gamma^5)v_{\bar{\nu}}\big] 2 k_e\cdot k_p \int^1_0dx\int^1_0dy\, 2
y\int^1_0dz\,3 z^2\nonumber\\
\hspace{-0.15in}&&\times \,\Big\{ -
\frac{1}{4}\,\frac{\eta_{\mu\nu}}{M^2_W(1 - z) + p^2(x)y^2z^2 -
  2q\cdot p(x)yz(1 - z) - q^2z(1- z)}\nonumber\\
\hspace{-0.15in}&& + \frac{1}{6}\,z^2\frac{(p(x)y + q)^2 \eta_{\mu\nu}
  - (p(x)y + q)_{\mu}(p(x)y + q)_{\nu}}{[M^2_W(1 - z) + p^2(x)y^2z^2 -
    2q\cdot p(x)yz(1 - z) - q^2z(1- z)]^2}\Big\}.
\end{eqnarray}
Neglecting dependence on $q$ giving the contributions smaller than
$10^{-5}$, i.e. setting formally $q = 0$, and having integrated over
$y$ we get
\begin{eqnarray}\label{eq:B.34}
\hspace{-0.3in}{\cal M}^{(\Box W\gamma)}_3 &=&
\frac{\alpha}{2\pi}\,\big[\bar{u}_p \gamma^{\mu}(1 - g_A \gamma^5)
  u_n\big]\big[\bar{u}_e \gamma^{\nu}(1 - \gamma^5)v_{\bar{\nu}}\big]
\,k_e\cdot k_p\, \Big\{ -
\frac{3}{2}\,\eta_{\mu\nu}\int^1_0\frac{dx}{p^2(x)}\int^1_0dz\,{\ell
  n}\Big(1 + \frac{p^2(x)}{M^2_W}\,\frac{z^2}{1 - z}\Big)\nonumber\\
\hspace{-0.15in}&& + \int^1_0\frac{dx}{p^2(x)}\Big(\eta_{\mu\nu} -
\frac{p(x)_{\mu} p(x)_{\nu}}{p^2(x)}\Big)\int^1_0dz\,\Big[{\ell
    n}\Big(1 + \frac{p^2(x)}{M^2_W}\,\frac{z^2}{1 - z}\Big) -
  \frac{M^2_W (1 - z)}{M^2_W(1 - z) + p^2(x)z^2} + 1\Big]\Big\}.
\end{eqnarray}
The integrals over $z$ are equal to
\begin{eqnarray}\label{eq:B.35}
\hspace{-0.3in}\int^1_0dz\,{\ell n}\Big(1 + a^2\,\frac{z^2}{1 -
  z}\Big) &=& -1 + \Big(1 - \frac{1}{2a^2}\Big)\,{\ell n}a^2 -
\frac{\sqrt{1 - 4 a^2}}{2a^2}\,{\ell n}\Big(\frac{1 - 2 a^2 + \sqrt{1
    - 4 a^2}}{1 - 2 a^2 - \sqrt{1 - 4 a^2}}\,\frac{1 - \sqrt{1 - 4
    a^2}}{1 + \sqrt{1 - 4 a^2}}\Big) = \nonumber\\
\hspace{-0.15in}&=&  - a^2 {\ell n} a^2 -
\frac{1}{2}\,a^2 + \ldots,\nonumber\\
\hspace{-0.3in}\int^1_0dz\,\frac{z - 1}{1 - z + a^2z^2} &=&
\frac{1}{2a^2}\,{\ell n} a^2 + \frac{1 - 2 a^2}{2 a^2\sqrt{1 - 4
    a^2}}\,{\ell n}\Big(\frac{1 - 2 a^2 + \sqrt{1 - 4 a^2}}{1 - 2 a^2
  - \sqrt{1 - 4 a^2}}\,\frac{1 - \sqrt{1 - 4 a^2}}{1 + \sqrt{1 - 4
    a^2}}\Big) = \nonumber\\
\hspace{-0.15in}&=& - 1 - a^2 {\ell n} a^2 - \frac{3}{2}\,a^2 +
\ldots,
\end{eqnarray}
where $a^2 = p^2(x)/M^2_W$. Keeping first the leading terms in the
large $M^2_W$ expansion and then the leading terms in the large $m_N$
expansion we obtain
\begin{eqnarray}\label{eq:B.36}
\hspace{-0.3in}{\cal M}^{(\Box W\gamma)}_3 =
\frac{\alpha}{2\pi}\,\big[\bar{u}_p \gamma^{\mu}(1 - g_A \gamma^5)
  u_n\big]\big[\bar{u}_e \gamma^{\nu}(1 - \gamma^5)v_{\bar{\nu}}\big]
\,(\eta_{\mu\nu} - 4\,\eta_{0\mu}\eta_{0\nu})\, \Big\{ \frac{E_e}{m_N}
\frac{m^2_N}{2 M^2_W}\,{\ell n}\Big(\frac{M^2_W}{m^2_N}\Big) +
\ldots\Big\}.
\end{eqnarray}
The contribution of ${\cal M}^{(\Box W\gamma)}_3$ to the radiative
corrections is of order $10^{-9}$ and can be neglected in comparison
with the corrections of order $10^{-5}$. In terms of the integrals over
the Feynman parameters the r.h.s. of Eq.(\ref{eq:B.30}) takes the form

\begin{eqnarray}\label{eq:B.37}
\hspace{-0.3in}&&{\cal M}^{(\Box W\gamma)}_4 = \frac{\alpha}{2\pi}
\int^1_0dx\int^1_0dy\, 2 y\int^1_0dz\, z^2\,\Big\{ -
\frac{1}{4}\,yz\,\frac{5 \eta_{\mu\nu} p(x)_{\alpha} -
  \eta_{\mu\alpha} p(x)_{\nu} - \eta_{\nu\alpha} p(x)_{\mu}}{M^2_W(1 -
  z) + p^2(x)y^2z^2}\nonumber\\
\hspace{-0.3in}&& + \frac{1}{2}\,y^3z^3\frac{(p^2(x)\eta_{\mu\nu} -
  p(x)_{\mu}p(x)_{\nu})p(x)_{\alpha}}{[M^2_W(1 - z) +
    p^2(x)y^2z^2]^2}\Big\} \Big\{\big[\bar{u}_p
  \hat{k}_e\gamma^{\alpha} \gamma^{\mu}(1 - g_A \gamma^5)
  u_n\big]\big[\bar{u}_e\,\gamma^{\nu}(1 - \gamma^5)v_{\bar{\nu}}\big]
\nonumber\\
\hspace{-0.3in}&& - \big[\bar{u}_p \gamma^{\mu}(1 - g_A \gamma^5)
  u_n\big]\big[\bar{u}_e \hat{k}_p\gamma^{\alpha}\gamma^{\nu}(1 -
  \gamma^5)v_{\bar{\nu}}\big]\Big\}.
\end{eqnarray}
Having integrated over $y$ we arrive at the expression
\begin{eqnarray}\label{eq:B.38}
\hspace{-0.15in}&&{\cal M}^{(\Box W\gamma)}_4 = \frac{\alpha}{2\pi}
\Big\{ - \frac{1}{2}\int^1_0\frac{dx}{p^2(x)}\big(5 \eta_{\mu\nu}
p(x)_{\alpha} - \eta_{\mu\alpha} p(x)_{\nu} - \eta_{\nu\alpha}
p(x)_{\mu}\big) \int^1_0dz\, z \Big(1 -
\sqrt{\frac{M^2_W}{p^2(x)}\frac{1 - z}{z^2}} \arctan
\sqrt{\frac{p^2(x)}{M^2_W} \frac{z^2}{1 - z}} \Big)\nonumber\\
\hspace{-0.15in}&& + \frac{3}{2}\int^1_0\frac{dx
  p(x)_{\alpha}}{p^2(x)}\Big(\eta_{\mu\nu} -
\frac{p(x)_{\mu}p(x)_{\nu}}{p^2(x)}\Big)\int^1_0dz z \Big[\Big(1 -
  \sqrt{\frac{M^2_W}{p^2(x)}\frac{1 - z}{z^2}} \arctan
  \sqrt{\frac{p^2(x)}{M^2_W} \frac{z^2}{1 - z}} \Big) -
  \frac{1}{3}\,\frac{p^2(x) z^2}{M^2_W(1 - z) + p^2(x)
    z^2}\Big]\Big\}\nonumber\\
\hspace{-0.15in}&& \times\Big\{\big[\bar{u}_p \hat{k}_e\gamma^{\alpha}
  \gamma^{\mu}(1 - g_A \gamma^5)
  u_n\big]\big[\bar{u}_e\,\gamma^{\nu}(1 - \gamma^5)v_{\bar{\nu}}\big]
- \big[\bar{u}_p \gamma^{\mu}(1 - g_A \gamma^5) u_n\big]\big[\bar{u}_e
  \hat{k}_p\gamma^{\alpha}\gamma^{\nu}(1 -
  \gamma^5)v_{\bar{\nu}}\big]\Big\}.
\end{eqnarray}
Keeping only the leading terms in the large $M_W$ expansion and then
the leading terms in the large $m_N$ expansion we get \cite{MathB39}
\begin{eqnarray}\label{eq:B.39}
\hspace{-0.15in}&&{\cal M}^{(\Box W\gamma)}_4 = \frac{\alpha}{2\pi}
\Big\{ - \frac{1}{12}\big(5 \eta_{\mu\nu}\eta_{0\alpha} -
\eta_{\mu\alpha}\eta_{0\nu} - \eta_{\nu\alpha}\eta_{0\mu}\big) +
\frac{1}{2}\,\eta_{0\alpha}\big(\eta_{\mu\nu} -
\eta_{0\mu}\eta_{0\nu}\big)\Big\}\,\Big\{\frac{m^2_N}{M^2_W}\,{\ell
  n}\Big(\frac{M^2_W}{m^2_N}\Big) + \ldots\Big\}\nonumber\\
\hspace{-0.15in}&&\times \Big\{\big[\bar{u}_p
  \frac{\hat{k}_e}{m_N}\gamma^{\alpha} \gamma^{\mu}(1 - g_A \gamma^5)
  u_n\big]\big[\bar{u}_e\,\gamma^{\nu}(1 - \gamma^5)v_{\bar{\nu}}\big]
- \big[\bar{u}_p \gamma^{\mu}(1 - g_A \gamma^5) u_n\big]\big[\bar{u}_e
  \frac{\hat{k}_p}{m_N}\gamma^{\alpha}\gamma^{\nu}(1 -
  \gamma^5)v_{\bar{\nu}}\big]\Big\}.
\end{eqnarray}
The r.h.s. of Eq.(\ref{eq:B.31}), expressed in terms of the integrals
over the Feynman parameters, takes the form
\begin{eqnarray}\label{eq:B.40}
\hspace{-0.3in}&&{\cal M}^{(\Box W\gamma)}_5 = - \frac{\alpha}{4\pi}
\int^1_0dx\int^1_0dy\, 2 y\int^1_0dz\, 3 z^2\,\Big\{\Big(\frac{5}{24}\,\eta_{\mu\nu}\eta_{\rho\omega} -
\frac{1}{24}\,(\eta_{\mu\rho}\eta_{\nu\omega} +
\eta_{\mu\omega}\eta_{\nu\rho})\Big)\Big[{\ell n}\frac{\Lambda^2}{m^2_N} - {\ell n}\Big(\frac{M^2_W}{m^2_N}\Big)\nonumber\\
\hspace{-0.15in}&& - {\ell n}\Big(1 - z + \frac{p^2(x)}{M^2_W}\,y^2
z^2\Big)\Big] - \frac{y^2 z^2}{M^2_W(1 - z) + p^2(x) y^2
  z^2}\Big(\frac{1}{12}\,\big(p^2(x)\eta_{\mu\nu} -
p(x)_{\mu}p(x)_{\nu}\big) \eta_{\rho\omega} + \eta_{\mu\nu}
p(x)_{\rho}p(x)_{\omega}\nonumber\\
\hspace{-0.15in}&& - \frac{1}{12}\,\big(\eta_{\mu\rho} p(x)_{\nu}p(x)_{\omega} +
\eta_{\nu\rho} p(x)_{\mu}p(x)_{\omega} + \eta_{\mu\omega}
p(x)_{\nu}p(x)_{\rho} + \eta_{\nu\omega} p(x)_{\mu}p(x)_{\rho}\big)
\Big) + \frac{1}{6}\,\frac{y^4 z^4}{(M^2_W(1 - z) + p^2(x) y^2
  z^2)^2}\nonumber\\
\hspace{-0.15in}&&\times \big(p^2(x)\eta_{\mu\nu} -
p(x)_{\mu}p(x)_{\nu}\big) p(x)_{\rho}p(x)_{\omega}\Big\}
\big[\bar{u}_p \gamma^{\alpha}\gamma^{\rho} \gamma^{\mu}(1 - g_A
  \gamma^5) u_n\big]\big[\bar{u}_e \gamma_{\alpha}\gamma^{\omega}
  \gamma^{\nu}(1 - \gamma^5)v_{\bar{\nu}}\big].
\end{eqnarray}
Keeping the leading contributions in the large $M_W$ and then in the
large $m_N$ expansion we get
\begin{eqnarray}\label{eq:B.41}
\hspace{-0.3in}&&{\cal M}^{(\Box W\gamma)}_5 = - \frac{\alpha}{4\pi}
\Big\{\Big(\frac{5}{8}\,\eta_{\mu\nu}\eta_{\rho\omega} -
\frac{1}{8}\,(\eta_{\mu\rho}\eta_{\nu\omega} +
\eta_{\mu\omega}\eta_{\nu\rho})\Big)\Big[{\ell n}\frac{\Lambda^2}{m^2_N}- {\ell n}\Big(\frac{M^2_W}{m^2_N}\Big) +
  \frac{11}{18} - \frac{1}{6}\,\frac{m^2_N}{M^2_W}\,{\ell
    n}\Big(\frac{M^2_W}{m^2_N}\Big) + \ldots\Big]\nonumber\\
\hspace{-0.15in}&& - \Big(\frac{1}{24}\,\big(\eta_{\mu\nu} -
\eta_{0\mu}\eta_{0\nu}\big) \eta_{\rho\omega} +
\frac{1}{2}\,\eta_{\mu\nu} \eta_{0\rho}\eta_{0\omega} -
\frac{1}{24}\,\big(\eta_{\mu\rho} \eta_{0\nu} \eta_{0\omega} +
\eta_{\nu\rho} \eta_{0\mu}\eta_{0\omega} + \eta_{\mu\omega}
\eta_{0\nu} \eta_{0\rho} +
\eta_{\nu\omega}\eta_{0\mu}\eta_{0\rho}\big) \Big) \Big[\frac{m^2_N}{M^2_W}\,{\ell
    n}\Big(\frac{M^2_W}{m^2_N}\Big)\nonumber\\
\hspace{-0.15in}&& + \ldots\Big] + \frac{1}{6}\, \big(\eta_{\mu\nu} -
\eta_{0\mu}\eta_{0\nu}\big)\eta_{0\rho}\eta_{0\omega}
\Big[\frac{m^2_N}{M^2_W} + \ldots\Big]\Big\} \big[\bar{u}_p
  \gamma^{\alpha}\gamma^{\rho} \gamma^{\mu}(1 - g_A \gamma^5)
  u_n\big]\big[\bar{u}_e \gamma_{\alpha}\gamma^{\omega} \gamma^{\nu}(1
  - \gamma^5)v_{\bar{\nu}}\big].
\end{eqnarray}
Summing up the contributions of Eq.(\ref{eq:B.36}), Eq.(\ref{eq:B.39})
and Eq.(\ref{eq:B.41}) we obtain the analytical expression for ${\cal
  M}^{(\Box W\gamma)}_{3 + 4 + 5}$ and, correspondingly, the
analytical expression for the second integral in Eq.(\ref{eq:B.19}).
\begin{eqnarray*}
\hspace{-0.15in}&&M^{(W\gamma)}(n\to p\,e^-\bar{\nu}_e)_{\rm
  Fig.\,\ref{fig:fig6a}a} = -
G_V\,\frac{\alpha}{2\pi}\Bigg\{\Big\{\Big[2 {\ell
    n}\frac{\Lambda^2}{m^2_N} + 4 + \frac{1}{2}\,{\ell
    n}\Big(\frac{m^2_N}{m^2_e}\Big) + 2 {\ell
    n}\Big(\frac{\mu}{m_e}\Big) + 2{\ell n}\Big(\frac{\mu}{m_e}\Big)
  \Big[\frac{1}{2\beta}\,{\ell n}\Big(\frac{1 + \beta}{1 - \beta}\Big)
    - 1 \Big]\nonumber\\
\hspace{-0.15in}&& - \frac{1}{4\beta}\,{\ell n}^2\Big(\frac{1 +
  \beta}{1 - \beta}\Big) - \frac{1}{\beta}\,{\rm
  Li}_2\Big(\frac{2\beta}{1 + \beta}\Big)\Big] + \frac{E_e}{m_N}\Big[1
  +\frac{1 - \beta^2}{\beta}\,{\ell n}\Big(\frac{1 + \beta}{1 -
    \beta}\Big)\Big]\Big\}[\bar{u}_p\gamma^{\mu}(1 - g_A
  \gamma^5)u_n][\bar{u}_e\gamma_{\mu}(1 - \gamma^5)v_{\bar{\nu}}] +
\Big\{(g_A - 1)\nonumber\\
\hspace{-0.15in}&&\times \, [\bar{u}_p\gamma^{\mu}(1 +
  \gamma^5)u_n][\bar{u}_e\gamma_{\mu}(1 - \gamma^5)v_{\bar{\nu}}]
\Big\{\frac{3}{4}\,{\ell n}\frac{\Lambda^2}{m^2_N}  + \frac{5}{4} -
\frac{3}{4}\,\frac{E_e}{m_N}\Big[{\ell n}\Big(\frac{m^2_N}{m^2_e}\Big)
  - \beta\,{\ell n}\Big(\frac{1 + \beta}{1 - \beta}\Big)\Big]\Big\} -
(g_A - 1) [\bar{u}_p\gamma^{\mu}(1 + \gamma^5)\nonumber\\
\end{eqnarray*}
\begin{eqnarray}\label{eq:B.42}
\hspace{-0.15in}&&\times u_n][\bar{u}_e\gamma^{\nu}(1 -
  \gamma^5)v_{\bar{\nu}}] \Big\{-
\frac{1}{4}\,\frac{k_{p\mu}k_{p\nu}}{m^2_N}\Big\{1 -
2\,\frac{E_e}{m_N}\Big[1 - \frac{1}{2}\,{\ell
    n}\Big(\frac{m^2_N}{m^2_e}\Big) + \frac{1 +
    \beta^2}{4\beta}\,{\ell n}\Big(\frac{1 + \beta}{1 -
    \beta}\Big)\Big]\Big\} - \frac{1}{4}\,\frac{k_{p\mu}k_{e\nu} +
  k_{p\nu}k_{e\mu}}{m^2_N}\nonumber\\
\hspace{-0.15in}&&\times \Big\{1 - \frac{1}{2}\,{\ell
  n}\Big(\frac{m^2_N}{m^2_e}\Big) - \frac{1}{2\beta}\,{\ell
  n}\Big(\frac{1 + \beta}{1 - \beta}\Big)\Big]\Big\} +
  \frac{1}{4}\,\frac{k_{e\mu}k_{e\nu}}{E_e
    m_N}\,\frac{1}{2\beta}\,{\ell n}\Big(\frac{1 + \beta}{1 -
    \beta}\Big) \Big\} + [\bar{u}_p \frac{\hat{k}_e}{m_N}
    \gamma^{\mu}(1 - g_A \gamma^5)u_n] [\bar{u}_e\gamma_{\mu}(1 -
    \gamma^5)v_{\bar{\nu}}]\nonumber\\
\hspace{-0.15in}&&\times \Big[\frac{1}{2}\,{\ell
    n}\Big(\frac{m^2_N}{m^2_e}\Big) - \frac{1}{2 \beta}\,{\ell
    n}\Big(\frac{1 + \beta}{1 - \beta}\Big)\Big] - \frac{m_e}{2E_e}\,
       [\bar{u}_p\gamma^{\mu}(1 -
         g_A\gamma^5)u_n][\bar{u}_e\frac{\hat{k}_p}{m_N}\gamma_{\mu}(1
         - \gamma^5)v_{\bar{\nu}}]\Big\{\frac{1}{\beta}\,{\ell
         n}\Big(\frac{1 + \beta}{1 - \beta}\Big) +
       \frac{E_e}{m_N}\,\Big[{\ell n}\Big(\frac{m^2_N}{m^2_e}\Big)
         \nonumber\\
\hspace{-0.15in}&& - \frac{1}{\beta}\,{\ell n}\Big(\frac{1 + \beta}{1
  - \beta}\Big) \Big]\Big\} + [\bar{u}_p \gamma^{\mu}(1 - g_A
         \gamma^5)u_n] [\bar{u}_e\gamma_{\mu}(1 -
         \gamma^5)v_{\bar{\nu}}]\,\Big\{\,\frac{1}{2\beta}\,{\ell
         n}\Big(\frac{1 + \beta}{1 - \beta}\Big) \Big\} +
       \Big\{\frac{1}{12}\big(5 \eta_{\mu\nu}\eta_{0\alpha} -
       \eta_{\mu\alpha}\eta_{0\nu} -
       \eta_{\nu\alpha}\eta_{0\mu}\big)\nonumber\\
\hspace{-0.15in}&& - \frac{1}{2}\,\eta_{0\alpha}\big(\eta_{\mu\nu} -
\eta_{0\mu}\eta_{0\nu}\big)\Big\}\,\Big\{\frac{m^2_N}{M^2_W}\,{\ell
  n}\Big(\frac{M^2_W}{m^2_N}\Big)\Big\} \big[\bar{u}_p \gamma^{\mu}(1
  - g_A \gamma^5) u_n\big]\big[\bar{u}_e
  \frac{\hat{k}_p}{m_N}\gamma^{\alpha}\gamma^{\nu}(1 -
  \gamma^5)v_{\bar{\nu}}\big] - \Big\{\Big(\frac{5}{16}\,
\eta_{\mu\nu}\eta_{\rho\omega} -
\frac{1}{16}\,(\eta_{\mu\rho}\eta_{\nu\omega} \nonumber\\
\hspace{-0.15in}&& +
\eta_{\mu\omega}\eta_{\nu\rho})\Big)\Big[\Gamma\Big(2 -
  \frac{n}{2}\Big) - {\ell n}\Big(\frac{M^2_W}{m^2_N}\Big) +
  \frac{11}{18} - \frac{1}{6}\,\frac{m^2_N}{M^2_W}\,{\ell
    n}\Big(\frac{M^2_W}{m^2_N}\Big)\Big] -
\Big(\frac{1}{48}\,\big(\eta_{\mu\nu} - \eta_{0\mu}\eta_{0\nu}\big)
\eta_{\rho\omega} + \frac{1}{4}\,\eta_{\mu\nu}
\eta_{0\rho}\eta_{0\omega} - \frac{1}{48}\,\big(\eta_{\mu\rho}
\eta_{0\nu}\nonumber\\
\hspace{-0.15in}&&\times\, \eta_{0\omega} + \eta_{\nu\rho}
\eta_{0\mu}\eta_{0\omega} + \eta_{\mu\omega} \eta_{0\nu} \eta_{0\rho}
+ \eta_{\nu\omega}\eta_{0\mu}\eta_{0\rho}\big) \Big)
\Big[\frac{m^2_N}{M^2_W}\,{\ell
    n}\Big(\frac{M^2_W}{m^2_N}\Big)\Big]\Big\} \,\big[\bar{u}_p
  \gamma^{\alpha}\gamma^{\rho} \gamma^{\mu}(1 - g_A \gamma^5)
  u_n\big]\big[\bar{u}_e \gamma_{\alpha}\gamma^{\omega} \gamma^{\nu}(1
  - \gamma^5)v_{\bar{\nu}}\big]\Bigg\},\nonumber\\
\hspace{-0.15in}&&
\end{eqnarray}
where we have neglected the terms of order smaller than $10^{-6}$.

\subsection*{Analytical expressions for the Feynman diagrams in Fig.\,
\ref{fig:fig6a}b - Fig.\,\ref{fig:fig6a}f}

Since the calculation of the Feynman diagrams in
Fig.\,\ref{fig:fig6a}b, Fig.\,\ref{fig:fig6a}c and
Fig.\,\ref{fig:fig6a}d, defined by the momentum integrals in
Eq.(\ref{eq:B.2}), Eq.(\ref{eq:B.3}) and Eq.(\ref{eq:B.4}),
respectively, runs parallel the calculation of the Feynman diagram in
Fig.\,\ref{fig:fig6a}a, defined by the momentum integral in
Eq.(\ref{eq:B.1}), we skip such standard intermediate calculations and
adduce only the final results
\begin{eqnarray}\label{eq:B.43}
\hspace{-0.15in}&&M(n \to p e^- \bar{\nu}_e)_{\rm
  Fig.\,\ref{fig:fig6a}b} =
G_V\big(\frac{\alpha}{2\pi}\Big)\Big\{\big(\eta^{\varphi}{}_{\mu} -
\eta^{\varphi}{}_0 \eta_{0\mu}\big)\Big[
  \frac{1}{4}\,\frac{m^2_N}{M^2_W}\,{\ell
    n}\Big(\frac{M^2_W}{m^2_N}\Big)\Big]\big[\bar{u}_p\gamma^{\mu}(1 -
  g_A \gamma^5) u_n]\, \frac{M^2_W}{M^2_W - q^2 - i0}\nonumber\\
\hspace{-0.15in}&&\times \, \Big(- \eta_{\varphi\nu} +
\frac{q_{\varphi}q_{\nu}}{M^2_W}\Big)\big[\bar{u}_e \gamma^{\nu}(1 -
  \gamma^5) v_{\bar{\nu}}\big] +
\Big\{\big(\eta^{\varphi}{}_{\rho}\eta_{\alpha \mu} - \eta_{\alpha
  \rho} \eta^{\varphi}{}_{\mu}\big)\Big[- \frac{1}{16} {\ell
    n}\frac{\Lambda^2}{m^2_N} + \frac{1}{16} {\ell
    n}\Big(\frac{M^2_W}{m^2_N}\Big) - \frac{3}{64} +
  \frac{1}{48}\frac{m^2_N}{M^2_W}\nonumber\\
\hspace{-0.15in}&&\times {\ell n}\Big(\frac{M^2_W}{m^2_N}\Big)\Big] +
\eta_{0\rho}\big(\eta^{\varphi}{}_0 \eta_{\alpha \mu} -
\eta_{0\alpha}\eta^{\varphi}{}_{\mu}\big)\Big[
  \frac{1}{12}\,\frac{m^2_N}{M^2_W}\,{\ell
    n}\Big(\frac{M^2_W}{m^2_N}\Big)\Big]\Big\}\big[\bar{u}_p
  \gamma^{\alpha}\gamma^{\rho}\gamma^{\mu}(1 - g_A \gamma^5)u_n]\,
\frac{M^2_W}{M^2_W - q^2 - i0}\nonumber\\
\hspace{-0.15in}&&\times\, \Big(- \eta_{\varphi\nu} +
\frac{q_{\varphi}q_{\nu}}{M^2_W}\Big) \, \big[\bar{u}_e \gamma^{\nu}(1
  - \gamma^5) v_{\bar{\nu}}\big]\Big\}
\end{eqnarray}
and 
\begin{eqnarray}\label{eq:B.44}
\hspace{-0.15in}&&M(n \to p e^- \bar{\nu}_e)_{\rm
  Fig.\,\ref{fig:fig6a}c} =
G_V\big(\frac{\alpha}{2\pi}\Big)\Big[\frac{3}{4}\,{\ell n}\frac{\Lambda^2}{m^2_N} - \frac{3}{4}\,{\ell
    n}\Big(\frac{M^2_W}{m^2_N}\Big) + \frac{9}{8}\Big]
\big[\bar{u}_p\gamma^{\mu}(1 - g_A \gamma^5) u_n]\, \frac{M^2_W}{M^2_W
  - q^2 - i0}\nonumber\\
\hspace{-0.15in}&&\times \, \Big(- \eta_{\mu\nu} +
\frac{q_{\mu}q_{\nu}}{M^2_W}\Big)\big[\bar{u}_e \gamma^{\nu}(1 -
  \gamma^5) v_{\bar{\nu}}\big].
\end{eqnarray}
After renormalization the contributions of the Feynman diagrams in
Fig.\,\ref{fig:fig6a}d, Fig.\,\ref{fig:fig6a}e and
Fig.\,\ref{fig:fig6a}f vanish. The renormalization constants of the
wave functions of the proton and electron are equal to
\begin{eqnarray}\label{eq:B.45}
Z^{(p)}_2 - 1 &=& \frac{\alpha}{2\pi}\,\Big[- \frac{1}{2}\,{\ell
    n}\frac{\Lambda^2}{m^2_N} - 1 - {\ell
    n}\Big(\frac{\mu^2}{m^2_N}\Big)\Big], \nonumber\\ Z^{(e)}_2 - 1
&=& \frac{\alpha}{2\pi}\,\Big[- \frac{1}{2}\,{\ell
    n}\frac{\Lambda^2}{m^2_N} - 1 - \frac{1}{2}\,{\ell
    n}\Big(\frac{m^2_N}{m^2_e}\Big) - {\ell
    n}\Big(\frac{\mu^2}{m^2_e}\Big)\Big],
\end{eqnarray}
where $\mu$ is an infinitesimal photon mass. The renormalization
constant $Z^{(W)}_3$ of the wave function of the electroweak
$W^-$--boson is equal to unity, i.e. $Z^{(W)}_3 = 1$. Indeed, a
non--trivial renormalization of the wave function of the electroweak
$W^-$--boson appears in the amplitude of the neutron $\beta^-$--decay
(see Eq.(\ref{eq:B.4})) to order $(\alpha/\pi)(q^2/M^2_W) \sim
10^{-12}$, which we neglect in our analysis of contributions of order
$10^{-5}$.  The contributions of the counter--terms to the
amplitude of the neutron $\beta^-$--decay are described by the
Lagrangian (see Eq.(\ref{eq:45}))
\begin{eqnarray}\label{eq:B.46}
\hspace{-0.15in}\delta {\cal L}^{(\rm CT)}_{\rm L \sigma M + SEM} &=&
\big(\tilde{Z}^{(N)}_1 Z_N \sqrt{Z^{(p)}_2} -
1\big)\,\frac{g^{(r)}}{2\sqrt{2}}\,\bar{\psi}^{(r)}_p\gamma^{\mu}(1 -
\gamma^5)\,\psi^{(r)}_n\,W^{(r)+}_{\mu}\nonumber\\
 \hspace{-0.15in}&-& \big(\tilde{Z}^{(\ell)}_1 \sqrt{Z^{(e)}_2} -
 1\big)\,\frac{g^{(r)}}{2\sqrt{2}}\,\bar{\psi}^{(r)}_e\gamma^{\mu}
 \big(1 - \gamma^5\big)\,\psi^{(r)}_{\nu L}\,W^{(r)-}_{\mu} + {\rm h.c.},
\end{eqnarray}
where the abbreviation ${\rm CT}$ means ``Counter-Terms'' and $Z_N =
1$, are given by
\begin{eqnarray}\label{eq:B.47}
\hspace{-0.15in}M(n \to p e^- \bar{\nu}_e)_{\,\rm CT} &=&
G_V\Big\{\big(\tilde{Z}^{(N)}_1 - 1\big) + \frac{Z^{(p)}_2 - 1}{2} +
\frac{Z^{(e)}_2 - 1}{2} + \big(\tilde{Z}^{(\ell)}_1 - 1\big)\Big\}
\big[\bar{u}_p\gamma^{\mu}(1 - g_A \gamma^5) u_n]\, \frac{M^2_W}{M^2_W
  - q^2 - i0}\nonumber\\
\hspace{-0.15in}&&\times\, \Big(- \eta_{\varphi\nu} +
\frac{q_{\varphi}q_{\nu}}{M^2_W}\Big)\big[\bar{u}_e \gamma^{\nu}(1 -
  \gamma^5) v_{\bar{\nu}}\big].
\end{eqnarray}
Plugging Eq.(\ref{eq:B.45}) into Eq.(\ref{eq:B.46}) we get the
following contribution of the counter--terms
\begin{eqnarray}\label{eq:B.48}
\hspace{-0.15in}M(n \to p e^- \bar{\nu}_e)_{\,\rm CT} &=&
G_V\Big\{\big(\tilde{Z}^{(N)}_1 - 1\big) + \big(\tilde{Z}^{(\ell)}_1 -
1\big) + \frac{\alpha}{2\pi}\,\Big[- \frac{1}{2}\, {\ell
    n}\frac{\Lambda^2}{m^2_N} - 1 + \frac{1}{4}\,{\ell
    n}\Big(\frac{m^2_N}{m^2_e}\Big) - 2{\ell
    n}\Big(\frac{\mu}{m_e}\Big)\Big]\Big\} \nonumber\\
\hspace{-0.15in}&&\times\,\big[\bar{u}_p\gamma^{\mu}(1 - g_A \gamma^5)
  u_n]\, \frac{M^2_W}{M^2_W - q^2 - i0} \Big(- \eta_{\mu\nu} +
\frac{q_{\mu}q_{\nu}}{M^2_W}\Big)\big[\bar{u}_e \gamma^{\nu}(1 -
  \gamma^5) v_{\bar{\nu}}\big].
\end{eqnarray}
Now we may proceed to the calculation of the total contribution of the
Feynman diagrams in Fig.\,\ref{fig:fig6a}, describing radiative
corrections of order $O(\alpha/\pi)$ to any order in the large nucleon
mass expansion, to the amplitude of the neutron $\beta^-$--decay.

\section*{Appendix C: Radiative corrections of order $O(\alpha/\pi)$ 
and $O(\alpha E_e/m_N)$ to the amplitude of the neutron
$\beta^-$--decay, caused by the Feynman diagrams in
Fig.\,\ref{fig:fig6a}}
\renewcommand{\theequation}{C-\arabic{equation}}
\setcounter{equation}{0}

Summing up the contributions of the Feynman diagrams in
Fig.\,\ref{fig:fig6a} and the counter--terms, given by
Eq.(\ref{eq:B.47}) we arrive at the following expression for the
radiative corrections to the amplitude of the neutron $\beta^-$--decay
of order $O(\alpha/\pi)$ and $O(\alpha E_e/m_N)$. We get
\begin{eqnarray*}
\hspace{-0.15in}&&M^{(W\gamma)}(n\to p\,e^-\bar{\nu}_e)_{\rm
  Fig.\,\ref{fig:fig6a}} = - G_V\,\Bigg\{\Big[\big(\tilde{Z}^{(N)}_1 -
  1\big) + \big(\tilde{Z}^{(\ell)}_1 -
  1\big)\Big][\bar{u}_p\gamma^{\mu}(1 - g_A
  \gamma^5)u_n][\bar{u}_e\gamma_{\mu}(1 - \gamma^5)v_{\bar{\nu}}] +
\frac{\alpha}{2\pi}\Bigg[\frac{9}{4}\,{\ell
    n}\frac{\Lambda^2}{m^2_N}\nonumber\\
\hspace{-0.15in}&& - \frac{3}{4}\,{\ell
  n}\Big(\frac{M^2_W}{m^2_N}\Big) + \frac{33}{8} + \frac{3}{4}\,{\ell
  n}\Big(\frac{m^2_N}{m^2_e}\Big) + 2{\ell n}\Big(\frac{\mu}{m_e}\Big)
\Big[\frac{1}{2\beta} {\ell n}\Big(\frac{1 + \beta}{1 - \beta}\Big) -
  1 \Big] + \frac{1}{2\beta} {\ell n}\Big(\frac{1 + \beta}{1 -
  \beta}\Big) - \frac{1}{4\beta} {\ell n}^2\Big(\frac{1 + \beta}{1 -
  \beta}\Big)\nonumber\\
\hspace{-0.15in}&& - \frac{1}{\beta}\,{\rm Li}_2\Big(\frac{2\beta}{1 +
  \beta}\Big) + \frac{E_e}{m_N}\Big[1 +\frac{1 -
    \beta^2}{\beta}\,{\ell n}\Big(\frac{1 + \beta}{1 -
    \beta}\Big)\Big]\Big\} [\bar{u}_p\gamma^{\mu}(1 - g_A
  \gamma^5)u_n][\bar{u}_e\gamma_{\mu}(1 - \gamma^5)v_{\bar{\nu}}] +
\Big\{(g_A - 1) [\bar{u}_p\gamma^{\mu}(1 +
  \gamma^5)u_n]\nonumber\\
\hspace{-0.15in}&&\times [\bar{u}_e\gamma_{\mu}(1 -
  \gamma^5)v_{\bar{\nu}}] \Big\{\frac{3}{4}\, {\ell n}\frac{\Lambda^2}{m^2_N} + \frac{5}{4} -
\frac{3}{4}\frac{E_e}{m_N}\Big[{\ell n}\Big(\frac{m^2_N}{m^2_e}\Big)
  - \beta\,{\ell n}\Big(\frac{1 + \beta}{1 - \beta}\Big)\Big]\Big\} -
(g_A - 1) [\bar{u}_p\gamma^{\mu}(1 + \gamma^5) u_n]\nonumber\\
\hspace{-0.15in}&&\times [\bar{u}_e\gamma^{\nu}(1 -
  \gamma^5)v_{\bar{\nu}}] \Big\{-
\frac{1}{4}\,\frac{k_{p\mu}k_{p\nu}}{m^2_N}\Big\{1 -
2\,\frac{E_e}{m_N}\Big[1 - \frac{1}{2}\,{\ell
    n}\Big(\frac{m^2_N}{m^2_e}\Big) + \frac{1 +
    \beta^2}{4\beta}\,{\ell n}\Big(\frac{1 + \beta}{1 -
    \beta}\Big)\Big]\Big\} - \frac{1}{4}\,\frac{k_{p\mu}k_{e\nu} +
  k_{p\nu}k_{e\mu}}{m^2_N}\nonumber\\
\hspace{-0.15in}&&\times \Big\{1 - \frac{1}{2}\,{\ell
  n}\Big(\frac{m^2_N}{m^2_e}\Big) - \frac{1}{2\beta}\,{\ell
  n}\Big(\frac{1 + \beta}{1 - \beta}\Big)\Big]\Big\} +
\frac{1}{4}\,\frac{k_{e\mu}k_{e\nu}}{E_e m_N}\,\frac{1}{2\beta}\,{\ell
  n}\Big(\frac{1 + \beta}{1 - \beta}\Big) \Big\}\nonumber\\
\hspace{-0.15in}&& + [\bar{u}_p \frac{\hat{k}_e}{m_N} \gamma^{\mu}(1 -
  g_A \gamma^5)u_n] [\bar{u}_e\gamma_{\mu}(1 - \gamma^5)v_{\bar{\nu}}]
\Big[\frac{1}{2}\,{\ell n}\Big(\frac{m^2_N}{m^2_e}\Big) - \frac{1}{2
    \beta}\,{\ell n}\Big(\frac{1 + \beta}{1 - \beta}\Big)\Big] -
\frac{m_e}{2E_e}\, [\bar{u}_p\gamma^{\mu}(1 -
  g_A\gamma^5)u_n]\nonumber\\
\hspace{-0.15in}&&\times [\bar{u}_e\frac{\hat{k}_p}{m_N}\gamma_{\mu}(1
  - \gamma^5)v_{\bar{\nu}}]\Big\{\frac{1}{\beta}\,{\ell n}\Big(\frac{1
  + \beta}{1 - \beta}\Big) + \frac{E_e}{m_N}\,\Big[{\ell
    n}\Big(\frac{m^2_N}{m^2_e}\Big) - \frac{1}{\beta}\,{\ell
    n}\Big(\frac{1 + \beta}{1 - \beta}\Big) \Big]\Big\} +
\Big\{\frac{1}{12}\big(5 \eta_{\mu\nu}\eta_{0\alpha} -
\eta_{\mu\alpha}\eta_{0\nu} -
\eta_{\nu\alpha}\eta_{0\mu}\big)\nonumber\\
\hspace{-0.15in}&& - \frac{1}{2}\,\eta_{0\alpha}\big(\eta_{\mu\nu} -
\eta_{0\mu}\eta_{0\nu}\big)\Big\}\,\Big\{\frac{m^2_N}{M^2_W}\,{\ell
  n}\Big(\frac{M^2_W}{m^2_N}\Big)\Big\} \big[\bar{u}_p \gamma^{\mu}(1
  - g_A \gamma^5) u_n\big]\big[\bar{u}_e
  \frac{\hat{k}_p}{m_N}\gamma^{\alpha}\gamma^{\nu}(1 -
  \gamma^5)v_{\bar{\nu}}\big] - \Big\{\Big(\frac{5}{16}\,
\eta_{\mu\nu}\eta_{\rho\omega} -
\frac{1}{16}\,(\eta_{\mu\rho}\eta_{\nu\omega} \nonumber\\
\hspace{-0.15in}&& + \eta_{\mu\omega}\eta_{\nu\rho})\Big)\Big[{\ell
    n}\frac{\Lambda^2}{m^2_N} - {\ell n}\Big(\frac{M^2_W}{m^2_N}\Big)
  + \frac{11}{18} - \frac{1}{6}\,\frac{m^2_N}{M^2_W}\,{\ell
    n}\Big(\frac{M^2_W}{m^2_N}\Big)\Big] -
\Big(\frac{1}{48}\,\big(\eta_{\mu\nu} - \eta_{0\mu}\eta_{0\nu}\big)
\eta_{\rho\omega} + \frac{1}{4}\,\eta_{\mu\nu}
\eta_{0\rho}\eta_{0\omega} - \frac{1}{48}\,\big(\eta_{\mu\rho}
\eta_{0\nu}\nonumber\\
\end{eqnarray*}
\begin{eqnarray}\label{eq:C.1}
\hspace{-0.15in}&&\times\, \eta_{0\omega} + \eta_{\nu\rho}
\eta_{0\mu}\eta_{0\omega} + \eta_{\mu\omega} \eta_{0\nu} \eta_{0\rho}
+ \eta_{\nu\omega}\eta_{0\mu}\eta_{0\rho}\big) \Big)
\Big[\frac{m^2_N}{M^2_W}\,{\ell
    n}\Big(\frac{M^2_W}{m^2_N}\Big)\Big]\Big\} \,\big[\bar{u}_p
  \gamma^{\alpha}\gamma^{\rho} \gamma^{\mu}(1 - g_A \gamma^5)
  u_n\big]\big[\bar{u}_e \gamma_{\alpha}\gamma^{\omega} \gamma^{\nu}(1
  - \gamma^5)v_{\bar{\nu}}\big]\nonumber\\
\hspace{-0.15in}&& + \big(\eta_{\mu\nu} - \eta_{0\nu}
\eta_{0\mu}\big)\Big\{ \frac{1}{4}\,\frac{m^2_N}{M^2_W}\,{\ell
  n}\Big(\frac{M^2_W}{m^2_N}\Big)\Big\}\big[\bar{u}_p\gamma^{\mu}(1 -
  g_A \gamma^5) u_n]\big[\bar{u}_e \gamma^{\nu}(1 - \gamma^5)
  v_{\bar{\nu}}\big] + \Big\{\big(\eta_{\nu\rho}\eta_{\alpha \mu} -
\eta_{\alpha \rho} \eta_{\nu\mu}\big)\Big[- \frac{1}{16}\, {\ell
    n}\frac{\Lambda^2}{m^2_N}\nonumber\\
\hspace{-0.15in}&& + \frac{1}{16} {\ell
  n}\Big(\frac{M^2_W}{m^2_N}\Big) - \frac{3}{64} +
\frac{1}{48}\frac{m^2_N}{M^2_W} {\ell
  n}\Big(\frac{M^2_W}{m^2_N}\Big)\Big] + \eta_{0\rho}\big(\eta_{0\nu}
\eta_{\alpha \mu} - \eta_{0\alpha}\eta_{\nu\mu}\big)\Big[
  \frac{1}{12}\,\frac{m^2_N}{M^2_W}\,{\ell
    n}\Big(\frac{M^2_W}{m^2_N}\Big)\Big]\Big\} \big[\bar{u}_p
  \gamma^{\alpha}\gamma^{\rho}\gamma^{\mu}(1 - g_A
  \gamma^5)u_n]\nonumber\\
\hspace{-0.15in}&&\times \big[\bar{u}_e \gamma^{\nu}(1 - \gamma^5)
  v_{\bar{\nu}}\big]\Bigg]\Bigg\}.
\end{eqnarray}
Following Sirlin \cite{Sirlin1967} (see also \cite{Ivanov2013}) we
rewrite the amplitude of the neutron $\beta^-$--decay in the
non--relativistic approximation for the neutron and proton. For this
aim we use the relation for the Dirac $\gamma$--matrices
\begin{eqnarray}\label{eq:C.2}
\hspace{-0.3in}\gamma^{\alpha}\gamma^{\nu}\gamma^{\mu} = \gamma^{\alpha}g^{\nu\mu} -
\gamma^{\nu}g^{\mu\alpha} + \gamma^{\mu}g^{\alpha\nu} +
i\,\varepsilon^{\alpha\nu\mu\beta}\,\gamma_{\beta}\gamma^5,
\end{eqnarray}
where $\varepsilon^{\alpha\nu\mu\beta}$ is the Levi--Civita tensor
defined by $\varepsilon^{0123} = 1$ and
$\varepsilon_{\alpha\nu\mu\beta}= - \varepsilon^{\alpha\nu\mu\beta}$
\cite{Itzykson1980}. In the non--relativistic approximation for the
neutron and proton the r.h.s. of Eq.(\ref{eq:C.1}) reads
\begin{eqnarray}\label{eq:C.3}
\hspace{-0.3in}&&M^{(W\gamma)}(n\to p\,e^-\bar{\nu}_e)_{\rm
  Fig.\,\ref{fig:fig6a}} = - 2 m_N G_V
\Bigg\{\Big[\big(\tilde{Z}^{(N)}_1 - 1\big) +
  \big(\tilde{Z}^{(\ell)}_1 -
  1\big)\Big]\Big\{[\varphi^{\dagger}_p\varphi_n][\bar{u}_e\gamma^0(1
  - \gamma^5)v_{\bar{\nu}}] + g_A
           [\varphi^{\dagger}_p\vec{\sigma}\,\varphi_n]\nonumber\\
\hspace{-0.3in}&&\times [\bar{u}_e\vec{\gamma} \,(1 -
  \gamma^5)v_{\bar{\nu}}]\Big\} + \frac{\alpha}{2\pi}\Big\{
       [\varphi^{\dagger}_p\varphi_n][\bar{u}_e\gamma^0(1 -
         \gamma^5)v_{\bar{\nu}}]\Big[\Big(\frac{11}{8} + (g_A -
         1)\,\frac{21}{8}\Big)\,{\ell n}\frac{\Lambda^2}{m^2_N}  +
         \Big(\frac{1}{8} - (g_A - 1)\,\frac{15}{8}\Big)\,{\ell
           n}\frac{M^2_W}{m^2_N}\nonumber\\
\hspace{-0.3in}&& + \frac{1445}{288} + (g_A - 1)\,\frac{127}{48} -
\Big(\frac{21}{16} + (g_A - 1)\,2\Big)\,\frac{m^2_N}{M^2_W}\,{\ell
  n}\frac{M^2_W}{m^2_N} + f_{\beta^-_c}(E_e,\mu) +
\frac{E_e}{m_N}\,f_V(E_e)\Big] + g_A
       [\varphi^{\dagger}_p\vec{\sigma}\,\varphi_n][\bar{u}_e\vec{\gamma}
         \,(1 - \gamma^5)v_{\bar{\nu}}]\nonumber\\
         \hspace{-0.3in}&&\times \Big[\Big(\frac{11}{8} - \frac{g_A -
    1}{g_A}\,\frac{21}{8}\Big)\,{\ell n}\frac{\Lambda^2}{m^2_N}  +
  \Big(\frac{1}{8} + \frac{g_A - 1}{g_A}\,\frac{15}{8}\Big)\,{\ell
    n}\frac{M^2_W}{m^2_N} + \frac{1445}{288} - \frac{g_A -
    1}{g_A}\,\frac{115}{48} - \Big(\frac{1}{8} - \frac{g_A -
    1}{g_A}\,\frac{2}{3}\Big)\,\frac{m^2_N}{M^2_W}\,{\ell
    n}\frac{M^2_W}{m^2_N}\nonumber\\
\hspace{-0.3in}&& + f_{\beta^-_c}(E_e,\mu) +
\frac{E_e}{m_N}\,f_A(E_e)\Big] +
       [\varphi^{\dagger}_p\varphi_n][\bar{u}_e\gamma^0(1 -
         \gamma^5)v_{\bar{\nu}}]\Big[- \frac{\sqrt{1 -
             \beta^2}}{2\beta}\,{\ell n}\Big(\frac{1 + \beta}{1 -
           \beta}\Big) + \frac{E_e}{m_N}\,f_S(E_e)\Big] + g_A
       [\varphi^{\dagger}_p\vec{\sigma}\,\varphi_n]\nonumber\\
\hspace{-0.3in}&&\times [\bar{u}_e\gamma^0 \vec{\gamma} \,(1 -
  \gamma^5)v_{\bar{\nu}}]\Big[- \frac{\sqrt{1 -
      \beta^2}}{2\beta}\,{\ell n}\Big(\frac{1 + \beta}{1 - \beta}\Big)
  + \frac{E_e}{m_N}\,f_T(E_e)\Big] +
 \frac{E_e}{m_N}\,\Big\{[\varphi^{\dagger}_p\frac{\vec{k}_e\cdot
    \vec{\sigma}}{E_e} \varphi_n]\,g_S(E_e) +
       [\varphi^{\dagger}_p\frac{\vec{k}_{\bar{\nu}}\cdot
           \vec{\sigma}}{E_e}
         \varphi_n]\,h_S(E_e)\Big\}\nonumber\\
\hspace{-0.3in}&&\times [\bar{u}_e(1 - \gamma^5)v_{\bar{\nu}}] +
       [\varphi^{\dagger}_p \frac{\vec{k}_e\cdot \vec{\sigma}}{E_e}
         \varphi_n][\bar{u}_e\gamma^0(1 -
         \gamma^5)v_{\bar{\nu}}]\,\frac{E_e}{m_N}\,g_V(E_e) +
       \frac{E_e}{m_N}\,[\varphi^{\dagger}_p\frac{(\vec{k}_e\cdot
           \vec{\sigma}\,)\vec{\sigma}}{E_e} \varphi_n]\cdot
            [\bar{u}_e \vec{\gamma} \,(1 -
              \gamma^5)v_{\bar{\nu}}]\,h_A(E_e)\Bigg\},
\end{eqnarray}
where we have denoted
\begin{eqnarray}\label{eq:C.4}
\hspace{-0.15in}f_{\beta^-_c}(E_e, \mu) &=& \frac{3}{4}\,{\ell
  n}\frac{m^2_N}{m^2_e} - \frac{11}{8} + 2{\ell
  n}\Big(\frac{\mu}{m_e}\Big) \Big[\frac{1}{2\beta} {\ell
    n}\Big(\frac{1 + \beta}{1 - \beta}\Big) - 1 \Big] +
\frac{1}{2\beta}\, {\ell n}\Big(\frac{1 + \beta}{1 - \beta}\Big) -
\frac{1}{4\beta}\, {\ell n}^2\Big(\frac{1 + \beta}{1 -
  \beta}\Big) - \frac{1}{\beta}\,{\rm Li}_2\Big(\frac{2\beta}{1 +
  \beta}\Big),\nonumber\\
\hspace{-0.15in}f_V(E_e) &=& 1 + \frac{1}{2}\,{\ell
  n}\frac{m^2_N}{m^2_e} + \frac{2 - 3\beta^2}{2 \beta}\,{\ell
  n}\Big(\frac{1 + \beta}{1 - \beta}\Big) + (g_A - 1)\,\Big[-
  \frac{1}{4} - \frac{5}{8}\,{\ell n}\frac{m^2_N}{m^2_e} - \frac{2 -
    5\beta^2}{8 \beta}\,{\ell n}\Big(\frac{1 + \beta}{1 -
    \beta}\Big)\Big],\nonumber\\
\hspace{-0.15in}f_A(E_e) &=&1 + \frac{1}{2}\,{\ell
  n}\frac{m^2_N}{m^2_e} + \frac{2 - 3\beta^2}{2 \beta}\,{\ell
  n}\Big(\frac{1 + \beta}{1 - \beta}\Big) + \frac{g_A - 1}{g_A}\,\Big[
  \frac{3}{4}\,{\ell n}\frac{m^2_N}{m^2_e} - \frac{3}{4}\,\beta\,
       {\ell n}\Big(\frac{1 + \beta}{1 - \beta}\Big)\Big],\nonumber\\
\hspace{-0.15in}f_S(E_e) &=&\sqrt{1 - \beta^2}\Big\{-
\frac{1}{2}\,{\ell n}\frac{m^2_N}{m^2_e} + \frac{2E_0 -
  E_e}{E_e}\,\frac{1}{2 \beta}\,{\ell n}\Big(\frac{1 + \beta}{1 -
  \beta}\Big) + (g_A - 1)\,\Big[ \frac{1}{4} - \frac{1}{8}\,{\ell
    n}\frac{m^2_N}{m^2_e} - \frac{1}{4 \beta}\,{\ell n}\Big(\frac{1 +
    \beta}{1 - \beta}\Big)\Big]\Big\},\nonumber\\
\hspace{-0.15in}f_T(E_e) &=&\sqrt{1 - \beta^2}\Big[-
  \frac{1}{2}\,{\ell n}\frac{m^2_N}{m^2_e} + \frac{1}{2 \beta}\,{\ell
    n}\Big(\frac{1 + \beta}{1 - \beta}\Big)\Big],\nonumber\\
\hspace{-0.15in}g_S(E_e) &=&(g_A - 1)\,\frac{\sqrt{1 - \beta^2}}{8
  \beta}\,{\ell n}\Big(\frac{1 + \beta}{1 - \beta}\Big),\nonumber\\
\hspace{-0.15in}h_S(E_e) &=& \frac{\sqrt{1 - \beta^2}}{\beta}\,{\ell
  n}\Big(\frac{1 + \beta}{1 - \beta}\Big),\nonumber\\
\hspace{-0.15in}g_V(E_e) &=&- \frac{1}{2}\,{\ell n}\frac{m^2_N}{m^2_e}
+ \frac{1}{2 \beta}\,{\ell n}\Big(\frac{1 + \beta}{1 - \beta}\Big) +
(g_A - 1)\,\Big[- \frac{1}{4} - \frac{3}{8}\,{\ell
    n}\frac{m^2_N}{m^2_e} + \frac{5}{8 \beta}\,{\ell n}\Big(\frac{1 +
    \beta}{1 - \beta}\Big)\Big],\nonumber\\
\hspace{-0.15in}h_A(E_e) &=& - \frac{1}{2}\,{\ell
  n}\frac{m^2_N}{m^2_e} + \frac{1}{2 \beta}\,{\ell n}\Big(\frac{1 +
  \beta}{1 - \beta}\Big).
\end{eqnarray}
The function $f_{\beta^-_c}(E_e, \mu)$ and the terms
\begin{eqnarray}\label{eq:C.5}
\frac{\alpha}{2\pi}\Big\{ [\varphi^{\dagger}_p\varphi_n][\bar{u}_e \,(1 -
  \gamma^5)v_{\bar{\nu}}] + g_A
       [\varphi^{\dagger}_p\vec{\sigma}\,\varphi_n]\cdot
       [\bar{u}_e\gamma^0 \vec{\gamma} \,(1 -
         \gamma^5)v_{\bar{\nu}}]\Big\} \Big[- \frac{\sqrt{1 -
             \beta^2}}{2\beta}\,{\ell n}\Big(\frac{1 + \beta}{1 -
           \beta}\Big)\Big]
\end{eqnarray}
have been calculated by Sirlin \cite{Sirlin1967} to leading order in
the large nucleon mass expansion (see also Appendix D of
Ref.\cite{Ivanov2013}). The other functions in Eq.(\ref{eq:C.4})
define radiative corrections of order $O(\alpha E_e/m_N)$.

\subsection*{Renormalization of the radiative corrections of order 
$O(\alpha/\pi)$ and $O(\alpha E_e/m_N)$}

For the calculation of observable radiative corrections of order
$O(\alpha/\pi)$ we have to delete ultra--violate divergent
contributions, which are defined by the terms ${\ell
  n}(\Lambda^2/m^2_N)$. For this aim we write down the amplitude of
the neutron $\beta^-$--decay as follows
\begin{eqnarray}\label{eq:C.6}
\hspace{-0.45in}&&M(n\to p\,e^-\bar{\nu}_e) = - 2 m_N G_V\Bigg\{\Big\{1
  + \big(\tilde{Z}^{(N)}_1 - 1\big) + \big(\tilde{Z}^{(\ell)}_1 -
  1\big) + \frac{\alpha}{2\pi}\Big[\Big(\frac{11}{8} + (g_A -
    1)\,\frac{21}{8}\Big)\,{\ell n}\frac{\Lambda^2}{m^2_N}  \nonumber\\
\hspace{-0.45in}&&+ \Big(\frac{1}{8} - (g_A -
1)\,\frac{15}{8}\Big)\,{\ell n}\frac{M^2_W}{m^2_N} + \frac{1445}{288}
+ (g_A - 1)\,\frac{127}{48} - \Big(\frac{21}{16} + (g_A -
1)\,2\Big)\,\frac{m^2_N}{M^2_W}\,{\ell n}\frac{M^2_W}{m^2_N} +
f_{\beta^-_c}(E_e,\mu)\nonumber\\
\hspace{-0.45in}&& + \frac{E_e}{m_N}\,f_V(E_e)\Big]\Big\}
  [\varphi^{\dagger}_p\varphi_n][\bar{u}_e\gamma^0(1 -
    \gamma^5)v_{\bar{\nu}}] + g_A \Big\{1 + \big(\tilde{Z}^{(N)}_1 -
  1\big) + \big(\tilde{Z}^{(\ell)}_1 - 1\big) +
  \frac{\alpha}{2\pi}\Big[\Big(\frac{11}{8} - \frac{g_A -
      1}{g_A}\,\frac{21}{8}\Big)\,{\ell n}\frac{\Lambda^2}{m^2_N} 
    \nonumber\\ 
\hspace{-0.45in}&&+ \Big(\frac{1}{8} + \frac{g_A -
  1}{g_A}\,\frac{15}{8}\Big)\,{\ell n}\frac{M^2_W}{m^2_N} +
\frac{1445}{288} - \frac{g_A - 1}{g_A}\,\frac{115}{48} -
\Big(\frac{1}{8} + \frac{g_A -
  1}{g_A}\,\frac{2}{3}\Big)\,\frac{m^2_N}{M^2_W}\,{\ell
  n}\frac{M^2_W}{m^2_N} + f_{\beta^-_c}(E_e,\mu) +
\frac{E_e}{m_N}\,f_A(E_e)\Big]\Big\}\nonumber\\
\hspace{-0.45in}&&\times [\varphi^{\dagger}_p\vec{\sigma}\,\varphi_n]
\cdot [\bar{u}_e\vec{\gamma}\, (1 - \gamma^5)v_{\bar{\nu}}] +
\frac{\alpha}{2\pi}\Big[ [\varphi^{\dagger}_p\varphi_n][\bar{u}_e (1 -
    \gamma^5)v_{\bar{\nu}}]\Big[- \frac{\sqrt{1 -
        \beta^2}}{2\beta}\,{\ell n}\Big(\frac{1 + \beta}{1 -
      \beta}\Big) + \frac{E_e}{m_N}\,f_S(E_e)\Big] + g_A
  [\varphi^{\dagger}_p\vec{\sigma}\,\varphi_n]\nonumber\\
\hspace{-0.45in}&&\cdot [\bar{u}_e\gamma^0 \vec{\gamma} \,(1 -
  \gamma^5)v_{\bar{\nu}}]\Big[- \frac{\sqrt{1 -
      \beta^2}}{2\beta}\,{\ell n}\Big(\frac{1 + \beta}{1 - \beta}\Big)
  + \frac{E_e}{m_N}\,f_T(E_e)\Big] +
\frac{E_e}{m_N}\,\Big\{[\varphi^{\dagger}_p\frac{\vec{k}_e\cdot
    \vec{\sigma}}{E_e} \varphi_n]\,g_S(E_e) +
     [\varphi^{\dagger}_p\frac{\vec{k}_{\bar{\nu}}\cdot
         \vec{\sigma}}{E_e} \varphi_n]\,h_S(E_e)\Big\}\nonumber\\
\hspace{-0.45in}&&\times [\bar{u}_e(1 - \gamma^5)v_{\bar{\nu}}] +
       [\varphi^{\dagger}_p \frac{\vec{k}_e\cdot \vec{\sigma}}{E_e}
         \varphi_n][\bar{u}_e\gamma^0(1 -
         \gamma^5)v_{\bar{\nu}}]\,\frac{E_e}{m_N}\,g_V(E_e) +
       [\varphi^{\dagger}_p\frac{(\vec{k}_e\cdot
           \vec{\sigma}\,)\vec{\sigma}}{E_e} \varphi_n]\cdot
       [\bar{u}_e \vec{\gamma}\,(1 -
         \gamma^5)v_{\bar{\nu}}]\frac{E_e}{m_N}\,
       h_A(E_e)\Big]\Bigg\}.
\end{eqnarray}
Because of conservation of the charged hadronic vector current
\cite{Feynman1958} and the Kinoshita--Lee--Nauenberg (KLN) theorem for
the radiative corrections to the neutron lifetime
\cite{Kinoshita1959,Kinoshita1962,Lee1964} (see also Appendix E of
Ref.\cite{Ivanov2013}) the renormalization constants are defined by
\begin{eqnarray}\label{eq:C.7}
\hspace{-0.3in}\big(\tilde{Z}^{(N)}_1 - 1\big) +
\big(\tilde{Z}^{(\ell)}_1 - 1\big) &+&
\frac{\alpha}{2\pi}\Big[\Big(\frac{11}{8} + (g_A -
  1)\,\frac{21}{8}\Big)\,{\ell n}\frac{\Lambda^2}{m^2_N}  +
  \Big(\frac{1}{8} - (g_A - 1)\,\frac{15}{8}\Big)\,{\ell
    n}\frac{M^2_W}{m^2_N}\nonumber\\
\hspace{-0.35in}&&~~~~~+ \frac{1445}{288} + (g_A - 1)\,\frac{127}{48} -
\Big(\frac{21}{16} + (g_A - 1)\,2\Big)\,\frac{m^2_N}{M^2_W}\,{\ell
  n}\frac{M^2_W}{m^2_N}\Big] = 0.
\end{eqnarray}
Plugging Eq.(\ref{eq:C.7}) into Eq.(\ref{eq:C.6}) we arrive at the
expression
\begin{eqnarray}\label{eq:C.8}
\hspace{-0.3in}&&M(n\to p\,e^-\bar{\nu}_e) = - 2 m_N G_V\Big\{\Big[1
  + \frac{\alpha}{2\pi}\Big(f_{\beta^-_c}(E_e,\mu) +
  \frac{E_e}{m_N}\,f_V(E_e)\Big)\Big]
       [\varphi^{\dagger}_p\varphi_n][\bar{u}_e\gamma^0(1 -
         \gamma^5)v_{\bar{\nu}}] + g_A \Big\{1 +
       \frac{\alpha}{2\pi}\,\frac{g_A - 1}{g_A}\nonumber\\
\hspace{-0.3in}&&\times\,\Big[- \frac{21}{4}\,{\ell
    n}\frac{\Lambda^2}{m^2_N} + \frac{15}{4}\,{\ell
    n}\frac{M^2_W}{m^2_N}- \frac{121}{24} + g_A\Big(\frac{11}{2} -
  \frac{g_A - 1}{g_A}\,\frac{2}{3}\Big)\,\frac{m^2_N}{M^2_W}\,{\ell
    n}\frac{M^2_W}{m^2_N}\Big] + \frac{\alpha}{2\pi}\Big[
  f_{\beta^-_c}(E_e,\mu) + \frac{E_e}{m_N}\,f_A(E_e) \nonumber\\
\hspace{-0.3in}&& + \frac{5}{2}\,\frac{m^2_N}{M^2_W}\, {\ell
  n}\frac{M^2_W}{m^2_N}\Big]\Big\}
       [\varphi^{\dagger}_p\vec{\sigma}\,\varphi_n]\cdot
       [\bar{u}_e\vec{\gamma}\, (1 - \gamma^5)v_{\bar{\nu}}] +
       \frac{\alpha}{2\pi}\Big[
         [\varphi^{\dagger}_p\varphi_n][\bar{u}_e (1 -
           \gamma^5)v_{\bar{\nu}}]\Big[- \frac{\sqrt{1 -
               \beta^2}}{2\beta}\,{\ell n}\Big(\frac{1 + \beta}{1 -
             \beta}\Big) + \frac{E_e}{m_N}\,f_S(E_e)\Big]\nonumber\\
\hspace{-0.3in}&& + g_A [\varphi^{\dagger}_p\vec{\sigma}\,\varphi_n]
\cdot [\bar{u}_e\gamma^0 \vec{\gamma} \,(1 -
  \gamma^5)v_{\bar{\nu}}]\Big[- \frac{\sqrt{1 -
      \beta^2}}{2\beta}\,{\ell n}\Big(\frac{1 + \beta}{1 - \beta}\Big)
  + \frac{E_e}{m_N}\,f_T(E_e)\Big] +
\Big\{[\varphi^{\dagger}_p\frac{\vec{k}_e\cdot \vec{\sigma}}{E_e}
  \varphi_n]\,\frac{E_e}{m_N}\,g_S(E_e) +
      [\varphi^{\dagger}_p\frac{\vec{k}_{\bar{\nu}}\cdot
          \vec{\sigma}}{E_e} \varphi_n ]\nonumber\\
\hspace{-0.3in}&&\times \,\frac{E_e}{m_N} \,h_S(E_e)\Big\}
       [\bar{u}_e(1 - \gamma^5)v_{\bar{\nu}}] + [\varphi^{\dagger}_p
         \frac{\vec{k}_e\cdot \vec{\sigma}}{E_e}
         \varphi_n][\bar{u}_e\gamma^0(1 -
         \gamma^5)v_{\bar{\nu}}]\,\frac{E_e}{m_N}\,g_V(E_e) +
       [\varphi^{\dagger}_p\frac{(\vec{k}_e\cdot
           \vec{\sigma}\,)\vec{\sigma}}{E_e} \varphi_n]\cdot
            [\bar{u}_e \vec{\gamma}\,(1 - \gamma^5)v_{\bar{\nu}}]\nonumber\\
\hspace{-0.3in}&&\times\,\frac{E_e}{m_N}\, h_A(E_e)\Big]\Big\}.
\end{eqnarray}
The calculation of the Feynman diagrams in Fig.\,\ref{fig:fig6a} in
the tree--approximation for strong low--energy interactions, described
by the Lagrangian Eq.(\ref{eq:44}), assumes the axial coupling
constant $g_A = 1$. Setting $g_A = 1$ in Eq.(\ref{eq:C.8}) we get

\begin{eqnarray}\label{eq:C.9}
\hspace{-0.35in}&&M(n\to p\,e^-\bar{\nu}_e) = - 2 m_N G_V\Big\{\Big[1
  + \frac{\alpha}{2\pi}\Big(f_{\beta^-_c}(E_e,\mu) +
  \frac{E_e}{m_N}\,f_V(E_e)\Big)\Big]
       [\varphi^{\dagger}_p\varphi_n][\bar{u}_e\gamma^0(1 -
         \gamma^5)v_{\bar{\nu}}] + \Big[1 + \frac{\alpha}{2\pi}\Big(
         f_{\beta^-_c}(E_e,\mu) \nonumber\\
\hspace{-0.35in}&&+ \frac{E_e}{m_N}\,f_A(E_e) +
\frac{5}{2}\,\frac{m^2_N}{M^2_W}\, {\ell
  n}\frac{M^2_W}{m^2_N}\Big)\Big]
       [\varphi^{\dagger}_p\vec{\sigma}\,\varphi_n]\cdot
       [\bar{u}_e\vec{\gamma}\, (1 - \gamma^5)v_{\bar{\nu}}] +
       \frac{\alpha}{2\pi}\Big[
         [\varphi^{\dagger}_p\varphi_n][\bar{u}_e (1 -
           \gamma^5)v_{\bar{\nu}}]\Big(- \frac{\sqrt{1 -
               \beta^2}}{2\beta}\,{\ell n}\Big(\frac{1 + \beta}{1 -
             \beta}\Big) \nonumber\\
\hspace{-0.35in}&&+ \frac{E_e}{m_N}\,f_S(E_e)\Big) +
         [\varphi^{\dagger}_p\vec{\sigma}\,\varphi_n] \cdot
         [\bar{u}_e\gamma^0 \vec{\gamma} \,(1 -
           \gamma^5)v_{\bar{\nu}}]\Big(- \frac{\sqrt{1 -
               \beta^2}}{2\beta}\,{\ell n}\Big(\frac{1 + \beta}{1 -
             \beta}\Big) + \frac{E_e}{m_N}\,f_T(E_e)\Big) +
         \Big([\varphi^{\dagger}_p\frac{\vec{k}_e\cdot
             \vec{\sigma}}{E_e} \varphi_n]\,\frac{E_e}{m_N}\,g_S(E_e)
         \nonumber\\
\hspace{-0.35in}&& +
       [\varphi^{\dagger}_p\frac{\vec{k}_{\bar{\nu}}\cdot
           \vec{\sigma}}{E_e} \varphi_n ]\,\frac{E_e}{m_N}
       \,h_S(E_e)\Big) [\bar{u}_e(1 - \gamma^5)v_{\bar{\nu}}] +
          [\varphi^{\dagger}_p \frac{\vec{k}_e\cdot \vec{\sigma}}{E_e}
            \varphi_n][\bar{u}_e\gamma^0(1 -
            \gamma^5)v_{\bar{\nu}}]\,\frac{E_e}{m_N}\,g_V(E_e) +
          [\varphi^{\dagger}_p\frac{(\vec{k}_e\cdot
              \vec{\sigma}\,)\vec{\sigma}}{E_e} \varphi_n]\nonumber\\
\hspace{-0.35in}&&\cdot [\bar{u}_e \vec{\gamma}\,(1 -
  \gamma^5)v_{\bar{\nu}}]\,\frac{E_e}{m_N}\, h_A(E_e)\Big]\Big\}.
\end{eqnarray}
This testifies gauge invariance and renormalizability of the quantum
field theory of strong low--energy and electroweak interactions
described by the Lagrangian Eq.(\ref{eq:44}). The use of the axial
coupling constant $g_A \neq 1$ does not violate gauge invariance. This
allows to take into account the contributions of strong low--energy
interactions at Sirlin's level \cite{Sirlin1967}. However, unlike
Sirlin \cite{Sirlin1967, Sirlin1978} gauge invariance of the radiative
corrections of order $O(\alpha/\pi)$ valid to any order in the large
nucleon mass expansion to the amplitude of the neutron
$\beta^-$--decay is not related to the contribution of hadronic
structure of the neutron and proton beyond the axial coupling
constant, but caused by the electroweak $W^-$--boson exchanges (see
Fig.\,\ref{fig:fig6a}b, c and d).

For complete renormalizability of the radiative corrections to the
amplitude of the neutron $\beta^-$--decay, given by Eq.(\ref{eq:C.8})
with $g_A > 1$, we may follow Sirlin \cite{Sirlin1967} (see also
\cite{Ivanov2013}) and renormalize the axial coupling constant
\begin{eqnarray}\label{eq:C.10}
\hspace{-0.15in}g_A \Big\{1 + \frac{\alpha}{2\pi}\,\frac{g_A -
  1}{g_A}\,\Big[- \frac{21}{4}\,{\ell n}\frac{\Lambda^2}{m^2_N}  +
  \frac{15}{4}\,{\ell n}\frac{M^2_W}{m^2_N}- \frac{121}{24} +
  g_A\Big(\frac{11}{2} - \frac{g_A -
    1}{g_A}\,\frac{2}{3}\Big)\Big]\Big\} \to g_A.
\end{eqnarray}
As a result we arrive at the following amplitude of the neutron
$\beta^-$--decay
\begin{eqnarray}\label{eq:C.11}
\hspace{-0.30in}&&M(n\to p\,e^-\bar{\nu}_e) = - 2 m_N G_V\Big\{\Big[1
  + \frac{\alpha}{2\pi}\Big(f_{\beta^-_c}(E_e,\mu) +
  \frac{E_e}{m_N}\,f_V(E_e)\Big)\Big]
       [\varphi^{\dagger}_p\varphi_n][\bar{u}_e\gamma^0(1 -
         \gamma^5)v_{\bar{\nu}}] + g_A \Big[1 +
         \frac{\alpha}{2\pi}\Big( f_{\beta^-_c}(E_e,\mu)\nonumber\\
\hspace{-0.30in}&& + \frac{E_e}{m_N}\,f_A(E_e) +
\frac{5}{2}\,\frac{m^2_N}{M^2_W}\, {\ell
  n}\frac{M^2_W}{m^2_N}\Big)\Big]
            [\varphi^{\dagger}_p\vec{\sigma}\,\varphi_n]\cdot
            [\bar{u}_e\vec{\gamma}\, (1 - \gamma^5)v_{\bar{\nu}}] +
            \frac{\alpha}{2\pi}\Big[
              [\varphi^{\dagger}_p\varphi_n][\bar{u}_e (1 -
                \gamma^5)v_{\bar{\nu}}]\Big(- \frac{\sqrt{1 -
                    \beta^2}}{2\beta}\,{\ell n}\Big(\frac{1 + \beta}{1
                  - \beta}\Big) \nonumber\\
\hspace{-0.30in}&&+ \frac{E_e}{m_N}\,f_S(E_e)\Big) + g_A
       [\varphi^{\dagger}_p\vec{\sigma}\,\varphi_n] \cdot
       [\bar{u}_e\gamma^0 \vec{\gamma} \,(1 -
         \gamma^5)v_{\bar{\nu}}]\Big(- \frac{\sqrt{1 -
           \beta^2}}{2\beta}\,{\ell n}\Big(\frac{1 + \beta}{1 -
         \beta}\Big) + \frac{E_e}{m_N}\,f_T(E_e)\Big) +
       \Big([\varphi^{\dagger}_p\frac{\vec{k}_e\cdot
           \vec{\sigma}}{E_e} \varphi_n] \frac{E_e}{m_N} \nonumber\\
\hspace{-0.30in}&&\times\, g_S(E_e) +
       [\varphi^{\dagger}_p\frac{\vec{k}_{\bar{\nu}}\cdot
           \vec{\sigma}}{E_e} \varphi_n ]\,\frac{E_e}{m_N}
       \,h_S(E_e)\Big) [\bar{u}_e(1 - \gamma^5)v_{\bar{\nu}}] +
          [\varphi^{\dagger}_p \frac{\vec{k}_e\cdot \vec{\sigma}}{E_e}
            \varphi_n][\bar{u}_e\gamma^0(1 -
            \gamma^5)v_{\bar{\nu}}]\,\frac{E_e}{m_N}\,g_V(E_e) +
          [\varphi^{\dagger}_p\frac{(\vec{k}_e\cdot
              \vec{\sigma}\,)\vec{\sigma}}{E_e} \varphi_n]\nonumber\\
\hspace{-0.30in}&&\cdot [\bar{u}_e \vec{\gamma}\,(1 -
  \gamma^5)v_{\bar{\nu}}]\,\frac{E_e}{m_N}\, h_A(E_e)\Big]\Big\},
\end{eqnarray}
where the contribution of strong low--energy interactions is described
by the axial coupling $g_A$. This amplitude takes into account
radiative corrections to order $O(\alpha/\pi)$ and $O(\alpha
E_e/m_N)$, respectively. The radiative corrections are calculated from
the set of Feynman diagrams in Fig.\,\ref{fig:fig6a}, which are gauge
invariant or independent of a gauge parameter $\xi$ of the photon
propagator. The amplitude is renormalized, i.e. all divergences are
absorbed by renormalization of i) the electroweak coupling $g$ or the
coupling constant $G_V$ and of ii) the axial coupling constant $g_A$
that agrees fully with Sirlin's elimination of ultra--violate
divergences \cite{Sirlin1967} (see also Appendix D of
Ref.\cite{Ivanov2013}). 

The radiative corrections of order $O(\alpha E_e/m_N)$ are calculated
as next--to--leading order corrections in the large nucleon mass
expansion to Sirlin's radiative corrections, calculated to leading
order in the large nucleon mass expansion. This confirms Sirlin's
confidence level of these corrections. Unlike Sirlin's analysis of
radiative corrections of order $O(\alpha/\pi)$ to the amplitude of the
neutron $\beta^-$--decay, where gauge invariance is caused by
one--virtual--photon exchanges with hadronic structure of the neutron
ans proton, gauge invariance of radiative corrections of order
$O(\alpha/\pi)$ and to any order in the large nucleon mass expansion
is fully due to the electroweak $W^-$--boson and photon exchanges (see
Fig.\,\ref{fig:fig6a}b, c and d). As a result, the contributions of
hadronic structure to radiative corrections of order $O(\alpha/\pi)$
to any order in the large nucleon mass expansion should be self-gauge
invariant. This is a separate problem, which we are planning to
investigate in our forthcoming publication.

\section*{Appendix D: Analytical expressions for the Feynman diagrams 
in Fig.\,\ref{fig:fig7a} for the amplitude of the neutron radiative
$\beta^-$--decay $n \to p + e^- + \bar{\nu}_e + \gamma$}
\renewcommand{\theequation}{D-\arabic{equation}}
\setcounter{equation}{0}

The Feynman diagrams, describing the amplitude of the neutron
radiative $\beta^-$--decay in the tree--approximation for electroweak,
electromagnetic and strong low--energy interactions , are shown in
Fig.\,\ref{fig:fig7a}. The Feynman diagrams are drawn to leading order
in the large mass $M_W$ of the electroweak $W^-$--boson expansion at
the neglect of the Feynman diagram with the vertex $W^-W^-\gamma$, the
contribution of which is suppressed by the factor $q\cdot k/M^2_W$,
where $k$ is a 4--momentum of a real photon. The amplitude of the
neutron radiative $\beta^-$--decay, described by the Feynman diagrams
in Fig.\,\ref{fig:fig7a}, can be written as follows \cite{Ivanov2018b}
\begin{eqnarray}\label{eq:D.1}
\hspace{-0.3in}M_{\rm Fig. \ref{fig:fig7a}}(n \to p e^- \bar{\nu}_e
\gamma)_{\lambda} &=& M_{\rm Fig. \ref{fig:fig7a}a +
  Fig. \ref{fig:fig7a}b}(n \to p e^- \bar{\nu}_e \gamma)_{\lambda} +
M_{\rm Fig. \ref{fig:fig7a}c + Fig. \ref{fig:fig7a}d}(n \to p e^-
\bar{\nu}_e \gamma)_{\lambda} \nonumber\\
\hspace{-0.3in}&+& M_{\rm Fig. \ref{fig:fig7a}e +
  Fig. \ref{fig:fig7a}f}(n \to p e^- \bar{\nu}_e \gamma)_{\lambda}.
\end{eqnarray}
The amplitude $M_{\rm Fig. \ref{fig:fig7a}a + Fig. \ref{fig:fig7a}b}(n
\to p e^- \bar{\nu}_e \gamma)_{\lambda}$, defined by the Feynman
diagrams in Fig.\ref{fig:fig7a}a and Fig.\ref{fig:fig7a}a, is equal to
\cite{Ivanov2013,Ivanov2017,Ivanov2017b}
\begin{eqnarray}\label{eq:D.2}
\hspace{-0.3in}&& M_{\rm Fig. \ref{fig:fig7a}a +
  Fig. \ref{fig:fig7a}b}(n \to p e^- \bar{\nu}_e \gamma)_{\lambda} = e
G_V\nonumber\\
\hspace{-0.3in}&&\times \Big\{\Big[\bar{u}_p(\vec{k}_p, \sigma_p)
  \gamma^{\mu}(1 - \gamma^5) u_n(\vec{k}_n, \sigma_n)\Big]
\Big[\bar{u}_e(\vec{k}_e,\sigma_e)\,\frac{1}{2k_e\cdot k}\,Q_{e,
    \lambda}\,\gamma_{\mu} (1 - \gamma^5) v_{\nu}(\vec{k}_{\nu}, +
  \frac{1}{2})\Big]\nonumber\\
\hspace{-0.3in}&&- \Big[\bar{u}_p(\vec{k}_p,
  \sigma_p)\,Q_{p, \lambda} \,\frac{1}{2k_p \cdot k}\,\gamma^{\mu}(1 -
  \gamma^5) u_n(\vec{k}_n,
  \sigma_n)\Big]\Big[\bar{u}_e(\vec{k}_e,\sigma_e) \gamma^{\mu} (1 -
  \gamma^5) v_{\nu}(\vec{k}_{\nu}, + \frac{1}{2})\Big]\Big\},
\end{eqnarray}
where $Q_{e,\lambda}$ and $Q_{p, \lambda}$ are given by
\cite{Ivanov2013,Ivanov2017,Ivanov2017b}
\begin{eqnarray}\label{eq:D.3}
\hspace{-0.3in}Q_{e, \lambda} = 2 \varepsilon^{*}_{\lambda}(k)\cdot k_e +
\hat{\varepsilon}^*_{\lambda}(k)\hat{k}\;,\; Q_{p,\lambda} = 2
\varepsilon^{*}_{\lambda}(k)\cdot k_p +
\hat{\varepsilon}^*_{\lambda}(k)\hat{k}.
\end{eqnarray}
Here $\varepsilon^*_{\lambda}(k)$ is the polarization vector of the
photon with the 4--momentum $k$ and in two polarization states
$\lambda = 1,2$, obeying the constraint $k\cdot
\varepsilon^*_{\lambda}(k) = 0$. For the derivation of
Eq.(\ref{eq:D.2}) we have used the Dirac equations for the free proton
and electron. Replacing $\varepsilon^{*}_{\lambda}(k) \to k$ and using
$k^2 = 0$ we get \cite{Ivanov2017,Ivanov2017b} (see also
\cite{Ivanov2013b})
\begin{eqnarray}\label{eq:D.4}
\hspace{-0.3in}M_{\rm Fig. \ref{fig:fig7a}a + Fig. \ref{fig:fig7a}b}(n
\to p e^- \bar{\nu}_e \gamma)_{\lambda}\Big|_{\varepsilon^{*}_{\lambda}(k) \to k} = 0.
\end{eqnarray}
This confirms invariance of the Feynman diagrams in
Fig.\,\ref{fig:fig7a}a and Fig.\,\ref{fig:fig7a}b  under a
gauge transformation $\varepsilon^*_{\lambda'}(k) \to
\varepsilon^*_{\lambda'}(k) + c\,k$, where $c$ is an arbitrary
constant.

The contributions of the Feynman diagrams in Fig.\,\ref{fig:fig7a}c -
Fig.\,\ref{fig:fig7a}f to the amplitude of the neutron radiative
$\beta^-$--decay take the form
\begin{eqnarray}\label{eq:D.5}
\hspace{-0.3in}&& M_{\rm Fig. \ref{fig:fig7a}c +
  Fig. \ref{fig:fig7a}d}(n \to p e^- \bar{\nu}_e \gamma)_{\lambda} = e
G_V\nonumber\\
\hspace{-0.3in}&&\times\,\Big\{\frac{2 m_N (q - k)_{\mu}}{m^2_{\pi} -
  (q - k)^2 - i 0}\,[\bar{u}_p(\vec{k}_p, \sigma_p)\gamma^5
  u_n(\vec{k}_n, \sigma_n)] \,\Big[\bar{u}_e(\vec{k}_e,\sigma_e)Q_{e,
    \lambda}\frac{1}{2 k_e\cdot k} \gamma^{\mu} (1 - \gamma^5)
  v_{\nu}(\vec{k}_{\nu}, +
  \frac{1}{2})\Big]\nonumber\\ \hspace{-0.3in}&& - \frac{2 m_N
  q_{\mu}}{m^2_{\pi} - q^2 - i 0}\,\Big[\bar{u}_p(\vec{k}_p,
  \sigma_p)\,Q_{p, \lambda} \,\frac{1}{2k_p \cdot k}\,\gamma^5
  u_n(\vec{k}_n, \sigma_n)\Big]\Big[\bar{u}_e(\vec{k}_e,\sigma_e)
  \gamma^{\mu} (1 - \gamma^5) v_{\nu}(\vec{k}_{\nu}, +
  \frac{1}{2})\Big]\Big\},\nonumber\\
\hspace{-0.3in}&& M_{\rm Fig. \ref{fig:fig7a}e +
  Fig. \ref{fig:fig7a}f}(n \to p e^- \bar{\nu}_e \gamma)_{\lambda} = e
G_V\nonumber\\
\hspace{-0.3in}&&\times\,\Big\{\frac{2 m_N q_{\mu}}{m^2_{\pi} - q^2 -
  i 0}\,\frac{(2 q - k)\cdot \varepsilon^*_{\lambda}(k)}{m^2_{\pi} -
  (q - k)^2 - i0}\,[\bar{u}_p(\vec{k}_p, \sigma_p)\gamma^5
  u_n(\vec{k}_n, \sigma_n)]\Big[\bar{u}_e(\vec{k}_e,\sigma_e)
  \gamma^{\mu} (1 - \gamma^5) v_{\nu}(\vec{k}_{\nu}, +
  \frac{1}{2})\Big]\nonumber\\\hspace{-0.3in}&& + \frac{2
  m_N}{m^2_{\pi} - (q - k)^2 - i0}\,[\bar{u}_p(\vec{k}_p,
  \sigma_p)\gamma^5 u_n(\vec{k}_n, \sigma_n)]\,
\Big[\bar{u}_e(\vec{k}_e,\sigma_e)\hat{\varepsilon}^*_{\lambda}(k) (1
  - \gamma^5) v_{\nu}(\vec{k}_{\nu}, + \frac{1}{2})\Big]\Big\},
\end{eqnarray}
where we have used the GT--relation $g_{\pi N} = m_N/f_{\pi}$. Making
a gauge transformation $\varepsilon^*_{\lambda'}(k) \to
\varepsilon^*_{\lambda'}(k) + c\,k$ one may show that
\begin{eqnarray}\label{eq:D.6}
\hspace{-0.3in}&&M_{\rm Fig. \ref{fig:fig7a}c +
  Fig. \ref{fig:fig7a}d}(n \to p e^- \bar{\nu}_e
\gamma)_{\lambda}\Big|_{\varepsilon^{*}_{\lambda}(k) \to k} = 0,
\nonumber\\
\hspace{-0.3in}&&M_{\rm Fig. \ref{fig:fig7a}e +
  Fig. \ref{fig:fig7a}f}(n \to p e^- \bar{\nu}_e
\gamma)_{\lambda}\Big|_{\varepsilon^{*}_{\lambda}(k) \to k} = 0.
\end{eqnarray}
One may be convinced that gauge invariance of the amplitude of the
neutron radiative $\beta^-$--decay is retained also for $g_A \neq
1$. The analytical expressions for the Feynman diagrams in
Fig.\ref{fig:fig7a}a -- Fig.\,\ref{fig:fig7a}f with the axial
coupling constant $g_A \neq 1$ are equal to
\begin{eqnarray}\label{eq:D.7}
\hspace{-0.3in}&& M_{\rm Fig. \ref{fig:fig7a}a +
  Fig. \ref{fig:fig7a}b}(n \to p e^- \bar{\nu}_e \gamma)_{\lambda} = e
G_V\nonumber\\
\hspace{-0.3in}&&\times \Big\{\Big[\bar{u}_p(\vec{k}_p, \sigma_p)
  \gamma^{\mu}(1 - g_A \gamma^5) u_n(\vec{k}_n, \sigma_n)\Big]
\Big[\bar{u}_e(\vec{k}_e,\sigma_e)\,\frac{1}{2k_e\cdot k}\,Q_{e,
    \lambda}\,\gamma_{\mu} (1 - \gamma^5) v_{\nu}(\vec{k}_{\nu}, +
  \frac{1}{2})\Big]\nonumber\\
\hspace{-0.3in}&&- \Big[\bar{u}_p(\vec{k}_p, \sigma_p)\,Q_{p, \lambda}
  \,\frac{1}{2k_p \cdot k}\,\gamma^{\mu}(1 - g_A \gamma^5)
  u_n(\vec{k}_n, \sigma_n)\Big]\Big[\bar{u}_e(\vec{k}_e,\sigma_e)
  \gamma^{\mu} (1 - \gamma^5) v_{\nu}(\vec{k}_{\nu}, +
  \frac{1}{2})\Big]\Big\},\nonumber\\
\hspace{-0.3in}&& M_{\rm Fig. \ref{fig:fig7a}c +
  Fig. \ref{fig:fig7a}d}(n \to p e^- \bar{\nu}_e \gamma)_{\lambda} = e
G_V\nonumber\\
\hspace{-0.3in}&&\times\,\Big\{\frac{2 m_N g_A (q -
  k)_{\mu}}{m^2_{\pi} - (q - k)^2 - i 0}\,[\bar{u}_p(\vec{k}_p,
  \sigma_p)\gamma^5 u_n(\vec{k}_n, \sigma_n)]
\,\Big[\bar{u}_e(\vec{k}_e,\sigma_e)Q_{e, \lambda}\frac{1}{2 k_e\cdot
    k} \gamma^{\mu} (1 - \gamma^5) v_{\nu}(\vec{k}_{\nu}, +
  \frac{1}{2})\Big]\nonumber\\ \hspace{-0.3in}&& - \frac{2 m_N g_A
  q_{\mu}}{m^2_{\pi} - q^2 - i 0}\,\Big[\bar{u}_p(\vec{k}_p,
  \sigma_p)\,Q_{p, \lambda} \,\frac{1}{2k_p \cdot k}\,\gamma^5
  u_n(\vec{k}_n, \sigma_n)\Big]\Big[\bar{u}_e(\vec{k}_e,\sigma_e)
  \gamma^{\mu} (1 - \gamma^5) v_{\nu}(\vec{k}_{\nu}, +
  \frac{1}{2})\Big]\Big\},\nonumber\\
\hspace{-0.3in}&& M_{\rm Fig. \ref{fig:fig7a}e +
  Fig. \ref{fig:fig7a}f}(n \to p e^- \bar{\nu}_e \gamma)_{\lambda} = e
G_V\nonumber\\
\hspace{-0.3in}&&\times\,\Big\{\frac{2 m_N g_A q_{\mu}}{m^2_{\pi} -
  q^2 - i 0}\,\frac{(2 q - k)\cdot
  \varepsilon^*_{\lambda}(k)}{m^2_{\pi} - (q - k)^2 -
  i0}\,[\bar{u}_p(\vec{k}_p, \sigma_p)\gamma^5 u_n(\vec{k}_n,
  \sigma_n)]\Big[\bar{u}_e(\vec{k}_e,\sigma_e) \gamma^{\mu} (1 -
  \gamma^5) v_{\nu}(\vec{k}_{\nu}, +
  \frac{1}{2})\Big]\nonumber\\\hspace{-0.3in}&& + \frac{2 m_N
  g_A}{m^2_{\pi} - (q - k)^2 - i0}\,[\bar{u}_p(\vec{k}_p,
  \sigma_p)\gamma^5 u_n(\vec{k}_n, \sigma_n)]\,
\Big[\bar{u}_e(\vec{k}_e,\sigma_e)\hat{\varepsilon}^*_{\lambda}(k) (1
  - \gamma^5) v_{\nu}(\vec{k}_{\nu}, + \frac{1}{2})\Big]\Big\}.
\end{eqnarray}
Since the contribution of the Feynman diagrams with one--pion--pole
exchanges to the rate of the neutron radiative $\beta^-$--decay is of
order $10^{-9}$ \cite{Ivanov2018b,Ivanov2018c}, one may take into
account only the contributions of the Feynman diagrams in
Fig\,\ref{fig:fig7a}a and Fig\,\ref{fig:fig7a}b. For the cancellation
of the infrared divergence in the rate $\lambda_{\beta^-_c}(\mu)$ one
may calculate the rate $\lambda_{\beta^-_c\gamma}(\mu)$ of the neutron
radiative $\beta^-$--decay to leading order in the large nucleon mass
expansion. The analytical expression for the rate
$\lambda_{\beta^-_c\gamma}(\mu)$, calculated to leading order in the
large nucleon mass expansion, we take from \cite{Ivanov2013}. It reads
\begin{eqnarray}\label{eq:D.8}
\hspace{-0.15in}&&\lambda_{\beta^-_c \gamma}(\mu) = (1 + 3
g^2_A)\,\frac{\alpha}{\pi}\,\frac{|G_V|^2}{2\pi^3}\int^{E_0}_{m_e}\,
\Big\{\Big[2{\ell n}\Big(\frac{2(E_0 - E_e)}{\mu}\Big) - 3 +
  \frac{2}{3}\,\frac{E_0 - E_e}{E_e}\, \Big(1 + \frac{1}{8} \frac{E_0
    - E_e}{E_e} \Big)\Big]\Big[\frac{1}{2\beta}\,{\ell n}\Big(\frac{1
    + \beta}{1 - \beta}\Big) - 1\Big] \nonumber\\
\hspace{-0.3in}&& + 1 + \frac{1}{12} \frac{(E_0 - E_e)^2}{E^2_e}+
\frac{1}{2\beta}\,{\ell n}\Big(\frac{1 + \beta}{1 - \beta}\Big) -
\frac{1}{4\beta}\,{\ell n}^2\Big(\frac{1 + \beta}{1 - \beta}\Big) -
\frac{1}{\beta}\,{\rm Li}_2\Big(\frac{2 \beta}{1 + \beta}
\Big)\Big\}\sqrt{E^2_e - m^2_e}\,E_e F(E_e, Z = 1)\,dE_e.
\end{eqnarray}
The calculation of the rate $\lambda_n = \lambda_{\beta^-_c}(\mu) +
\lambda_{\beta^-_c\gamma}(\mu)$ related to the neutron lifetime
$\tau_n = 1/\lambda_n$, where $\lambda_{\beta^-_c}(\mu)$ is the rate
of the neutron $\beta^-$--decay $n \to p + e^- + \bar{\nu}_e$, defined
by the amplitude in Eq.(\ref{eq:C.11}) (see Eq.(\ref{eq:73}) of section
\ref{sec:oneewloop}).


\begin{thebibliography}{9}
\bibitem{Lee1969a} The assertion that the Lagrangian Eq.(\ref{eq:7})
  with $b = f_{\pi}$ cannot be renormalizable in the normal ordered
  form (see the Appendix of Ref.\cite{Lee1969}) is valid only for the
  L$\sigma$M with a finite mass of the scalar isoscalar
  $\sigma$--meson. However, it is not valid in the limit of the
  infinite scalar isoscalar $\sigma$--meson mass.
\bibitem{Ivanov2019} In the dimensional regularization for $n < 4$ the
  momentum integral with the integrand $1/(p^2 + i0)^2$ can be
  calculated as follows \cite{Gerstein1969}
\[\int \frac{d^np}{(2\pi)^ni}\frac{1}{(p^2 + i0)^2} = \frac{1}{n - 4} 
\int \frac{d^np}{(2\pi)^ni}\frac{d}{d p_{\alpha}}
\frac{p_{\alpha}}{(p^2 + i0)^2} = \ \frac{1}{n -
  4}\,\frac{\Omega_n}{(2\pi)^n}\,\lim_{R\to \infty}\frac{R^n}{R^4} =
0,\] where $\Omega_n = 2 \pi^{n/2}/\Gamma(n/2)$ is the n--dimensional
solid angle.
\bibitem{MathB39} The analytical calculation of the integrals over $z$
  and the $1/M_W$--expansion has been carried out with Wolfram
  Mathematica 8.
\end{thebibliography}
\end{document}